\documentclass{aa}  

\usepackage{graphicx}
\usepackage{txfonts}
\usepackage{pifont}

\usepackage{natbib,twoopt}
\usepackage[breaklinks=true]{hyperref} 
\bibpunct{(}{)}{;}{a}{}{,}             
\makeatletter
  \newcommandtwoopt{\citeads}[3][][]{\href{http://adsabs.harvard.edu/abs/#3}%
    {\def\hyper@linkstart##1##2{}%
     \let\hyper@linkend\@empty\citealp[#1][#2]{#3}}}
  \newcommandtwoopt{\citepads}[3][][]{\href{http://adsabs.harvard.edu/abs/#3}%
    {\def\hyper@linkstart##1##2{}%
     \let\hyper@linkend\@empty\citep[#1][#2]{#3}}}
  \newcommandtwoopt{\citetads}[3][][]{\href{http://adsabs.harvard.edu/abs/#3}%
    {\def\hyper@linkstart##1##2{}%
     \let\hyper@linkend\@empty\citet[#1][#2]{#3}}}
  \newcommandtwoopt{\citeyearads}[3][][]%
    {\href{http://adsabs.harvard.edu/abs/#3}
    {\def\hyper@linkstart##1##2{}%
     \let\hyper@linkend\@empty\citeyear[#1][#2]{#3}}}
\makeatother

\usepackage[T1]{fontenc}
\usepackage[utf8]{inputenc}

\hypersetup{linkcolor=red,citecolor=green,filecolor=cyan,urlcolor=magenta}

\usepackage{soul}

\usepackage{graphicx}
\usepackage{subfigure,placeins}
\usepackage{booktabs}
\usepackage{multirow}
\usepackage{siunitx}

\usepackage{gensymb}

\usepackage{makecell}

\begin{document}

   \title{A JWST/MIRI view of \object{$\kappa$\,Andromedae\,b}: Refining its mass, age, and physical parameters
   }

   \author{N.~ Godoy
          \inst{1}
          \and
          E.~ Choquet
          \inst{1}
          \and
          E.~ Serabyn
          \inst{2}
          \and
          M.~ Mâlin
          \inst{3,4,5}
          \and
          P.~ Tremblin
          \inst{6}
          \and
          C.~ Danielski
          \inst{7}
          \and
          P.~O.~ Lagage
          \inst{8}
          \and
          A.~ Boccaletti
          \inst{5}
          \and
          B.~ Charnay
          \inst{5}
          \and
          M.~E.~ Ressler
          \inst{2}
          }

   \institute{ 
              Aix Marseille Universit\'e, CNRS, CNES, LAM, Marseille, France\\
              \email{nicolas.godoy@lam.fr}
             \and 
              Jet Propulsion Laboratory, California Institute of Technology, Pasadena, CA 91109, USA
              \and 
              Department of Physics \& Astronomy, Johns Hopkins University, 3400 N. Charles Street, Baltimore, MD 21218, USA
              \and 
              Space Telescope Science Institute, 3700 San Martin Drive, Baltimore, MD 21218, USA
              \and 
              LIRA, Observatoire de Paris, Université PSL, Sorbonne Université, Université Paris Cité, CY Cergy Paris Université, CNRS, 92190 Meudon, France
              \and 
              Université Paris-Saclay, UVSQ, CNRS, CEA, Maison de la Simulation, 91191 Gif-sur-Yvette, France
              \and 
              INAF – Osservatorio Astrofisico di Arcetri, Largo E. Fermi 5, 50125, Firenze, Italy
              \and 
              Université Paris-Saclay, Université Paris Cité, CEA, CNRS, AIM, 91191 Gif-sur-Yvette, France
             }

   \date{Received 19 March 2025 / Accepted 04 August 2025}

 
  \abstract
   { 
   $\kappa$\,And\,b is a substellar companion with a mass near the planet--brown dwarf boundary orbiting a B9IV star at $\sim50$-$100$\,au. Estimates of its age and mass vary, which has fueled a decade-long debate. Additionally, the atmospheric parameters ($\mathrm{T_{eff}}$ $1650$-$2050$\,K and log(g) $3.5$-$5.5$\,dex) remain poorly constrained. The differences in atmospheric models and inhomogeneous datasets contribute to the varied interpretations. 
   }
   {
   We aim to refine the characterization of $\kappa$\,And\,b by using mid-infrared data to capture its full bolometric emission. Combined with near-infrared (NIR) measurements, we aim to constrain $\mathrm{T_{eff}}$, log(g), and the radius to narrow down the uncertainties in age and mass.
   }
   { 
   We obtained JWST/MIRI coronagraphic data in the \texttt{F1065C}, \texttt{F1140C}, and \texttt{F1550C} filters and recalibrated existing NIR photometry using an updated ATLAS stellar model. We used MIRI color-magnitude diagrams to probe the likelihood of species (e.g., $\mathrm{CH_4}$, $\mathrm{NH_3}$, and silicates). We compared the H and \texttt{F1140C} colors and magnitudes of the companion to isochrones to constrain the age and mass. We then modeled its spectral energy distribution with atmospheric models to refine the estimates of $\mathrm{T_{eff}}$, radius, and log(g) and to constrain age and mass using evolutionary models.
   }
   {
   Cloudy atmosphere models fit the spectral energy distribution of $\kappa$\,And\,b best. This is consistent with its L0/L2 spectral type and its position near silicate-atmosphere field objects in the MIRI color-magnitude diagram. We derived an age of $47\pm7$\,Myr and a mass of $17.3\pm1.8\,\mathrm{M_{Jup}}$ by weight-mean combining the models. Atmospheric modeling yielded $\mathrm{T_{eff}}$\,=\,$1791\pm68$\,K and a radius of\\$1.42\pm0.06\,\mathrm{R_{Jup}}$. This improves the precision by $\sim30\%$ over previous estimates. Log(g) was constrained to $4.35\pm0.07$\,dex, which is an improvement in the precision by $\sim70\%$ relative to the most precise literature value of $4.75 \pm 0.25$ dex.
   }
   { 
   Our new mass estimate places $\kappa$\,And\,b slightly above the planet--brown dwarf boundary determined by the deuterium-burning limit. Our age estimate is $\sim75\%$ more precise than previous values and aligns the object with the Columba association ($42$\,Myr). The derived $\mathrm{T_{eff}}$ suggests silicate clouds, but this needs to be confirmed spectroscopically. MIRI data were crucial to refine the radius and temperature, which led to stronger constraints on the age and mass (both dependent on the model) and improved the overall characterization of $\kappa$\,And\,b.
   }

   \maketitle
%

\section{Introduction} \label{sec:intro}

Planetary-mass companions observed with direct imaging such as HR\,2562\,b (\citealt{Mesa+2018}), HR\,8799\,bcde (\citealt{Marois+2008}), and $\beta$\,Pic\,b (\citealt{Lagrange+2010}) provide unique insights into planet formation mechanisms at wide orbits, including disk instability and core accretion (\citealt{Spiegel+Burrows-2012}. These insights are further supported by statistical analyses from dedicated surveys (e.g., \citealt{Vigan+2021}). These companions offer us the opportunity to study planetary evolution in dynamically complex environments where host star radiation, disk chemistry, and gravitational interactions play a critical role in shaping their atmospheric and orbital properties (e.g., \citealt{Fortney+2008}; \citealt{Oberg+2011}; \citealt{Marleau+Cumming-2014}; \citealt{Bowler-2016B}; \citealt{Nielsen+2019}). Unlike isolated brown dwarfs, which are thought to form through cloud fragmentation, planetary-mass companions, particularly those at separations below a few hundred AU, likely form within circumstellar disks and evolve under the influence of their host star and the circumstellar disk. This provides a contrasting perspective on substellar atmospheres and formation histories.

L-type objects, in particular, early L types L0/L2, represent a critical transition regime between low-mass stars, brown dwarfs, and massive planets. Their effective temperatures ($\sim$ 1400–2000 K) and atmospheric conditions make them key for understanding the giant planet formation and atmospheric evolution (e.g., \citealt{Tremblin+2016}), especially during the L to T transition. This phase is characterized by dramatic changes in the cloud structure and spectral appearance because silicate and iron clouds condense below the visible/NIR photosphere (e.g., \citealt{Cushing+2005}; \citealt{Burrows+2006}; \citealt{Saumon+Marley-2008}; \citealt{Marley+2012}; \citealt{Suarez+Metchev-2022}). Thick silicate clouds and iron and alkali metals dominate their spectra. This introduces complexities in modeling their atmospheric properties (e.g., \citealt{Miles+2023}). These challenges make L-type objects, especially planetary companions, pivotal for studying cloud dynamics and chemical disequilibria.

Direct-imaging observations of young planetary-mass companions further highlight their differences from field brown dwarfs, particularly among L-type objects. The low surface gravity, enhanced chemical mixing, and highly cloudy or dusty atmospheres set them apart from brown dwarfs that formed through cloud fragmentation (e.g., \citealt{Currie+2011}; \citealt{Allers+2013}; \citealt{DeRosa+2016}; \citealt{Rajan+2017}; \citealt{Chauvin+2017}). By capturing their infrared spectra, direct imaging enables us to access their properties while they are still warm and luminous. NIR observations are crucial for constraining bulk parameters such as effective temperature, surface gravity, luminosity, and mass. The limited spectral coverage of ground-based instruments often restricts our ability to fully characterize these objects, but particularly L-types, whose spectral energy distributions (SEDs) challenge atmospheric models because their cloud and chemical processes are complex (e.g., \citealt{Saumon+Marley-2008}, \citealt{Tremblin+2016}).

The combination of NIR and mid-infrared (MIR) observations has proven essential for a more comprehensive characterization of these low-mass objects. MIR observations, particularly from space-based telescopes and observatories such as the \textit{James Webb} Space Telescope (JWST; \citealt{Rigby+2023}), improve the constraints on effective temperature, surface gravity, and radius (e.g., \citealt{Carter+2022}; \citealt{Boccaletti+2024}; \citealt{Godoy+2024}). These constraints are further enhanced when the system ages are well determined, as was shown for objects from well-dated moving groups such as $\beta$\,Pic (e.g., $20.4\pm2.5$\,Myr, \citealt{Couture+2023}) and Columba (e.g., $42\pm8$\,Myr, \citealt{Bell+2015}). The uncertainties in the ages can significantly affect the inferred physical properties, however, including the mass and radius. This complicates the differentiation between planetary-mass objects and brown dwarfs (e.g., \citealt{Carson+2013} and \citealt{Hinkley+2013}). It is essential to address these age-related uncertainties to refine evolutionary models and understand the formation pathways.

The atmospheric characterization of young planetary companions is often hindered by heterogeneous datasets and varying modeling assumptions. The conversion of contrast into absolute magnitudes relies on stellar spectral models that differ from the stellar parameters, which introduces systematic offsets in the companion properties and flux normalization (e.g., \citealt{Hinkley+2013} vs. \citealt{Currie+2018}). Additionally, differences in the spectral coverage, calibration, and method of the instruments can result in significant discrepancies in the retrieved spectra and physical parameters (e.g., \citealt{Sutlieff+2021}; \citealt{Nasedkin+2023}; \citealt{Xuan+2022}). This highlights the need for uniform analysis pipelines and multi-instrument integration for robust atmospheric constraints.

These challenges are well illustrated by $\kappa$\,Andromedae\,b. This planetary-mass companion (L0/L2; \citealt{Uyama+2020}) orbits a B9IV star and has been the focus of debate since its discovery (\citealt{Carson+2013}) because of the large uncertainties in the system age. Combined with varying model assumptions and calibration discrepancies, these uncertainties yield a wide range of estimates for its effective temperature, surface gravity, and mass.

We present the first JWST coronagraphic observations of $\kappa$\,And\,b using the Mid-Infrared Instrument (MIRI; \citealt{Rieke+2015}, \citealt{Wright+2015}) at wavelengths of 10.65\,$\mu m$, 11.40\,$\mu m$, and 15.50\,$\mu m$. These observations were conducted with the advanced coronagraphic abilities of JWST/MIRI (\citealt{Boccaletti+2015}; \citealt{Danielski+2018}; \citealt{Boccaletti+2022}) and provide unprecedented insights into the atmosphere and physical properties of $\kappa$\,And\,b. They address some of the limitations posed by earlier ground-based observations.

The structure of this paper is as follows: Section\,\ref{sec:system} provides an overview of the $\kappa$\,Andromedae system and summarizes its key properties and contextual relevance. Section\,\ref{sec:obs} describes the observations and data processing, including the recalibration of archival ground-based observations. Section\,\ref{sec:res} delves into the analysis of the new data and presents updated estimates of the physical parameters of the companion. Finally, we discuss in Section\,\ref{sec:dis} the broader implications of these findings for understanding planetary-mass companions and their formation mechanisms, which is followed by a summary of our conclusions in Section\,\ref{sec:sac}.

\section{The $\kappa$\,Andromedae system}\label{sec:system}

Kappa Andromedae ($\kappa$\,And), also known as HD\,222439, HIP\,116805, HR\,8976, and TYC\,3244-1530-1, is a high proper motion and young B9IVn-type star (\citealt{Garrison+Gray-1994}) at $\sim51$\,pc (\citealt{Gaia-EDR3}, \citealt{Gaia+2016}). In Table\,\ref{table:star_parameters} we summarize the main stellar parameters known to date, and in Table\,\ref{table:star_mags} the magnitudes and fluxes. The star hosts a planetary-mass companion, $\kappa$\,And\,b, discovered by \cite{Carson+2013}, at an angular separation of 1\arcsec.058 (on 2012 July 8), and semi-major axis of $103.6$\,AU ($57.4$\,AU-$107.4$\,AU $68\%$ confidence interval, \citealt{Uyama+2020}). We summarize the main properties of the companion later in this section.

\begin{table}[]
\centering
\caption{\label{table:star_parameters}Main properties and stellar parameters of the star $\kappa$\,And.}
\begin{tabular}{l c l}
\hline \hline
Parameter & Value & Reference \\
\hline
Spectral Type &  B9IVn  & (1) \\
Right Ascension (J2000) & 23:40:24.5076 & (2,3) \\
Declination (J2000) & +44:20:02.1566 & (2,3) \\
Parallax (mas) & $19.406 \pm 0.210$ & (2,3) \\
Distance (pc) & $51.530 \pm 0.558$ & (2,3) \\
$\mu_{\alpha}\,(mas\,yr^{-1})$ & $79.998 \pm 0.156$ & (2,3) \\
$\mu_{\delta}\,(mas\,yr^{-1})$ & $-19.011 \pm 0.128$ & (2,3) \\
Age (Myr) & $47^{+27}_{-40}$ & (4) \\
$v\,sin(i)$ (km\,s$^{-1}$) & $142.2^{+13.1}_{-21.1}$ & (4) \\
Mass ($M_{\odot}$) & $2.768^{+0.1}_{-0.109}$ & (4) \\
Radius ($R_{\odot}$) & $2.109^{+0.032}_{-0.018}$ & (4) \\
$\mathrm{T_{eff}}$ (Kelvin) & $11327^{+412}_{-44}$ & (4) \\
log(g) (dex) & $4.174^{+0.019}_{-0.012}$ & (4) \\
$[M/H]$ & $0.0$\tablefootmark{a} & (4) \\
$log_{10}(L_{total}/L_{\odot})$ & $1.7966^{+0.0682}_{-0.0155}$ & (4)\\
$log_{10}(L_{app}/L_{\odot})$ & $1.8574^{+0.0674}_{-0.0090}$ & (4) \\
$B - V$ (mag) & $ -0.074\pm 0.007$ & (5) \\
\hline
\end{tabular}
\tablefoot{References: ($1$) \cite{Garrison+Gray-1994}; (2) \cite{Gaia-EDR3}; (3) \textit{Gaia} DR3: \cite{Gaia+2016}; (4) \cite{Jones+2016}; (5) Johnson UBV photometry: \cite{Mermilliod-1997}, \cite{Johnson-UBV}.
\tablefoottext{a}{We used a solar metallicity as adopted by \cite{Jones+2016}.}
 }
\end{table}

\begin{table}[]
\centering
\caption{\label{table:star_mags} Archival and measured $\kappa$\,And magnitudes and fluxes.}
\begin{tabular}{l l c l}
\hline \hline
Parameter & $\lambda_{cen}$\tablefootmark{a} & Value & Ref \\
\hline
$B - V$ (mag) & & $ -0.074\pm 0.007$ & (1) \\
$.2395$ (mag) & $0.24\mu m$ & $3.656 \pm 0.042$ & (2) \\
$.2900$ (mag) & $0.30\mu m$ & $3.676 \pm 0.018$ & (2) \\
$U$ (mag)  & $0.36\mu m$ & $4.152 \pm 0.012$ & (1) \\
$V$ (mag) & $0.41\mu m$ & $4.152 \pm 0.012$ & (3) \\
$B$ (mag) & $0.43\mu m$ & $4.051 \pm 0.014$ & (4) \\
$B$ (mag) & $0.44\mu m$ & $4.067 \pm 0.007$ & (1) \\
$b$ (mag) & $0.47\mu m$ & $4.061 \pm 0.006$ & (3) \\
$H_{p}$ (mag) & $0.49\mu m$ & $4.1257 \pm 0.0004$ & (5) \\
$GB_{p}$ (mag) & $0.50 \mu m$ & $4.090 \pm 0.003$ & (6) \\
$V$ (mag) & $0.53 \mu m$ & $4.126 \pm 0.009$ & (4) \\
$y$ (mag) & $0.54 \mu m$ & $3.970 \pm 0.006$ & (3) \\
$V$ (mag) & $0.55 \mu m$ & $4.138 \pm 0.003$ & (1) \\
$G$ (mag) & $0.58 \mu m$ & $4.148 \pm 0.003$ & (6) \\
$GR_{p}$ (mag) & $0.76 \mu m$ & $4.168 \pm 0.005$ & (6) \\
$J$ (mag) & $1.24 \mu m$ & $4.26 \pm 0.04$ & (7) \\
$H$ (mag) & $1.66 \mu m$ & $4.31 \pm 0.05$ & (7) \\
$K_{s}$ (mag) & $2.16 \mu m$ & $4.32 \pm 0.05$ & (7) \\
$W1$ (mag) & $3.35 \mu m$ & $4.282 \pm 0.283$ & (8) \\
$W2$ (mag) & $4.60 \mu m$ & $4.079 \pm 0.138$ & (8) \\
$S9W$ (mag) & $8.23 \mu m$ & $4.163 \pm 0.011$ & (9) \\
$12\mu m$ (mag) & $10.16 \mu m$ & $4.105 \pm 0.076$ & (10) \\
$W3$ (mag) & $11.56 \mu m$ & $4.400 \pm 0.015$ & (8) \\
$L18W$ (mag) & $17.61 \mu m$ & $4.260 \pm 0.157$ & (9) \\
$25\mu m$ (mag) & $21.74 \mu m$ & $4.052 \pm 0.206$ & (10) \\
$W4$ (mag) & $22.09 \mu m$ & $4.337 \pm 0.027$ & (8) \\
$\mathrm{m_{F1065C}}$ (mag) & $10.58 \mu m$ & $4.356 \pm 0.019$ & This paper  \\
$\mathrm{m_{F1140C}}$ (mag) & $11.30 \mu m$ & $4.342 \pm 0.019$ & This paper  \\
$\mathrm{m_{F1550C}}$ (mag) & $15.50 \mu m$ & $4.346 \pm 0.019$ & This paper  \\
$\mathrm{F_{F1065C}}$ (mJy) & & $615 \pm 28$ & This paper \\
$\mathrm{F_{F1140C}}$ (mJy) & & $542 \pm 25$ & This paper \\
$\mathrm{F_{F1550C}}$ (mJy) & & $290 \pm 14$ & This paper \\
\hline
\end{tabular}
\tablefoot{
\tablefoottext{a}{$\lambda_{cen}$ refers to the central wavelength.}
References: (1) Johnson UBV photometry: \cite{Mermilliod-1997}, \cite{Johnson-UBV}; (2) \cite{IEU}; (3) Str{\"o}mgren-Crawford uvby$\beta$ photometry: \cite{Paunzen+2015P}; (4) Tycho-2 catalog: \cite{TYCHO+0}; (5) HIPPARCOS (HIgh Precision PARallax COllecting Satellite) catalog: \cite{HIPPARCOS+0}; (6) \textit{Gaia} DR3: \cite{Gaia+2016}; (7) Maunakea Observatories (MKO) bandpasses: \cite{Currie+2018}; (8) Wide-field Infrared Survey Explorer (WISE) catalog: \cite{WISE}; (9) AKARI catalog: \cite{AKARI+0}. \cite{AKARI+1}, \cite{AKARI+2}; (10) Infrared Astronomical Satellite (IRAS) catalog: \cite{IRAS+0}. \cite{IRAS+1}. The catalogs were taken from The VizieR database of astronomical catalogues \cite{Vizier-data}.
}
\end{table}

The age of the $\kappa$\,And system has been a topic of significant discussion over the past decade. \citet{Zuckerman+2011} first proposed that the system is a member of the Columba association ($\sim30$\,Myr), adopted by \citet{Carson+2013} in their analysis of $\kappa$\,And\,b. \citet{Hinkley+2013} questioned the membership of the system in the Columba association, however, and proposed an older isochronal age of $220 \pm 100$\,Myr for the host star. Subsequent studies aimed to refine the system’s age. \citet{Bonnefoy+2014} supported a younger age of $30^{+120}_{-10}$\,Myr based on kinematics and color-magnitude diagram (CMD) coupled with isochrones. \citet{Jones+2016} used long-baseline optical interferometry to measure the star’s size, oblateness, and rotation velocity, enabling strong constraints on its age and mass. Combining these with evolutionary models, they derived an age of $47^{+27}_{-40}$\,Myr, assuming solar metallicity ([M/H] = 0.0). This estimate is consistent with the revised age of the Columba association of $42^{+6}_{-4}$\,Myr from \citet{Bell+2015} \citet{Currie+2018} estimated an age of $40^{+34}_{-19}$\,Myr from empirical comparisons, kinematics, and the properties of the host star and companion. \citet{Stone+2020} used hot-start evolutionary models and companion data to suggest a broader 10–100\,Myr range, while \citet{Hoch+2020} set an upper limit of $50$\,Myr based on evolutionary models. Despite differing methods, these studies consistently support a young system age.

The ongoing uncertainty in the system’s age has significantly affected the inferred properties of $\kappa$\,And\,b, in particular its mass estimate. Assuming a younger age of $\sim$30 Myr, \citet{Carson+2013} estimated a mass of $12.8^{+2}_{-1}\,M_{\mathrm{Jup}}$ for the companion, consistent with a planetary-mass object. In contrast, the older age proposed by \citet{Hinkley+2013} implied a more massive object ($50^{+16}_{-13}\,M_{\mathrm{Jup}}$) with a surface gravity consistent with a brown dwarf classification. \cite{Jones+2016}, using the derived atmospheric models fit parameters from \cite{Hinkley+2013}, infer a mass of $22^{+6}_{-7}\,M_{\mathrm{Jup}}$. Recent studies, including \citet{Currie+2018} and \citet{Uyama+2020}, found that $\kappa$\,And\,b exhibits low-gravity features in its spectrum, aligning with an L0–L1 spectral type and a mass of $13^{+12}_{-2}\,M_{\mathrm{Jup}}$. \cite{Gratton+2024}, using photometry measurements from \cite{Uyama+2020} and an average age of $36\pm8$\,Myr, estimate a mass of $15.03\pm0.66\,M_{\mathrm{Jup}}$.

Previous studies inferred the effective temperature and surface gravity of $\kappa$\,And\,b from its SED using various atmospheric models, datasets, and calibrations, leading to discrepancies in parameter estimates. For instance, temperatures range from $\sim1600$\,K \citep{Carson+2013} up to 1900–2040\,K \citep{Hinkley+2013, Bonnefoy+2014, Hoch+2020}, with several studies suggesting intermediate values around 1680–1800\,K \citep{Todorov+2016, Currie+2018, Stone+2020, Uyama+2020, Morris+2024, Gratton+2024, Xuan+2024}. Surface gravity is less well constrained, with log(g) estimates spanning from $\sim3.8$ to $4.7$ dex \citep{Hoch+2020, Bonnefoy+2014, Currie+2018, Stone+2020, Uyama+2020, Morris+2024}, but with large uncertainties (0.25–1.0 dex) spanning a broad range (3.5–5.5 dex).

The atomic and molecular abundances have also been measured for $\kappa$\,And\,b, first by \cite{Hoch+2020}, who found $C/O=0.70^{+0.09}_{-0.24}$, and later by \cite{Xuan+2024}, who reported $C/O=0.58^{+0.05}_{-0.04}$ and [C/H]$=-0.12^{+0.28}_{-0.19}$. The projected rotational velocity $v\sin(i)$ was also measured by \citealt{Morris+2024} as $38.42 \pm 0.05$~km\,s$^{-1}$, and by \cite{Xuan+2024} as $39.4 \pm 0.9$~km\,s$^{-1}$.

\begin{figure}
\centering
\includegraphics[width=8.5cm]{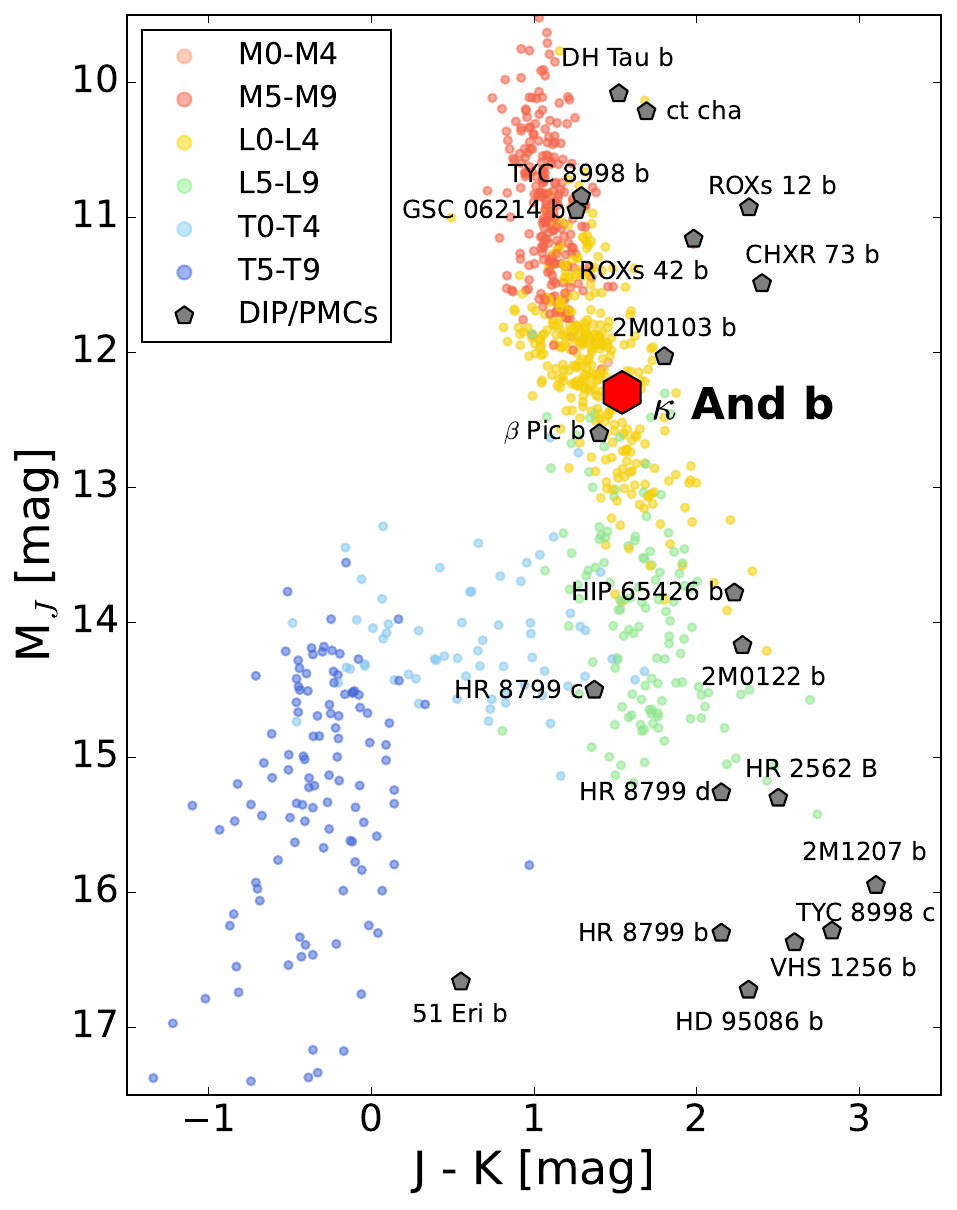}
\caption{CMD showing the position of $\kappa$\,And\,b (red hexagon marker) relative to the population of low-mass stars and brown dwarfs (colored circles) as obtained from \cite{Best+2021}. The different colors highlight the different spectral types. The gray pentagons correspond to selected directly imaged planets (DIP) and planetary-mass companions (PMCs).}
\label{fig:CMD_JK}%
\end{figure}

The spectral type of $\kappa$\,And\,b is better constrained, ranging consistently from L0 to L2 across multiple studies \citep{Carson+2013, Hinkley+2013, Bonnefoy+2014, Currie+2018, Stone+2020, Uyama+2020}, with L3 occasionally included \citep{Stone+2020}. This spectral classification helps us to limit the plausible surface gravity range to approximately 3.9–4.5 dex, thereby placing useful constraints on the companion’s atmospheric parameters. Figure\,\ref{fig:CMD_JK} shows $\kappa$\,And\,b’s position on a J-K CMD consistent with the spectral class derived by \citet[L0/L2]{Uyama+2020}.

These studies suggest that the mass and surface gravity estimates of the companion are closely tied to the system age and the wavelength coverage that was used in observations. By probing the system in the MIR, we can gain more accurate insights into these properties, helping to refine the physical characteristics of the companion.

\section{Observations and data reduction}\label{sec:obs}

\subsection{Observations and strategy}

The observations of the $\kappa$\,Andromedae system were carried out under the guarantee time observation program 1241 (PI: M. Ressler), using the JWST/MIRI instrument. This program is part of the MIRICLE collaboration (\citealt{Boccaletti+2024}; \citealt{Godoy+2024}; \citealt{Malin+2024}), an agreement between the US and European MIRI team to explore the properties of the known substellar companions with MIRI coronagraphic imaging. The observations were taken on September 19, 2023, within $3.5$ hours of execution time. We used the four-quadrant phase mask coronagraphs (4QPMs; \citealt{Rouan+2000}) with the narrow-band filters \texttt{F1065C} ($10.57\mu$m), \texttt{F1140C} ($11.3\mu$m), and \texttt{F1550C} ($15.5\mu$m). The observing strategy is identical to that used for HR\,2562\,b, as detailed in \cite{Godoy+2024}, while the exposure parameters were independently optimized using the MIRI Coronagraphic Simulation pipeline \citep{Danielski+2018}.

We used reference differential imaging (RDI; \citealt{Smith+Terrile-1984}), given that the roll subtraction (\citealt{Burrows+1995}) would induce excessive self-subtraction of the companion (separation of $\sim$$0.7$\arcsec, \citealt{Currie+2018}). We selected HD\,222389, a K5 (\citealt{Cannon+Pickering-1993}) star classified as a long-period variable (P=$92.6 \pm 2.6$\,days, $\Delta V$=$166$\,mmag, $\Delta G$=$168$\,mmag; \citealt{Burggraaff+2018}, \citealt{Gaia-variables+2023}), located at a distance of $\sim$$539$\,pc (\citealt{Gaia+2016}, \citealt{Gaia-EDR3}), as the reference star. We chose this star based on the {JWST User Documentation}\footnote{\url{https://jwst-docs.stsci.edu/jwst-mid-infrared-instrument/miri-observing-strategies/miri-coronagraphic-recommended-strategies}} and following the same consideration presented in \cite{Godoy+2024}. We used the dither-pattern technique with the small-grid dither pattern \texttt{5-point-small-grid} (see \citealt{Lajoie+2016}) in a cross configuration, to accurately subtract the science star coronagraphic point spread function (PSF; e.g., \citealt{Carter+2022}).

Given the presence of the ``glow stick'' stray light (\citealt{Boccaletti+2022}), we added background observations for the science target and the reference star to allow the subtraction of this feature. The observing setups are shown in Table\,\ref{table:obs_setup} for the two stars $\kappa$\,And and HD\,222389.

\begin{table*}[]
\caption{Observational setup.}
\centering
\begin{tabular}{ c c c c c c c c c c c c c }
\hline\hline
Target  & Filter & $\lambda_{\mathrm{cen}}$\tablefootmark{a} & BW\tablefootmark{a} & $\lambda_{\mathrm{eff}}$\tablefootmark{b} & $\mathrm{W_{eff}}$\tablefootmark{b} & FWHM\tablefootmark{b}  & $\mathrm{N_{groups}}$ & $\mathrm{N_{int}}$ & $\mathrm{t_{exp}}$ & $\mathrm{N_{dither}}$ & $\mathrm{t_{total}}$ & Total\\
        &    & [$\mu m$] & [$\mu m$] & [$\mu m$] & [$\mu m$] & [$\mu m$] & & & [sec] & & [sec] & integration\\
\hline
$\kappa$\,And   & F1065C & 10.575 & 0.75 & 10.554 & 0.567 & 0.568 & 28   & 3 & 6.71 & 1 & 20.61 & 3 \\
                & F1140C & 11.30  & 0.80 & 11.301 & 0.604 & 0.587 & 28   & 3 & 6.71 & 1 & 20.61 & 3 \\ 
                & F1550C & 15.50  & 0.90 & 15.508 & 0.704 & 0.734 & 1250 & 3 & 299.6 & 1 & 899.28 & 3 \\
\hline
HD\,222389  & F1065C & 10.575 & 0.75 & 10.554 & 0.567 & 0.568 & 7    & 3 & 1.68  & 5 & 5.51 & 15 \\
            & F1140C & 11.30  & 0.80 & 11.301 & 0.604 & 0.587 & 7    & 3 & 1.68  & 5 & 5.51 & 15 \\
            & F1550C & 15.50  & 0.90 & 15.508 & 0.704 & 0.734 & 277  & 3 & 66.39 & 5 & 199.65 & 15 \\
\hline
\end{tabular}
\tablefoot{
\tablefoottext{a}{Central wavelength ($\lambda_{\mathrm{cen}}$) and bandwidth (BW) were taken from \cite{Boccaletti+2022}. Note that these definitions and values reported by \cite{Boccaletti+2022} are different than the ones used in the Spanish Virtual Observatory (SVO) Filter Profile Service (\citealt{Rodrigo+2012}; \citealt{Rodrigo+2020}).}
\tablefoottext{b}{Effective wavelength ($\lambda_{\mathrm{eff}}$), effective width ($\mathrm{W_{eff}}$), and full width at half maximum (FWHM) were taken from SVO Filter Profile Service (\citealt{Rodrigo+2012}; \citealt{Rodrigo+2020}). }
}
\label{table:obs_setup}
\end{table*}

\subsection{Data reduction and preprocessing}\label{sec:cosmetics}

We reduced the data using the \texttt{JWST}\footnote{\url{https://jwst-pipeline.readthedocs.io/}}\footnote{\url{https://jwst-pipeline.readthedocs.io/en/latest/jwst/jump/description.html}} pipeline routines (version 1.10.0, \citealt{Bushouse+2022}; CRDS version 11.16.21), with the \texttt{spaceKLIP}\footnote{\url{https://github.com/kammerje/spaceKLIP/}} pipeline (version 0.1, \citealt{Kammerer+2022}). We modified some minor aspects in the pipeline to optimize the preprocessing and cosmetics calibrations (see \citealt{Godoy+2024} for more details). We emphasize that, at the time of our data reduction and post-processing, these improvements were not yet included in the official pipelines, although they have since been incorporated into recent releases. We proceed with stages 1 and 2 as in \cite{Godoy+2024}, optimizing the initial parameters for the best detection and correction of cosmic rays and bad pixels.

\begin{figure}
\centering
\includegraphics[width=4.3cm]{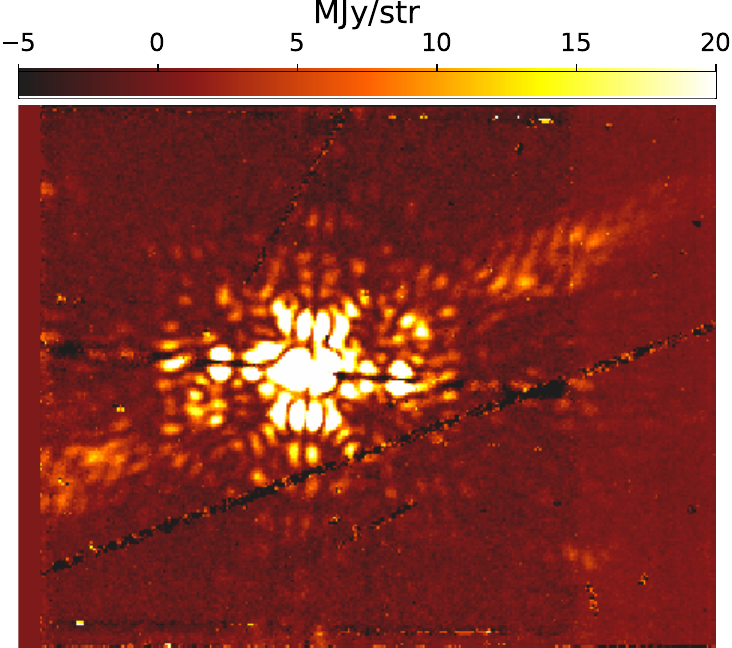}
\includegraphics[width=4.3cm]{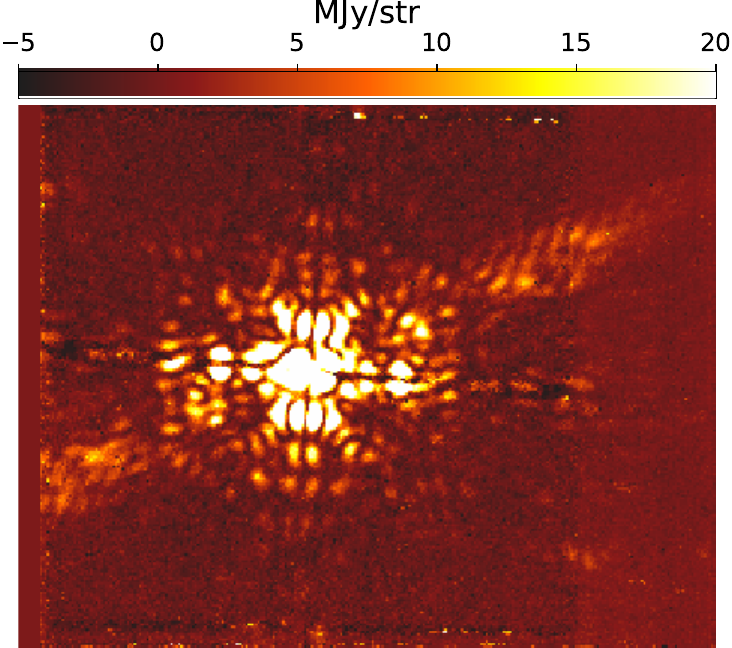}
\caption{Second integration of the third dither position of the reference star HD\,222389 after the background subtraction at the \texttt{F1550C} filter. \textit{Left}: Frame with the standard data reduction parameters using \texttt{spaceKLIP}. \textit{Right}: Same frame, but directly applying our cosmic-ray and bad-pixel corrections in the raw frame. }
\label{fig:cosmic_example}%
\end{figure}

The hit of energetic cosmic rays affected mostly the integrations at \texttt{F1550C}, making the post-processing challenging. \cite{Godoy+2024} addressed this issue as a post-reduction by applying custom techniques to identify bad pixels and cosmic ray remnants, which are then corrected using interpolation routines. Instead, we solve this problem as a pre-reduction step, correcting the \texttt{*uncal*} files before starting the data reduction. In short, we compute the flux difference between each consecutive pair of groups within each ramp, and we identify cosmic ray events from any departure above 5 times the median absolute deviation (MAD) flux difference value. The ramp is then corrected by replacing the flux difference between groups affected by the event with the median flux difference along the ramp. In Appendix \,\ref{apx:cosmic}, we explain in more detail this procedure. Figure\,\ref{fig:cosmic_example} shows an example of the energetic cosmic ray treatment, using an optimized \texttt{spaceKLIP} setup (left) and using our method (right). We observe that certain artifacts, such as the black line in the left image, caused by a cosmic ray, are effectively eliminated using our correction method. We applied the post-reduction correction method described by \cite{Godoy+2024} to correct the residual and persistent bad pixels in the reduced images.

For the background subtraction, we first computed the combined background for each target and filter. We employed the ``cube structure'' method described in \cite{Godoy+2024}, which achieves superior subtraction compared to mean-combining all background observations. We also used the same procedure as \cite{Godoy+2024} for removing the background and the ``glow stick'' artifacts, which has a considerable impact at \texttt{F1550C} filter.

We note that despite the application of pre- and post-corrections during data reduction, a fringing structure remains in the reference star observations at \texttt{F1065C} and \texttt{F1140C}, characterized by horizontal lines, which significantly impacts the post-processing stage (see Fig.\,\ref{fig:bad_PCAsub}). We attribute the presence of this structure to the small number of groups (seven; see Table\,\ref{table:obs_setup}), selected to prevent saturation. It limits the ability of the pipeline to adequately correct for the detector structure, however. We developed a step-by-step procedure to remove this fringing, starting with the identification of the structure and then applying row normalization to the image. In our example, we reduced the flux dispersion from $\sim$$100$ to $\sim$$50$ MJy/str. The complete procedure is detailed in Appendix\,\ref{Apx:fringing}. Figure\,\ref{fig:fringing_example} shows an example of an image at \texttt{F1065C} before (left) and after (right) fringing correction. We emphasize that the ``fringing structure'' is only present in the reference frames ($N_{groups}$=$7$ vs. $N_{groups}$=$28$ for the science star), so the correction does not affect the science data after PSF subtraction. Since we processed the data, this ``fringing'' has been associated with the 390\,Hz pattern noise seen in many MIRI sub-array observations, described in the JWST documentation website\footnote{ \url{https://jwst-docs.stsci.edu/known-issues-with-jwst-data/miri-known-issues\#MIRIKnownIssues-emiElectromagneticinterference(EMI)patternnoise} }. A pipeline correction is now available as part of the \texttt{\textit{emicorr}} step of the \texttt{calwebb\_detector1} module of the pipeline.

\begin{figure}
\centering
\includegraphics[width=4.3cm]{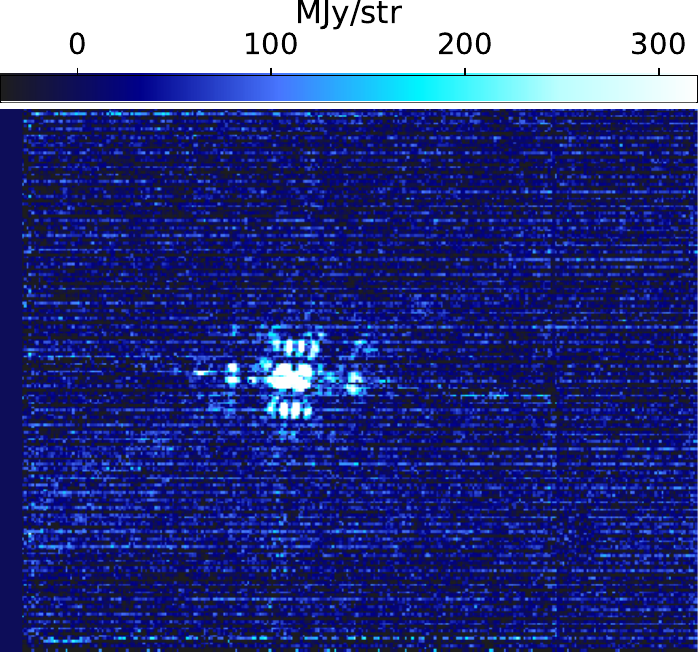}
\includegraphics[width=4.3cm]{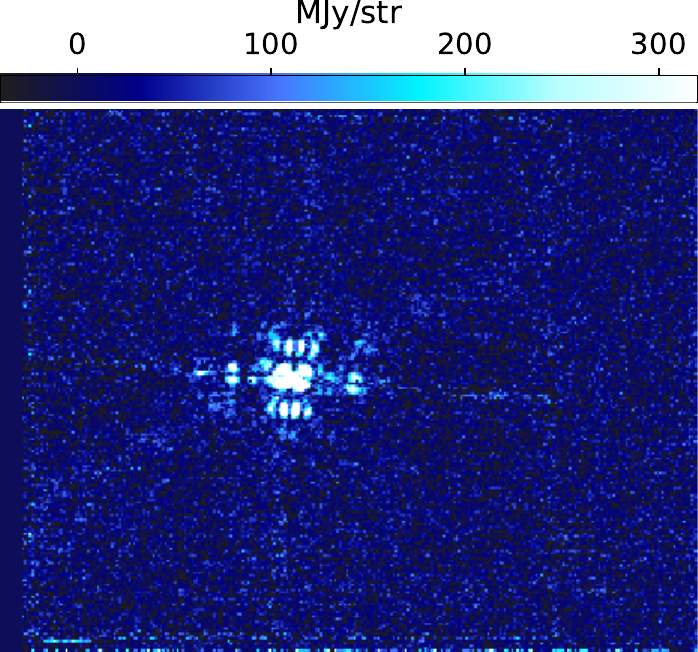}
\caption{First integration of the third dither position of the reference star HD\,222389 after the background subtraction at the \texttt{F1065C} filter. \textit{Left}: Frame with the fringing structure (horizontal lines) after applying the normal reduction. \textit{Right}: Same frame after applying our additional correction procedure, in which we removed the fringing structure.  }
\label{fig:fringing_example}%
\end{figure}

\subsection{Post-processing, photometry, and contrast}\label{sec:raw_contrast}

\begin{figure}
\centering
\includegraphics[width=8.7cm]{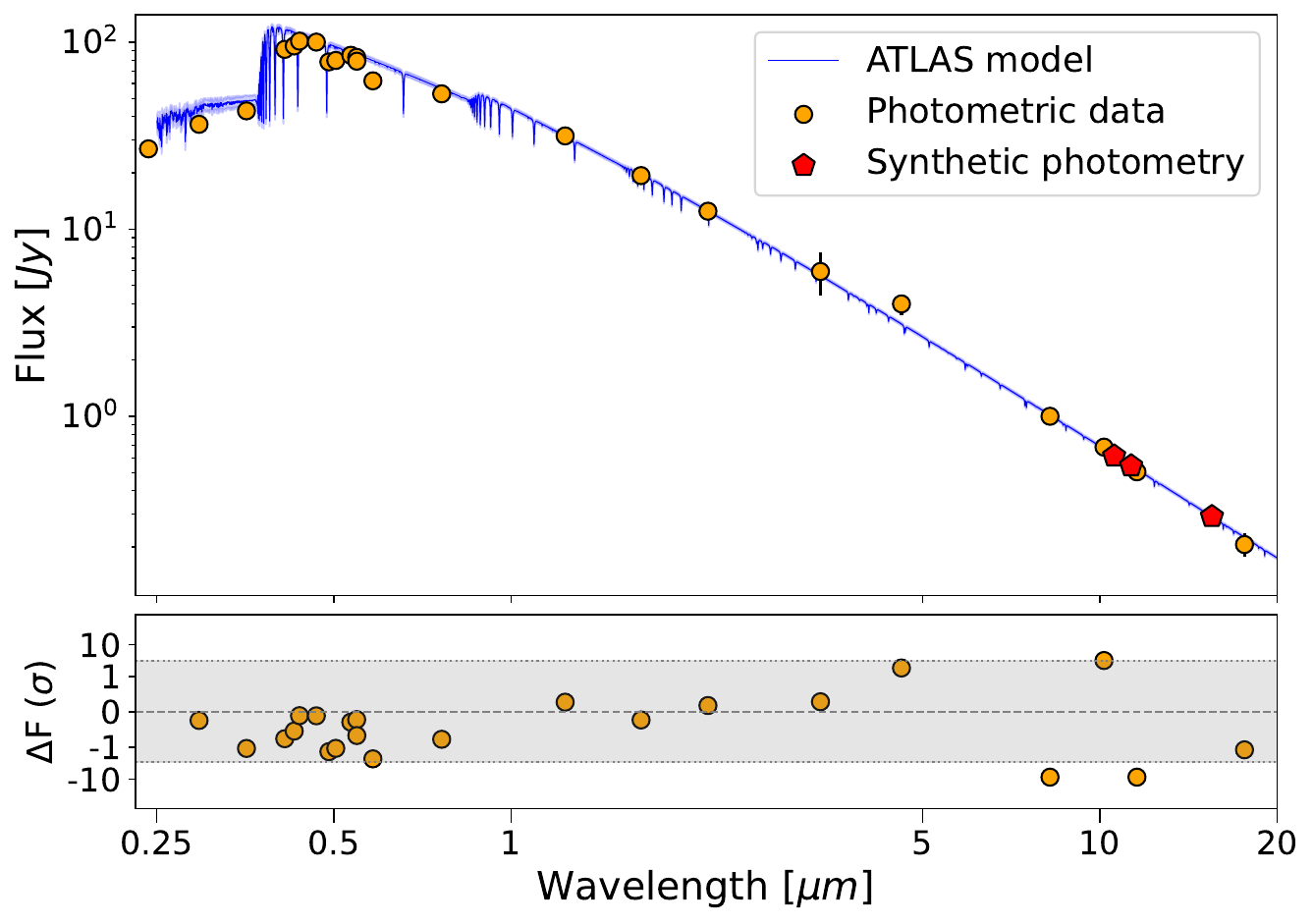} 
\caption{\emph{Top: }SED of the star $\kappa$\,Andromedae. The orange dots correspond to the photometric data from Table\,\ref{table:star_mags}. Red dots correspond to our synthetic photometry using the \texttt{F1065C}, \texttt{F1140C}, and \texttt{F1550C} bandpasses. The blue line corresponds to the mean between the 1\,000 realizations of \texttt{ATLAS} models with the respective uncertainties. \emph{Bottom: } Residuals between the model and the data. The gray region corresponds to $3\sigma$.}
\label{fig:Stellar_SED}%
\end{figure}

We used the \texttt{spaceKLIP} pipeline to optimally subtract the starlight with principal component analysis (PCA). \texttt{spaceKLIP} uses the stellar spectrum and magnitude to calibrate the contrast and extract the companion photometry. We used the \texttt{ATLAS/SYNTHE} atmospheric models (\citealt{ATLAS12}, \citealt{ATLAS12+code}) from \texttt{VidmaPy}\footnote{\url{https://github.com/RozanskiT/vidmapy}} to generate a synthetic spectrum\footnote{We used a microturbulence of 2\,km/s.} of the star. The stellar parameters and their uncertainties were taken from Table\,\ref{table:star_parameters} based on \cite{Jones+2016}, and the spectrum was used as input for \texttt{spaceKLIP}. The model includes the obliquity and the stellar rotation interferometric measurements, which significantly affect the stellar spectrum and spectral lines. We rescaled the spectrum using the new \texttt{2MASS} magnitudes estimated by \cite{Bonnefoy+2014}. To validate the stellar model, we collected archival photometry data from \texttt{Vizier} (\citealt{Vizier-data}), listed in Table\,\ref{table:star_mags}. We proceeded to generate 1\,000 synthetic spectra considering all the parameters and uncertainties using a Monte Carlo approach. At the same time, we calculated the synthetic photometry at each MIRI filter (\texttt{F1065C}, \texttt{F1140C}, and \texttt{F1550C}; see Table\,\ref{table:star_mags}). From the 1\,000 synthetic spectra, we computed the mean stellar spectrum that is used in \texttt{spaceKLIP} for contrast calibration, as well as the standard deviation as a representation of the uncertainty. We computed the synthetic photometry in the same manner. Figure\,\ref{fig:Stellar_SED} shows the SED of $\kappa$\,And, featuring the archival and synthetic photometry and the synthetic spectrum.

For the starlight subtraction, we proceeded in the same way as \cite{Godoy+2024}. We used the RDI technique to suppress the starlight using PCA. The reference star has 15 integrations coming from the dithering and integrations (see Table\,\ref{table:obs_setup}), so we explored the range of 1 to 15 components. We performed separate optimizations of the starlight subtraction to maximize the detection and flux extraction of the companion, and to derive the contrast limits for each filter. For the companion, we determined that the best extraction is using the ``blur'' (smooth using a Gaussian kernel) option in the \texttt{spaceKLIP} pipeline. We analyze this point in more detail in Appendix\,\ref{Apx:best_param}. Figure\,\ref{fig:Extraction_comp} shows the best extraction of the companion, obtained with 10, 7, and 8 components for the \texttt{F1065C}, \texttt{F1140C}, and \texttt{F1550C}, respectively. We note that at \texttt{F1550C} there are some hot and/or bad pixels at the location of the companion. To determine if our extraction is biased by these pixels, we injected a fake planet with and without bad pixels rotated 180\degree and at the same angular separation and same brightness (see Appendix\,\ref{Apx:Bad_Pix}). We conclude that these hot and/or bad pixels do not bias our measurements, since there is a small deviation within the uncertainties of around $\sim6\%$ compared to the $\sim17\%$ uncertainty (see Fig.\,\ref{fig:BP_check}). The measured fluxes of the companion $\kappa$\,And\,b in the MIRI filters are presented in Table\,\ref{table:Fluxes}. Figure\,\ref{fig:images} shows the pre- and post-processed frames using the optimal number of components for RDI.

\begin{figure}
\centering 
\includegraphics[width=9.1cm]{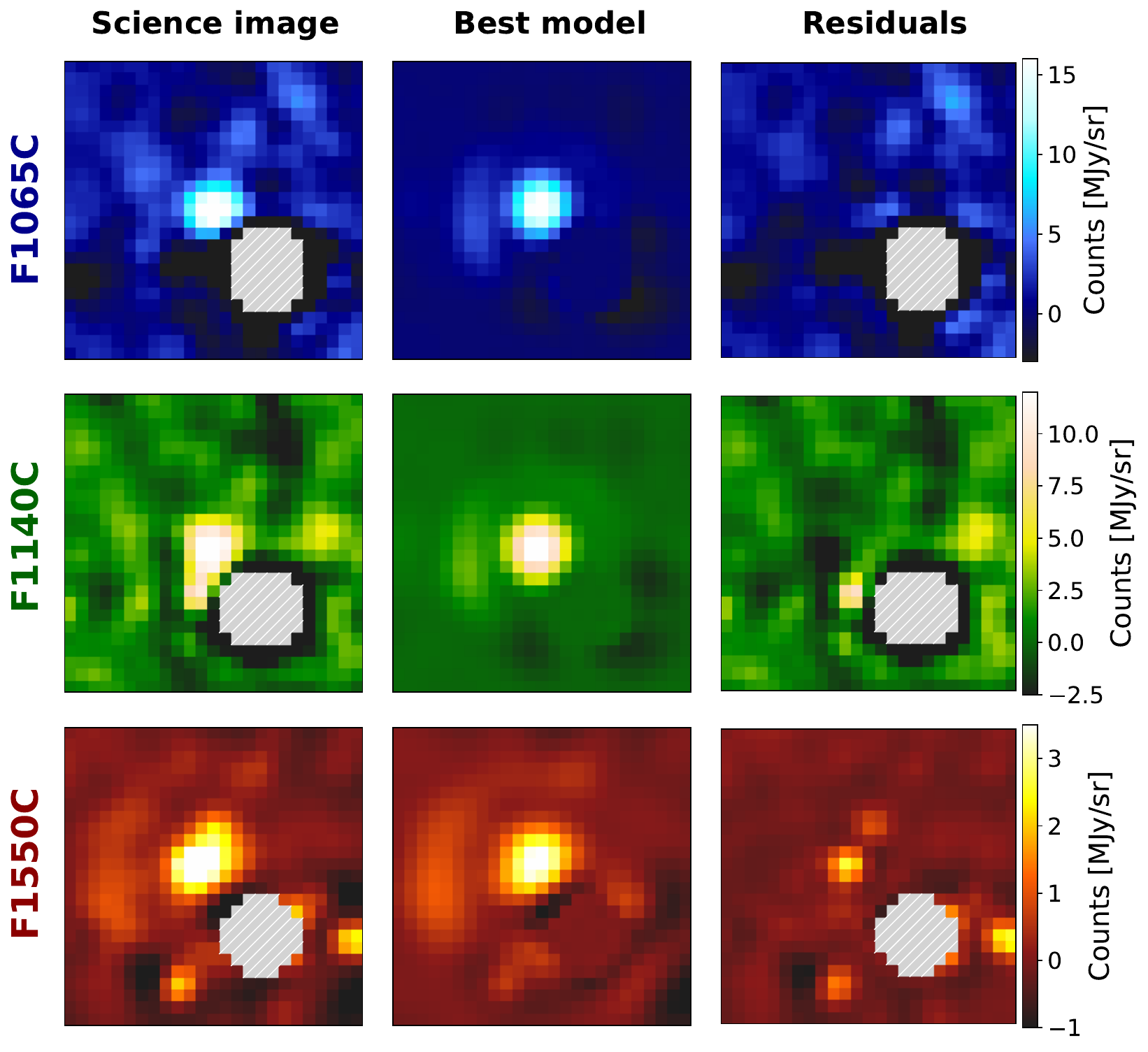}
\caption{Modeling and extraction of the companion $\kappa$\,And\,b. 
From \textit{top} to \textit{bottom}: Filters \texttt{F1065C}, \texttt{F1140C}, and \texttt{F1550C}. \textit{Left}: Post-processed science images, corresponding to 10, 7, and 8 components for the filters \texttt{F1065C}, \texttt{F1140C}, and \texttt{F1550C}, respectively. \textit{Middle}: Best \texttt{spaceKLIP} model of the companion using \texttt{webb\_psf}. \textit{Right}: Residuals after subtracting the model from the science data. The field of view corresponds to $25\times25$ pixels ($\sim2.75\arcsec\times2.75\arcsec$). The images in each row have the same color scale. The gray region with white lines represents a hatched mask corresponding to the star behind the coronagraph.}
\label{fig:Extraction_comp}
\end{figure}

\begin{table*}[]
\centering
\caption{Photometry and astrometry of $\kappa$\,And\,b derived from the MIRI coronagraphic observations.}
\begin{tabular}{ l c c c c c  }
\hline\hline
\addlinespace[3.5pt]
   & \texttt{F1065C} & \texttt{F1140C} & \texttt{F1550C} & Mean value \\
\addlinespace[3.5pt]
\hline
\addlinespace[4pt]
Flux $\kappa$\,And [mJy]   & $615\pm 28$ & $542\pm 25$ & $290\pm 14$ & --- \\
\addlinespace[4pt]
\hline
\addlinespace[4pt]
Flux $\kappa$\,And\,b [mJy] & $0.304\pm0.055$ & $0.247\pm0.045$ & $0.169\pm0.030$ & --- \\ 
\addlinespace[4pt]
PA [$^{\circ}$] & $+42.39\pm0.74$ & $+41.81\pm0.84$ & $+41.75\pm0.57$ & $+41.98\pm0.42$ \\
\addlinespace[4pt]
Angular Sep. [\arcsec] &  $0.766\pm0.089$ & $0.729\pm0.094$ & $0.766\pm0.068$ &  $0.753\pm0.049$ \\
\addlinespace[4pt]
\hline
\end{tabular}
\tablefoot{The mean angular separation and position angle were estimated by combining all three filters at each best number of components determined previously. The angular separation was estimated assuming a pixel scale of $0.11$\arcsec.}
\label{table:Fluxes}
\end{table*}

\begin{figure*}[htb]
    \centering
 \begin{subfigure}{}
  \includegraphics[width=4.8cm]{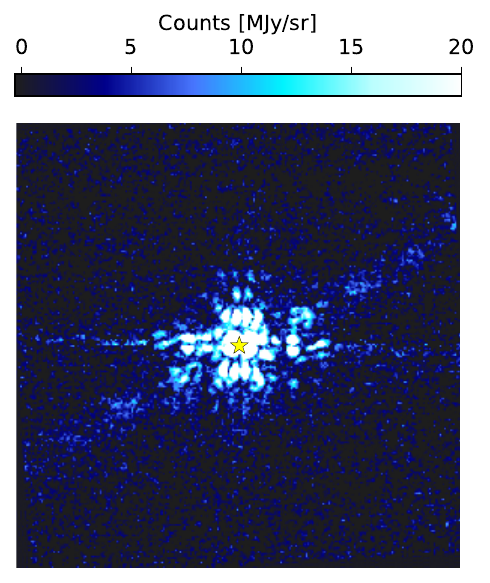}
 \end{subfigure}\hfil 
 \begin{subfigure}{}
  \includegraphics[width=4.8cm]{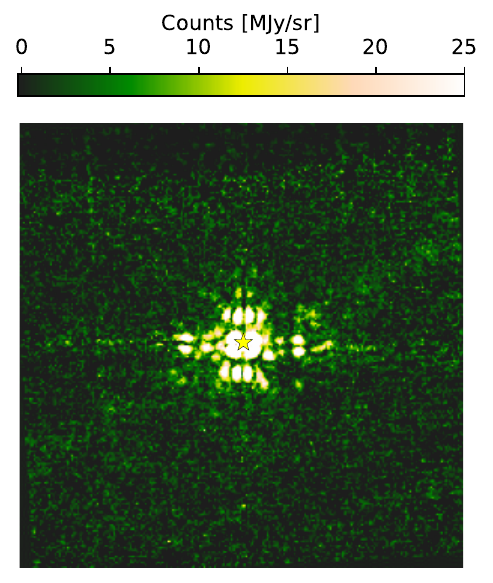}
 \end{subfigure}\hfil 
 \begin{subfigure}{}
  \includegraphics[width=4.8cm]{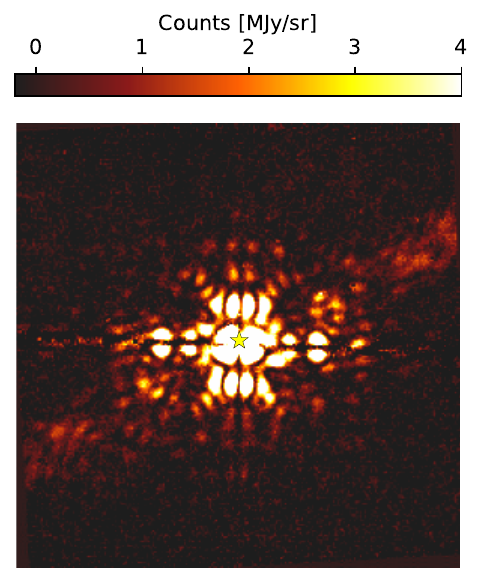}
 \end{subfigure}
 \medskip
 \begin{subfigure}{}
  \includegraphics[width=4.8cm]{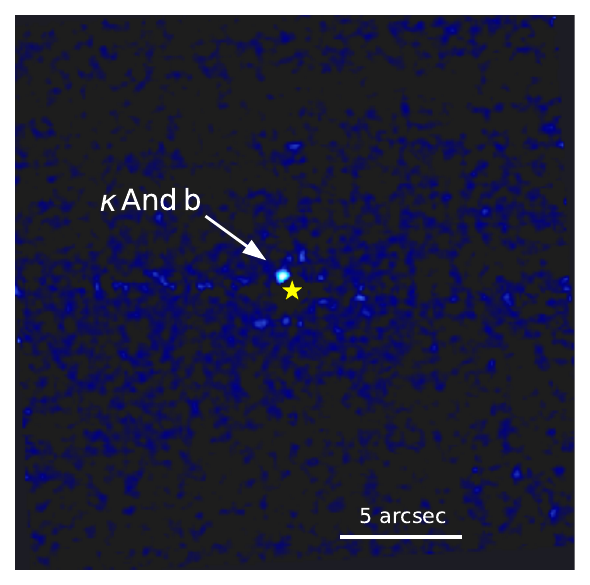}
 \end{subfigure}\hfil 
 \begin{subfigure}{}
  \includegraphics[width=4.8cm]{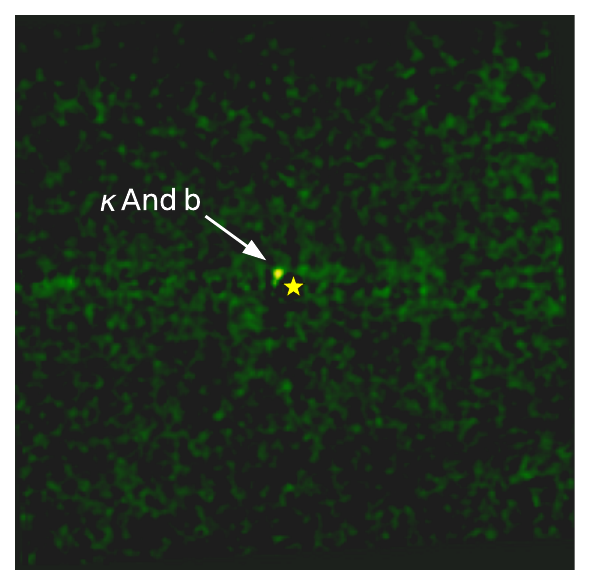}
 \end{subfigure}\hfil 
 \begin{subfigure}{}
  \includegraphics[width=4.8cm]{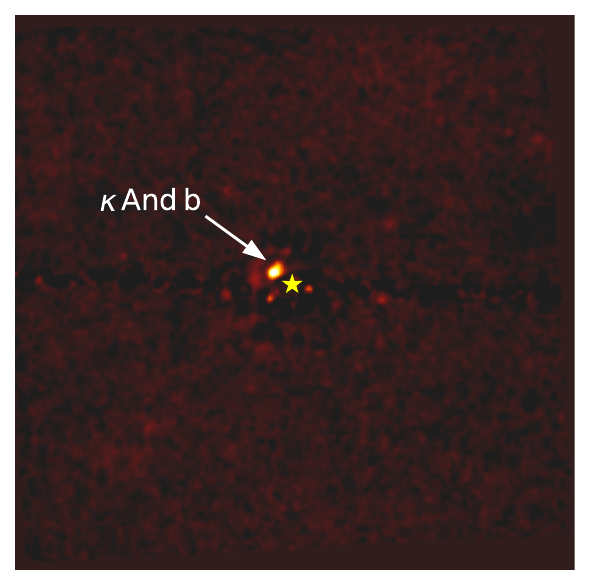}
 \end{subfigure}
\caption{Pre- and post-processing frames of $\kappa$\,And. \textit{Top}: Background-subtracted and stacked science frames. The stacking is only for visualization purposes. \textit{Bottom}: Starlight-subtracted image using RDI and \texttt{KLIP} at the optimal number of components. From left to right: \texttt{F1065C}, \texttt{F1140C}, and \texttt{F1550C} filters. The horizontal white arrow highlights the position of $\kappa$\,And\,b. Each of the columns (i.e., filters) has the same color scale. The yellow star in each subplot marks the position of the star $\kappa$\,And. North is up and east to the left.}
\label{fig:images}
\end{figure*}

\begin{figure}[]
\centering
\includegraphics[width=8.4cm]{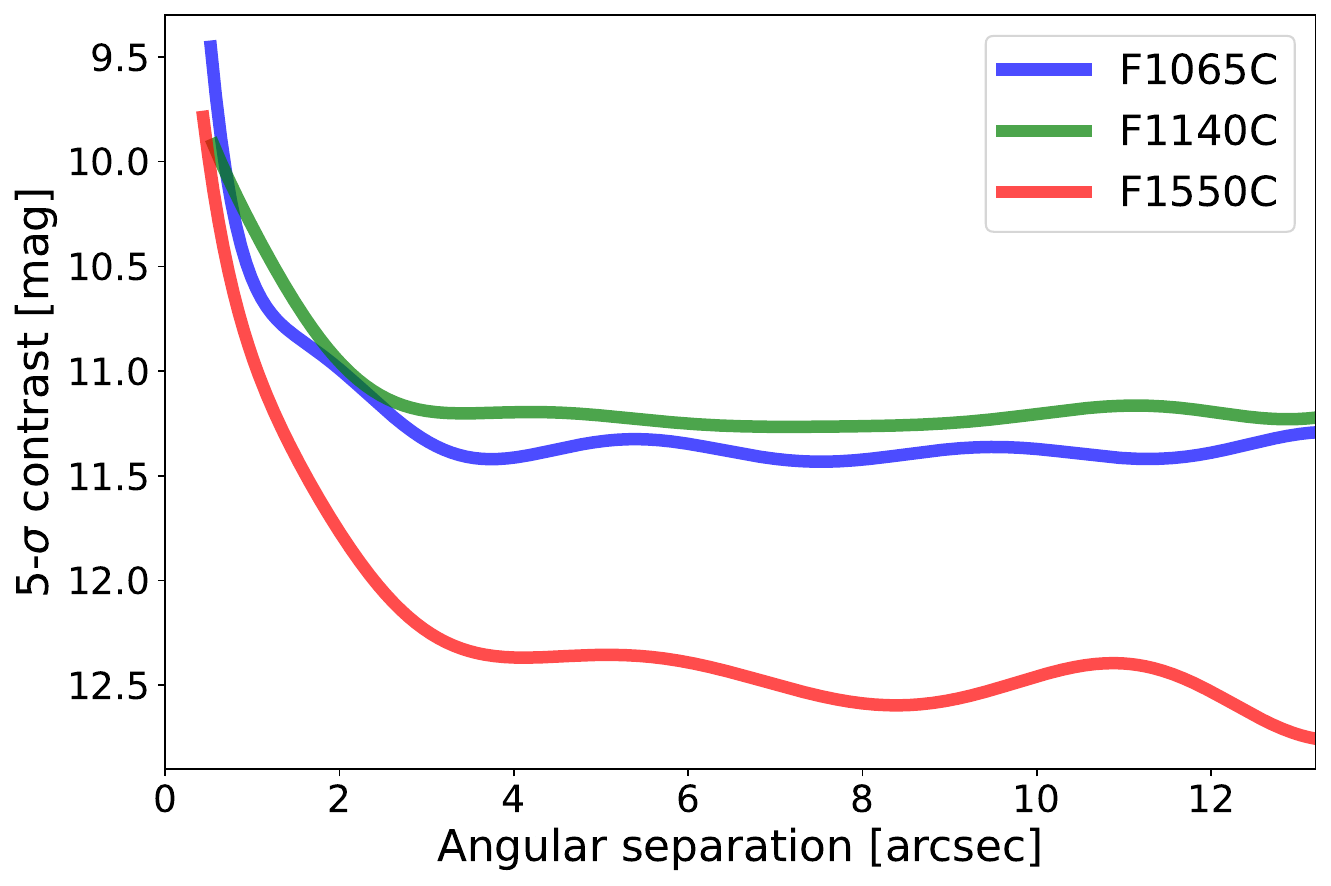} 
\caption{
Smoothed contrast limits for all JWST/MIRI $\kappa$\,And observations in the \texttt{F1065C} (blue), \texttt{F1140C} (green), and \texttt{F1550C} (red) bandpasses.}
\label{fig:Contrast_mag}%
\end{figure}

We estimated the contrast limits for each filter using an adaptive approach that optimizes performance at all angular separations. Instead of fixing the number of components, we evaluated contrasts for 1 to 15 components and retained the best contrast at each separation to produce a combined contrast curve (\citealt{Xuan+2018}). The contrast limits were derived directly from post-processed images after extracting the flux of $\kappa$\,And\,b. These limits were computed accounting for small sample statistics (\citealt{Mawet+2014}). Corrections were applied for coronagraphic transmission and over-subtraction biases introduced by the Karhunen–Loève Image Projection (KLIP), which were quantified by injecting and recovering calibration point sources in the raw and processed images. To reduce noise fluctuations (due to the small-step angular separation used), we first extended the contrast data by stacking it symmetrically and applied a Gaussian window function in the frequency domain using a Fourier transform. Afterward, we performed an inverse Fourier transform to bring the data back to the spatial domain and truncated it to the original length. This allowed us to remove the noise in the contrast curve. Finally, we fitted a polynomial to the smoothed contrast values (reducing the high-frequency fluctuations) to obtain the final contrast curve. The final combined and smoothed contrast limits are presented in Figure\,\ref{fig:Contrast_mag}.

\subsection{Uniform recalibration of archival data}

All available observations of the companion $\kappa$\,And\,b are compiled in Table\,\ref{table:archival_data}. To avoid biases from inconsistent calibrations and ensure a homogeneous flux calibration across all datasets, we reprocessed and uniformly recalibrated all existing flux measurements of the companion with a unique and state-of-the-art stellar atmospheric model. For all the archival observations, we retrieved the raw, uncalibrated contrast of the companion, along with the instrument filter transmission. We then used the precise stellar model of $\kappa$\,And (see Fig.\,\ref{fig:Stellar_SED}), to compute the stellar flux in the archival data filter bandpasses, and consistently calibrate the raw companion fluxes. Below, we summarize the recalibration for each observation, and Appendix\,\ref{Apx:data} provides additional details about each calibration. We highlight that the uncertainty associated with our \texttt{ATLAS} spectrum was considered in the calibration of all the datasets.

\begin{itemize}

\item \textbf{\citealt{Carson+2013}:} Carson et al. observed $\kappa$\,And with the \texttt{Subaru} telescope on 2012 January-July using the \texttt{HiCIAO} (High Contrast Instrument for the Subaru Next Generation Adaptive Optics; \citealt{Hodapp+2008}) and \texttt{IRCS} (Infrared Camera and Spectrograph; \citealt{Tokunaga+1998}) instruments in the J, H, $\mathrm{K_{s}}$, and $L^{\prime}$ bands (see Table\,\ref{table:archival_data}). They used the \texttt{2MASS} J, H, and $\mathrm{K_{s}}$ magnitudes for flux calibration to convert the $\kappa$\,And\,b contrast into apparent magnitudes. We note a difference of about $3\%$ ($0.03$ in magnitude) between the \texttt{2MASS} and \texttt{Subaru/HiCIAO} filters. Additionally, as highlighted by \cite{Bonnefoy+2014}, the \texttt{2MASS} images are saturated, which leads to incorrect flux measurements for the star $\kappa$\,And. We used our stellar model to compute the \texttt{2MASS} magnitudes, which agree with the ones reported by \cite{Bonnefoy+2014}. We then computed the \texttt{Subaru/HiCIAO} J, H, and $\mathrm{K_{s}}$ magnitudes using our stellar models to convert the $\kappa$\,And\,b contrast into apparent magnitudes.

\item \textbf{\citealt{Hinkley+2013}:} Hinkley et al. observed $\kappa$\,And using the P1640 instrument at Palomar on 2012 December 23 (\citealt{Hinkley+2011c}, \citealt{Oppenheimer+2012}), integral field spectroscopy (IFS) mode (\citealt{Hinkley+2008}) covering the YJH bands. They flux-calibrated the observations using a B9V stellar spectrum model from the \texttt{Pickles Stellar Library} (\citealt{Pickles-1998}). As noted by \cite{Currie+2018}, the Pickles B9 spectrum shows significant deviations in the NIR with respect to other stellar models (e.g., Kurucz atmospheric models). To correct this, we applied a correction factor by dividing the B9V-calibrated model by our \texttt{ATLAS} stellar spectrum model. Note that the calibrator factor has a dependence on wavelength and it is not a single value. The differences between the \texttt{Pickles} and \texttt{ATLAS} models result in a change in the shape of the IFS spectrum of $\kappa$\,And\,b (see Figure\,\ref{fig:Spec_star_2}).

\item \textbf{\citealt{Bonnefoy+2014}:} Bonnefoy et al. observed $\kappa$\,And on 2012 October 6 and 30, and November 3 using the \texttt{Keck} and the Large Binocular Telescope Interferometer (\texttt{LBTI}) observatories in the $\mathrm{K_{s}}$, $L^{\prime}$, Br$\alpha$, and $M^{\prime}$ bands. They also re-estimated the \texttt{2MASS} magnitudes of $\kappa$\,And using a calibrator star on 2012 October 30 with the Mimir instrument (\citealt{Clemens+2007}) mounted on the 1.8m \texttt{Perkins} telescope at Lowell Observatory. We calculated the contrast magnitudes of $\kappa$\,And\,b and derived the magnitudes in each filter using our stellar spectrum model. The resulting magnitudes are consistent with those from \cite{Bonnefoy+2014}, with about $0.01$\,mag differences. We did not re-estimate the $L^{\prime}$ magnitude, as it was flux-calibrated using HR\,8799, which we consider an unbiased measurement.

\item \textbf{\citealt{Currie+2018}:} Currie et al. observed $\kappa$\,And using the \texttt{Subaru} \texttt{SCExAO} (Subaru Coronagraphic Extreme AO)/\texttt{CHARIS} integral field spectrograph in the JHK bands (\citealt{Peters+2012}; \citealt{Groff+2014}; \citealt{Groff+2015}) on 2017 September 8. To flux-calibrate the IFS data, they used a Kurucz stellar atmosphere model (\citealt{Castelli+Kurucz-2003}) with\footnote{We adopted a metalicity equal to zero, as \cite{Jones+2016}} $\mathrm{T_{eff}} = 11\,400$,K and log(g)$=4.0$\,dex. As with the \cite{Hinkley+2013} observations, we computed a correction factor by comparing the Kurucz models with our \texttt{ATLAS} spectrum. The correction is small and within the uncertainties, primarily affecting the H-band peak (see Figure\,\ref{fig:Spec_comp_corr}). \cite{Currie+2018} also highlights that they did not include a $5\%$ uncertainty from the flux calibration, so we incorporated this uncertainty in our correction.

\item \textbf{\citealt{Kuhn+2018}:} They observed $\kappa$\,And as part of the first-light and on-sky commissioning results of the H-band Vector Vortex Coronagraph for the SCExAO System on 2016 November 12. They reported only the $\kappa$\,And\,b contrast, which is $10.35$. This value differs from those reported by \cite{Carson+2013} and \cite{Currie+2018}. We chose not to use this contrast, as it may be biased due to data processing or calibration issues, and it was not reported the related uncertainty which is crucial to determining the possibility of biases.

\item \textbf{\citealt{Hoch+2020}:} They observed $\kappa$\,And with the \texttt{OSIRIS} (OH-Suppressing Infrared Imaging Spectrograph) integral field spectrograph at the Keck Observatory (R$\sim4\,000$, \citealt{Larkin+2006}) in the K broadband mode on 2016 November 6-8 and 2017 November 4. \cite{Hoch+2020} did not perform direct flux calibration. To remove telluric absorptions in $\kappa$\,And,b, however, they masked the H lines in $\kappa$\,And and normalized the continuum using a blackbody with the same temperature as determined by \cite{Jones+2016}. We did not apply any additional corrections at this step since they used the same temperature as our \texttt{ATLAS} model. Then, they calibrated the resulting $\kappa$\,And\,b spectrum using the K-band apparent magnitude computed by \cite{Currie+2018} and a distance of $50.0\pm0.1$\,pc. To renormalize the \texttt{OSIRIS} spectrum, we calculated the \texttt{WIYN/WHIRC}\footnote{\texttt{WIYN:} Wisconsin-Indiana-Yale-NOIRLab. \texttt{WHIRC:} WIYN High-Resolution Infrared Camera.} $\mathrm{Ks_{MKO}}$ magnitude in the \texttt{OSIRIS} spectrum and in the corrected Currie et al. CHARIS IFS data. Using both, we computed the correction factor to renormalize the \texttt{OSIRIS} spectrum of $\kappa$\,And\,b.

\item \textbf{\citealt{Stone+2020}:} Stone et al. observed $\kappa$\,And using the \texttt{LBTI/ALES} (Arizona Lenslets for Exoplanet Spectroscopy) $2.8-4.1\mu\,m$ mode on 2016 November 13. They flux-calibrated the spectrum using the \texttt{NEXGEN} atmospheric library model (\citealt{Hauschildt+1999}) for an A0 star. Since they did not specify the key atmospheric properties, and since \texttt{NEXGEN} models are available for temperatures up to $10\,000$\,K, we adopted $\mathrm{T_{eff}} = 10\,000$,K, log(g) = $4.0$\,dex, and solar metallicity\footnote{The differences between \texttt{NEXGEN} models with $\mathrm{T_{eff}} = 10\,000$\,K, log(g) values of $4.0$ and $4.5$\,dex, and different metallicities are minor compared to our \texttt{ATLAS} spectrum and within uncertainties.}. Stone et al. normalized the stellar spectrum using the \texttt{Keck/NIRC2} (Near-infrared Camera, Second Generation) $L^{\prime}$ magnitude from \citet[][$4.32 \pm 0.05$]{Bonnefoy+2014}. We calculated the $L^{\prime}$ magnitude using our \texttt{ATLAS} spectrum ($4.32 \pm 0.02$) to normalize the A0 stellar spectrum, propagating the uncertainties. Then, we computed the correction factor to adjust the $\kappa$\,And\,b spectrum. The differences between \texttt{ALES} spectrum and our corrected version are small, with slight deviations around 3.7$\mu m$ and $4.0\mu m$.

\item \textbf{\citealt{Uyama+2020}:} They observed $\kappa$\,And with \texttt{Subaru/SCExAO}+\texttt{HiCIAO} using H- and Y-band broadband filters on 2016 July 18. The Y-band flux calibration was performed with HIP\,118133 ($6.60\pm 0.06$\,mag) and the H-band with HIP\,79977 ($7.85\pm 0.03$\,mag). The magnitudes of $\kappa$\,And and $\kappa$\,And\,b were determined to be $4.28\pm 0.09$ and $17.04\pm 0.15$\,mag in the Y-band, and $15.18\pm 0.56$\,mag in the H-band for $\kappa$\,And\,b. Due to the larger uncertainties in the H-band, they excluded it from the atmospheric analysis. We decided not to use magnitude in the H-band for the same reason. Since the flux calibration was done using different stars (HIP\,118133 for the Y-band), this measurement can be considered unbiased. As a check on our stellar spectrum, we calculated the magnitude in the Y-band of $\kappa$\,And, obtaining $4.25\pm0.01$\,mag, consistent with \citet[][$4.28\pm 0.09$\,mag]{Uyama+2020}.

\end{itemize}

\begin{figure}
\centering
\includegraphics[width=8.8cm]{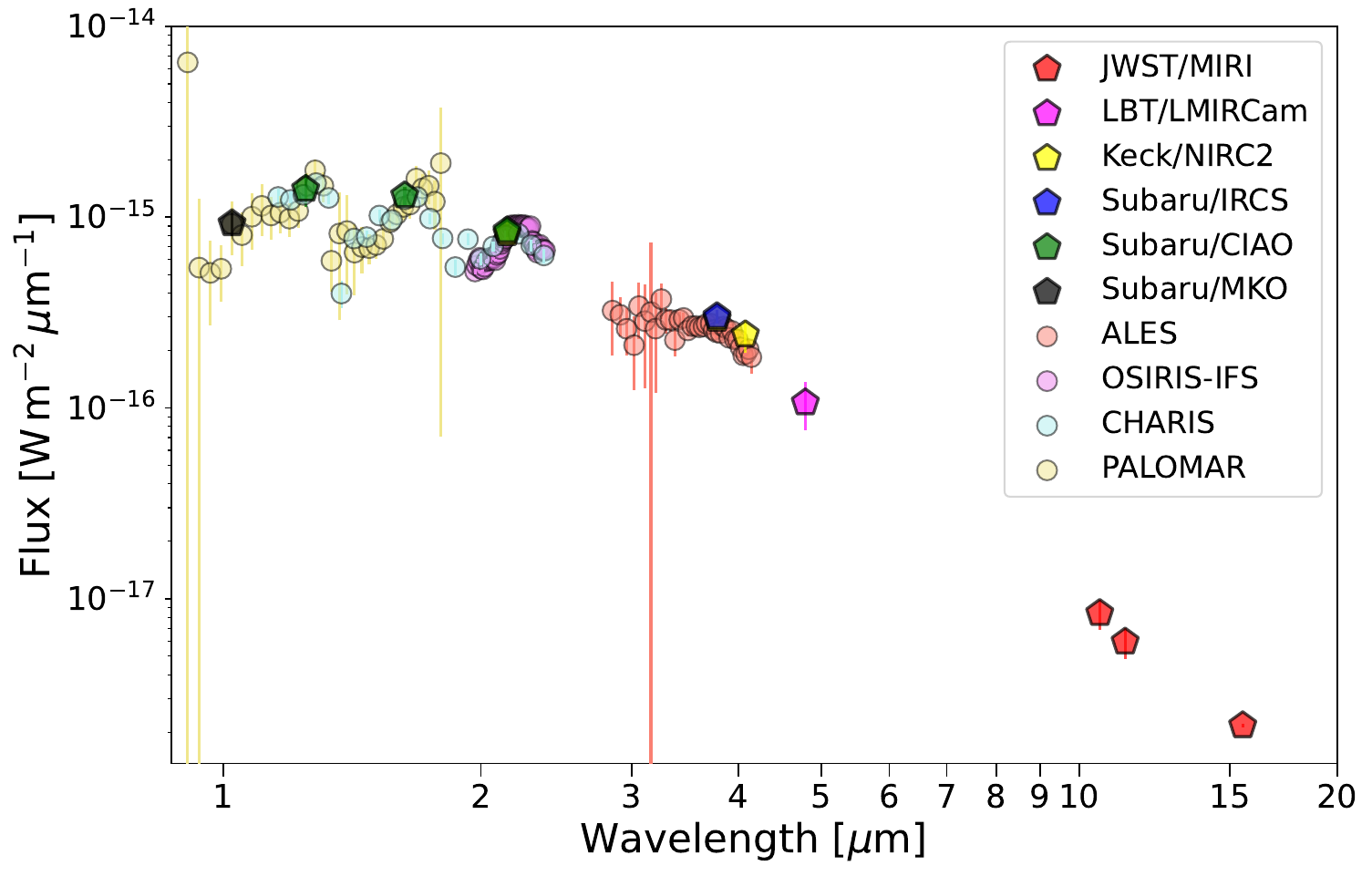}
\caption{SED of $\kappa$\,And\,b. The colored circles correspond to the IFS/spectrum data, while the colored pentagons are the photometric data. Note that we reduced the OSIRIS resolution spectrum to a low-resolution, IFS-like data (hereafter ``OSIRIS-IFS''). }
\label{fig:data_alone}%
\end{figure}

Figure\,\ref{fig:data_alone} shows all the corrected data plus our JWST/MIRI observations. We note that all the datasets are consistent within $1$-$2\sigma$.

\begin{table*}[]
\centering
\caption{\label{table:archival_data}Corrected magnitudes archival data of $\kappa$\,And\,b using $\kappa$\,And \texttt{ATLAS} atmospheric model.}
\begin{tabular}{l l c c c} 
\hline \hline
Date & Instrument & Band & magnitude\tablefootmark{a} & Reference \\
\hline
2012-01/2012-07 & Subaru/HiCIAO & J & $15.850 \pm 0.211$  &  \cite{Carson+2013} \\
                & Subaru/HiCIAO & H & $14.923 \pm 0.122$  &  \cite{Carson+2013}\\
                & Subaru/HiCIAO & Ks & $14.304 \pm 0.083$ &  \cite{Carson+2013}\\
                & Subaru/IRCS & $L^{\prime}$ & $13.12 \pm 0.09$\tablefootmark{b} & \cite{Carson+2013} \\
2012-10-06      & LBT/LMIRCam & $M^{\prime}$ & $13.23 \pm 0.30$ &  \cite{Bonnefoy+2014} \\
2012-10-30      & Keck/NIRC2 & $L^{\prime}$ & $13.14 \pm 0.09$ &  \cite{Bonnefoy+2014} \\
2012-11-03      & Keck/NIRC2 & Ks & $14.34 \pm 0.15$ &  \cite{Bonnefoy+2014} \\
                & Keck/NIRC2 & $NB\_4.05$\tablefootmark{c} & $13.00 \pm 0.21$ &  \cite{Bonnefoy+2014} \\ 
2016-11-12      & Subaru/CIAO & H & $14.63 \pm 0.06$\tablefootmark{d} &  \cite{Kuhn+2018} \\
2016-07-18      & Subaru/CIAO & Y & $17.04 \pm 0.15$\tablefootmark{b} &  \cite{Uyama+2020} \\
\hline
\end{tabular}
\tablefoot{
\tablefoottext{a}{Apparent magnitudes in the Vega system.}
\tablefoottext{b}{We did not re-estimate these magnitudes since they were flux-calibrated using a different star.}
\tablefoottext{c}{The filter corresponds to the $Br\alpha$ narrow band with $\lambda_{c}=4.05\mu m$.}
\tablefoottext{d}{We did not include this magnitude in our study, but we included the transformed contrast to magnitude and only included the uncertainty of our \texttt{ATLAS} spectrum.}
}
\end{table*}

\section{Analysis and results}\label{sec:res}

In this section we analyze the main properties of the companion $\kappa$\,And\,b. In Section\,\ref{sec:CMD} we qualitative describe the companion from the CMD from the MIRI filter CMD point of view. Then, in Section\,\ref{sec:EtCMD}, we use two main families, cloudy and cloud-free, of atmospheric and evolutionary tracks models to infer the properties of $\kappa$\,And\,b. We first derive the age and mass by combining the MIRI CMD with isochrones. Then, by using atmospheric models, we fit the SED to obtain main physical parameters such as effective temperature and radius in Section\,\ref{sec:atmosp}. We combine these results with evolutionary tracks to derive the age and mass of the companion in Section\,\ref{sec:atm-tracks}. Finally, we used our age estimate to generate our MIRI sensitivity maps in Section\,\ref{sec:PMD}.

\subsection{MIRI color-magnitude diagram}\label{sec:CMD}

Color-magnitude diagrams have long been essential for studying the atmospheric and physical properties of substellar objects across the M-to-Y spectral range (e.g., \citealt{Kirkpatrick+1999}, \citealt{Cushing+2008}). These diagrams provide a framework to disentangle the influence of effective temperature (T$_{\mathrm{eff}}$), surface gravity (log (g)), and atmospheric composition on the photometric and spectral properties of these objects. In the NIR (e.g., J–K vs. J, Fig.\,\ref{fig:CMD_JK}), M-type objects are dominated by water vapor and metal hydrides (red colors), while cooler T- and Y-dwarfs exhibit strong methane and ammonia absorption features displaying bluer colors (e.g., \citealt{Leggett+2010L}, \citealt{Bonnefoy+2013}). MIRI/JWST \texttt{F1065C}-\texttt{F1140C} CMD reveal similar trends with greater sensitivity to condensates and minor species, enabling more precise differentiation between objects based on their atmospheric chemistry and thermal structure (e.g., \citealt{Carter+2022}; \citealt{Matthews+2024}; \citealt{Malin+2024}; \citealt{Godoy+2024}; \citealt{Malin+2025}).

\begin{figure*}
\centering
\includegraphics[width=5.8cm]{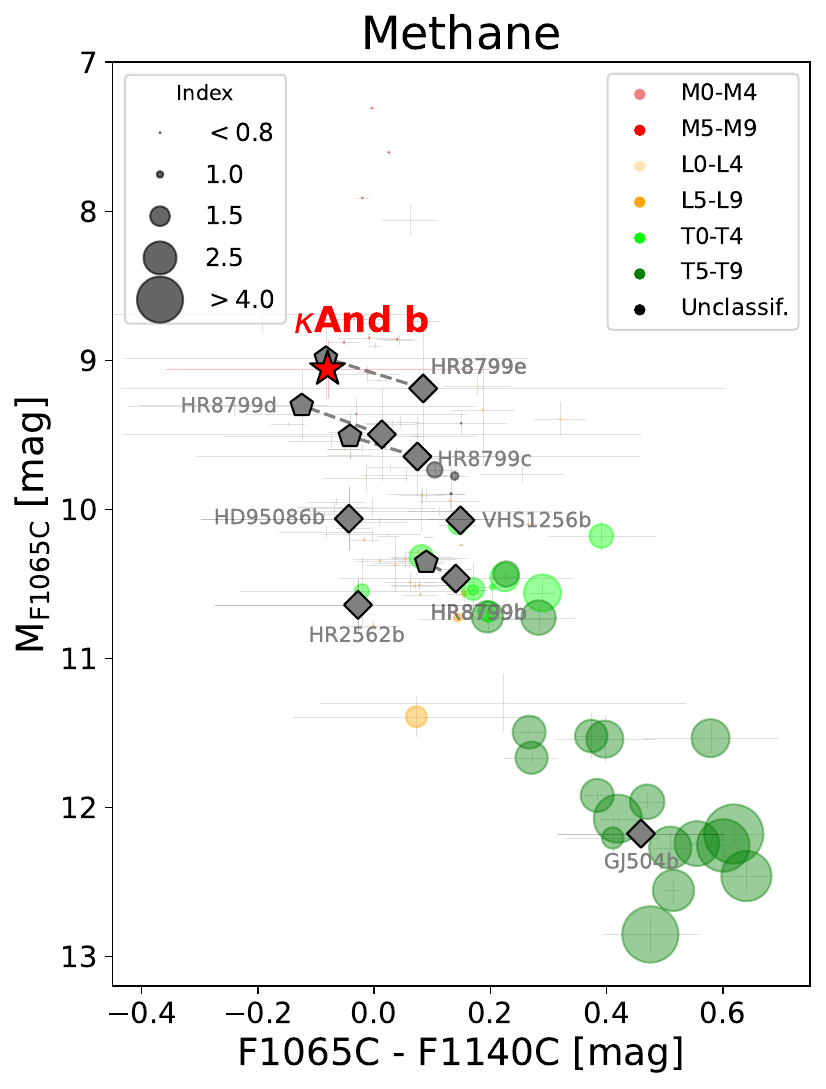}
\includegraphics[width=5.8cm]{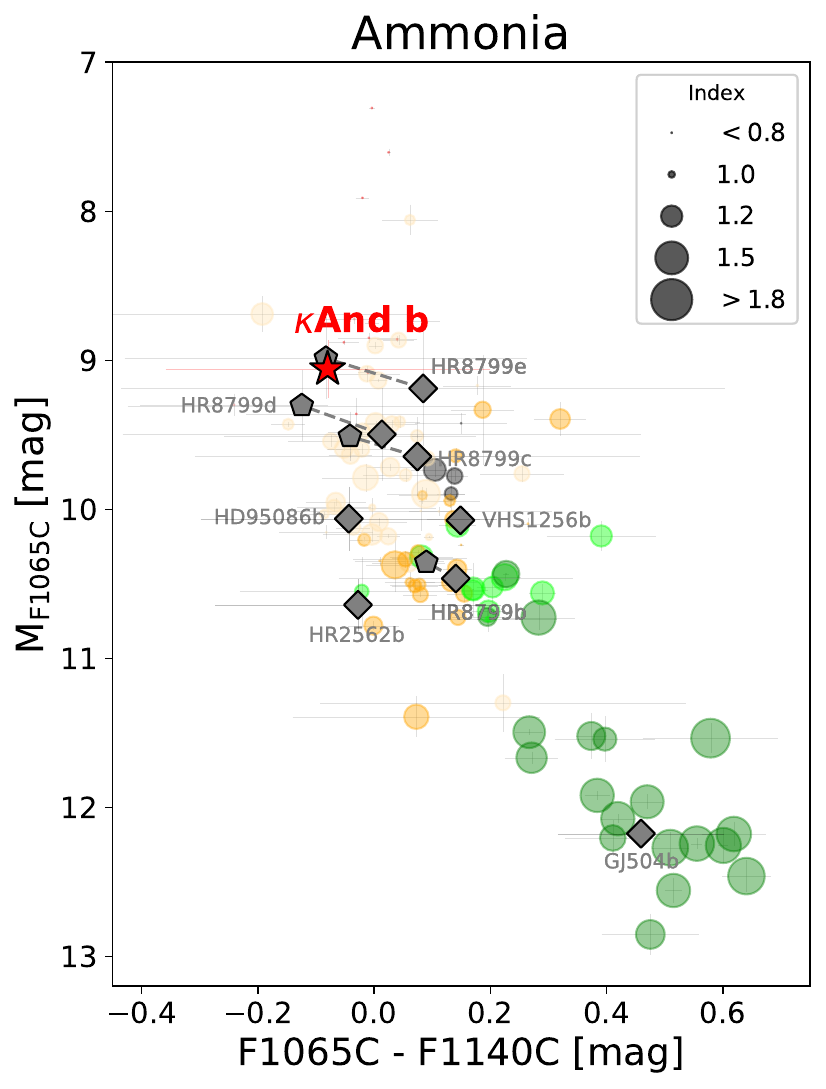}
\includegraphics[width=5.8cm]{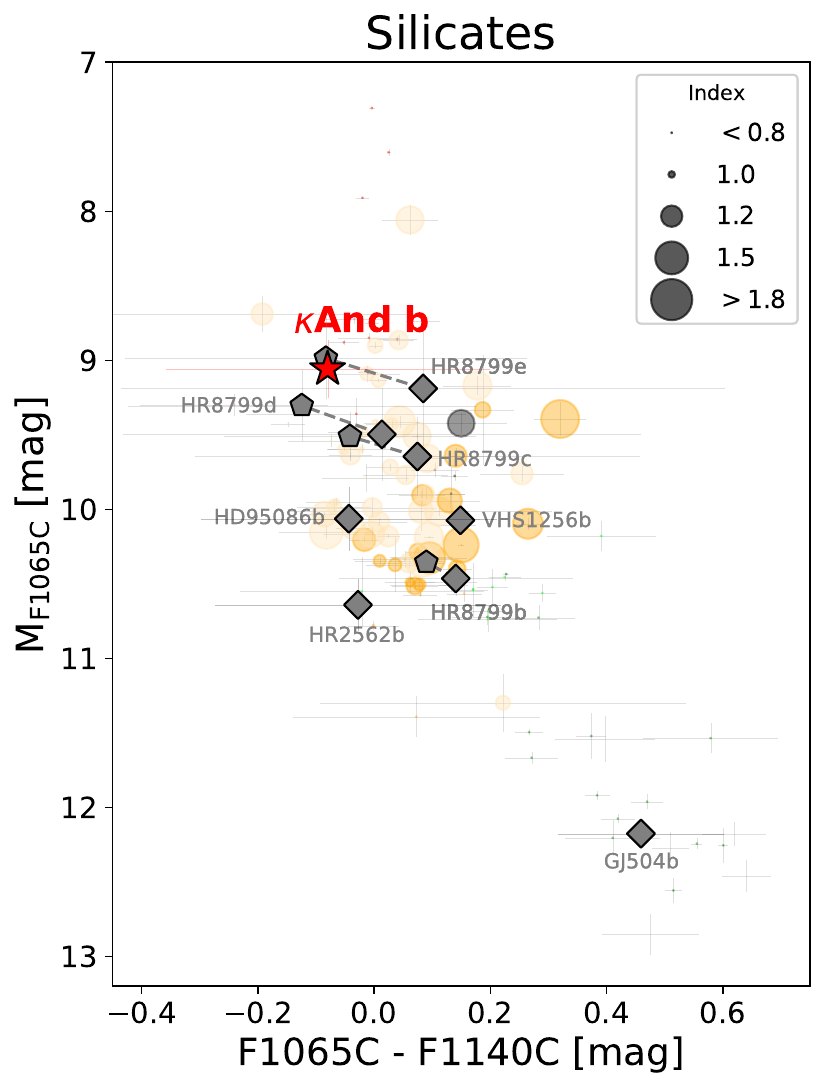}
\caption{CMD using the \texttt{F1065C} ad \texttt{F1140C} filters from MIRI. The dots correspond to the photometry obtained from the Spitzer spectra sample (\citealt{Suarez+Metchev-2022}), while the colors refer to the spectral type. The subplots correspond to the same CMD but show the methane (\textit{left}), ammonia (\textit{middle}), and silicates (\textit{right}) spectral indices, as defined in \cite{Suarez+Metchev-2022}, related to the depth of the absorption feature of each component. The circle sizes in each subplot refer to the value of the spectral index, as defined by \cite{Suarez+Metchev-2022}. The gray dots in each panel correspond to species non-detection (i.e., plotted with index=0, not highlighted sources). The black dots refer to unclassified spectral types. The red star corresponds to $\kappa$\,And\,b, while the gray diamonds correspond to known planetary-mass companions (HR\,2562\,b from \citealt{Godoy+2024}; VHS\,1256\,b from \citealt{Miles+2023}; HD\,95086\,b from \citealt{Malin+2024}; GJ\,504\,b from \citealt{Malin+2025}; HR\,8799\,bcde from \citealt{Boccaletti+2024}). The location and size of each planetary-mass companion are not related to the spectral index. Note that the Y-axis (i.e., $M_{\mathrm{\texttt{F1065C}}}$) refers to the absolute magnitude. Sample data taken from \cite{Godoy+2024}}
\label{fig:CMD_index}%
\end{figure*}

Figure\,\ref{fig:CMD_index} shows the \texttt{F1065C}–\texttt{F1140C} CMD, plotting with colored dots the field sources with different spectral types. The CMDs from left to right highlight (with marker sizes) the spectral index (from \citealt{Suarez+Metchev-2022}) related to the presence of different chemical species (methane, ammonia, and silicates, respectively). Figure\,\ref{fig:CMD_index} highlights the pivotal role of silicate and metal oxide clouds in early L-type atmospheres, especially for planetary-mass companions ( \citealt{Kirkpatrick-2005}; \citealt{Visscher+2010}; \citealt{Morley+2012}; \citealt{Morley+2014}). In these objects, T$_{\mathrm{eff}}$ typically ranges from $1500$\,K to $2100$\,K, with surface gravities lower than those of field dwarfs due to their younger ages (\citealt{Burgasser+2002}; \citealt{Leggett+2002}). These differences result in thicker cloud decks and distinct molecular absorption features that set them apart from their field counterparts (e.g., \citealt{Saumon+Marley-2008}, \citealt{Burgasser-2011}). Early L dwarfs in particular serve as benchmarks for understanding the transition from cloud-dominated M-dwarfs to the clear atmospheres of T-dwarfs. Their location on the MIRI CMD is influenced by the presence of silicate clouds in the upper atmosphere, which gradually condense and disappear as the temperature drops, leaving behind clouds composed of different molecules and leading to a transition towards cloud-free atmospheres (e.g., \citealt{Lodders+2002}; \citealt{Zhang+2018a}; \citealt{Zhang+2018b}; \citealt{Suarez+Metchev-2022}).

$\kappa$\,And\,b is situated at the top of the early L-type sequence in the \texttt{F1065C}–\texttt{F1140C} CMD (red star marker in Fig.\,\ref{fig:CMD_index}), consistent with a $\mathrm{T_{eff}}$ between $1600$\,K and $2000$\,K. This location suggests the presence of silicate clouds, consistent with its spectral classification of L0–L2 (e.g., \citealt{Marley+2002}, \citealt{Suarez+Metchev-2022}). Compared to field sources, the younger age and lower surface gravity of the target imply a thicker, more optically active cloud layer (\citealt{Allers+2013}). Its position is notably distinct from that of other planetary-mass companions shown with gray diamonds in the CMD, which span a broader range of spectral types and $\mathrm{T_{eff}}$. The classification of our target as L0–L2 highlights its importance as a probe for understanding the early L-type sequence and its relation to the field and companion objects.

Although the CMD provides key insights into the atmospheric properties of the target, further progress requires a detailed MIR spectrum (e.g., \citealt{Daemgen+2017} for $\lambda<5\mu$m; \citealt{Miles+2023} covering the NIR to MIR). Such observations would allow for the precise identification of species like silicates and metal oxides, as well as an assessment of their concentration and vertical distribution in the $\kappa$\,And\,b atmosphere (e.g., \citealt{Skemer+2014}). A spectrum would also enable robust comparisons between this target and other benchmark objects, advancing our understanding of the early L-type sequence and the broader context of planetary-mass companions in the MIR (e.g., \citealt{Lodieu+2018}). Proven techniques for obtaining such spectra of high-contrast planets with JWST, in particular, using NIRSpec, have already been demonstrated (e.g., \citealt{Ruffio+2024}), paving the way for similar analyzes of $\kappa$\,And\,b.

\subsection{Evolutionary tracks and color-magnitude diagram}\label{sec:EtCMD}

When combined with theoretical evolutionary tracks, CMDs provide a robust method to derive key properties such as mass or age, directly from the photometric measurements, minimizing dependence on model fitting. This approach has been widely used in the literature to constrain the ages of stellar populations and substellar objects (e.g., \citealt{Bonatto-2019}), and giant planet companions such as AF\,Lep\,b (\citealt{Gratton+2024}), and HIP\,65426\,b (\citealt{Chauvin+2017}), enabling precise placement of objects on evolutionary grids and facilitating comparisons across a broad range of ages and masses.

For this analysis, we employed five evolutionary track models: ATMO chemical equilibrium (\citealt{Phillips+2020}), AMES-COND (\citealt{Allard+2001}), AMES-DUSTY (\citealt{Chabrier+2000}), BT-Settl (\citealt{Allard+2012}), and Sonora (\citealt{Karalidi+2021}, \citealt{Marley+2021-son}). These models span a range of atmospheric and cloud physics assumptions, which can display differences in the properties of substellar objects. To better understand the properties of $\kappa$\,And\,b, we used photometry in the H-band and the JWST/MIRI \texttt{F1140C} filter. The H-band is affected by cloud properties, changing the shape of the spectrum (e.g., \citealt{Samland+2017}, \citealt{Cheetham+2019}), while \texttt{F1140C} is affected by the presence of specific chemical species (e.g., \citealt{Suarez+Metchev-2022}, \citealt{Malin+2025}). These bands cover key insights of the atmosphere properties, making these filters ideal to test different atmospheric models with different physics (cloudy and cloud-free properties).

\begin{figure*}
\centering
\sidecaption
\includegraphics[width=6.3cm]{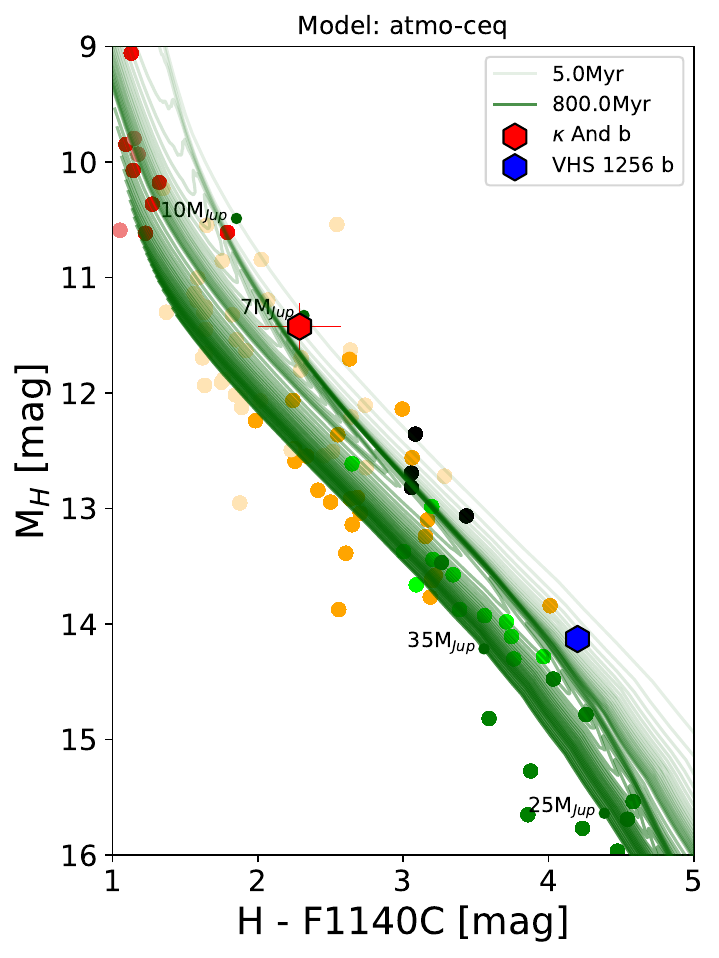}
\includegraphics[width=6.3cm]{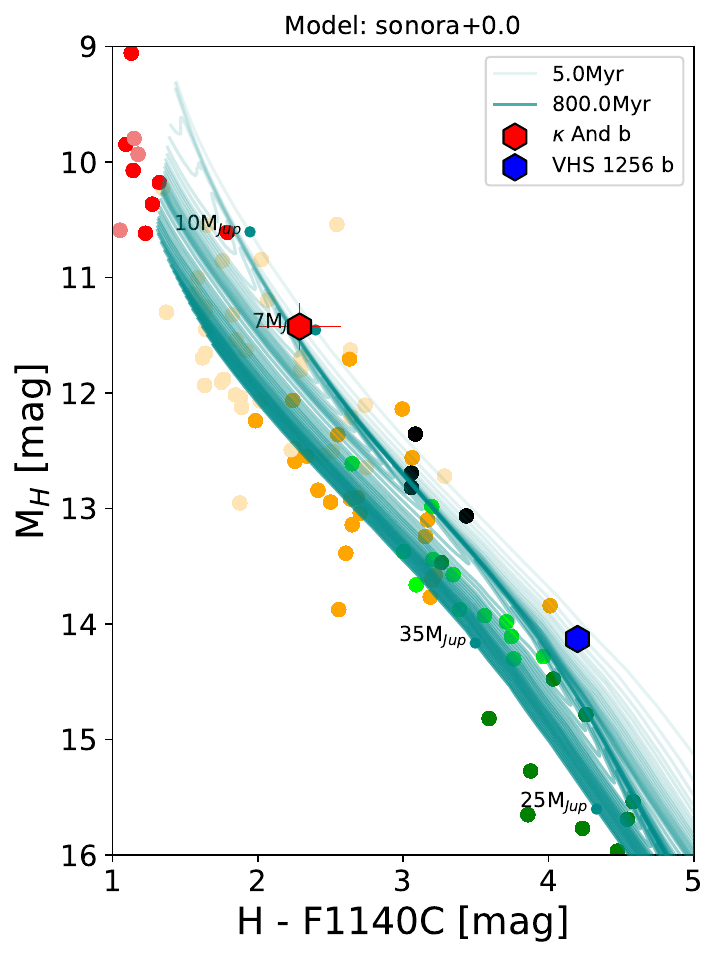}
\caption{CMD using $\mathrm{H_{2MASS}}$ and \texttt{F1140C}. The blue pentagon corresponds to $\kappa$\,And\,b, while the red one to VHS\,1256\,b (from \citealt{Miles+2023} and \citealt{Godoy+2024}). The colored circles correspond to field sources shown in Fig.\,\ref{fig:CMD_index}. \textit{Left:} ATMO chemical equilibrium isochrones from 5 to 800 Myr, highlighting the 7 and 10 $\mathrm{M_{Jup}}$ for the isochrone of 5\,Myr, also 25 and 35 $\mathrm{M_{Jup}}$ for the one of 800\,Myr. \textit{Right:} Same as the left panel but for Sonora solar metallicity isochrones. }
\label{fig:CMD_tracks}%
\end{figure*}

To estimate the age and mass of $\kappa$\,And\,b for each evolutionary track, we calculated the compatibility between the observed photometry and each isochrone. Assuming Gaussian uncertainties for the magnitude and color of $\kappa$\,And\,b, we evaluated a score based on the proximity of the photometry to each isochrone in magnitude-color space. We first normalized the 2D Gaussian to have 1 at the maximum. Then, we evaluate each point of the isochrone in the Gaussian, obtaining a value in the range between 0 to 1, depending on the proximity to the Gaussian center. This score reflects the probability that the photometry for the object is consistent with a given isochrone. We associated masses from the isochrone with their respective scores for each age, constructing a probability-weighted distribution of masses and ages separately. Figure\,\ref{fig:CMD_tracks} presents the H-\texttt{F1140C} CMD, showing the isochrones spanning ages from 5 to 800 Myr for two evolution models (ATMO and Sonora). The colored circles represent field objects as described in Fig.\,\ref{fig:CMD_index}.

We derived the most probable age and mass for $\kappa$\,And\,b for each evolutionary track. Since we also have the mass-age relation (each age associated with a score and mass), we calculated the age that corresponds to the derived mass (age as a function of mass). To ensure the robustness of our method, we validated it using VHS\,1256\,b, a well-characterized benchmark object. A detailed description of the procedure and results for VHS\,1256\,b is provided in Appendix\,\ref{Apx:cmd-tracks}. The derived properties for $\kappa$\,And\,b are summarized in Table\,\ref{table:cmd-iso} for each evolutionary model and our final estimates for cloudy and cloud-free families. The cloud-free family has a final estimate in mass and age of $15.7\pm2.3\,\mathrm{M_{Jup}}$ and $50\pm8$\,Myr, while the cloudy family $18.1\pm4.8\,\mathrm{M_{Jup}}$ and $50\pm13$\,Myr, respectively. Both families align well with the ranges reported in the literature for the age ($47^{+27}_{-40}$\,Myr \citealt{Jones+2016}) and mass (e.g., $13^{+12}_{-2}\mathrm{M_{Jup}}$, \citealt{Currie+2018}), further supporting the reliability of the derived parameters.

\begin{table}[h!]
\centering
\caption{Age and mass estimates for $\kappa$\,And\,b from the H-\texttt{F1140C} CMD and isochrones.}\label{table:cmd-iso}
\begin{tabular}{l c c c}
    \hline
    \hline
    Isochrone & Age & Mass & Age(mass) \\
              & [Myr] & [$\mathrm{M_{Jup}}$] & [Myr] \\
    \hline
    AMES-COND & $49\pm6$ & $17.0\pm3.4$ & $55\pm12$ \\
    ATMO-ceq & $42\pm3$ & $14.5\pm2.0$ & $45\pm6$ \\
    BT-Settl & $33\pm3$ & $14.5\pm2.0$ & $42\pm7$ \\
    Sonora+0.0 & $39\pm4$ & $15.3\pm2.8$ & $40\pm8$ \\
    AMES-DUSTY & $78\pm16$ & $24.5\pm3.7$ & $68\pm7$ \\
    \hline
    \hline
    cloud-free & $45\pm5$ & $15.7\pm2.3$ & $50\pm8$ \\
    cloudy   & $50\pm21$ & $18.1\pm4.8$ & $50\pm13$ \\
    \hline
    \hline
\end{tabular}
\tablefoot{ The age and mass were calculated separately using the weighted distribution from the CMD (see text). The age (mass) refers to the age calculated from the mass-age relation given the estimated mass.
}
\end{table}

\subsection{Atmospheric characterization of $\kappa$\,And\,b}\label{sec:atmosp}

\begin{figure}
\centering
\includegraphics[width=8.5cm]{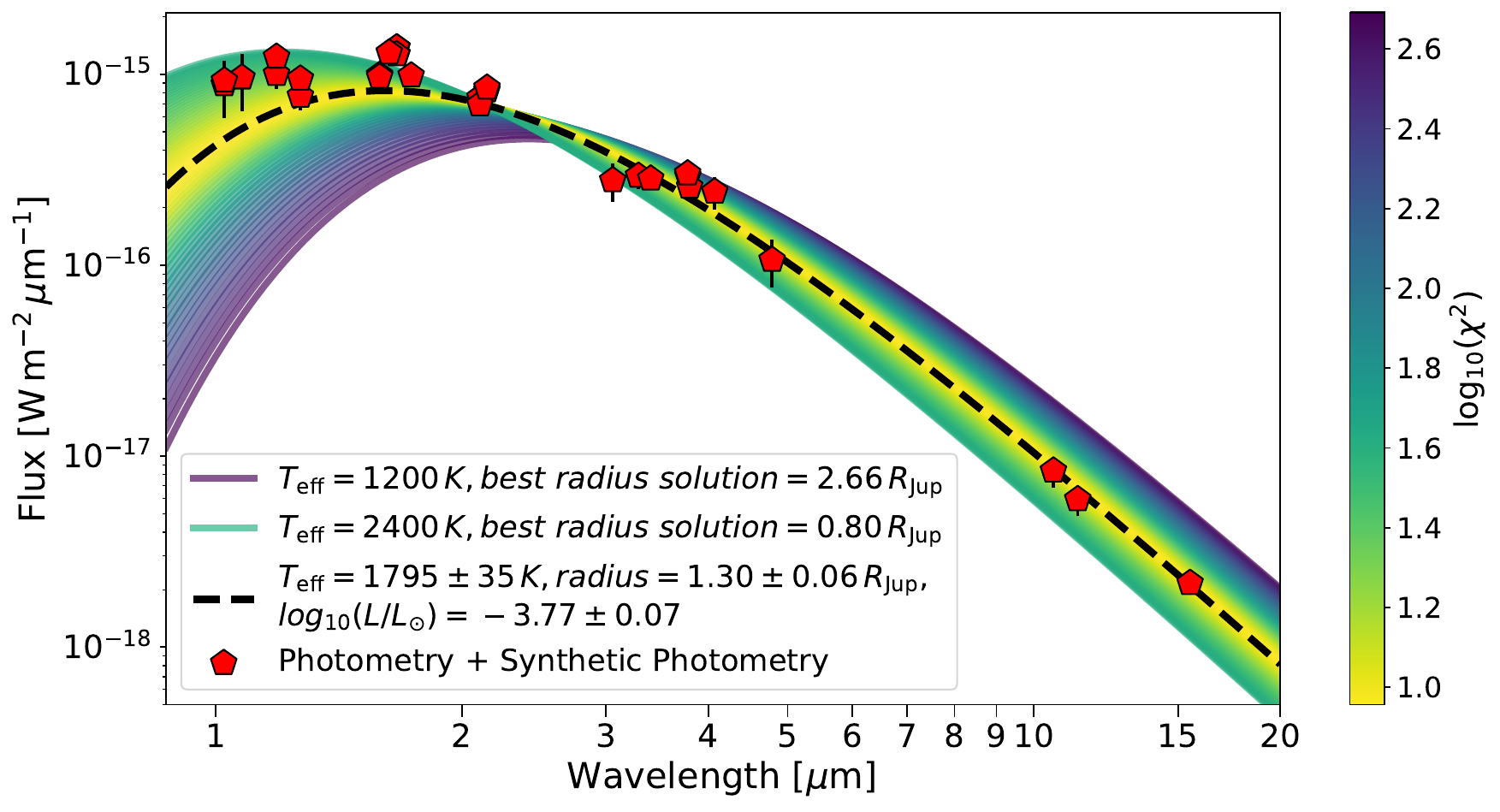}
\caption{SED of $\kappa$\,And\,b and the blackbody fitting results. The red circles correspond to observed photometry and calculated photometry from spectra observations (synthetic photometry). The colored lines correspond to the black body fit with temperatures from 1200\,K to 2400\,K (fixed) and the respective best-fit radius. The color gradient corresponds to the goodness-of-fit. The dashed black line corresponds to the best-fit model (temperature and radius as free parameters).}
\label{fig:black_body}%
\end{figure}

Our MIRI measurements cover the Rayleigh-Jeans tail of the companion SED, providing for the first time a complete coverage of its bolometric emission. We thus started our analysis with a simple black body fit, which allows us to provide first-order constraints on its temperature and radius. Figure\,\ref{fig:black_body} shows the best-fit results in the dashed black line, and the range of $\mathrm{T_{eff}}$-radius and the respective $\chi^2$ values in color-scale lines. The best parameters are $\mathrm{T_{eff}}$=$1795\pm35$\,K, radius=$1.30\pm0.06\,\mathrm{R_{Jup}}$, and log($\mathrm{L/L_{\odot}}$)=$-3.77\pm0.07$. These values are in agreement with the values in the literature (see Table\,\ref{table:summary_properties}). For these fits, we used a reduced number of photometric and synthetic photometric data points compared to the following more complex atmospheric model fits (see text below), but with adequate wavelength coverage to ensure a reliable fit. We also fit the black body excluding the MIRI data, obtaining a $\mathrm{T_{eff}}$=$1949\pm78$\,K, radius=$1.21\pm0.08\,\mathrm{R_{Jup}}$, and log($\mathrm{L/L_{\odot}}$)=$-3.70\pm0.06$.

We then employed a variety of atmospheric models to gain a comprehensive understanding of the main characteristics of $\kappa$\,And\,b. These included ATMO in chemical equilibrium (ATMO-ceq, \citealt{Phillips+2020}, \citealt{Petrus+2023}), AMES-COND (\citealt{Allard+2001}), AMES-DUSTY (\citealt{Allard+2001}), BT-COND (\citealt{Allard+2012}), BT-DUSTY (\citealt{Allard+2012}), BT-Settl, DRIFT-PHOENIX (\citealt{Helling+2008}), and EXO-REM (\citealt{Charnay+2018})), configured for the cloudy (resolutions of 500 and 20\,000)\footnote{Although generated with the same code and parameters, the two EXO-REM cloudy models show small intrinsic differences ($\sim$0.2\%) before and after resolution degradation, possibly due to differences in the numerical processes requiring more or less detailed atmospheric modeling.} and cloud-free (resolution of $500$) conditions. Each model incorporates different physical assumptions and computational approaches, which are crucial to consider when comparing the best-fit results. Below, we briefly summarize the key features of each model.

The ATMO-ceq model simulates planetary atmospheres under the assumption of chemical equilibrium, where gas-phase abundances are determined by thermochemical equilibria without accounting for cloud formation. In contrast, EXO-REM is a versatile model that accommodates the cloud-free and cloud-inclusive configurations, offering low- and high-resolution setups suitable for a range of exoplanetary environments. The AMES-COND and AMES-DUSTY models both focus on cloud formation but differ in their treatment of particulate matter: AMES-COND models cloud condensates, while AMES-DUSTY incorporates the role of dust grains, which significantly affect radiative transfer and energy balance. Similarly, BT-COND models atmospheric cloud condensates in radiative-convective equilibrium for brown dwarfs, while BT-DUSTY explicitly accounts for the scattering and absorption properties of dust particles. The BT-Settl model builds upon BT-COND by incorporating dynamic cloud-settling processes for a more realistic representation of cloud and dust behavior over time. Finally, PHOENIX, widely used for stellar and substellar objects, provides detailed radiative transfer calculations.

These models can be broadly categorized as cloud-free (ATMO-ceq, EXO-REM without clouds, BT-COND, AMES-COND) and cloudy (AMES-DUSTY, BT-DUSTY, BT-Settl, EXO-REM with clouds, and PHOENIX). From previous studies, we expect that the cloudy family fit better the SED of $\kappa$\,And\,b (e.g., \citealt{Bonnefoy+2014}; \citealt{Stone+2020}; \citealt{Uyama+2020}).

We performed atmospheric fitting considering various data combinations and scenarios. Our dataset includes extensive NIR observations using different setups (narrow- and broad-band filters, IFS, and low- and medium-resolution spectra). These were grouped into different subsets to identify the best-fit models based on the lowest $\chi^2$ values. For the OSIRIS spectrum, we reduced the resolution to $\sim$50 to have the same order of magnitude resolution as the other IFS data (R of$\sim$20-70). The groups were all data combined, IFS + MIRI, photometry (NIR + MIR) alone, photometry (NIR + MIR) + synthetic photometry from IFS, IFS with priors from photometric (NIR + MIR) fitting, and IFS with priors from photometric (NIR + MIR) + synthetic photometry. 
Each atmospheric model was fitted to each data subset, resulting in 60 atmospheric fits. In addition, we did the same exercise, excluding the MIRI observations, to highlight the impact of the MIR data. We applied two critical considerations during the fitting process as listed below.

\begin{itemize}
\item Wavelength weighting: Data were weighted across wavelengths to prevent biases due to regions with an over-density of observations in the NIR. This is, weighting by the number of observations in a given wavelength bin or band. Note that the standard fitting procedure assigns equal weight to all data points, ensuring that IFS observations do not carry more weight than photometric ones, but does not prevent over-density data in wavelength.
\item Parameter priors: Gaussian priors were applied to log(g) ($N(4.5, 0.25)$), effective temperature ($N(1750, 350)$ K), and radius ($N(1.4, 0.15)$ , $\mathrm{R_{Jup}}$). These priors were informed by results from the black body fit and ensured that solutions remained physically plausible and did not converge to the edge of the parameter grid.
\end{itemize}

\begin{figure*}
\centering
\includegraphics[width=8.7cm]{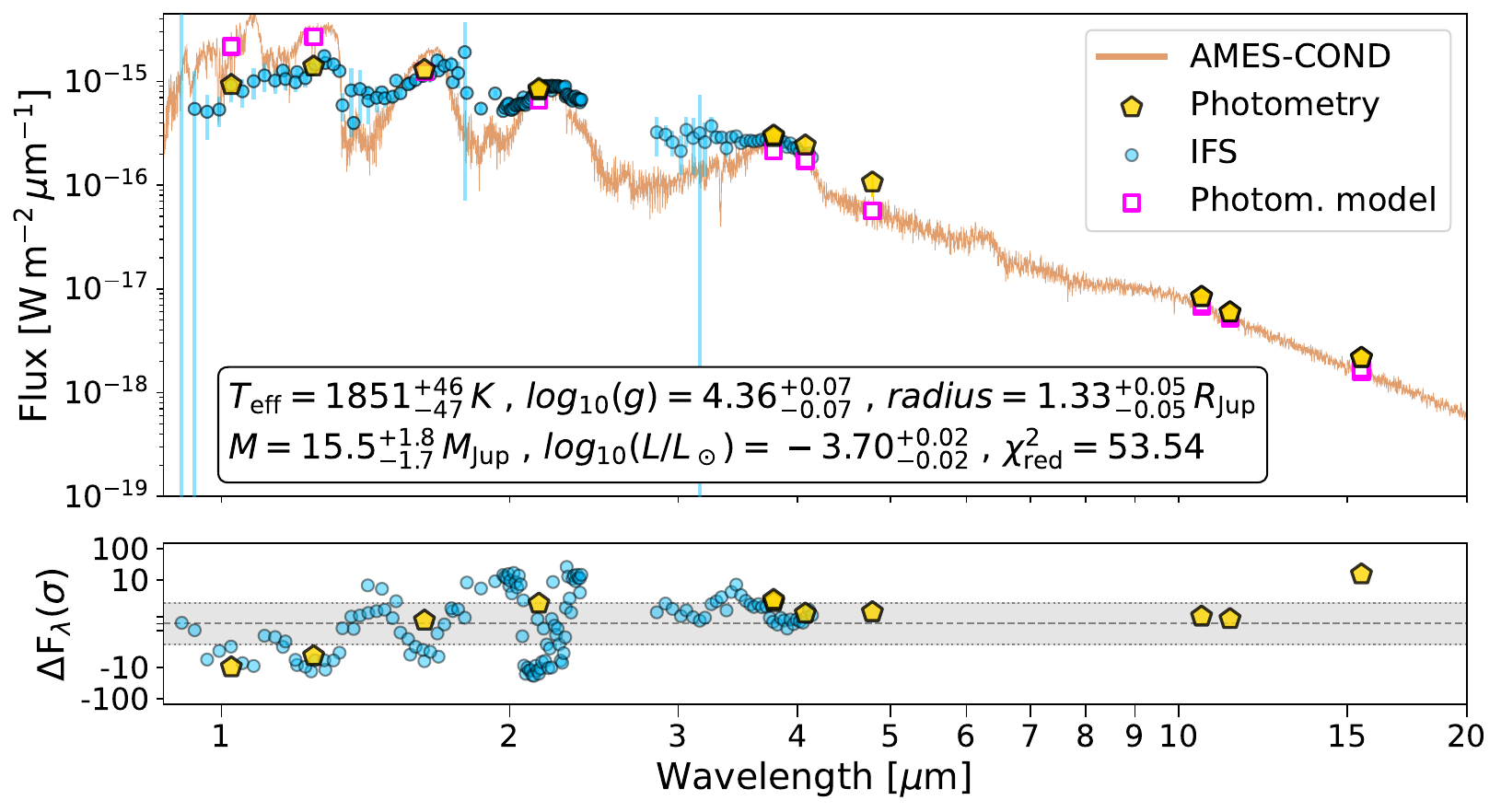}
\includegraphics[width=8.7cm]{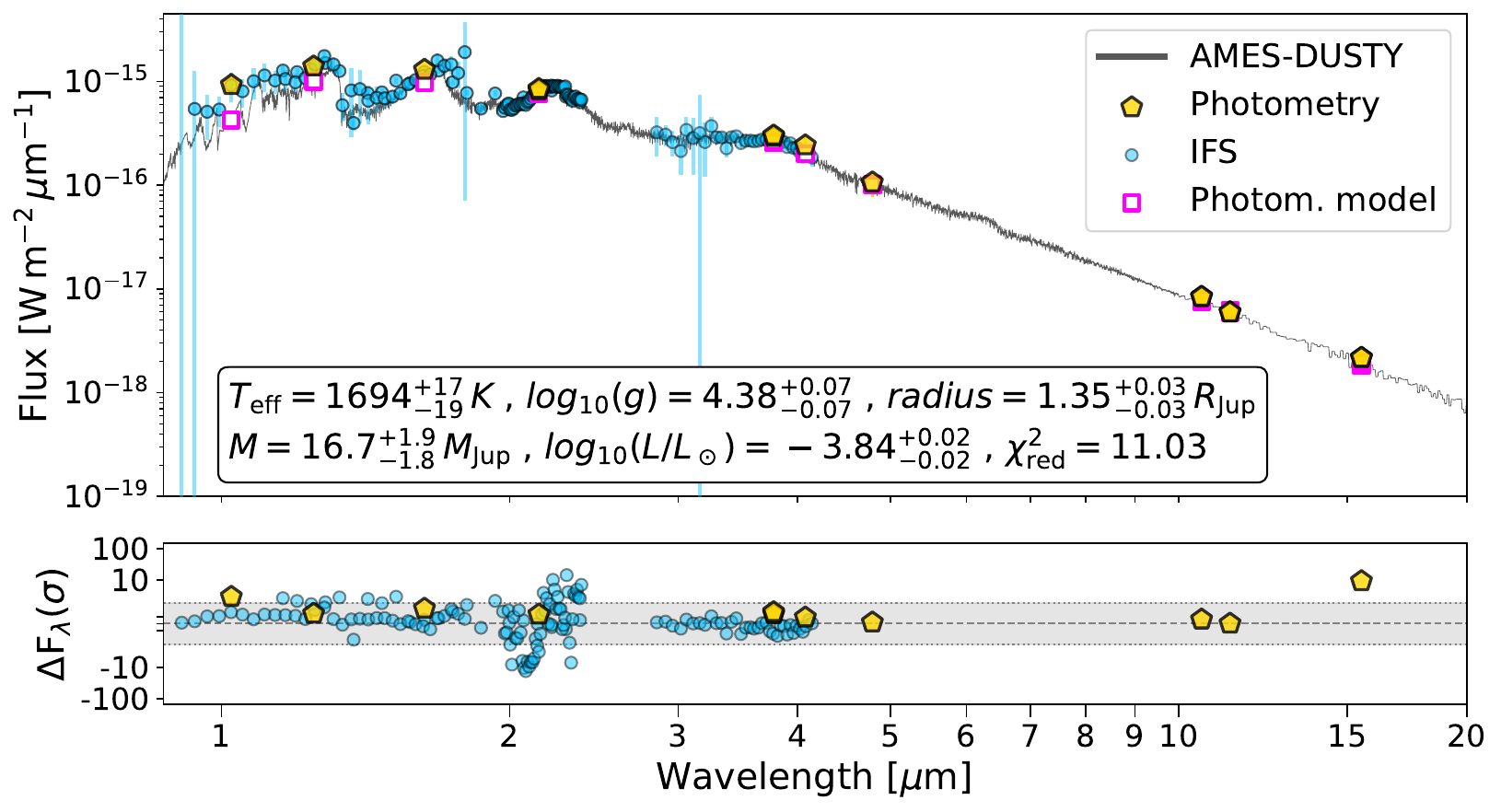}
\includegraphics[width=8.7cm]{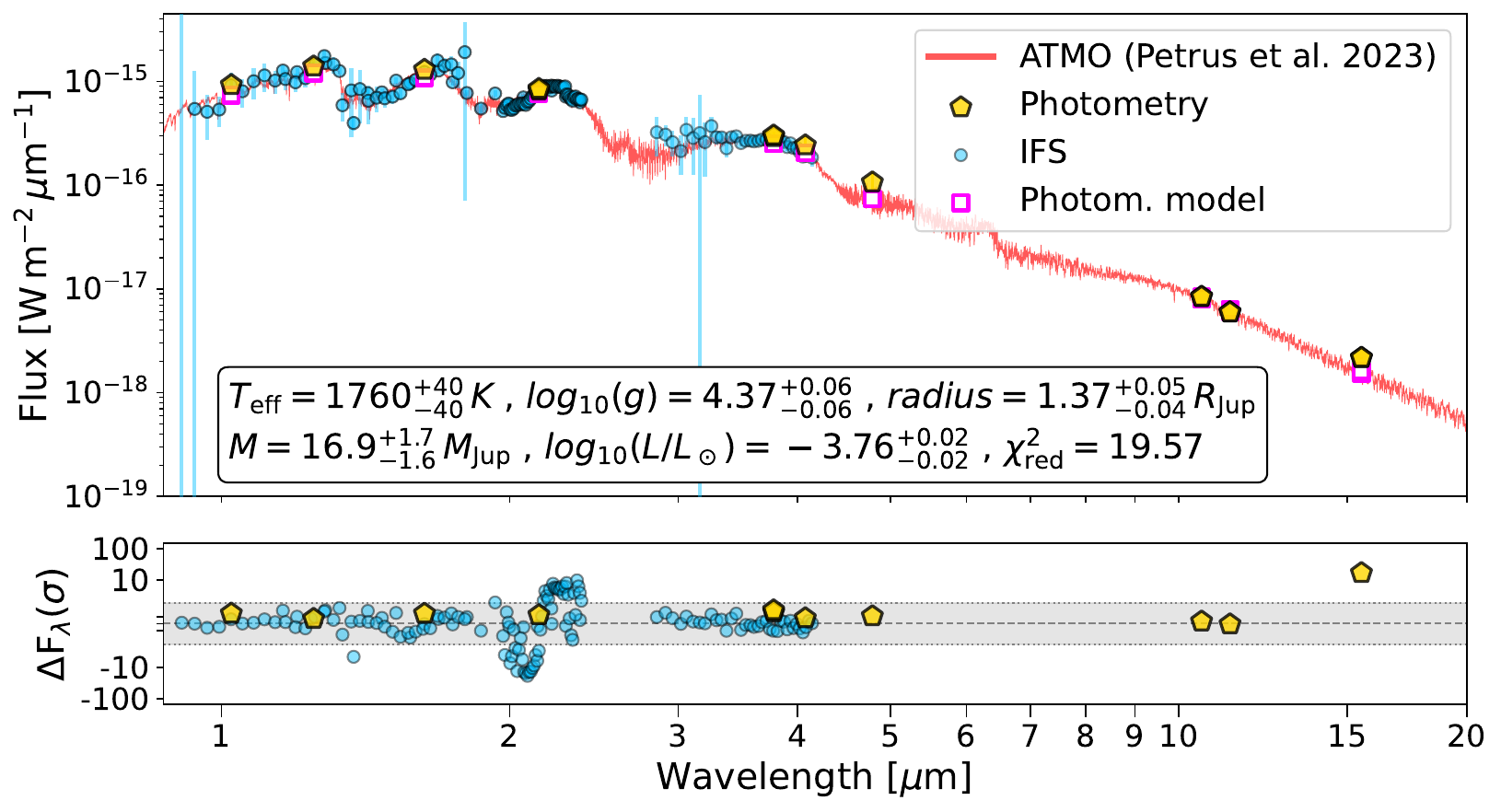}
\includegraphics[width=8.7cm]{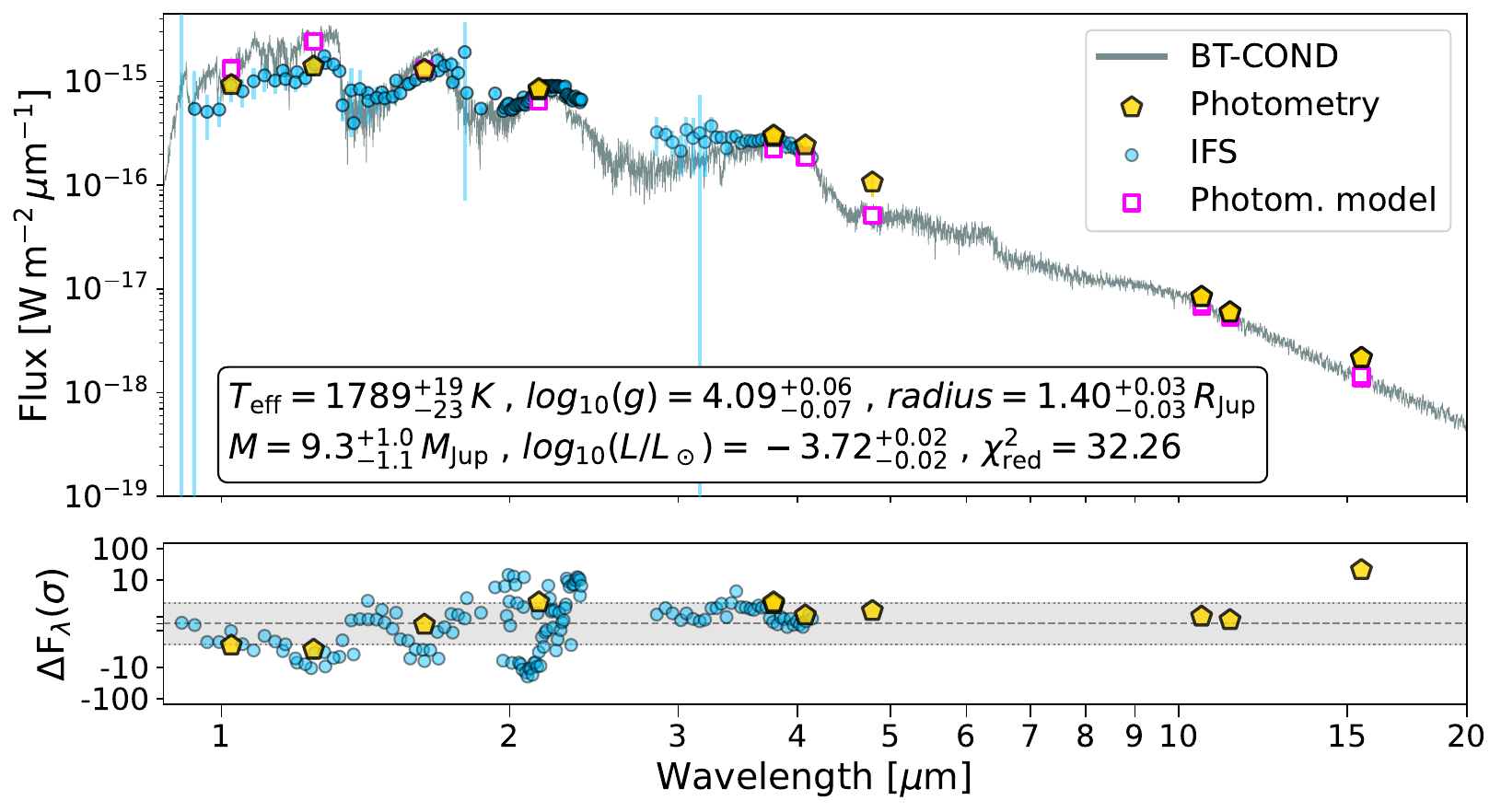}
\includegraphics[width=8.7cm]{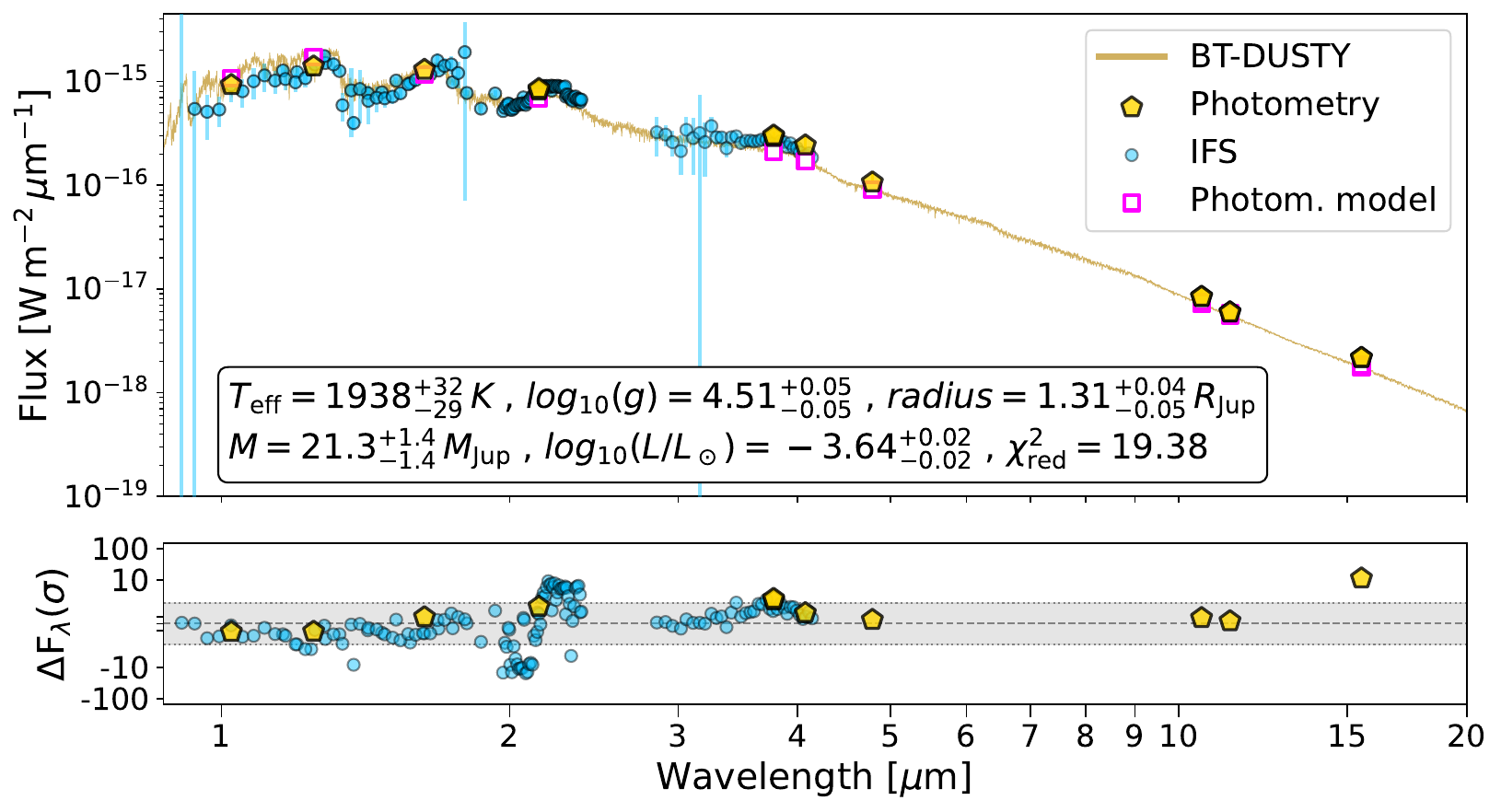}
\includegraphics[width=8.7cm]{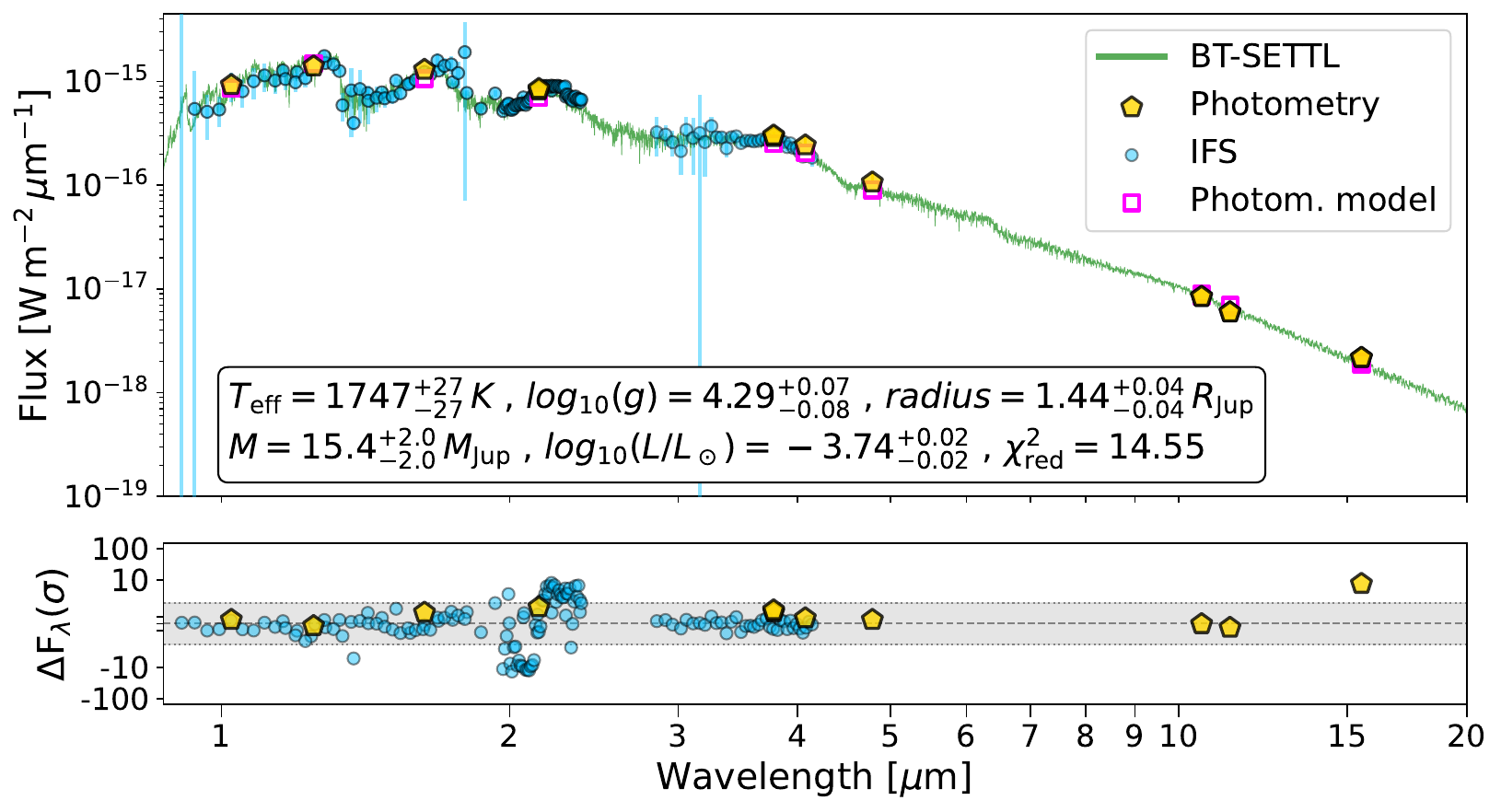}
\includegraphics[width=8.7cm]{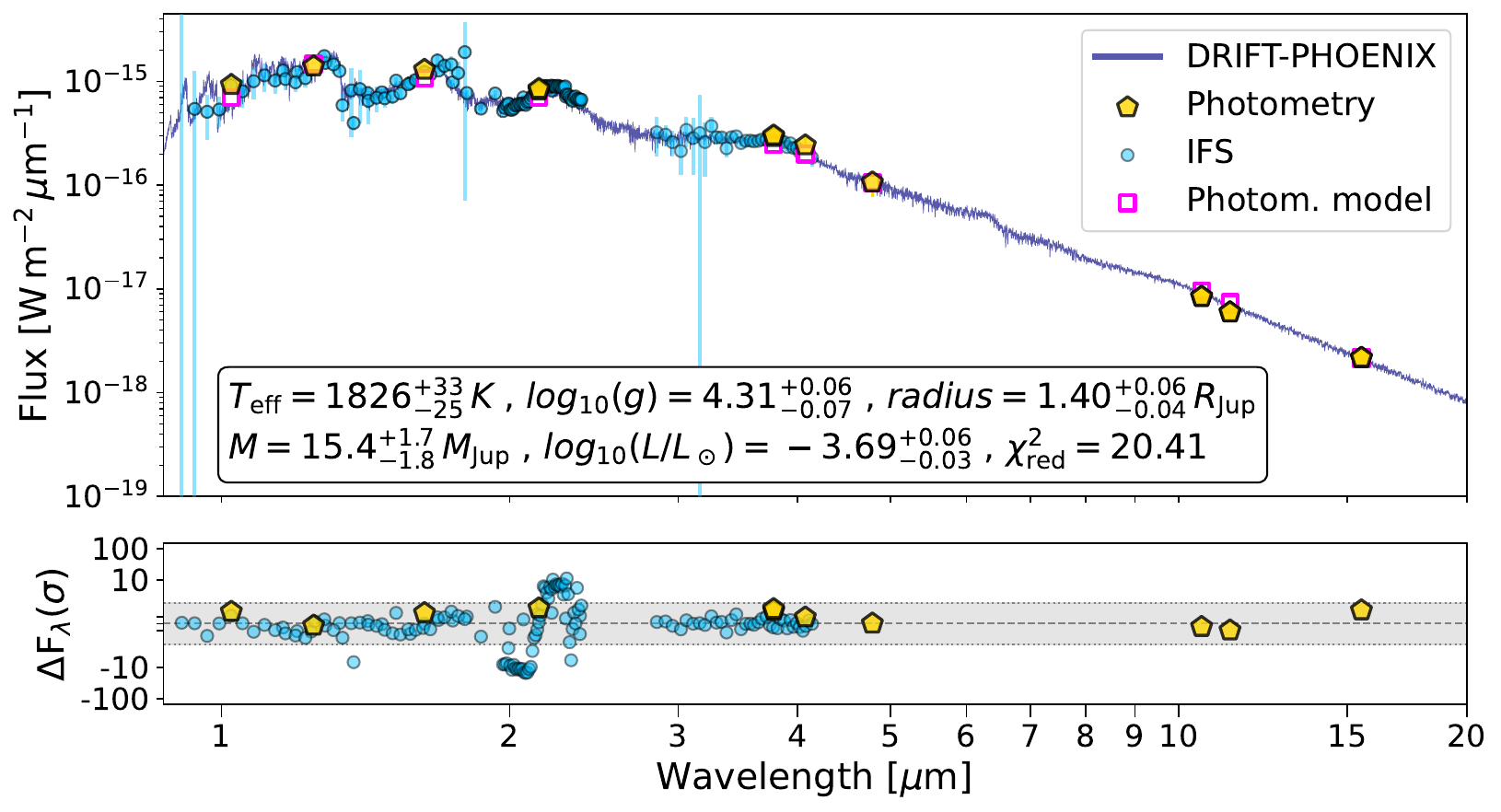}
\includegraphics[width=8.7cm]{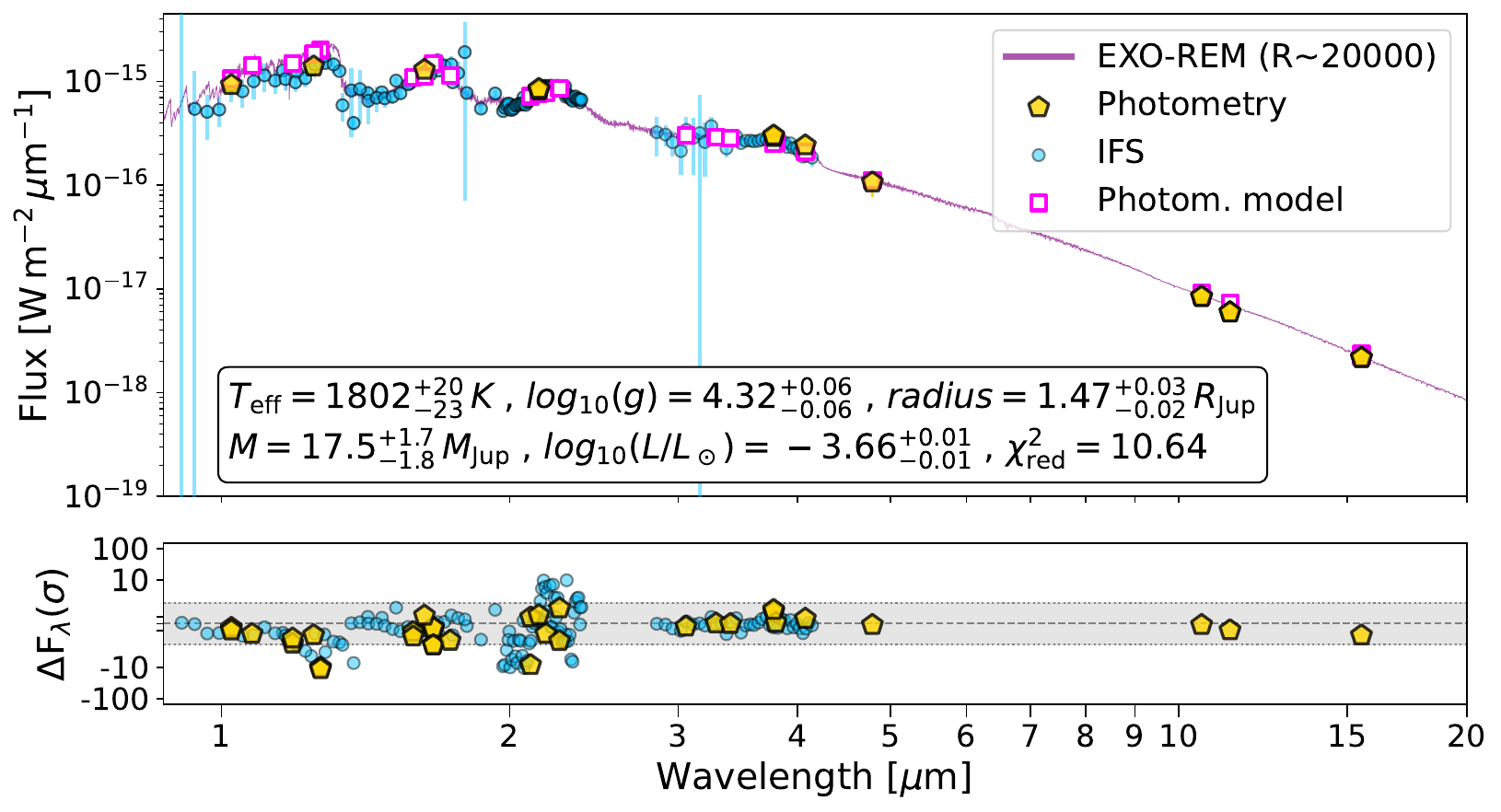}
\includegraphics[width=8.7cm]{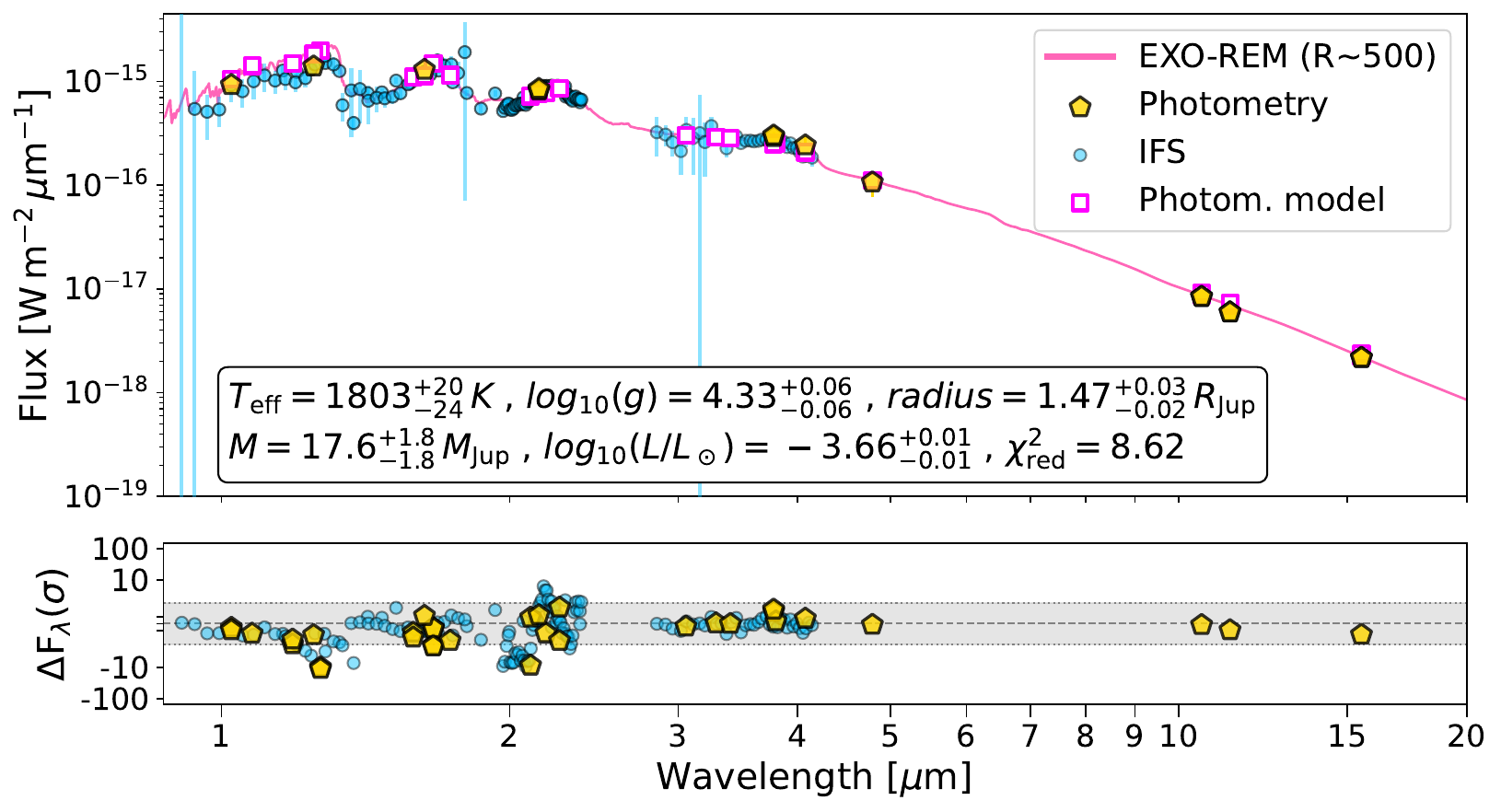}
\includegraphics[width=8.7cm]{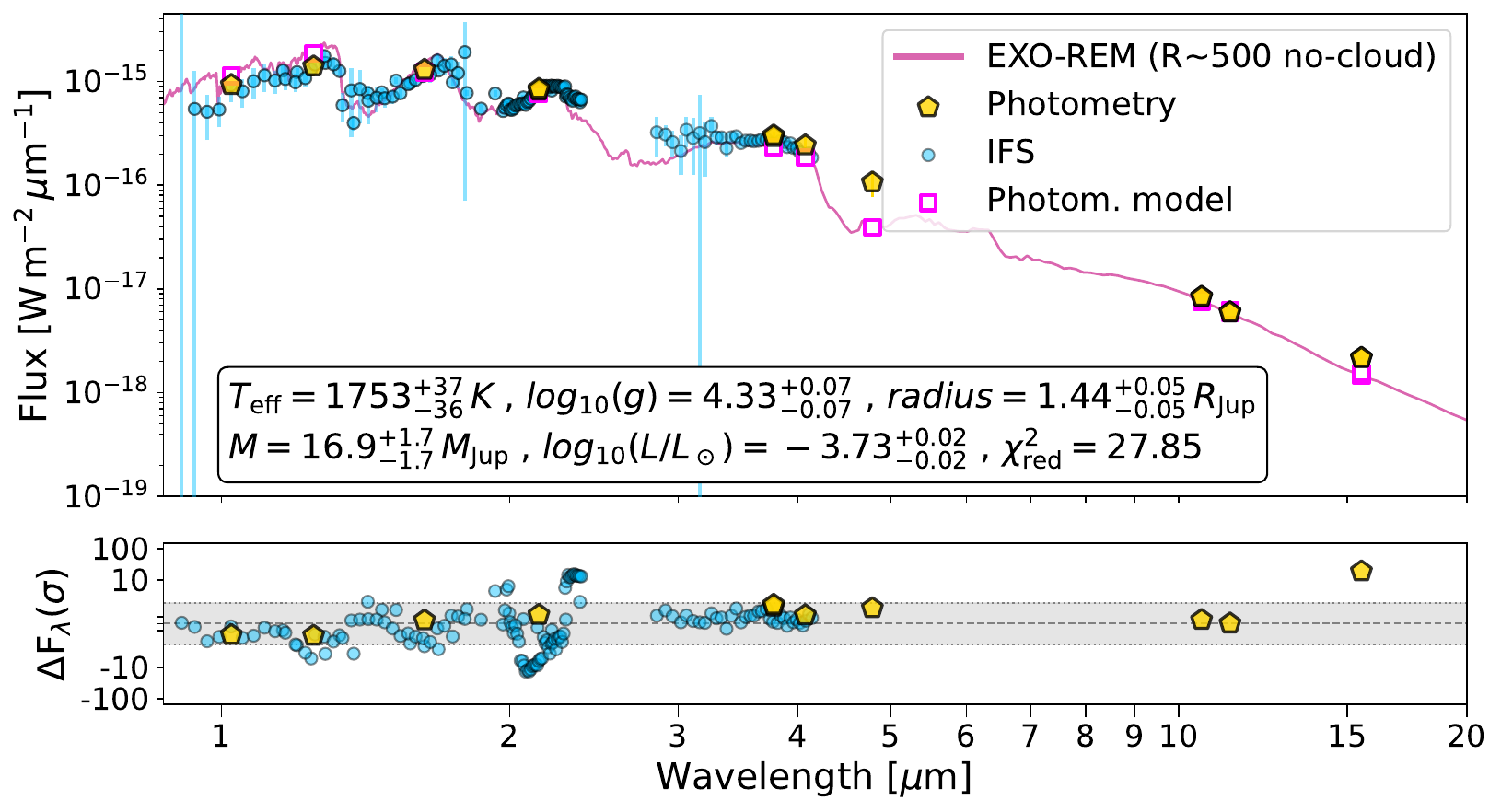}
\caption{
Best-fit results from different atmospheric models. Each panel shows the best-fit model from a different combination of datasets, selected from all possible combinations. Each panel shows the photometric data (yellow pentagons), IFS (cyan circles), photometry from atmospheric model (magenta squares), and the best-fit model (colored line), and the residuals are shown in the bottom panel. Table\,\ref{table:summary_properties_v2} shows the data subset that corresponds to the best fit for each atmospheric model.}
\label{fig:best_atmosph_models_black}%
\end{figure*}

We selected the best-fit models based on constraints requiring log(g) $>3.9$, radius $>1.0\,\mathrm{R_{Jup}}$, and temperatures $> 1650$\,K. We choose these restrictions based on previous estimates and on the L0/L2 spectral types properties. To account for known modeling biases due to post-processing choices (\citealt{Nasedkin+2023}), we adopt a 1\% systematic uncertainty on all parameters. This level is consistent with typical $1\sigma$ biases reported by \cite{Nasedkin+2023}, and ensures realistic error estimates, especially in cases where the fit fails to converge properly (e.g., underestimated errors in cloud-free models). The added uncertainty remains conservative without overcoming the formal statistical errors. Figure\,\ref{fig:best_atmosph_models_black} shows the best-fit model for the data-subset with the best results for each atmospheric model considered. Cloudy models provided significantly better fits to the data, while cloud-free models exhibited larger discrepancies, particularly in the NIR. The EXO-REM model yielded the lowest $\chi^2$ values. 

We note that when excluding MIRI observations, the model fitting converges, in general, to slightly higher effective temperatures, similar surface gravities, and smaller radii, regardless of the atmospheric family used. This suggests that without MIR constraints, models tend to over-estimate the effective temperature while compensating with lower radii to match the observed luminosity. Conversely, the MIRI data allow us to remove the degeneracy between these parameters. Regarding the improvement in uncertainties when using MIRI observations, we obtained for the cloud-free family an improvement of $6\%$ for $\mathrm{T_{eff}}$, $2\%$ for log(g), $6\%$ for radius, and $3\%$ for the spectroscopic mass, with no improvement for the luminosity. For the cloudy family, the improvements were $22\%$ in $\mathrm{T_{eff}}$, $9\%$ in log(g), $33\%$ in radius, $1\%$ in the spectroscopic mass, and $35\%$ in luminosity.

We also tested including a circumplanetary blackbody disk component in the cloudy and cloud-free families fits. Cloud-free models are compatible with a disk flux contribution ranging from $0.2\%$ to $35\%$ across the MIRI bands, with flux-to-error ratios $>30$, yet still provide poor fits to the data. In contrast, cloudy models, especially the best-fit \texttt{EXO-REM}, indicate negligible disk contributions ($<0.02\%$) with low significance ($<1$). Since the best-fitting models do not require a disk to explain the data, we conclude that any circumplanetary disk contribution is not statistically significant.

Table\,\ref{table:summary_properties_v2} shows the results for the cloudy and cloud-free models in terms of effective temperature, surface gravity, radius, spectroscopic mass, luminosity, and the respective $\chi^2$. We selected the data combination with the best $\chi^2_{\mathrm{red}}$. The table also shows the results without considering the MIRI data. We also calculated the weighted mean for the cloudy and cloud-free models for each parameter (fit including MIRI data) considering the natural dispersion in the uncertainties.

\begin{table*}[]
\centering
\caption{Atmospheric and physical properties of $\kappa$\,And\,b.}\label{table:summary_properties_v2}
\begin{tabular}{l c c c c c c c} 
\hline \hline
   Atmospheric model   & $\mathrm{T_{eff}}$ & log(g)   & Radius               &  Mass                  & $\mathrm{log_{10}}(L/L_{\odot})$ & $\chi^2_{red}$ & Data\\
       & [Kelvin]           &  [dex]   & [$\mathrm{R_{Jup}}$] &  [$\mathrm{M_{Jup}}$]  &  [dex]                       &  & combination\tablefootmark{*}  \\
\hline
\\[-1.8ex]
\multicolumn{7}{c}{\textbf{with MIRI data}} \\
\\[-1.8ex]
\hline 
\\[-1.7ex]

\\[-1.7ex]

ATMO    &  $1760^{+40}_{-40}$ & $4.37^{+0.06}_{-0.06}$ & $1.37^{+0.05}_{-0.04}$ & $16.9^{+1.7}_{-1.6}$ & $-3.76^{+0.02}_{-0.02}$ & 19.57 & (IP) \\
\\[-1.7ex]
AMES-COND  & $1851^{+46}_{-47}$ & $4.36^{+0.07}_{-0.07}$ & $1.33^{+0.05}_{-0.05}$ & $15.5^{+1.8}_{-1.7}$ & $-3.70^{+0.02}_{-0.02}$ & 53.54 & (IP) \\
\\[-1.7ex]
BT-COND & $1789^{+19}_{-23}$ & $4.09^{+0.06}_{-0.07}$ & $1.40^{+0.03}_{-0.03}$ & $9.3^{+1.0}_{-1.1}$ & $-3.72^{+0.02}_{-0.02}$ & 32.26 & (IPSP)  \\
\\[-1.7ex]
EXO-REM ($\mathrm{R_{500}}$ - no-cloud) & $1756^{+37}_{-36}$ & $4.33^{+0.07}_{-0.07}$ & $1.44^{+0.05}_{-0.05}$ & $16.9^{+1.7}_{-1.7}$ & $-3.73^{+0.02}_{-0.02}$ & 27.85 & (IP)  \\
\\[-1.7ex]
\cline{2-8}
\\[-1.7ex]
BT-DUSTY & $1938^{+32}_{-29}$ & $4.51^{+0.05}_{-0.05}$ & $1.31^{+0.04}_{-0.05}$ & $21.3^{+1.4}_{-1.4}$ & $-3.64^{+0.02}_{-0.02}$ & 19.38 & (IPSP)  \\
\\[-1.7ex]
AMES-DUSTY & $1694^{+17}_{-19}$ & $4.38^{+0.07}_{-0.07}$ & $1.35^{+0.03}_{-0.03}$ & $16.7^{+1.9}_{-1.8}$ & $-3.84^{+0.02}_{-0.02}$ & 11.03 & (IPSP)  \\
\\[-1.7ex]
BT-Settl & $1747^{+27}_{-27}$ & $4.29^{+0.07}_{-0.08}$ & $1.44^{+0.04}_{-0.04}$ & $15.4^{+2.0}_{-2.0}$ & $-3.74^{+0.02}_{-0.02}$ & 14.55 & (IPSP) \\
\\[-1.7ex]
DRIFT-PHOENIX & $1826^{+33}_{-25}$ & $4.31^{+0.06}_{-0.07}$ & $1.40^{+0.06}_{-0.04}$ & $15.4^{+1.7}_{-1.8}$ & $-3.69^{+0.06}_{-0.03}$ & 20.41 & (IPSP)  \\
\\[-1.7ex]
EXO-REM ($\mathrm{R_{20000}}$) & $1802^{+20}_{-23}$ & $4.32^{+0.06}_{-0.06}$ & $1.47^{+0.03}_{-0.02}$ & $17.5^{+1.7}_{-1.8}$ & $-3.66^{+0.01}_{-0.01}$ & 10.64 & (PSP)  \\
\\[-1.7ex]
EXO-REM ($\mathrm{R_{500}}$) & $1803^{+20}_{-24}$ & $4.33^{+0.06}_{-0.06}$ & $1.47^{+0.03}_{-0.02}$ & $17.6^{+1.8}_{-1.8}$ & $-3.66^{+0.01}_{-0.01}$ & \textbf{8.62} & (PSP)  \\
\\[-1.7ex]
\hline
\\[-1.8ex]
\multicolumn{7}{c}{\textbf{without MIRI data}} \\
\\[-1.8ex]
\hline 
\\[-1.7ex]

\\[-1.7ex]

ATMO &  $1725^{+37}_{-36}$ & $4.34^{+0.07}_{-0.07}$ & $1.40^{+0.04}_{-0.04}$ & $16.4^{+1.7}_{-1.7}$ & $-3.78^{+0.01}_{-0.01}$ & 13.69 & (PSP) \\
\\[-1.7ex]
AMES-COND & $1982^{+71}_{-76}$ & $4.42^{+0.10}_{-0.10}$ & $1.18^{+0.08}_{-0.07}$ & $14.2^{+2.2}_{-2.2}$ & $-3.69^{+0.02}_{-0.02}$ & 43.66 & (IPSP) \\
\\[-1.7ex]
BT-COND & $1918^{+31}_{-23}$ & $4.27^{+0.05}_{-0.06}$ & $1.24^{+0.03}_{-0.04}$ & $10.8^{+0.8}_{-1.1}$ & $-3.70^{+0.02}_{-0.02}$ & 27.34 & (IPSP) \\
\\[-1.7ex]
EXO-REM ($\mathrm{R_{500}}$ - no-cloud) & $1709^{+39}_{-30}$ & $4.29^{+0.06}_{-0.07}$ & $1.51^{+0.04}_{-0.05}$ & $17.2^{+1.7}_{-1.7}$ & $-3.73^{+0.01}_{-0.01}$ & 17.46 & (PSP) \\
\\[-1.7ex]
\cline{2-8}
\\[-1.7ex]
BT-DUSTY & $1940^{+33}_{-29}$ & $4.51^{+0.05}_{-0.05}$ & $1.30^{+0.05}_{-0.05}$ & $21.1^{+1.4}_{-1.5}$ & $-3.64^{+0.02}_{-0.02}$ & 18.95 & (IPSP)  \\
\\[-1.7ex]
AMES-DUSTY & $1805^{+72}_{-57}$ & $4.40^{+0.07}_{-0.08}$ & $1.29^{+0.04}_{-0.04}$ & $15.9^{+2.0}_{-2.0}$ & $-3.77^{+0.05}_{-0.04}$ & 10.51 & (IPSP) \\
\\[-1.7ex]
BT-Settl & $1746^{+28}_{-27}$ & $4.29^{+0.07}_{-0.08}$ & $1.44^{+0.04}_{-0.04}$ & $15.4^{+1.9}_{-2.0}$ & $-3.74^{+0.02}_{-0.02}$ & 14.45 & (IPSP) \\
\\[-1.7ex]
DRIFT-PHOENIX & $1910^{+25}_{-20}$ & $4.29^{+0.10}_{-0.08}$ & $1.41^{+0.12}_{-0.13}$ & $15.3^{+2.0}_{-1.9}$ & $-3.59^{+0.06}_{-0.09}$ & 18.75 & (IPSP) \\
\\[-1.7ex]
EXO-REM ($\mathrm{R_{20000}}$) & $1822^{+29}_{-41}$ & $4.35^{+0.06}_{-0.07}$ & $1.42^{+0.06}_{-0.04}$ & $17.2^{+1.7}_{-1.6}$ & $-3.68^{+0.03}_{-0.03}$ & 15.78 & (AD) \\
\\[-1.7ex]
EXO-REM ($\mathrm{R_{500}}$) & $1803^{+43}_{-42}$ & $4.35^{+0.07}_{-0.07}$ & $1.41^{+0.06}_{-0.05}$ & $17.1^{+1.7}_{-1.6}$ & $-3.70^{+0.02}_{-0.02}$ & 12.14 & (IF) \\

\\[-1.8ex]
\hline
\hline
\\[-1.8ex]
Cloud-free models & $1778\pm37$ & $4.24\pm0.14$ & $1.39\pm0.04$ & $15.0\pm3.2$ & $-3.73\pm0.02$ & -- & (WMC) \\
\\[-1.7ex]
Cloudy models\tablefootmark{**}  & $1791\pm69$ & $4.35\pm0.07$ & $1.42\pm0.06$ & $17.3\pm1.8$ & $-3.71\pm0.07$ & -- & (WMC) \\
\\[-1.7ex]
\hline
\hline
\end{tabular}
\tablefoot{ Mass refers to spectroscopic mass derived from log(g) and radius. (PSP): Photometry plus synthetic photometry. (IPSP): IFS plus prior using photometry plus synthetic photometry. (IP): IFS plus prior using photometry alone. (IF): IFS no prior. (AD): using the entire dataset. (WMC): Weighted-mean combined values only for the results including MIRI observations. 
\tablefoottext{*}{Here, data combination refers to the best fits between all the subgroups of datasets (see text for more details).}
\tablefoottext{**}{We considered these values as our final estimates.}
}
\end{table*}

\subsection{Evolutionary tracks and physical parameters from atmospheric modeling}\label{sec:atm-tracks}

We used the results from atmospheric models combined with evolutionary tracks to assess whether the derived properties of $\kappa$\,And\,b align with theoretical predictions. Four evolutionary tracks were selected: ATMO-ceq, Saumon-2008 (\citealt{Saumon+2008}) Hybrid, Saumon-2008 No-Clouds, and Sonora solar metalicity (\citealt{Marley+2021-son}). These models can be grouped into two families based on cloud treatment: cloudy (Saumon Hybrid and Sonora solar) and cloud-free (ATMO-ceq and Saumon No-Clouds). The atmospheric models were consistently paired with their respective evolutionary track families for analysis.

\begin{figure*}
\centering
\sidecaption
\includegraphics[width=7.5cm]{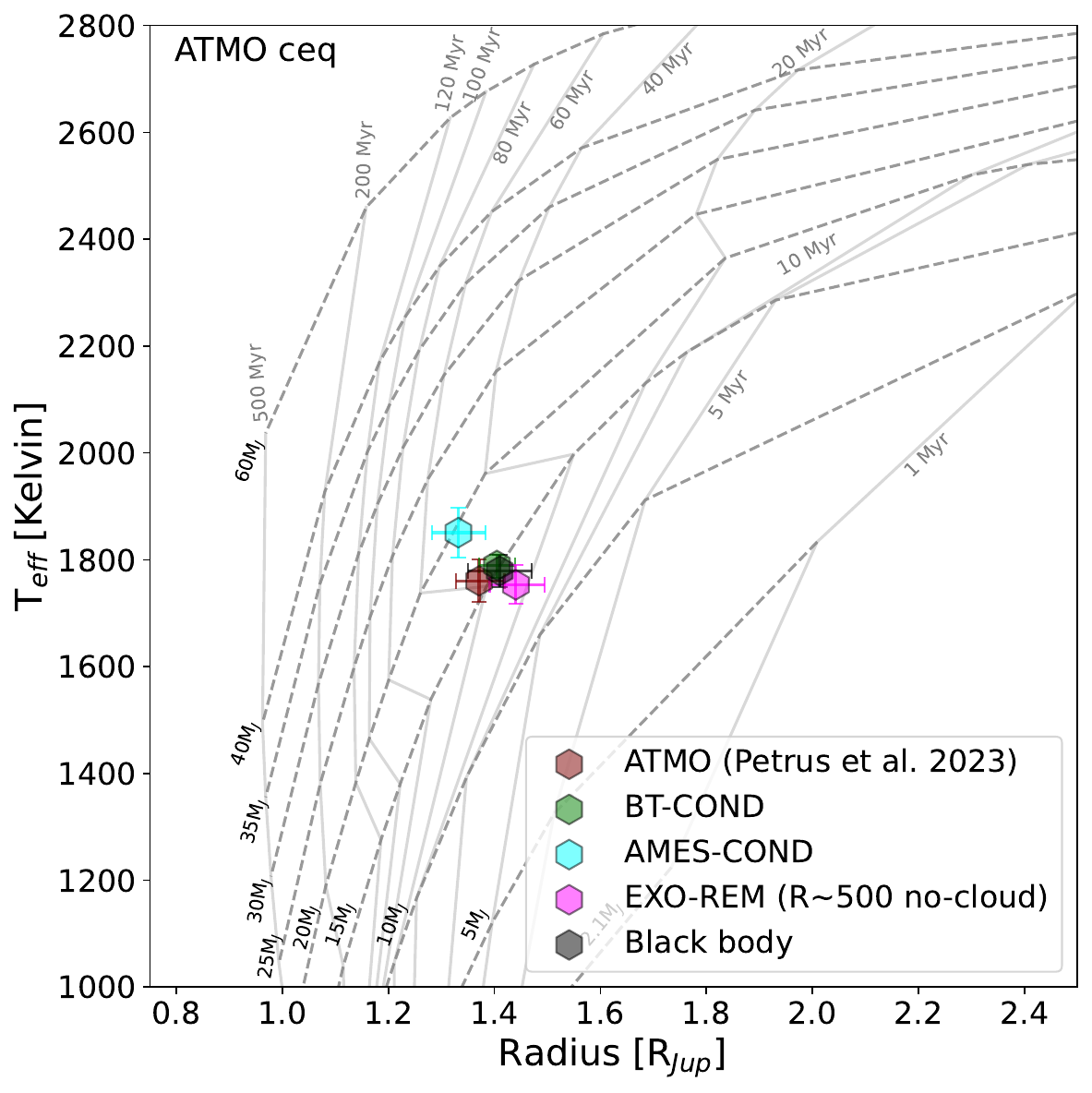}
\includegraphics[width=7.5cm]{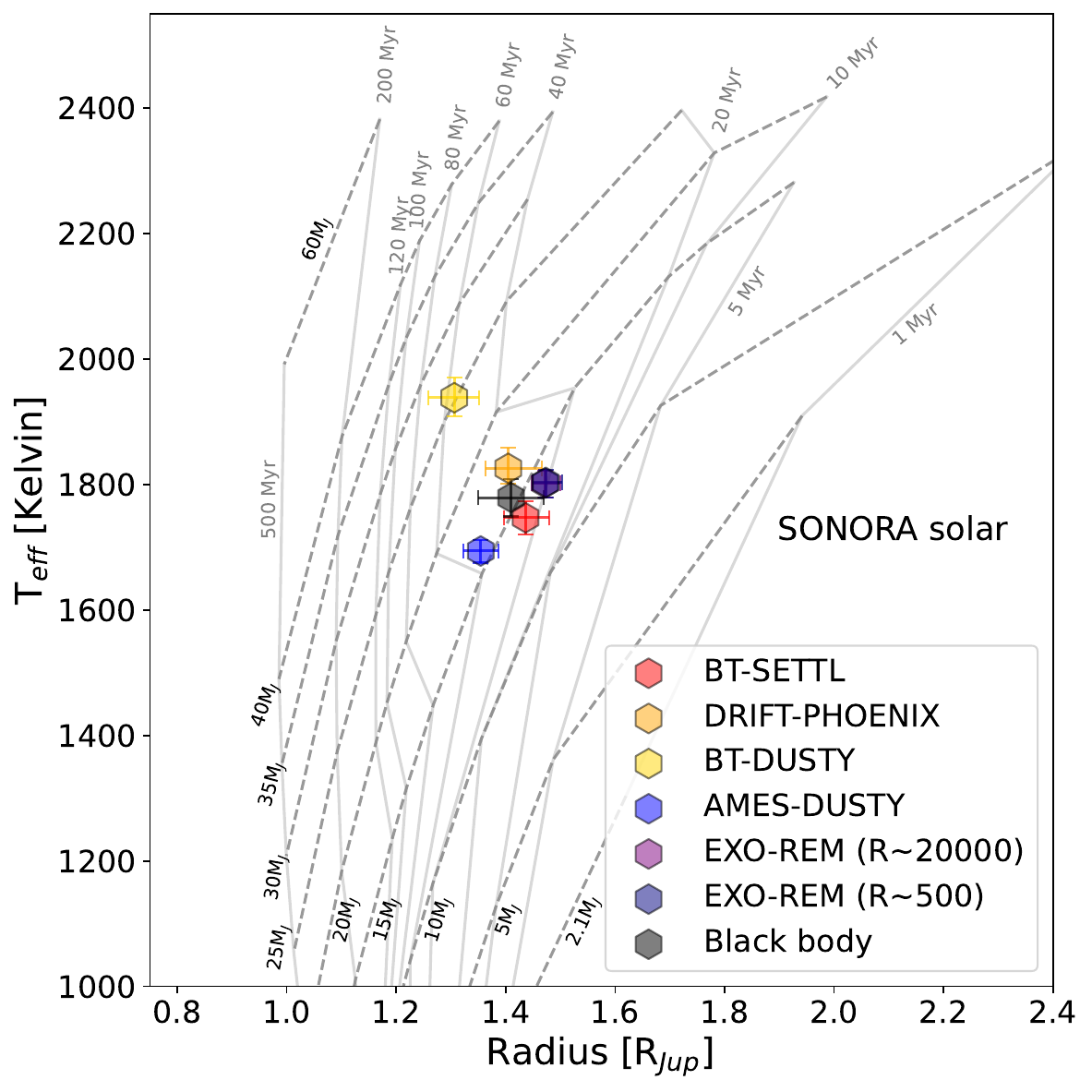}
\includegraphics[width=7.5cm]{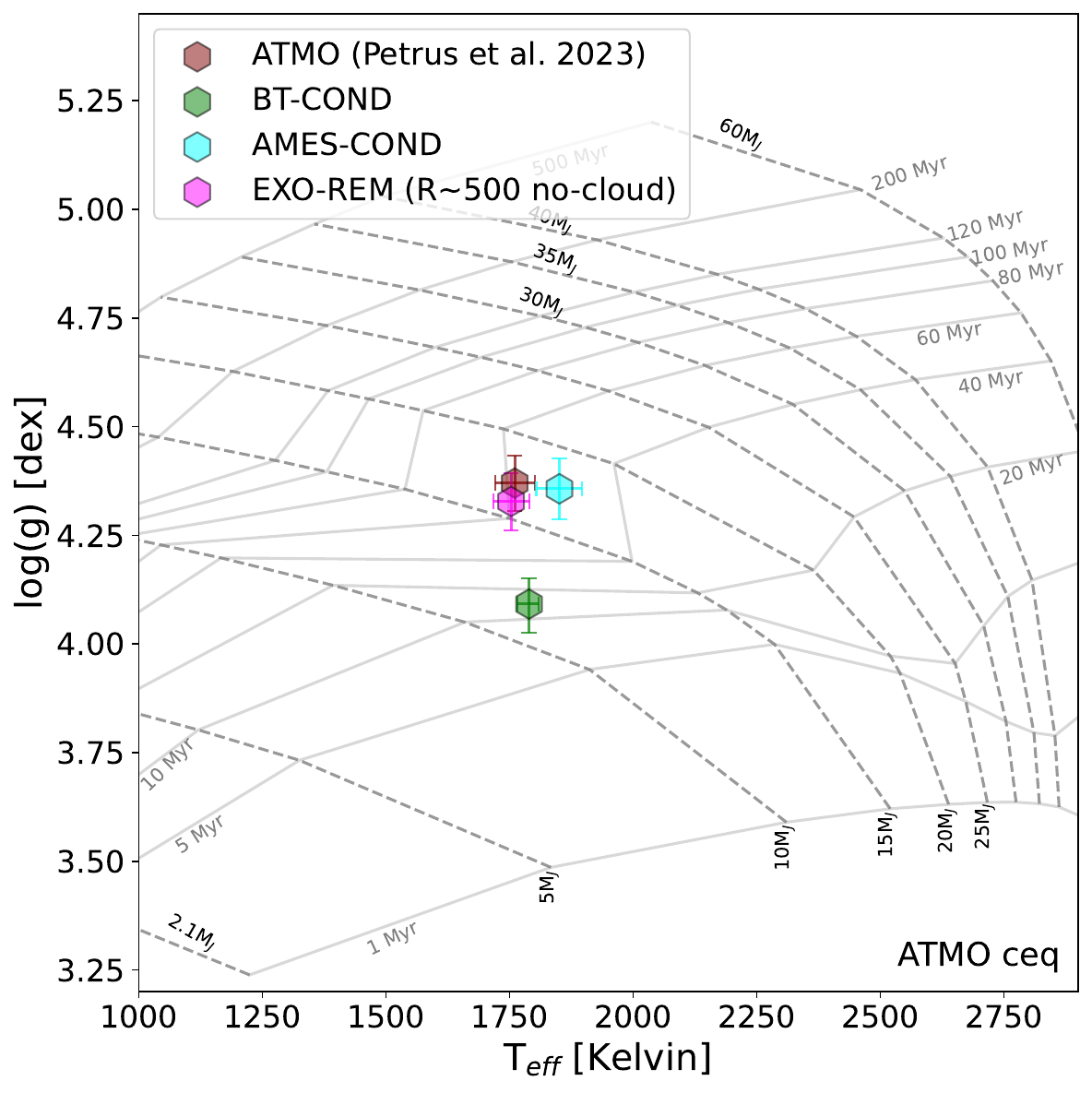}
\includegraphics[width=7.5cm]{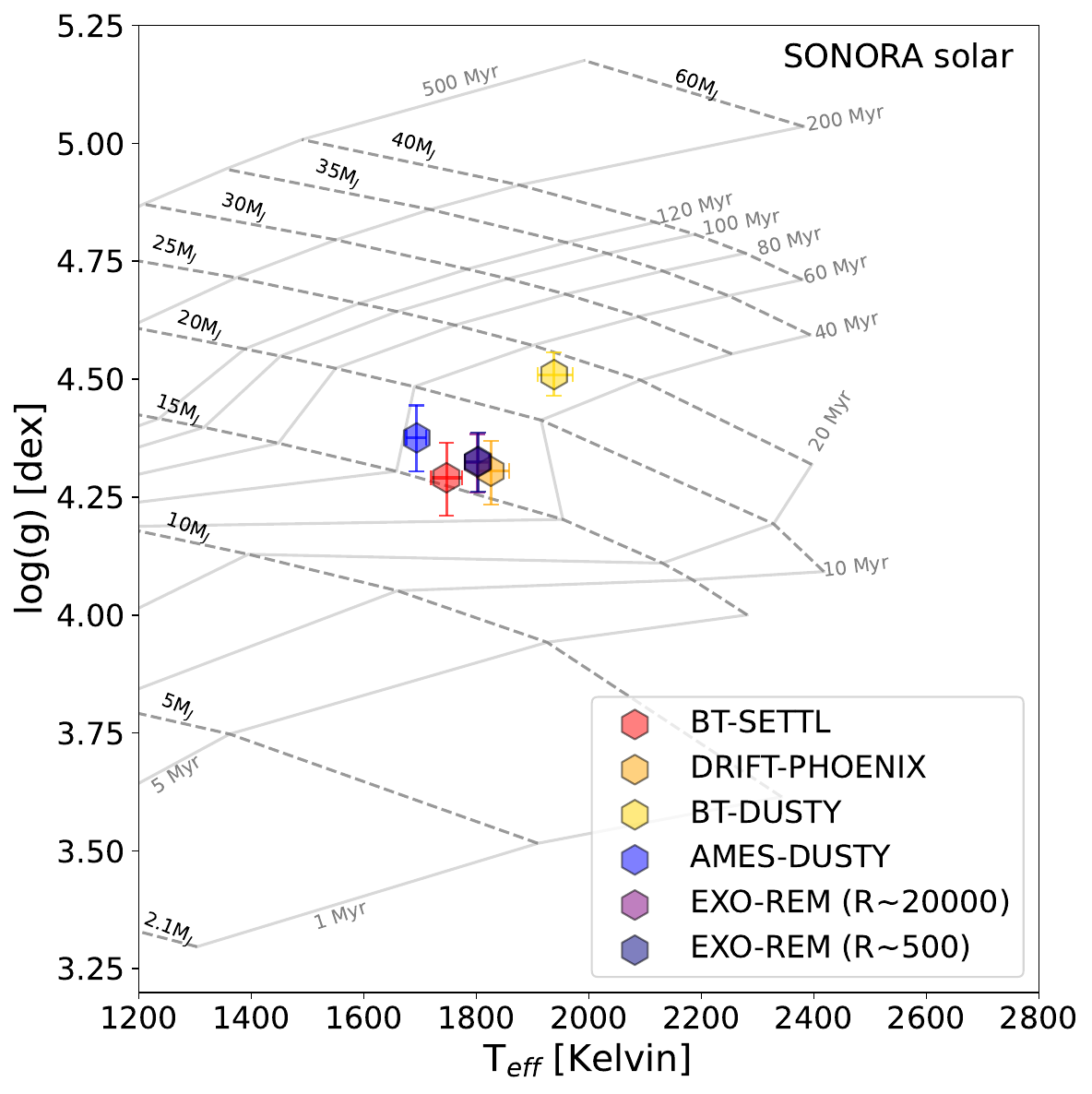}
\caption{Evolutionary tracks showing the age (continuum gray lines) and the mass (dashed dark gray lines) for a cloudy (Sonora with solar metallicity, right panels), and a cloud-free model (ATMO chemical equilibrium, left panels). \textit{Top:} Evolutionary tracks using the radius and effective temperature. \textit{Bottom:} Evolutionary tracks using effective temperature and log(g). The differently colored pentagons correspond to the values obtained from the atmospheric models. }
\label{fig:evol_tracks}%
\end{figure*}

Using the effective temperature, surface gravity, and radius values listed in Table\,\ref{table:summary_properties_v2}, we derived the mass and age of $\kappa$\,And\,b by interpolating within the evolutionary track grids. The interpolation was performed for three parameter combinations: $\mathrm{T_{eff}-log(g)}$, $\mathrm{T_{eff}-radius}$, and $\mathrm{radius-log(g)}$. Figure\,\ref{fig:evol_tracks} illustrates the evolutionary tracks for the ATMO-ceq and Sonora solar models, showing the grid combinations for $\mathrm{T_{eff}-radius}$ and $\mathrm{T_{eff}-log(g)}$.

We computed weighted averages of the derived mass and age values from the three-parameter grids for each evolutionary track family, using the $\chi^2$ values from the atmospheric model fits as weights. We also consider the natural dispersion of these measurements in the uncertainties. Table\,\ref{table:evol_tracks} presents the combined estimates for each evolutionary track and the final values for the cloudy and cloud-free families.

\begin{table}[h!]
\centering
\caption{Age and mass estimates of $\kappa$\,And\,b using the evolutionary tracks and the atmospheric model fit results.}\label{table:evol_tracks}
\begin{tabular}{|c|cc|cc|cc|}
    \hline
    \hline
    & \multicolumn{6}{c|}{Grid Track} \\ 
    \hline
    Evolut. & \multicolumn{2}{c|}{$\mathrm{T_{eff}}$-$\mathrm{log(g)}$} & \multicolumn{2}{c|}{$\mathrm{T_{eff}}$-$\mathrm{radius}$} & \multicolumn{2}{c|}{$\mathrm{radius}$-$\mathrm{log(g)}$} \\ 
    Track  & age & mass & age & mass & age & mass \\  
    \hline
    ATMO & $60^{+16}_{-16}$ & $16^{+3}_{-3}$ & $59^{+16}_{-16}$ & $14^{+3}_{-3}$ & $51^{+14}_{-15}$ & $17^{+5}_{-6}$ \\ 
    S08-nc & $58^{+14}_{-14}$ & $19^{+3}_{-4}$ & $57^{+14}_{-14}$ & $17^{+3}_{-3}$ & $53^{+15}_{-16}$ & $19^{+4}_{-4}$ \\ 
    S08-Hy & $44^{+5}_{-6}$ & $19^{+3}_{-3}$ & $49^{+12}_{-12}$ & $20^{+5}_{-5}$ & $56^{+12}_{-12}$ & $18^{+2}_{-3}$ \\ 
    SA+0.0 & $49^{+12}_{-11}$ & $17^{+1}_{-1}$ & $44^{+12}_{-12}$ & $16^{+2}_{-2}$ & $40^{+8}_{-7}$ & $18^{+1}_{-1}$ \\ 
    \hline
    \hline
    cloud-free & $59^{+11}_{-11}$ & $17^{+2}_{-2}$ & $58^{+11}_{-11}$ & $16^{+2}_{-2}$ & $52^{+11}_{-11}$ & $18^{+3}_{-4}$ \\
    cloudy & $\mathbf{47^{+7}_{-7}}$\tablefootmark{a} & $17^{+2}_{-2}$ & $47^{+8}_{-8}$ & $17^{+3}_{-3}$ & $48^{+7}_{-7}$ & $19^{+2}_{-2}$ \\
    \hline
    \hline
\end{tabular}
\tablefoot{ Age in Myr and mass in $\mathrm{M_{Jup}}$. \textit{ATMO} to ATMO-ceq, \textit{S08-nc} refers to Saumon 2008 no-clouds, \textit{S08-Hy} to Saumon 2008 Hybrid, and \textit{SA+0.0} to Sonora-solar.
\tablefoottext{a}{This corresponds to the finally adopted age estimate.}
}
\end{table}

A key observation from Table\,\ref{table:summary_properties_v2} and Figure\,\ref{fig:evol_tracks} is that log(g) exhibits larger uncertainties than $\mathrm{T_{eff}}$ and radius, but this does not affect the uncertainties strongly because they are dominated by the data value dispersion in each grid (see Table\,\ref{table:evol_tracks}). We adopted the mass and age estimates with the lowest uncertainties, corresponding to the $\mathrm{T_{eff}-log(g)}$ grid, as our final results. 

Interestingly, the comparison between cloudy and cloud-free models revealed similar trends. The cloud-free models predict slightly older ages (between 51\,Myr and 60\,Myr), while the cloudy models suggest younger ages (from 40\,Myr to 56\,Myr). The masses are similar and within the uncertainties. The precision of the cloudy family is better, however, which highlights the importance of the cloud treatment in atmospheric and evolutionary modeling for substellar companions such as $\kappa$\,And\,b.

\subsection{Final derived properties of $\kappa$\,And\,b}\label{sec:prop}

Table\,\ref{table:summary_properties} presents the final estimates for the physical and atmospheric properties of $\kappa$\,And\,b, derived from a combination of CMDs with isochrones, atmospheric model fitting, and evolutionary tracks. We provide the results for the cloudy and cloud-free models, summarizing the findings into distinct categories for clarity. Additionally, we compare these estimates to values from the literature.

\begin{figure*}
\centering
\includegraphics[width=9.1cm]{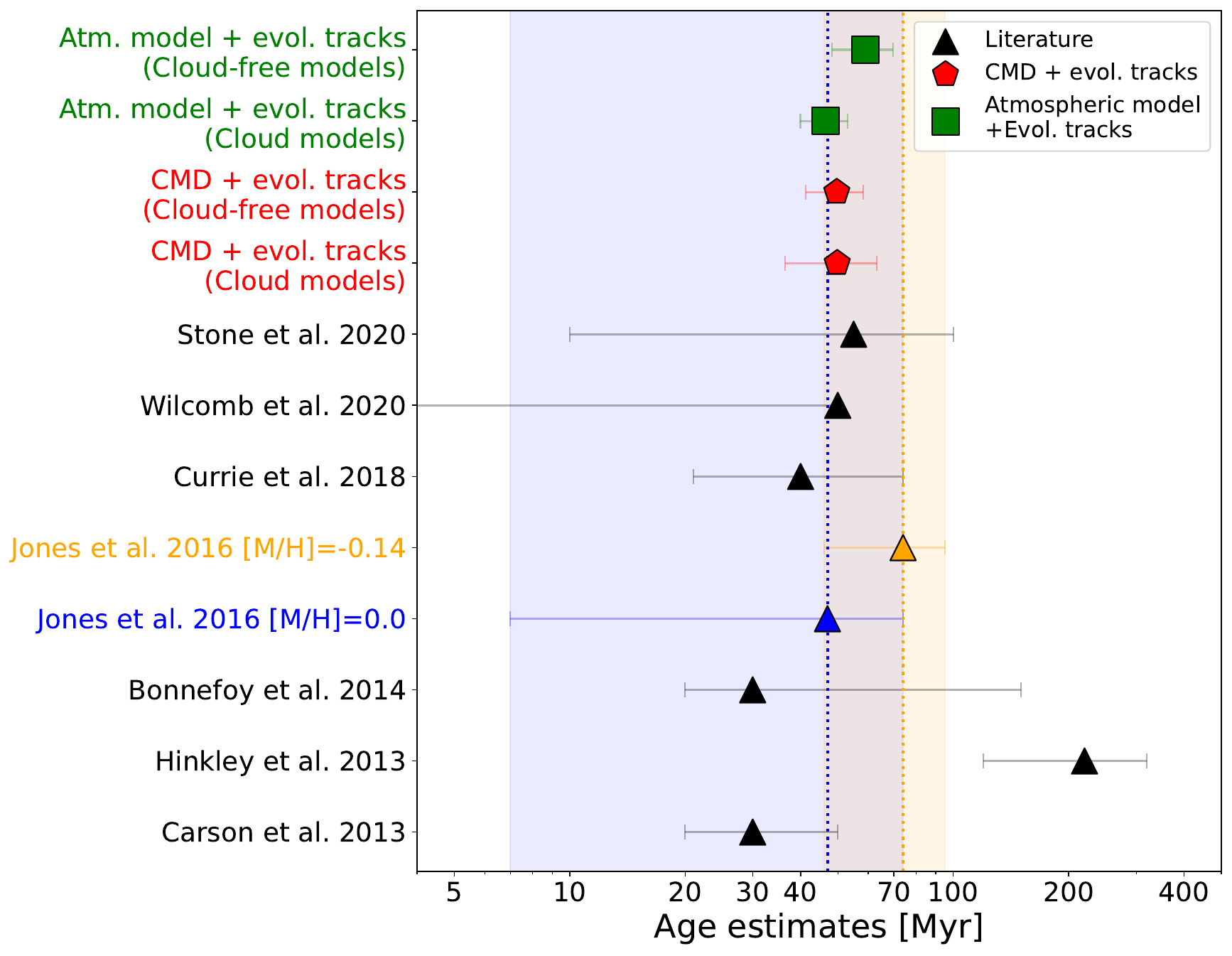}
\includegraphics[width=9.1cm]{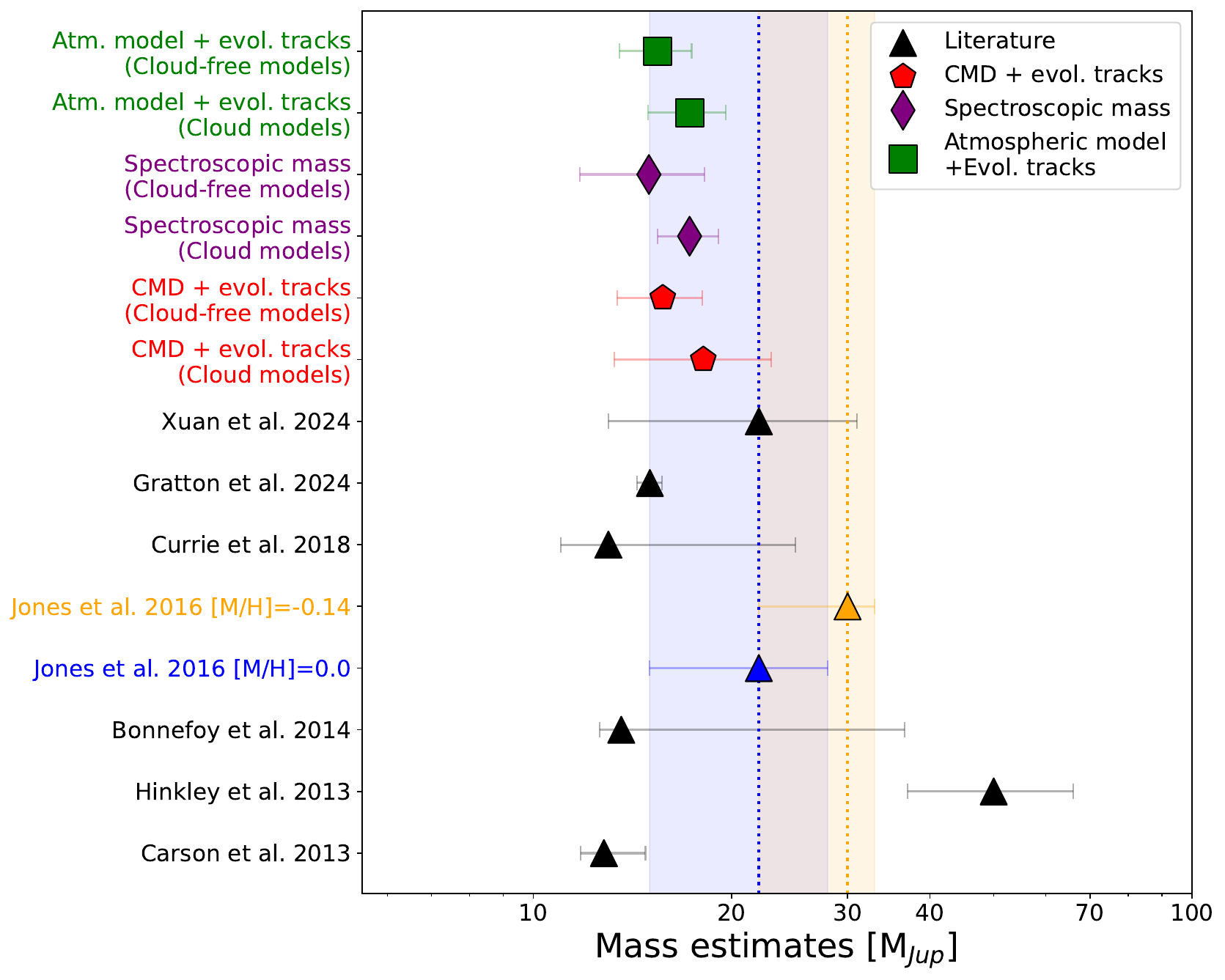}
\caption{Ages and masses estimated from literature and this work. \textit{Left:} Age estimates from the literature (black triangles), CMD plus isochrones (red pentagons), and evolutionary tracks plus atmospheric models (green squares). \textit{Right:} Mass estimates from the literature (black triangles), CMD plus isochrones (red pentagons), spectroscopic masses (purple diamond), and evolutionary tracks plus atmospheric models (green squares). For the estimates in this study, each method was sub-grouped into cloud and cloud-free models. The vertical dashed lines and filled area highlight the values obtained by \cite{Jones+2016} using solar (blue) and subsolar (orange) metallicities. }
\label{fig:final_values}%
\end{figure*}

To offer a more comprehensive view, Figure\,\ref{fig:final_values} plots the derived, model-dependent age and mass from this study alongside literature values, highlighting the range of possible solutions and the convergence or divergence with previous studies. This summary provides a foundation for further discussion in Section\,\ref{sec:dis}, where we compare our findings with those of previous studies and discuss their implications for the formation and evolution of $\kappa$\,And\,b.

\begin{table*}[]
\centering
\caption{Atmospheric and physical properties of $\kappa$\,And\,b.}\label{table:summary_properties}
\begin{tabular}{l c c c c c c} 
\hline \hline
   Data    & $\mathrm{T_{eff}}$ & log(g) & Radius             &  Mass                & $\mathrm{log_{10}}(L/L_{\odot})$ & Age\tablefootmark{*} \\
           & [K]                & [dex]  & [$\mathrm{R_{Jup}}$] &  [$\mathrm{M_{Jup}}$]  &  [dex]                       &  [Myr]  \\
\hline
\\[-1.4ex]
\multicolumn{7}{c}{\textbf{Literature}} \\
\\[-1.4ex]
\hline 

\cite{Carson+2013} & $1680^{+30}_{-20}$     &  --                     & --                  & $12.8^{+2.0}_{-1.0}$  & --      &  $30^{+20}_{-10}$ \\
\cite{Hinkley+2013}& $2040\pm60$            & $4.33^{+0.88}_{-0.79}$  & --                  & $50^{+16}_{-13}$      & --      & $220\pm100$ \\
\cite{Bonnefoy+2014} & $1900^{+100}_{-200}$ & $4.5\pm1.0$             & --                  & $14_{+25}^{-2}$ & $-3.76\pm0.06$ & $30^{+120}_{-10}$   \\
\cite{Jones+2016}    & $2040\pm 60$         &  --                     & --                  & $22^{+6}_{-7}$        & --        & $47^{+27}_{-40}$ \\
\cite{Todorov+2016}  & $1733\pm 467$        &  --                     & --                  & --                    & -- & -- \\
                     & $2050\pm 230$        &  --                     & --                  & --                    & -- & -- \\
\cite{Currie+2018}   & $1850\pm 150$\tablefootmark{a}     &  $4.5\pm0.5$            & --                  & $13^{+12}_{-2}$       & $-3.81\pm 0.06$ & $40^{+34}_{-19}$ \\
\cite{Hoch+2020}     & $ 2050\pm100$\tablefootmark{b}     &  $3.8^{+0.7}_{-0.3}$    & $1.2^{+0.3}_{-0.2}$ & --                    & $-3.8^{-0.1}_{+0.3}$\tablefootmark{c} & $<50$ \\
\cite{Stone+2020}    & $1850\pm 50$\tablefootmark{d}       &  $4.75\pm 0.25$\tablefootmark{e}       & --                  & --                    & $-3.735\pm 0.045$\tablefootmark{f} & $10-100$ \\
\cite{Uyama+2020}    & $1850\pm 50$\tablefootmark{g}      &  $4.25\pm 0.25$\tablefootmark{h}         & $1.45\pm 0.15$\tablefootmark{i}      & --                    & -- & $47^{+27}_{-40}$ \\
                     & $1700$               &  $4.00$                 & --                  & --                    & -- & -- \\
\cite{Morris+2024}   & $1700\pm 100$        &  $4.7\pm 0.5$           & --                  & --                    & -- & $47^{+27}_{-40}$ \\
\cite{Gratton+2024}  & $1776\pm61$          & --                      & --                  & $15.03\pm0.66$        & -- & $36\pm8$\tablefootmark{j} \\
\cite{Xuan+2024}     & $1680^{+60}_{-50}$\tablefootmark{k}   & --                      & $1.35\pm 0.25$\tablefootmark{l}      & $22\pm9$\tablefootmark{l}           & -- & -- \\
\hline
\hline
\\[-1.4ex]
\multicolumn{7}{c}{\textbf{This work}} \\
\\[-1.4ex]
\hline
\\[-1.7ex]
\multicolumn{7}{c}{Cloud-free family} \\

\\[-1.7ex]
\cline{2-7}
\\[-1.7ex]
Isochrone+CMD & -- & -- & -- & $15.7\pm2.3$ & -- & $50\pm8$ \\
Atmospheric models & $1778\pm37$ & $4.24\pm0.14$ & $1.39\pm0.04$ & $15.0\pm3.2$ & $-3.73\pm0.02$ & -- \\
Evol. tracks+AM & -- & -- & -- & $16\pm2$ & -- & $52\pm11$ \\
\hline
\\[-1.7ex]
\multicolumn{7}{c}{Cloudy family} \\

\\[-1.7ex]
\cline{2-7}
\\[-1.7ex]
Isochrone+CMD & -- & -- & -- & $18.1\pm4.8$ & -- & $50\pm13$ \\
Atmospheric models & $1791\pm69$ & $4.35\pm0.07$ & $1.42\pm0.06$ & $17.3\pm1.8$ & $-3.71\pm0.07$ & -- \\
Evol. tracks + AM & -- & -- & -- & $19\pm2$ & -- & $47\pm7$ \\
\hline
\hline 
\end{tabular}
\tablefoot{ 
\tablefoottext{*}{Estimated or assumed age.}
\tablefoottext{a}{The $\pm150$\,K does not refer to the uncertainty in the measured effective temperature, but to the range $1950$\,K-$2150$\,K.}
\tablefoottext{b}{The $\pm100$\,K does not refer to the uncertainty in the measured effective temperature, but to the range $1950$\,K-$2150$\,K.}
\tablefoottext{c}{The $+0.3$ and $-0.1$ refer to the luminosity range of $log_{}10(\mathrm{L/L_{\odot}})$=$-3.5$ to $-3.9$.}
\tablefoottext{d}{The $\pm50$\,K does not refer to the uncertainty in the measured effective temperature, but to the range $1800$\,K-$1900$\,K.}
\tablefoottext{e}{The $\pm25$\,dex does not refer to the uncertainty in the surface gravity, but to the range $4.5$-$5.0$\,dex.}
\tablefoottext{f}{The $\pm0.045$ uncertainty refers to the range $-3.78$ to $-3.69$ in  $log_{}10(\mathrm{L/L_{\odot}})$.}
\tablefoottext{g}{The $\pm50$\,K does not refer to the uncertainty in the measured effective temperature, but to the range $1800$\,K-$1900$\,K.}
\tablefoottext{h}{The $\pm0.25$\,dex does not refer to the uncertainty in the measured surface gravity, but to the range $4.0$-$4.5$\,dex.}
\tablefoottext{i}{The $\pm0.15$\,K does not refer to the uncertainty in the measured radius, but to the range $1.3$-$1.6\,\mathrm{R_{Jup}}$.}
\tablefoottext{j}{Corresponds to the mean and standard deviation values from literature.}
\tablefoottext{k}{Corresponds to the retrieved effective temperature.}
\tablefoottext{l}{Values obtained using the bolometric luminosity from \cite{Currie+2018}, a uniform prior in age (5 to 100\,Myr), and a series of evolutionary models (see \citealt{Xuan+2024} for further details). }
}
\end{table*}

\subsection{Mass limits and sensitivity}\label{sec:PMD}

We used the synthetic stellar magnitudes for MIRI coronagraphic observations (see Table\,\ref{table:star_mags}), the parallax of $19.406\pm0.210$\,mas (Table\,\ref{table:star_parameters}), and an age of $47\pm7$\,Myr from Table\,\ref{table:summary_properties}, combined with the contrast limits obtained in Section\,\ref{sec:raw_contrast}, to compute the sensitivity limits to additional companions, in Mass units.

\begin{figure}
\centering
\includegraphics[width=7.1cm]{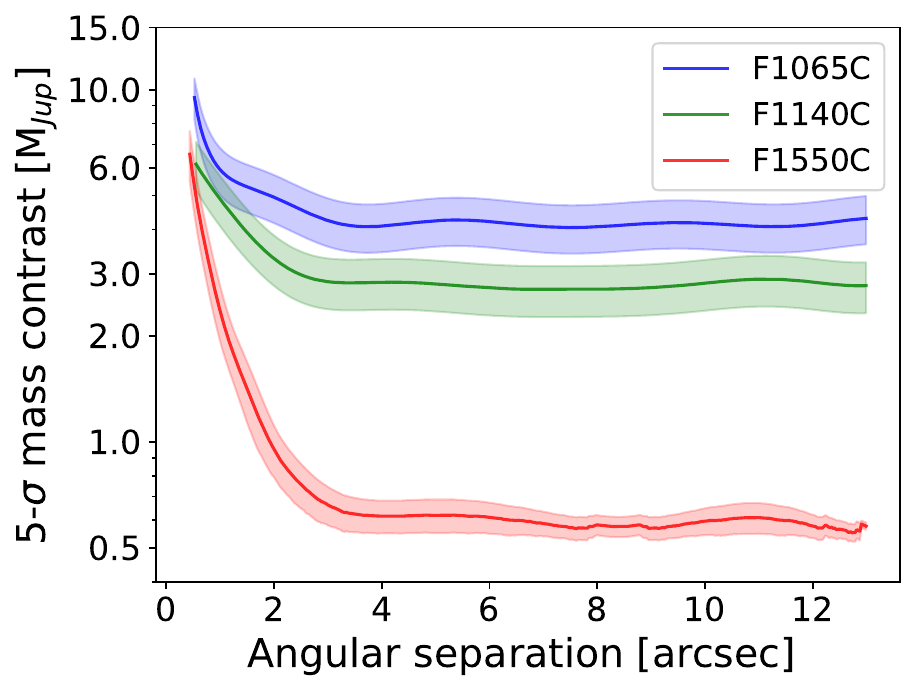}
\includegraphics[width=7.1cm]{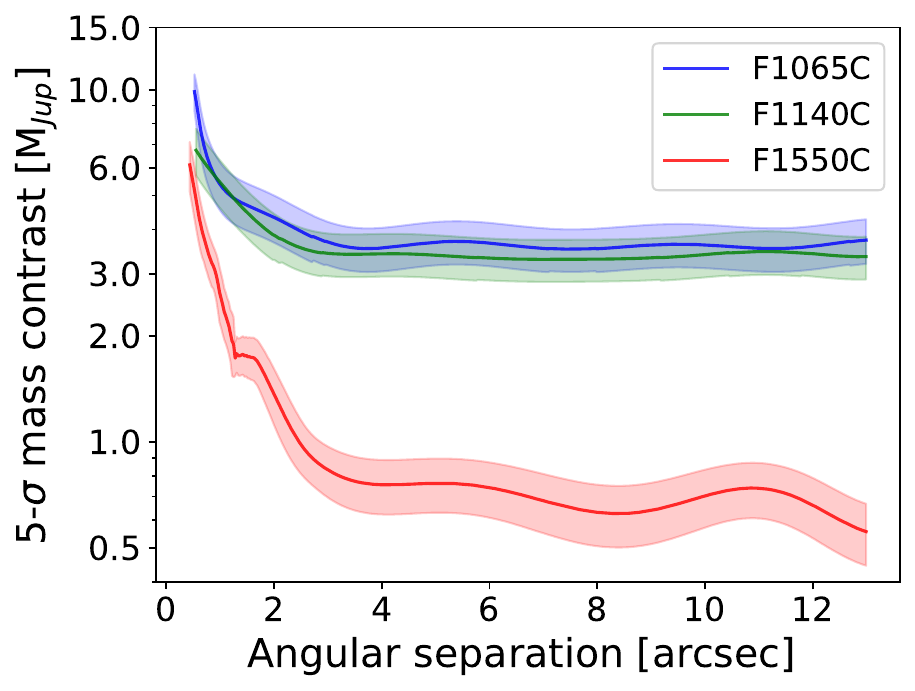}
\caption{Mass (from a $5$$\sigma$ sensitivity contrast) as a function of angular separation. \textit{Top:} ATMO chemical equilibrium mass limits for \texttt{F1065C}, \texttt{F1140C}, and \texttt{F1550C}. \textit{Bottom:} Same as in the left panel but for DUSTY-Linder2009. }
\label{fig:contrast_mass}%
\end{figure}

We compared the detection limits with two evolutionary models: ATMO chemical equilibrium and AMES-DUSTY, a cloud-free and a cloudy model, respectively. Since the sensitivity magnitude for the \texttt{F1550C} data goes to faint sources outside of the DUSTY grid, we coupled it with the ``Linder2019'' (BEX models; \citealt{Linder+2019}) evolutionary models (hereafter DUSTY-Linder2019). The Linder2019 models, in particular, account for the opacity and radiative effects of clouds, including their impact on atmospheric temperature gradients and emergent spectra. This makes them better suited for studying young or massive exoplanets, where cloud processes are significant. Figure\,\ref{fig:contrast_mass} shows the sensitivity as a function of angular separation for the ATMO and DUSTY-Linder2019 models. The mass sensitivity is around $5\,\mathrm{M_{Jup}}$ above 4\arcsec and below $10\,\mathrm{M_{Jup}}$ in the inner working angle region for both models.

\begin{figure*}
\centering
\includegraphics[width=6.0cm]{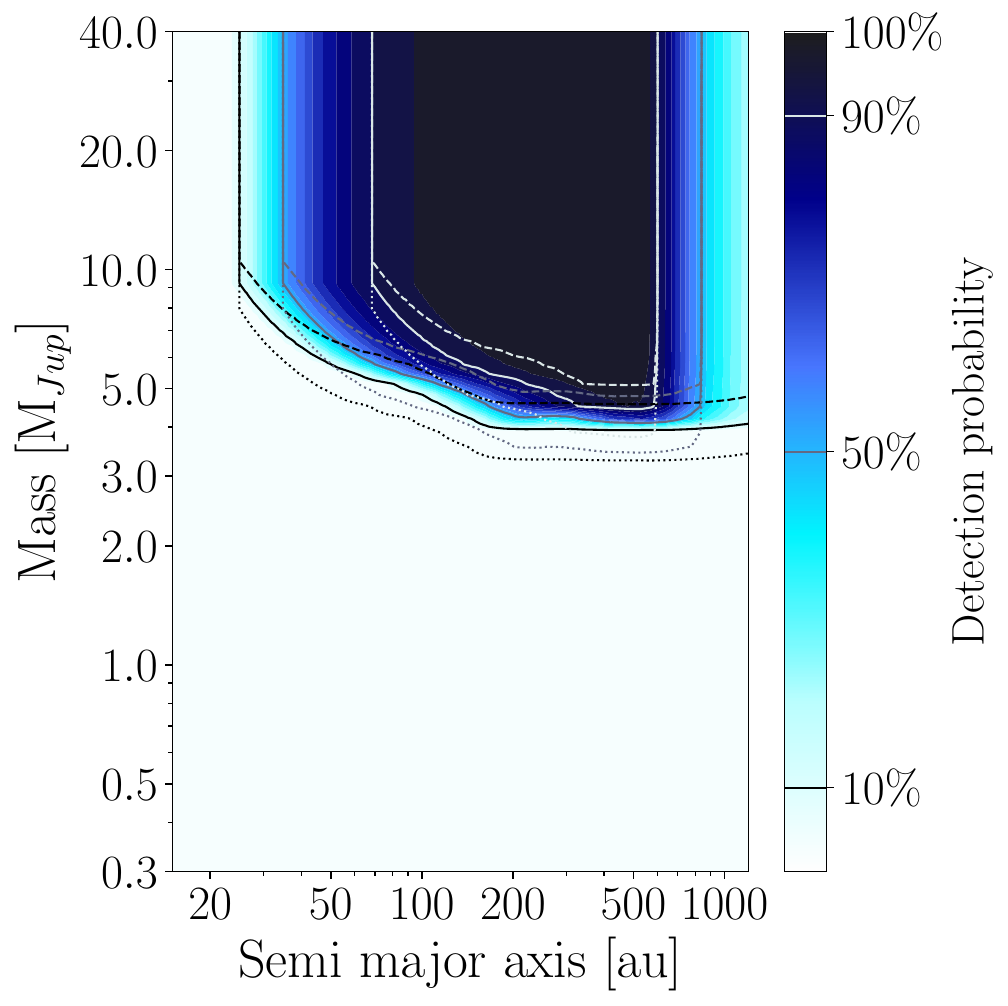}
\includegraphics[width=6.0cm]{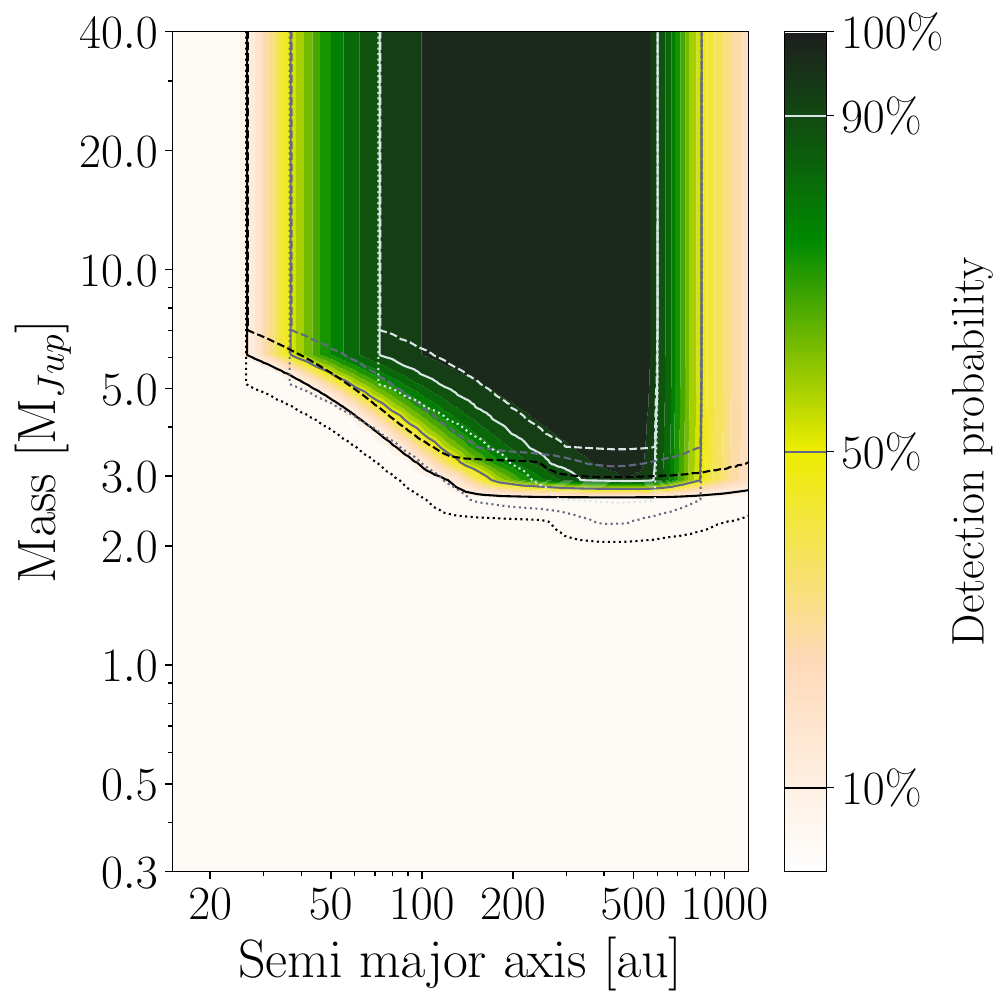}
\includegraphics[width=6.0cm]{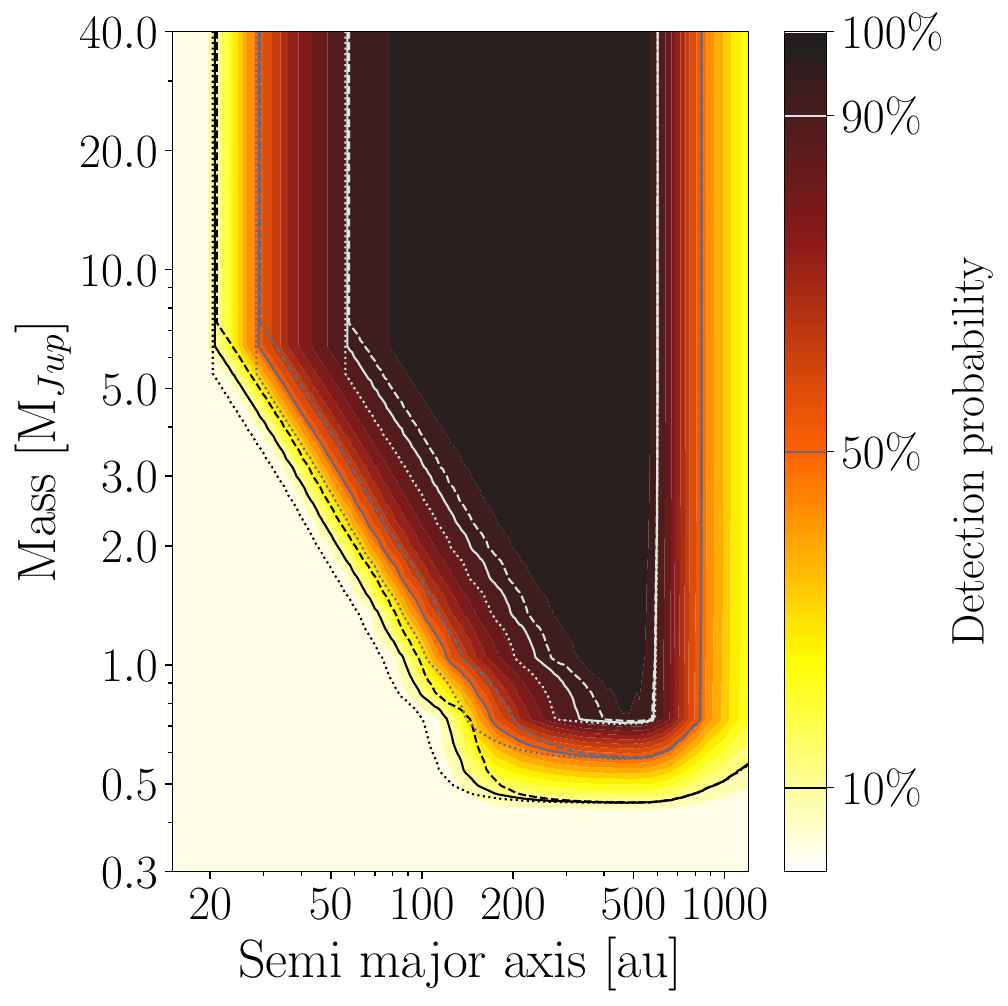}
\caption{Sensitivity of JWST/MIRI observations using the ATMO chemical equilibrium evolutionary track and an age of $47\pm7$\,Myr. From left to right: \texttt{F1065C}, \texttt{F1140C}, and \texttt{F1550C}. The color bar in each plot means the detection probability, and the solid lines highlight
the $10\%$, $50\%$, and $90\%$ detection thresholds. The dotted and dashed lines correspond to $1$$\sigma$ uncertainties, respectively. }
\label{fig:PMD_atmo}%
\end{figure*}

\begin{figure*}
\centering
\includegraphics[width=6.0cm]{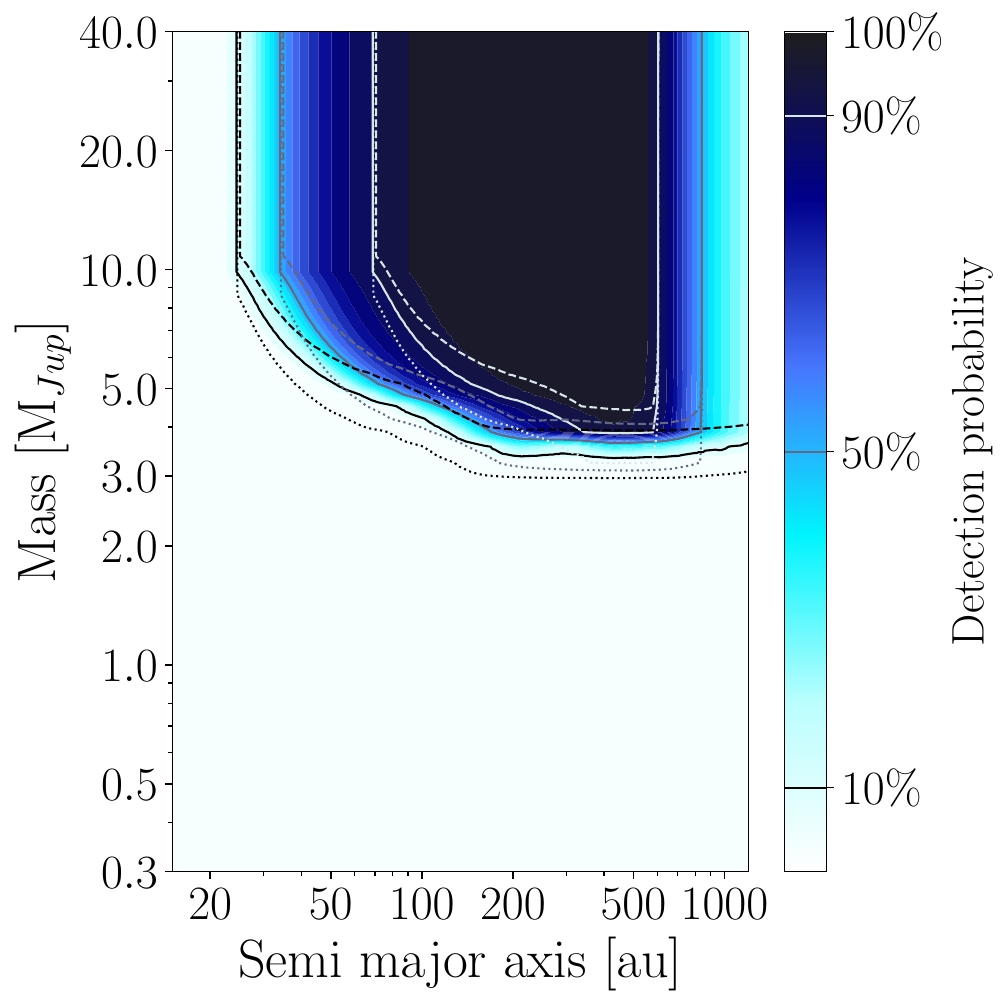}
\includegraphics[width=6.0cm]{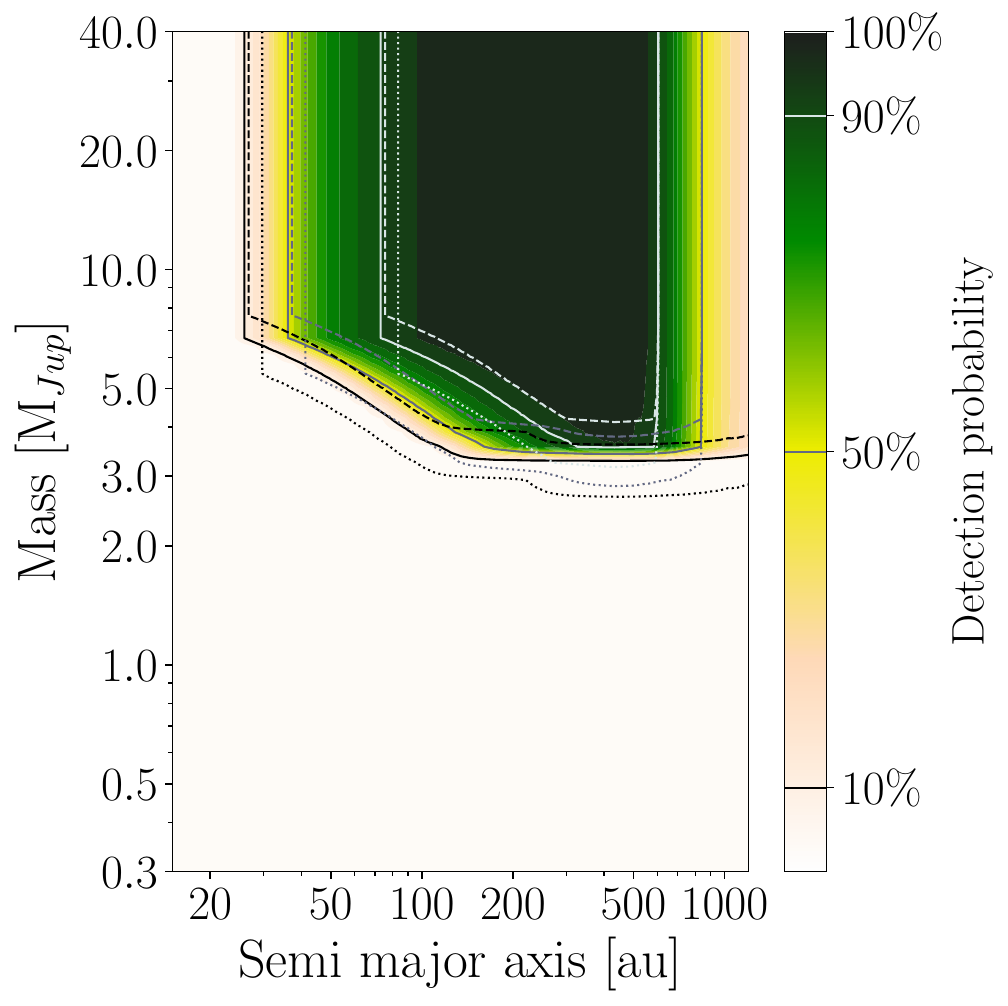}
\includegraphics[width=6.0cm]{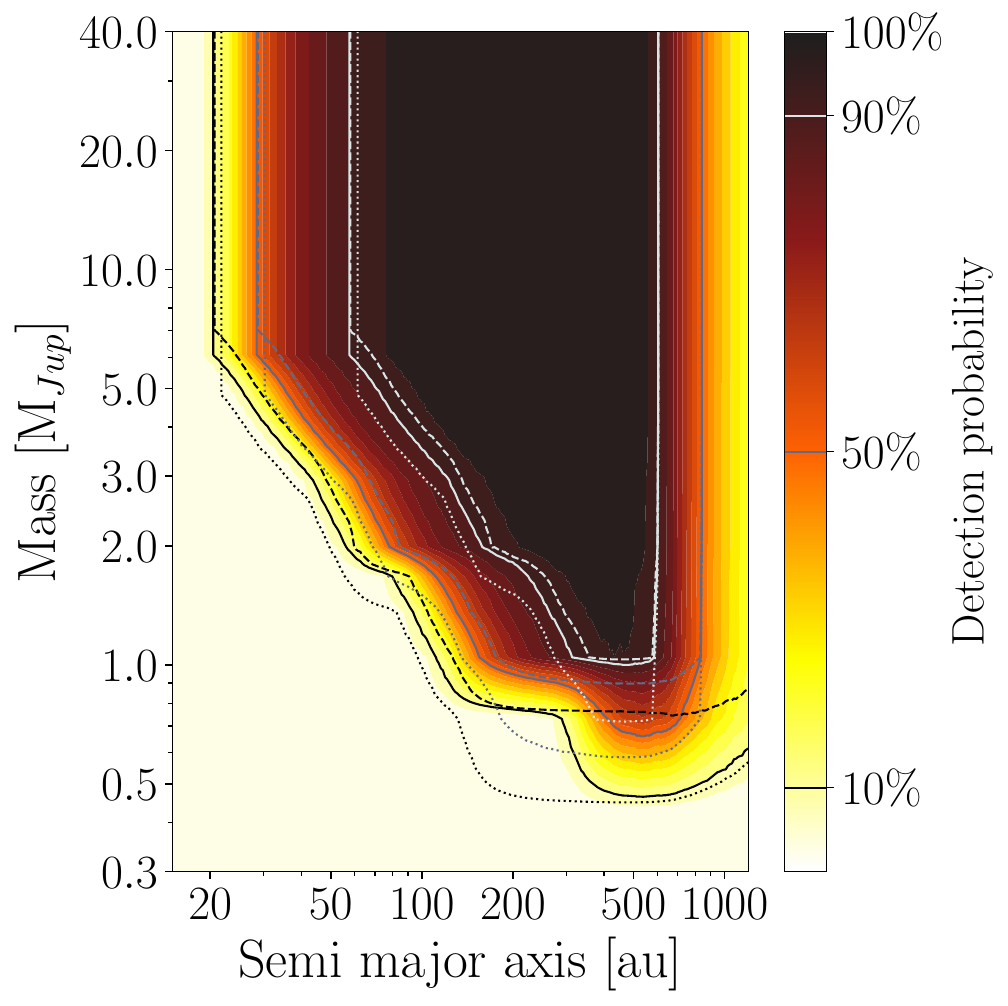}
\caption{Same as figure\,\ref{fig:PMD_atmo} but using the DUSTY-Linder2009 evolutionary tracks.}
\label{fig:PMD_linder}%
\end{figure*}

We determined the sensitivity maps using the Exoplanet Detection Map Calculator (\texttt{Exo-DMC}\footnote{\url{https://github.com/mbonav/Exo_DMC}}, \citealt{Exo-DMC}) code. We followed the same approach described in \cite{Godoy+2024} to compute the sensitivity map and propagate uncertainties, varying the mass between $0.1$–$110\,\mathrm{M_{Jup}}$ and the semi-major axis between $1$ and $1200$\,AU. We simulated $\sim1.2\times 10^{8}$ orbits, with all the orbital parameters uniformly distributed, except for the eccentricity (see \citealt{Exo-DMC}). Figures\,\ref{fig:PMD_atmo} and \ref{fig:PMD_linder} present the results for the ATMO and DUSTY-Linder2019 mass limits.

The factor that most influences the uncertainty in the sensitivity map is age. By reducing the age uncertainties by a factor of $\sim 3$ with respect to the previous estimates (e.g., \citealt{Jones+2016}), we better constrain the sensitivity. The reduced age uncertainty is reflected in the small deviation of masses in the sensitivity map (colored area in Figure\,\ref{fig:contrast_mass}, and dashed and dotted lines in Figures\,\ref{fig:PMD_atmo} and \ref{fig:PMD_linder}). The small differences between the predictions from ATMO and DUSTY-Linder2019 come from the relatively old age of $47$\,Myr, meaning an advanced atmospheric stage.

Our observations can detect, with $50\%$ probability, objects with masses around $8\,\mathrm{M_{Jup}}$ at 40 AU ($6.5\,\mathrm{M_{Jup}}$ at -$1$$\sigma$ uncertainty) for both models. At 200 AU, the ATMO model has a sensitivity of $0.65\,\mathrm{M_{Jup}}$ ($50\%$ probability), while the DUSTY-Linder2019 model reaches $0.95\,\mathrm{M_{Jup}}$, improving to $0.6\,\mathrm{M_{Jup}}$ and $0.67\,\mathrm{M_{Jup}}$ at $1$$\sigma$, respectively. The semi-major axis of $\kappa$\,And\,b ranges between $55$–$125$ AU (\citealt{Uyama+2020}), and its clear detection in MIRI data implies a mass above $4\,\mathrm{M_{Jup}}$, consistent with the literature and our measurements.

\section{Discussion}\label{sec:dis}

\subsection{Physical parameters of $\kappa$\,And\,b}

The large uncertainties in the age, the wide range of possible temperatures, and the model dependence in the derived mass, at the top of a nonuniform dataset, have generated discrepancies between past studies. The leverage provided by our long-wavelength observations has allowed us to contain the main properties of $\kappa$\,And\,b. Below we briefly discuss the implications of our new measurements on each parameter, and the meaning in the context of $\kappa$\,And\,b:

\paragraph{New age constraints:} Age is perhaps the most important parameter for interpreting the physical characteristics of $\kappa$\,And\,b. It directly impacts mass, effective temperature, radius, and evolutionary state estimates, which in turn inform models of formation and atmospheric dynamics. The age of $\kappa$\,And\,b has been a source of significant controversy, however. Combining all the literature ages, the age ranges from $10$ to $250$\,Myr (see Fig.\,\ref{fig:final_values}), based on different techniques such as isochrone fitting (\citealt{Bonnefoy+2014}), host stellar properties (\citealt{Jones+2016}), or membership in moving groups (\citealt{Hinkley+2013}, \citealt{Bonnefoy+2014}). 

\cite{Jones+2016} derived an age of $47$\,Myr ($+27$/$-40$\,Myr) based on interferometric observations and isochrone fitting using the inferred star properties, and this estimate has often been adopted as a reference. We employed two complementary methods based on the observations of the companion $\kappa$\,And\,b: CMD fitting with isochrones, and evolutionary track combined with atmospheric modeling. For the cloud-based atmospheric family, we derived a model-dependent age of $47\pm7$\,Myr, perfectly consistent with \cite{Jones+2016} and previous estimates but with an improvement in precision by a factor of $\sim4.8$ and better uncertainty by $\sim79$\%. This reduced uncertainty represents an important advance and provides a better constraint for deriving other parameters such as mass.

Furthermore, our age estimate supports the membership of $\kappa$\,And\,b in the Columba moving group (\citealt{Carson+2013}; \citealt{Bonnefoy+2014}; \citealt{Stone+2020}). Also, a well-constrained age enables better comparisons with other planetary-mass companions (e.g., those near to $\kappa$\,And\,b in the CMD; Figures\,\ref{fig:CMD_JK} and \ref{fig:CMD_index}), and more precise evolutionary history track, making $\kappa$\,And\,b a valuable benchmark for future atmospheric studies.\newline

\paragraph{Mass estimates}: The companion’s mass remains a key point of discussion. Estimates range from $13\,\mathrm{M_{Jup}}$ (\citealt{Carson+2013}, \citealt{Gratton+2024}) to nearly $50\,\mathrm{M_{Jup}}$ (\citealt{Hinkley+2013}), depending on the assumed age and choice of evolutionary or atmospheric models (Fig.\,\ref{fig:final_values}). Dynamical constraints (when available) provide the most robust mass estimates, especially when combining direct imaging with absolute astrometry (e.g., \citealt{Snellen+Brown-2018}; \citealt{Brandt+2019}; \citealt{Franson+2022}; \citealt{Franson+2023}; \citealt{Currie+2023}). The wide separation of $\kappa$\,And\,b (57–107 AU; \citealt{Uyama+2020}) and the long orbital period limit the current effectiveness of this approach, however. Relative astrometry currently covers less than a decade and shows only linear motion, providing limited constraints on the orbit. Although no significant proper motion anomaly is seen in \textit{Gaia} DR3, future analyzes combining longer-baseline astrometry and high-precision imaging, similar to those applied to $\beta$\,Pic\,b (\citealt{Brandt+2021}) and HR\,2562\,b (\citealt{Zhang+2023}), may provide stronger dynamical constraints in the coming decades.

We derived two primary mass estimates: $17$–$19\,\mathrm{M_{Jup}}$ for the cloud-based family and $15$–$16\,\mathrm{M_{Jup}}$ for the cloud-free family. Given that the cloudy models better match the observed spectrum, the mass from this type of model is more likely accurate, so our best mass estimate from the overall spectroscopy is $17.3\pm1.8\,\mathrm{M_{Jup}}$, with errors bars smaller than previous estimates by 20\%. This intermediate mass reconciles the long-standing debate between low-mass and high-mass solutions, situating $\kappa$\,And\,b in the transition between planets and brown dwarfs.

Notably, the mass estimate converges with values derived independently using three different methods (isochrones+CMD, atmospheric modeling, evolutionary tracks), all within the $1$$\sigma$ range: $18.1\pm4.8$, $17.3\pm 1.8$, and $19\pm2\,M_\mathrm{Jup}$, respectively. This convergence underscores the robustness of our results and the impact of MIRI observations. \newline

\paragraph{Effective temperature:} The $\mathrm{T_{eff}}$ of $\kappa$\,And\,b has been challenging to constrain due to its strong dependence on the assumed age, atmospheric models, and data used. Previous estimates ranged from $1650$\,K (\citealt{Carson+2013}) to $2040$\,K (\citealt{Hinkley+2013}). More recent studies (e.g., \citealt{Uyama+2020}, \citealt{Gratton+2024}) have suggested temperatures between $1700$\,K and $1850$\,K. It is important to note that the data used in each study may show discrepancies, either minor or significant, primarily due to differences in calibration procedures and in the stellar spectrum used to transform the contrast to absolute magnitude (see Section\,\ref{sec:obs}).

We find an effective temperature of $1791\pm69$\,K using cloudy models, consistent with previous estimates and at similar precision (e.g., $69$\,K vs. $50$\,K, \citealt{Stone+2020}). Cloud-free family yield similar averaged temperatures ($1778$\,K). Our results highlight the importance of considering cloud-based physics when modeling such objects and the fitting procedure. Also, this temperature reflects well the position of $\kappa$\,And\,b in the CMD (see Figures\,\ref{fig:CMD_JK} and \ref{fig:CMD_index}), at the L0-L4 objects with high temperature and, qualitatively, low concentration of silicates.

We observe that the black body fits yield similar temperature results ($1795\pm35$\,K), highlighting the importance and impact of covering most ($\sim90$\%) of the bolometric flux from substellar companions (from the NIR to the MIR). Even with a simple black body model, we can obtain a reliable estimate of the effective temperature of $\kappa$\,And\,b, showing the importance of MIRI in characterizing the atmosphere of substellar companions.\newline

\paragraph{Radius:} Previous radius estimates for $\kappa$\,And\,b were sparse and poorly constrained, with \cite{Hoch+2020} suggesting $1.2\pm0.3\,\mathrm{R_{Jup}}$ and \cite{Uyama+2020} reporting $1.45\pm0.15\,\mathrm{R_{Jup}}$. Our analysis yields a more precise value of $1.42\pm0.07\,\mathrm{R_{Jup}}$, improving the precision by over $45\%$.

The cloud-free models predict a slightly reduced radius of $1.39\pm0.04\,\mathrm{R_{Jup}}$, consistent with the absence of prominent clouds and the need to compensate for slightly lower temperatures to match the observed luminosity. The radius from the cloud-free family is consistent with the radius from the cloudy family at 1$\sigma$, however. The gain of covering the NIR and MIR is that we can obtain a good estimate of the temperature and radius regardless of the atmospheric family we use. Again, the black body fit provides consistent results at $2$$\sigma$ level ($1.30\pm0.06\,\mathrm{R_{Jup}}$), highlighting the importance of MIRI observations to constrain this parameter.\newline

\paragraph{Surface gravity:} The log(g) estimates in the literature have varied between $3.5$ and $5.5$ dex, with the most precise value being $4.75\pm0.25$\,dex (\citealt{Uyama+2020}). Our measurement of $4.35\pm0.07$\,dex is notably lower but still within the $2$$\sigma$ range of previous estimates. This represents a $\sim70\%$ improvement in precision.

The lower surface gravity has important implications for the classification of $\kappa$\,And\,b, supporting an L0–L2 spectral type (\citealt{Uyama+2020}). A lower log(g) suggests a less compact atmosphere, which may influence cloud formation and atmospheric dynamics, and a younger age (e.g., log(g)=4.75 and 10-100\,Myr, versus log(g)=4.35 and an age of 47\,Mr). Our results highlight the need for medium- and high-resolution spectroscopy to refine and constrain the log(g) using the new constrained $\mathrm{T_{eff}}$ as prior in the modeling.\newline

\paragraph{Luminosity:} Among the physical parameters of $\kappa$\,And\,b, luminosity estimates have shown minimal discrepancies across different studies. Previous estimates have ranged from $\mathrm{log_{10}(L/L_{\odot})}$ = $-3.76$ to $-3.81$, with most values trending toward lower luminosities. The most precise measurement in the literature is $-3.735 \pm 0.045$.

Our new measurement of $-3.71 \pm 0.07$ is in agreement with previous estimates at the $2$$\sigma$ level, demonstrating the robustness of current models when incorporating new observational data. We note that the cloudy and cloud-free families have small differences, at the $1$$\sigma$ level ($-3.76\pm0.05$ vs. $-3.71\pm0.07$, respectively). As we mentioned before, the NIR observations coupled with MIRI cover most of the bolometric emission of $\kappa$\,And\,b, and this helps us to be less biased in the luminosity measurement. This is also reflected in the black body estimates ($-3.77\pm0.07$), which are also in agreement with the cloudy, cloud-free families and literature.\newline

The luminosity and age are key for studying the evolutionary pathway and dynamics of the atmosphere of exoplanets in direct imaging. These new constraints on the $\kappa$\,And\,b properties can help in further comparison with the directly imaged exoplanets and, in particular, with those reported in associations of similar ages such as the Tucana-Horologium Association (e.g., HR\,8799 system). With the uncertainties of the main parameters of $\kappa$\,And\,b narrowed down, new studies related to the metallicity, chemical disequilibrium, presence of different species, and dust sedimentation will be keys to better understanding the atmospheric evolution of this companion and dynamic history of the $\kappa$\,And system.

\subsection{Impact of variability}

We emphasize the importance of a self-consistent calibration of all data in the analysis because this helps us to reduce bias sources. This recalibration does not eliminate other biases, however, such as internal calibration uncertainties or instrument performance variations. Another factor that we did not consider in our analysis is the potential variability due to a patchy atmosphere. Significant variability can impact atmospheric modeling, especially when combining datasets from different epochs, and it may vary across wavelengths. In addition, strong variability is very unlikely outside of the L to T transition (\citealt{Radigan+2014b}).

We estimate the rotational period of $\kappa$\,And\,b using the projected rotational velocity $v\,sin(i)=38.42\pm0.05$\,km/s from \cite{Morris+2024} and our radius estimate of $1.42\pm0.06\,\mathrm{R_{Jup}}$. \cite{Morris+2024} highlight that $\kappa$\,And\,b is a fast rotator, which suggests a short rotational period. We estimate the lower limit for the period to be $P\geq 4.61\pm0.19$\,hours, slightly longer than the MIRI science target execution time of $\sim1.5$ hours. In case of strong variability, the \texttt{F1550C} data would be the most impacted as it has the longest exposure time. The 11 years of observations, taken at different times and wavelengths, provide a broad temporal baseline that helps us to mitigate the impact of potential variability, if any. Since the data sample different epochs, the best-fit model effectively represents an average state of the source, with the residual dispersion naturally accounting for possible minor variations over time. To account for the impact of variability on the parameters, we have included 1\% uncertainty in the retrieved parameters.

\subsection{Brief comparison with other planetary-mass companions}

$\kappa$\,And\,b lies in the L0–L4 spectral type range. This is a key phase for probing atmospheric processes in young self-luminous exoplanets. L-type companions offer insights into cloud formation, gravity effects, and thermal structure during early evolution. Comparable objects include $\beta$\,Pic\,b \citep{Lagrange+2009}, 2MASS\,J0219–3925\,b \citep{Artigau+2015}, and AB\,Pic\,b \citep{Chauvin+2005}, all with ages between 21–37 Myr \citep{Gratton+2024}, close to $\kappa$\,And\,b’s estimated age of $47^{+27}_{-40}$ Myr \citep{Jones+2016}. These companions share similar atmospheric parameters: $\mathrm{T_{eff}} \approx 1680$–$1740$\,K, log(g) $\approx 4.2$–$4.5$, and masses of $12$–$15,\mathrm{M_{Jup}}$. The close match to $\beta$\,Pic\,b particularly supports the derived values for $\kappa$\,And\,b ($\mathrm{T_{eff}}$\,$\approx$\,$1790$,K, log(g)\,$\approx$\,$4.35$, M\,$\approx$\,$17\,\mathrm{M_{Jup}}$), reinforcing its reliability and placing it among well-characterized young L-type companions.

Combining NIR and MIR data offers a more complete view of $\kappa$\,And\,b’s atmosphere. Our models indicate that cloudy atmospheres fit the spectrum better than cloud-free ones, consistent with condensate clouds inferred from high-resolution NIR spectra \citep{Xuan+2024}. MIRI/MRS (Medium Resolution Spectroscopy) observations would enable direct detection of the $8$–$11,\mu$m silicate absorption feature – seen in other L dwarfs \citep[e.g.,][]{Cushing+2006, Miles+2023} — and provide deeper constraints on molecular abundances and the P–T structure. With its brightness and wide separation, $\kappa$\,And\,b is an ideal benchmark for studying cloud properties and atmospheric chemistry in young planetary-mass companions.

\section{Summary and conclusions}\label{sec:sac}

We presented a refined characterization of the planetary-mass companion $\kappa$\,And\,b based on new JWST/MIRI observations in the \texttt{F1065C}, \texttt{F1140C}, and \texttt{F1550C} bandpasses. By combining MIR and NIR data, we captured most of the bolometric flux and significantly improved the constraints on the atmospheric parameters. Our analysis reduced the uncertainties in key properties such as effective temperature and radius and provided new estimates for the companion mass and age of the system. These improvements helped us to resolve a decade-long debate surrounding $\kappa$\,And\,b, whose inferred mass ($13$–$50$\,$\mathrm{M_{Jup}}$), surface gravity ($3.5$–$5.5$\,dex), and effective temperature ($1680$–$2050$\,K) have varied widely in the literature. The discrepancies were driven by inhomogeneous treatments of the observed dataset, the different atmospheric models that were used, and the large uncertainties in the age ($10$–$250$\,Myr). Our study addressed these issues through a consistent reanalysis that offered a clearer picture of the nature of the companion.

Our first step was to recalibrate the archival data to have homogenous and consistent flux measurements of $\kappa$\,And\,b. To achieve this, we used the precise stellar parameters of $\kappa$\,And from \citet{Jones+2016} to compute a stellar spectrum with a higher accuracy using the ATLAS/SYNTHE models. We conducted a study using four different methods to constrain the physical parameters of $\kappa$\,And\,b as listed below.

\begin{itemize}
    \item The MIRI CMD using the spectral index from \cite{Suarez+Metchev-2022} related to the presence of species in field brown dwarfs was used to understand the atmosphere stage and possible presence of specific species. From this comparison, we concluded that $\kappa$\,And\,b has a high temperature between $1600$\,K and $2000$\,K, with a possible small amount of silicates in the atmosphere. The presence of other species (methane, and ammonia) is completely ruled out because of the location of this companion in the CMD and the high-temperature restriction.
    \item The combination of CMD with evolutionary tracks to derive the mass and age of the companion $\kappa$\,And\,b using the H and \texttt{F1140C} bands. We used two families of isochrones based on the physics used in the models: the cloudy family (AMES-DUSTY, BT-Settl, and Sonora solar metallicity), and the cloud-free family (AMES-COND and ATMO-ceq). We obtained an age and mass of $50\pm8$\,Myr and $15.7\pm2.3\,\mathrm{M_{Jup}}$ for the cloud-free family, and $50\pm13$\,Myr and $18.1\pm4.8\,\mathrm{M_{Jup}}$ for the cloudy family. This agrees with the values in the literature (e.g., $13^{+12}_{-2}\,\mathrm{M_{Jup}}$, \citealt{Currie+2018}, and $47^{+27}_{-40}$\,Myr, \citealt{Jones+2016}).
    \item We modeled the atmosphere of $\kappa$\,And\,b using recalibrated archival data and new MIRI observations,  and we tested cloudy (BT-DUSTY, BT-Settl, DRIFT-PHOENIX, and EXO-REM) and cloud-free (ATMO, AMES-COND, BT-COND, and EXO-REM no clouds) models. Physical priors and wavelength weighting were applied to ensure realistic solutions and to balance the NIR and MIR contributions. Cloudy models, especially EXO-REM, provided the best match to the full SED, while cloud-free models failed to reproduce the NIR spectrum. From the best fit, we derived a radius of $1.42\pm0.06\,\mathrm{R_{Jup}}$, a surface gravity of $log(g)$=$4.35\pm0.07$\,dex, and an effective temperature of $1791\pm69$\,K. This improves the precision by $\sim$22\% for $T_\mathrm{eff}$ and by $\sim$33\% for the radius compared to previous estimates.
   \item We combined the atmospheric results with evolutionary tracks to constrain the age and mass of $\kappa$\,And\,b. We considered two model families: cloudy (Saumon2008 hybrid and Sonora solar metallicity) and cloud-free (ATMO and Saumon2008 no cloud) using the $\mathrm{T_{eff}}$, log(g), and radius solutions from all atmospheric models grouped accordingly. Since the cloudy models match the SED of $\kappa$\,And\,b best, we adopted them for our final estimate. The $\mathrm{T_{eff}}$–log(g) combination yields the smallest uncertainties and leads to an age of $47\pm7$\,Myr and a mass of $17.3\pm1.8\,\mathrm{M_{Jup}}$.

\end{itemize}

With these results and analysis, we consider the following values as our final estimates of the physical parameters of $\kappa$\,And\,b: a {spectroscopic mass} of ${17.3\pm1.8\,\mathrm{M_{Jup}}}$, a ${\mathrm{T_{eff}}}$ of ${1791\pm68\,\mathrm{K}}$, a {radius} of ${1.42\pm0.06\,\mathrm{R_{Jup}}}$, a {surface gravity} of ${4.35\pm0.07\,\mathrm{dex}}$, ${\mathrm{log_{10}(L/L_{\odot})}}$ of ${-3.73\pm0.02}$, and an {age} of ${47\pm7\,\mathrm{Myr}}$. These are all model-dependent estimates.

The MIRI observations have been key to obtaining precise measurements and improving the uncertainties of these parameters, especially in the effective temperature ($22\%$) and radius ($33\%$). This shows the impact of properly covering most of the bolometric flux of planetary-mass companions. Without MIRI data, all model classes were found to overestimate the effective temperatures and to underestimate the radii. The MIRI data allowed us to remove the degeneracy between these parameters, however. Our (model-dependent) age estimates of $47\pm7$\,Myr agrees with the adopted age from \citet[$47^{+27}_{-40}$\,Myr]{Jones+2016}, but with significantly smaller uncertainties. This age is also consistent with the age of the Columba association ($42^{+6}_{-4}$\,Myr; \citealt{Bell+2015}), which indicates that the $\kappa$\,And system might be a member of Columba, as previously suggested by \cite{Carson+2013} and \cite{Bonnefoy+2014}. Our results lie in the middle range of the earlier measurements and also significantly reduce the error bars by $80\%$.

We obtained a final value of $17.3\pm1.8\,\mathrm{M_{Jup}}$ for the mass. This estimated value lies between the values reported in the literature. Consequently, $\kappa$\,And\,b is at the transition between giant planets and brown dwarfs. The \textit{Gaia} mission, combined with new observations in the coming decades, will place strong constraints on the dynamical mass and help us to corroborate this estimate. The effective temperature and radius are consistent with the spectral classification (L0–L2; \citealt{Uyama+2020}), suggesting a low abundance of certain metals and silicates in the atmosphere. The new constraints on age, mass, effective temperature, and radius will help us to understand the dynamical evolution of the system. 
Furthermore, while previous studies (e.g., \citealt{Xuan+2024}, \citealt{Hoch+2020}) have measured $\mathrm{CO}$ and $\mathrm{H_2O}$, constrained the C/O ratio, and retrieved a P–T profile for $\kappa$\,And\,b, the uncertainties remain large. New MIRI/MRS observations can confirm the presence of silicates, detect additional species, and refine our understanding of its atmospheric composition, cloud properties, and sedimentation state.

\begin{acknowledgements}
This project is funded/Co-funded by the European Union (ERC, ESCAPE, project No 101044152). Views and opinions expressed are however those of the author(s) only and do not necessarily reflect those of the European Union or the European Research Council Executive Agency. Neither the European Union nor the granting authority can be held responsible for them. These observations are associated with program \#1241. This research has made use of computing facilities operated by CeSAM ( Centre de données Astrophysiques de Marseille) data center at LAM (Laboratoire Astrophysique de Marseille), Marseille, France. We thank T. Stolker for providing support using the Species python package. We thank the authors Carson et al., Hinkley et al., Bonnefoy et al., Currie et al., Kuhn et al., Wilcomb et al., Stone et al., and Uyama et al., for providing and/or publishing the observational data used in this study. This work has made use of data from the European Space Agency (ESA) mission {\it Gaia} (\url{https://www.cosmos.esa.int/gaia}), processed by the {\it Gaia} Data Processing and Analysis Consortium (DPAC, \url{https://www.cosmos.esa.int/web/gaia/dpac/consortium}). Funding for the DPAC has been provided by national institutions, in particular, the institutions participating in the {\it Gaia} Multilateral Agreement. This research has made use of the Spanish Virtual Observatory (\url{https://svo.cab.inta-csic.es}) project funded by MCIN/AEI/10.13039/501100011033/ through grant PID2020-112949GB-I00. This publication makes use of VOSA, developed under the Spanish Virtual Observatory (\url{https://svo.cab.inta-csic.es}) project funded by MCIN/AEI/10.13039/501100011033/ through grant PID2020-112949GB-I00. VOSA has been partially updated by using funding from the European Union's Horizon 2020 Research and Innovation Programme, under Grant Agreement No 776403 (EXOPLANETS-A). This publication makes use of data products from the Two Micron All Sky Survey, which is a joint project of the University of Massachusetts and the Infrared Processing and Analysis Center/California Institute of Technology, funded by the National Aeronautics and Space Administration and the National Science Foundation. This publication makes use of data products from the Wide-field Infrared Survey Explorer, which is a joint project of the University of California, Los Angeles, and the Jet Propulsion Laboratory/California Institute of Technology, funded by the National Aeronautics and Space Administration. This work has benefited from The UltracoolSheet, maintained by Will Best, Trent Dupuy, Michael Liu, Rob Siverd, and Zhoujian Zhang, and developed from compilations by Dupuy \& Liu (2012, ApJS, 201, 19), Dupuy \& Kraus (2013, Science, 341, 1492), Liu et al. (2016, ApJ, 833, 96), Best et al. (2018, ApJS, 234, 1), and Best et al. (2020b, AJ, in press). This research has made use of the Washington Double Star Catalog maintained at the U.S. Naval Observatory. M.M., A.B., P.-O.L. acknowledge funding support by CNES.

\end{acknowledgements}

%
%

\bibliographystyle{aa} 

\begin{thebibliography}{147}
\expandafter\ifx\csname natexlab\endcsname\relax\def\natexlab#1{#1}\fi

\bibitem[{HIP(1997)}]{HIPPARCOS+0}
 1997, ESA Special Publication, Vol. 1200, {The HIPPARCOS and TYCHO catalogues.
  Astrometric and photometric star catalogues derived from the ESA HIPPARCOS
  Space Astrometry Mission}

\bibitem[{{Allard} {et~al.}(2001){Allard}, {Hauschildt}, {Alexander},
  {Tamanai}, \& {Schweitzer}}]{Allard+2001}
{Allard}, F., {Hauschildt}, P.~H., {Alexander}, D.~R., {Tamanai}, A., \&
  {Schweitzer}, A. 2001, \apj, 556, 357

\bibitem[{{Allard} {et~al.}(2012){Allard}, {Homeier}, \&
  {Freytag}}]{Allard+2012}
{Allard}, F., {Homeier}, D., \& {Freytag}, B. 2012, Philosophical Transactions
  of the Royal Society of London Series A, 370, 2765

\bibitem[{{Allers} \& {Liu}(2013)}]{Allers+2013}
{Allers}, K.~N. \& {Liu}, M.~C. 2013, \apj, 772, 79

\bibitem[{{Anderson} \& {Gordon}(2011)}]{Anderson+Gordon-2011}
{Anderson}, R.~E. \& {Gordon}, K.~D. 2011, \pasp, 123, 1237

\bibitem[{{Artigau} {et~al.}(2015){Artigau}, {Gagn{\'e}}, {Faherty}, {Malo},
  {Naud}, {Doyon}, {Lafreni{\`e}re}, \& {Beletsky}}]{Artigau+2015}
{Artigau}, {\'E}., {Gagn{\'e}}, J., {Faherty}, J., {et~al.} 2015, \apj, 806,
  254

\bibitem[{{Beichman} {et~al.}(1988){Beichman}, {Neugebauer}, {Habing}, {Clegg},
  \& {Chester}}]{IRAS+0}
{Beichman}, C.~A., {Neugebauer}, G., {Habing}, H.~J., {Clegg}, P.~E., \&
  {Chester}, T.~J., eds. 1988, {Infrared Astronomical Satellite (IRAS) Catalogs
  and Atlases.Volume 1: Explanatory Supplement.}, Vol.~1

\bibitem[{{Bell} {et~al.}(2015){Bell}, {Mamajek}, \& {Naylor}}]{Bell+2015}
{Bell}, C. P.~M., {Mamajek}, E.~E., \& {Naylor}, T. 2015, \mnras, 454, 593

\bibitem[{Best {et~al.}(2021)Best, Dupuy, Liu, Siverd, \& Zhang}]{Best+2021}
Best, W. M.~J., Dupuy, T.~J., Liu, M.~C., Siverd, R.~J., \& Zhang, Z. 2021,
  {The UltracoolSheet: Photometry, Astrometry, Spectroscopy, and Multiplicity
  for 3000+ Ultracool Dwarfs and Imaged Exoplanets}

\bibitem[{{Boccaletti} {et~al.}(2022){Boccaletti}, {Cossou}, {Baudoz},
  {Lagage}, {Dicken}, {Glasse}, {Hines}, {Aguilar}, {Detre}, {Nickson},
  {Noriega-Crespo}, {G{\'a}sp{\'a}r}, {Labiano}, {Stark}, {Rouan}, {Reess},
  {Wright}, {Rieke}, {Garcia Marin}, {Lajoie}, {Girard}, {Perrin}, {Soummer},
  \& {Pueyo}}]{Boccaletti+2022}
{Boccaletti}, A., {Cossou}, C., {Baudoz}, P., {et~al.} 2022, \aap, 667, A165

\bibitem[{{Boccaletti} {et~al.}(2015){Boccaletti}, {Lagage}, {Baudoz},
  {Beichman}, {Bouchet}, {Cavarroc}, {Dubreuil}, {Glasse}, {Glauser}, {Hines},
  {Lajoie}, {Lebreton}, {Perrin}, {Pueyo}, {Reess}, {Rieke}, {Ronayette},
  {Rouan}, {Soummer}, \& {Wright}}]{Boccaletti+2015}
{Boccaletti}, A., {Lagage}, P.~O., {Baudoz}, P., {et~al.} 2015, \pasp, 127, 633

\bibitem[{{Boccaletti} {et~al.}(2024){Boccaletti}, {M{\^a}lin}, {Baudoz},
  {Tremblin}, {Perrot}, {Rouan}, {Lagage}, {Whiteford}, {Molli{\`e}re},
  {Waters}, {Henning}, {Decin}, {G{\"u}del}, {Vandenbussche}, {Absil},
  {Argyriou}, {Bouwman}, {Cossou}, {Coulais}, {Gastaud}, {Glasse}, {Glauser},
  {Kamp}, {Kendrew}, {Krause}, {Lahuis}, {Mueller}, {Olofsson}, {Patapis},
  {Pye}, {Royer}, {Serabyn}, {Scheithauer}, {Colina}, {van Dishoeck}, {Ostlin},
  {Ray}, \& {Wright}}]{Boccaletti+2024}
{Boccaletti}, A., {M{\^a}lin}, M., {Baudoz}, P., {et~al.} 2024, \aap, 686, A33

\bibitem[{{Bonatto}(2019)}]{Bonatto-2019}
{Bonatto}, C. 2019, \mnras, 483, 2758

\bibitem[{{Bonavita}(2020)}]{Exo-DMC}
{Bonavita}, M. 2020, {Exo-DMC: Exoplanet Detection Map Calculator},
  Astrophysics Source Code Library, record ascl:2010.008

\bibitem[{{Bonnefoy} {et~al.}(2013){Bonnefoy}, {Boccaletti}, {Lagrange},
  {Allard}, {Mordasini}, {Beust}, {Chauvin}, {Girard}, {Homeier}, {Apai},
  {Lacour}, \& {Rouan}}]{Bonnefoy+2013}
{Bonnefoy}, M., {Boccaletti}, A., {Lagrange}, A.~M., {et~al.} 2013, \aap, 555,
  A107

\bibitem[{{Bonnefoy} {et~al.}(2014){Bonnefoy}, {Currie}, {Marleau},
  {Schlieder}, {Wisniewski}, {Carson}, {Covey}, {Henning}, {Biller}, {Hinz},
  {Klahr}, {Marsh Boyer}, {Zimmerman}, {Janson}, {McElwain}, {Mordasini},
  {Skemer}, {Bailey}, {Defr{\`e}re}, {Thalmann}, {Skrutskie}, {Allard},
  {Homeier}, {Tamura}, {Feldt}, {Cumming}, {Grady}, {Brandner}, {Helling},
  {Witte}, {Hauschildt}, {Kandori}, {Kuzuhara}, {Fukagawa}, {Kwon}, {Kudo},
  {Hashimoto}, {Kusakabe}, {Abe}, {Brandt}, {Egner}, {Guyon}, {Hayano},
  {Hayashi}, {Hayashi}, {Hodapp}, {Ishii}, {Iye}, {Knapp}, {Matsuo}, {Mede},
  {Miyama}, {Morino}, {Moro-Martin}, {Nishimura}, {Pyo}, {Serabyn}, {Suenaga},
  {Suto}, {Suzuki}, {Takahashi}, {Takami}, {Takato}, {Terada}, {Tomono},
  {Turner}, {Watanabe}, {Yamada}, {Takami}, \& {Usuda}}]{Bonnefoy+2014}
{Bonnefoy}, M., {Currie}, T., {Marleau}, G.~D., {et~al.} 2014, \aap, 562, A111

\bibitem[{{Bowler}(2016)}]{Bowler-2016B}
{Bowler}, B.~P. 2016, \pasp, 128, 102001

\bibitem[{{Brandt} {et~al.}(2021){Brandt}, {Brandt}, {Dupuy}, {Li}, \&
  {Michalik}}]{Brandt+2021}
{Brandt}, G.~M., {Brandt}, T.~D., {Dupuy}, T.~J., {Li}, Y., \& {Michalik}, D.
  2021, \aj, 161, 179

\bibitem[{{Brandt} {et~al.}(2019){Brandt}, {Dupuy}, \& {Bowler}}]{Brandt+2019}
{Brandt}, T.~D., {Dupuy}, T.~J., \& {Bowler}, B.~P. 2019, \aj, 158, 140

\bibitem[{{Burgasser}(2011)}]{Burgasser-2011}
{Burgasser}, A.~J. 2011, in Astronomical Society of the Pacific Conference
  Series, Vol. 450, Molecules in the Atmospheres of Extrasolar Planets, ed.
  J.~P. {Beaulieu}, S.~{Dieters}, \& G.~{Tinetti}, 113

\bibitem[{{Burgasser} {et~al.}(2002){Burgasser}, {Kirkpatrick}, {Brown},
  {Reid}, {Burrows}, {Liebert}, {Matthews}, {Gizis}, {Dahn}, {Monet}, {Cutri},
  \& {Skrutskie}}]{Burgasser+2002}
{Burgasser}, A.~J., {Kirkpatrick}, J.~D., {Brown}, M.~E., {et~al.} 2002, \apj,
  564, 421

\bibitem[{{Burggraaff} {et~al.}(2018){Burggraaff}, {Talens}, {Spronck},
  {Lesage}, {Stuik}, {Otten}, {Van Eylen}, {Pollacco}, \&
  {Snellen}}]{Burggraaff+2018}
{Burggraaff}, O., {Talens}, G.~J.~J., {Spronck}, J., {et~al.} 2018, \aap, 617,
  A32

\bibitem[{{Burrows} {et~al.}(2006){Burrows}, {Sudarsky}, \&
  {Hubeny}}]{Burrows+2006}
{Burrows}, A., {Sudarsky}, D., \& {Hubeny}, I. 2006, \apj, 640, 1063

\bibitem[{{Burrows} {et~al.}(1995){Burrows}, {Krist}, {Stapelfeldt}, \& {WFPC2
  Investigation Definition Team}}]{Burrows+1995}
{Burrows}, C.~J., {Krist}, J.~E., {Stapelfeldt}, K.~R., \& {WFPC2 Investigation
  Definition Team}. 1995, in American Astronomical Society Meeting Abstracts,
  Vol. 187, American Astronomical Society Meeting Abstracts, 32.05

\bibitem[{{Bushouse} {et~al.}(2022){Bushouse}, {Eisenhamer}, {Dencheva},
  {Davies}, {Greenfield}, {Morrison}, {Hodge}, {Simon}, {Grumm}, {Droettboom},
  {Slavich}, {Sosey}, {Pauly}, {Miller}, {Jedrzejewski}, {Hack}, {Davis},
  {Crawford}, {Law}, {Gordon}, {Regan}, {Cara}, {MacDonald}, {Bradley},
  {Shanahan}, \& {Jamieson}}]{Bushouse+2022}
{Bushouse}, H., {Eisenhamer}, J., {Dencheva}, N., {et~al.} 2022,
  {spacetelescope/jwst: JWST 1.6.2}, Zenodo

\bibitem[{{Cannon} \& {Pickering}(1993)}]{Cannon+Pickering-1993}
{Cannon}, A.~J. \& {Pickering}, E.~C. 1993, {VizieR Online Data Catalog: Henry
  Draper Catalogue and Extension (Cannon+ 1918-1924; ADC 1989)}, VizieR On-line
  Data Catalog: III/135A. Originally published in: Harv. Ann. 91-100
  (1918-1924)

\bibitem[{{Carson} {et~al.}(2013){Carson}, {Thalmann}, {Janson}, {Kozakis},
  {Bonnefoy}, {Biller}, {Schlieder}, {Currie}, {McElwain}, {Goto}, {Henning},
  {Brandner}, {Feldt}, {Kandori}, {Kuzuhara}, {Stevens}, {Wong}, {Gainey},
  {Fukagawa}, {Kuwada}, {Brandt}, {Kwon}, {Abe}, {Egner}, {Grady}, {Guyon},
  {Hashimoto}, {Hayano}, {Hayashi}, {Hayashi}, {Hodapp}, {Ishii}, {Iye},
  {Knapp}, {Kudo}, {Kusakabe}, {Matsuo}, {Miyama}, {Morino}, {Moro-Martin},
  {Nishimura}, {Pyo}, {Serabyn}, {Suto}, {Suzuki}, {Takami}, {Takato},
  {Terada}, {Tomono}, {Turner}, {Watanabe}, {Wisniewski}, {Yamada}, {Takami},
  {Usuda}, \& {Tamura}}]{Carson+2013}
{Carson}, J., {Thalmann}, C., {Janson}, M., {et~al.} 2013, \apjl, 763, L32

\bibitem[{{Carter} {et~al.}(2023){Carter}, {Hinkley}, {Kammerer}, {Skemer},
  {Biller}, {Leisenring}, {Millar-Blanchaer}, {Petrus}, {Stone}, {Ward-Duong},
  {Wang}, {Girard}, {Hines}, {Perrin}, {Pueyo}, {Balmer}, {Bonavita},
  {Bonnefoy}, {Chauvin}, {Choquet}, {Christiaens}, {Danielski}, {Kennedy},
  {Matthews}, {Miles}, {Patapis}, {Ray}, {Rickman}, {Sallum}, {Stapelfeldt},
  {Whiteford}, {Zhou}, {Absil}, {Boccaletti}, {Booth}, {Bowler}, {Chen},
  {Currie}, {Fortney}, {Grady}, {Greebaum}, {Henning}, {Hoch}, {Janson},
  {Kalas}, {Kenworthy}, {Kervella}, {Kraus}, {Lagage}, {Liu}, {Macintosh},
  {Marino}, {Marley}, {Marois}, {Matthews}, {Mawet}, {McElwain}, {Metchev},
  {Meyer}, {Molliere}, {Moran}, {Morley}, {Mukherjee}, {Pantin}, {Quirrenbach},
  {Rebollido}, {Ren}, {Schneider}, {Vasist}, {Worthen}, {Wyatt},
  {Briesemeister}, {Bryan}, {Calissendorff}, {Cantalloube}, {Cugno}, {De
  Furio}, {Dupuy}, {Factor}, {Faherty}, {Fitzgerald}, {Franson}, {Gonzales},
  {Hood}, {Howe}, {Kuzuhara}, {Lagrange}, {Lawson}, {Lazzoni}, {Lew}, {Liu},
  {Llop-Sayson}, {Lloyd}, {Martinez}, {Mazoyer}, {Palma-Bifani}, {Quanz},
  {Redai}, {Samland}, {Schlieder}, {Tamura}, {Tan}, {Uyama}, {Vigan}, {Vos},
  {Wagner}, {Wolff}, {Ygouf}, {Zhang}, {Zhang}, \& {Zhang}}]{Carter+2022}
{Carter}, A.~L., {Hinkley}, S., {Kammerer}, J., {et~al.} 2023, \apjl, 951, L20

\bibitem[{{Castelli} \& {Kurucz}(2003)}]{Castelli+Kurucz-2003}
{Castelli}, F. \& {Kurucz}, R.~L. 2003, in IAU Symposium, Vol. 210, Modelling
  of Stellar Atmospheres, ed. N.~{Piskunov}, W.~W. {Weiss}, \& D.~F. {Gray},
  A20

\bibitem[{{Chabrier} {et~al.}(2000){Chabrier}, {Baraffe}, {Allard}, \&
  {Hauschildt}}]{Chabrier+2000}
{Chabrier}, G., {Baraffe}, I., {Allard}, F., \& {Hauschildt}, P. 2000, \apj,
  542, 464

\bibitem[{{Charnay} {et~al.}(2018){Charnay}, {B{\'e}zard}, {Baudino},
  {Bonnefoy}, {Boccaletti}, \& {Galicher}}]{Charnay+2018}
{Charnay}, B., {B{\'e}zard}, B., {Baudino}, J.~L., {et~al.} 2018, \apj, 854,
  172

\bibitem[{{Chauvin} {et~al.}(2017){Chauvin}, {Desidera}, {Lagrange}, {Vigan},
  {Gratton}, {Langlois}, {Bonnefoy}, {Beuzit}, {Feldt}, {Mouillet}, {Meyer},
  {Cheetham}, {Biller}, {Boccaletti}, {D'Orazi}, {Galicher}, {Hagelberg},
  {Maire}, {Mesa}, {Olofsson}, {Samland}, {Schmidt}, {Sissa}, {Bonavita},
  {Charnay}, {Cudel}, {Daemgen}, {Delorme}, {Janin-Potiron}, {Janson},
  {Keppler}, {Le Coroller}, {Ligi}, {Marleau}, {Messina}, {Molli{\`e}re},
  {Mordasini}, {M{\"u}ller}, {Peretti}, {Perrot}, {Rodet}, {Rouan}, {Zurlo},
  {Dominik}, {Henning}, {Menard}, {Schmid}, {Turatto}, {Udry}, {Vakili}, {Abe},
  {Antichi}, {Baruffolo}, {Baudoz}, {Baudrand}, {Blanchard}, {Bazzon}, {Buey},
  {Carbillet}, {Carle}, {Charton}, {Cascone}, {Claudi}, {Costille}, {Deboulbe},
  {De Caprio}, {Dohlen}, {Fantinel}, {Feautrier}, {Fusco}, {Gigan}, {Giro},
  {Gisler}, {Gluck}, {Hubin}, {Hugot}, {Jaquet}, {Kasper}, {Madec}, {Magnard},
  {Martinez}, {Maurel}, {Le Mignant}, {M{\"o}ller-Nilsson}, {Llored}, {Moulin},
  {Orign{\'e}}, {Pavlov}, {Perret}, {Petit}, {Pragt}, {Puget}, {Rabou},
  {Ramos}, {Rigal}, {Rochat}, {Roelfsema}, {Rousset}, {Roux}, {Salasnich},
  {Sauvage}, {Sevin}, {Soenke}, {Stadler}, {Suarez}, {Weber}, {Wildi},
  {Antoniucci}, {Augereau}, {Baudino}, {Brandner}, {Engler}, {Girard}, {Gry},
  {Kral}, {Kopytova}, {Lagadec}, {Milli}, {Moutou}, {Schlieder},
  {Szul{\'a}gyi}, {Thalmann}, \& {Wahhaj}}]{Chauvin+2017}
{Chauvin}, G., {Desidera}, S., {Lagrange}, A.~M., {et~al.} 2017, \aap, 605, L9

\bibitem[{{Chauvin} {et~al.}(2005){Chauvin}, {Lagrange}, {Zuckerman}, {Dumas},
  {Mouillet}, {Song}, {Beuzit}, {Lowrance}, \& {Bessell}}]{Chauvin+2005}
{Chauvin}, G., {Lagrange}, A.~M., {Zuckerman}, B., {et~al.} 2005, \aap, 438,
  L29

\bibitem[{{Cheetham} {et~al.}(2019){Cheetham}, {Samland}, {Brems}, {Launhardt},
  {Chauvin}, {S{\'e}gransan}, {Henning}, {Quirrenbach}, {Avenhaus}, {Cugno},
  {Girard}, {Godoy}, {Kennedy}, {Maire}, {Metchev}, {M{\"u}ller}, {Musso
  Barcucci}, {Olofsson}, {Pepe}, {Quanz}, {Queloz}, {Reffert}, {Rickman}, {van
  Boekel}, {Boccaletti}, {Bonnefoy}, {Cantalloube}, {Charnay}, {Delorme},
  {Janson}, {Keppler}, {Lagrange}, {Langlois}, {Lazzoni}, {Menard}, {Mesa},
  {Meyer}, {Schmidt}, {Sissa}, {Udry}, \& {Zurlo}}]{Cheetham+2019}
{Cheetham}, A.~C., {Samland}, M., {Brems}, S.~S., {et~al.} 2019, \aap, 622, A80

\bibitem[{{Clemens} {et~al.}(2007){Clemens}, {Sarcia}, {Grabau}, {Tollestrup},
  {Buie}, {Dunham}, \& {Taylor}}]{Clemens+2007}
{Clemens}, D.~P., {Sarcia}, D., {Grabau}, A., {et~al.} 2007, \pasp, 119, 1385

\bibitem[{{Couture} {et~al.}(2023){Couture}, {Gagn{\'e}}, \&
  {Doyon}}]{Couture+2023}
{Couture}, D., {Gagn{\'e}}, J., \& {Doyon}, R. 2023, \apj, 946, 6

\bibitem[{{Currie} {et~al.}(2023){Currie}, {Brandt}, {Brandt}, {Lacy},
  {Burrows}, {Guyon}, {Tamura}, {Liu}, {Sagynbayeva}, {Tobin}, {Chilcote},
  {Groff}, {Marois}, {Thompson}, {Murphy}, {Kuzuhara}, {Lawson}, {Lozi}, {Deo},
  {Vievard}, {Skaf}, {Uyama}, {Jovanovic}, {Martinache}, {Kasdin}, {Kudo},
  {McElwain}, {Janson}, {Wisniewski}, {Hodapp}, {Nishikawa}, {He{\l}miniak},
  {Kwon}, \& {Hayashi}}]{Currie+2023}
{Currie}, T., {Brandt}, G.~M., {Brandt}, T.~D., {et~al.} 2023, Science, 380,
  198

\bibitem[{{Currie} {et~al.}(2018){Currie}, {Brandt}, {Uyama}, {Nielsen},
  {Blunt}, {Guyon}, {Tamura}, {Marois}, {Mede}, {Kuzuhara}, {Groff},
  {Jovanovic}, {Kasdin}, {Lozi}, {Hodapp}, {Chilcote}, {Carson}, {Martinache},
  {Goebel}, {Grady}, {McElwain}, {Akiyama}, {Asensio-Torres}, {Hayashi},
  {Janson}, {Knapp}, {Kwon}, {Nishikawa}, {Oh}, {Schlieder}, {Serabyn},
  {Sitko}, \& {Skaf}}]{Currie+2018}
{Currie}, T., {Brandt}, T.~D., {Uyama}, T., {et~al.} 2018, \aj, 156, 291

\bibitem[{{Currie} {et~al.}(2011){Currie}, {Burrows}, {Itoh}, {Matsumura},
  {Fukagawa}, {Apai}, {Madhusudhan}, {Hinz}, {Rodigas}, {Kasper}, {Pyo}, \&
  {Ogino}}]{Currie+2011}
{Currie}, T., {Burrows}, A., {Itoh}, Y., {et~al.} 2011, \apj, 729, 128

\bibitem[{{Cushing} {et~al.}(2008){Cushing}, {Marley}, {Saumon}, {Kelly},
  {Vacca}, {Rayner}, {Freedman}, {Lodders}, \& {Roellig}}]{Cushing+2008}
{Cushing}, M.~C., {Marley}, M.~S., {Saumon}, D., {et~al.} 2008, \apj, 678, 1372

\bibitem[{{Cushing} {et~al.}(2005){Cushing}, {Rayner}, \&
  {Vacca}}]{Cushing+2005}
{Cushing}, M.~C., {Rayner}, J.~T., \& {Vacca}, W.~D. 2005, \apj, 623, 1115

\bibitem[{{Cushing} {et~al.}(2006){Cushing}, {Roellig}, {Marley}, {Saumon},
  {Leggett}, {Kirkpatrick}, {Wilson}, {Sloan}, {Mainzer}, {Van Cleve}, \&
  {Houck}}]{Cushing+2006}
{Cushing}, M.~C., {Roellig}, T.~L., {Marley}, M.~S., {et~al.} 2006, \apj, 648,
  614

\bibitem[{{Daemgen} {et~al.}(2017){Daemgen}, {Todorov}, {Silva}, {Hand},
  {Garcia}, {Currie}, {Burrows}, {Stassun}, {Ratzka}, {Debes}, {Lafreniere},
  {Jayawardhana}, \& {Correia}}]{Daemgen+2017}
{Daemgen}, S., {Todorov}, K., {Silva}, J., {et~al.} 2017, \aap, 601, A65

\bibitem[{{Danielski} {et~al.}(2018){Danielski}, {Baudino}, {Lagage},
  {Boccaletti}, {Gastaud}, {Coulais}, \& {B{\'e}zard}}]{Danielski+2018}
{Danielski}, C., {Baudino}, J.-L., {Lagage}, P.-O., {et~al.} 2018, \aj, 156,
  276

\bibitem[{{De Rosa} {et~al.}(2016){De Rosa}, {Rameau}, {Patience}, {Graham},
  {Doyon}, {Lafreni{\`e}re}, {Macintosh}, {Pueyo}, {Rajan}, {Wang},
  {Ward-Duong}, {Hung}, {Maire}, {Nielsen}, {Ammons}, {Bulger}, {Cardwell},
  {Chilcote}, {Galvez}, {Gerard}, {Goodsell}, {Hartung}, {Hibon}, {Ingraham},
  {Johnson-Groh}, {Kalas}, {Konopacky}, {Marchis}, {Marois}, {Metchev},
  {Morzinski}, {Oppenheimer}, {Perrin}, {Rantakyr{\"o}}, {Savransky}, \&
  {Thomas}}]{DeRosa+2016}
{De Rosa}, R.~J., {Rameau}, J., {Patience}, J., {et~al.} 2016, \apj, 824, 121

\bibitem[{{Ducati}(2002)}]{Johnson-UBV}
{Ducati}, J.~R. 2002, {VizieR Online Data Catalog: Catalogue of Stellar
  Photometry in Johnson's 11-color system.}, CDS/ADC Collection of Electronic
  Catalogues, 2237, 0 (2002)

\bibitem[{{Dupuy} {et~al.}(2023){Dupuy}, {Liu}, {Evans}, {Best}, {Pearce},
  {Sanghi}, {Phillips}, \& {Bardalez Gagliuffi}}]{Dupuy+2023}
{Dupuy}, T.~J., {Liu}, M.~C., {Evans}, E.~L., {et~al.} 2023, \mnras, 519, 1688

\bibitem[{{Fortney} {et~al.}(2008){Fortney}, {Marley}, {Saumon}, \&
  {Lodders}}]{Fortney+2008}
{Fortney}, J.~J., {Marley}, M.~S., {Saumon}, D., \& {Lodders}, K. 2008, \apj,
  683, 1104

\bibitem[{{Franson} \& {Bowler}(2023)}]{Franson+2023}
{Franson}, K. \& {Bowler}, B.~P. 2023, \aj, 165, 246

\bibitem[{{Franson} {et~al.}(2022){Franson}, {Bowler}, {Brandt}, {Dupuy},
  {Tran}, {Brandt}, {Li}, \& {Kraus}}]{Franson+2022}
{Franson}, K., {Bowler}, B.~P., {Brandt}, T.~D., {et~al.} 2022, \aj, 163, 50

\bibitem[{{Gaia Collaboration}(2020)}]{Gaia-EDR3}
{Gaia Collaboration}. 2020, VizieR Online Data Catalog, I/350

\bibitem[{{Gaia Collaboration} {et~al.}(2016){Gaia Collaboration}, {Prusti},
  {de Bruijne}, {Brown}, {Vallenari}, {Babusiaux}, {Bailer-Jones}, {Bastian},
  {Biermann}, {Evans}, {Eyer}, {Jansen}, {Jordi}, {Klioner}, {Lammers},
  {Lindegren}, {Luri}, {Mignard}, {Milligan}, {Panem}, {Poinsignon},
  {Pourbaix}, {Randich}, {Sarri}, {Sartoretti}, {Siddiqui}, {Soubiran},
  {Valette}, {van Leeuwen}, {Walton}, {Aerts}, {Arenou}, {Cropper}, {Drimmel},
  {H{\o}g}, {Katz}, {Lattanzi}, {O'Mullane}, {Grebel}, {Holland}, {Huc},
  {Passot}, {Bramante}, {Cacciari}, {Casta{\~n}eda}, {Chaoul}, {Cheek}, {De
  Angeli}, {Fabricius}, {Guerra}, {Hern{\'a}ndez}, {Jean-Antoine-Piccolo},
  {Masana}, {Messineo}, {Mowlavi}, {Nienartowicz}, {Ord{\'o}{\~n}ez-Blanco},
  {Panuzzo}, {Portell}, {Richards}, {Riello}, {Seabroke}, {Tanga},
  {Th{\'e}venin}, {Torra}, {Els}, {Gracia-Abril}, {Comoretto},
  {Garcia-Reinaldos}, {Lock}, {Mercier}, {Altmann}, {Andrae}, {Astraatmadja},
  {Bellas-Velidis}, {Benson}, {Berthier}, {Blomme}, {Busso}, {Carry},
  {Cellino}, {Clementini}, {Cowell}, {Creevey}, {Cuypers}, {Davidson}, {De
  Ridder}, {de Torres}, {Delchambre}, {Dell'Oro}, {Ducourant}, {Fr{\'e}mat},
  {Garc{\'\i}a-Torres}, {Gosset}, {Halbwachs}, {Hambly}, {Harrison}, {Hauser},
  {Hestroffer}, {Hodgkin}, {Huckle}, {Hutton}, {Jasniewicz}, {Jordan},
  {Kontizas}, {Korn}, {Lanzafame}, {Manteiga}, {Moitinho}, {Muinonen},
  {Osinde}, {Pancino}, {Pauwels}, {Petit}, {Recio-Blanco}, {Robin}, {Sarro},
  {Siopis}, {Smith}, {Smith}, {Sozzetti}, {Thuillot}, {van Reeven}, {Viala},
  {Abbas}, {Abreu Aramburu}, {Accart}, {Aguado}, {Allan}, {Allasia},
  {Altavilla}, {{\'A}lvarez}, {Alves}, {Anderson}, {Andrei}, {Anglada Varela},
  {Antiche}, {Antoja}, {Ant{\'o}n}, {Arcay}, {Atzei}, {Ayache}, {Bach},
  {Baker}, {Balaguer-N{\'u}{\~n}ez}, {Barache}, {Barata}, {Barbier}, {Barblan},
  {Baroni}, {Barrado y Navascu{\'e}s}, {Barros}, {Barstow}, {Becciani},
  {Bellazzini}, {Bellei}, {Bello Garc{\'\i}a}, {Belokurov}, {Bendjoya},
  {Berihuete}, {Bianchi}, {Bienaym{\'e}}, {Billebaud}, {Blagorodnova},
  {Blanco-Cuaresma}, {Boch}, {Bombrun}, {Borrachero}, {Bouquillon}, {Bourda},
  {Bouy}, {Bragaglia}, {Breddels}, {Brouillet}, {Br{\"u}semeister},
  {Bucciarelli}, {Budnik}, {Burgess}, {Burgon}, {Burlacu}, {Busonero}, {Buzzi},
  {Caffau}, {Cambras}, {Campbell}, {Cancelliere}, {Cantat-Gaudin}, {Carlucci},
  {Carrasco}, {Castellani}, {Charlot}, {Charnas}, {Charvet}, {Chassat},
  {Chiavassa}, {Clotet}, {Cocozza}, {Collins}, {Collins}, {Costigan}, {Crifo},
  {Cross}, {Crosta}, {Crowley}, {Dafonte}, {Damerdji}, {Dapergolas}, {David},
  {David}, {De Cat}, {de Felice}, {de Laverny}, {De Luise}, {De March}, {de
  Martino}, {de Souza}, {Debosscher}, {del Pozo}, {Delbo}, {Delgado},
  {Delgado}, {di Marco}, {Di Matteo}, {Diakite}, {Distefano}, {Dolding}, {Dos
  Anjos}, {Drazinos}, {Dur{\'a}n}, {Dzigan}, {Ecale}, {Edvardsson}, {Enke},
  {Erdmann}, {Escolar}, {Espina}, {Evans}, {Eynard Bontemps}, {Fabre},
  {Fabrizio}, {Faigler}, {Falc{\~a}o}, {Farr{\`a}s Casas}, {Faye}, {Federici},
  {Fedorets}, {Fern{\'a}ndez-Hern{\'a}ndez}, {Fernique}, {Fienga}, {Figueras},
  {Filippi}, {Findeisen}, {Fonti}, {Fouesneau}, {Fraile}, {Fraser}, {Fuchs},
  {Furnell}, {Gai}, {Galleti}, {Galluccio}, {Garabato}, {Garc{\'\i}a-Sedano},
  {Gar{\'e}}, {Garofalo}, {Garralda}, {Gavras}, {Gerssen}, {Geyer}, {Gilmore},
  {Girona}, {Giuffrida}, {Gomes}, {Gonz{\'a}lez-Marcos},
  {Gonz{\'a}lez-N{\'u}{\~n}ez}, {Gonz{\'a}lez-Vidal}, {Granvik}, {Guerrier},
  {Guillout}, {Guiraud}, {G{\'u}rpide}, {Guti{\'e}rrez-S{\'a}nchez}, {Guy},
  {Haigron}, {Hatzidimitriou}, {Haywood}, {Heiter}, {Helmi}, {Hobbs},
  {Hofmann}, {Holl}, {Holland}, {Hunt}, {Hypki}, {Icardi}, {Irwin}, {Jevardat
  de Fombelle}, {Jofr{\'e}}, {Jonker}, {Jorissen}, {Julbe}, {Karampelas},
  {Kochoska}, {Kohley}, {Kolenberg}, {Kontizas}, {Koposov}, {Kordopatis},
  {Koubsky}, {Kowalczyk}, {Krone-Martins}, {Kudryashova}, {Kull}, {Bachchan},
  {Lacoste-Seris}, {Lanza}, {Lavigne}, {Le Poncin-Lafitte}, {Lebreton},
  {Lebzelter}, {Leccia}, {Leclerc}, {Lecoeur-Taibi}, {Lemaitre}, {Lenhardt},
  {Leroux}, {Liao}, {Licata}, {Lindstr{\o}m}, {Lister}, {Livanou}, {Lobel},
  {L{\"o}ffler}, {L{\'o}pez}, {Lopez-Lozano}, {Lorenz}, {Loureiro},
  {MacDonald}, {Magalh{\~a}es Fernandes}, {Managau}, {Mann}, {Mantelet},
  {Marchal}, {Marchant}, {Marconi}, {Marie}, {Marinoni}, {Marrese},
  {Marschalk{\'o}}, {Marshall}, {Mart{\'\i}n-Fleitas}, {Martino}, {Mary},
  {Matijevi{\v{c}}}, {Mazeh}, {McMillan}, {Messina}, {Mestre}, {Michalik},
  {Millar}, {Miranda}, {Molina}, {Molinaro}, {Molinaro}, {Moln{\'a}r},
  {Moniez}, {Montegriffo}, {Monteiro}, {Mor}, {Mora}, {Morbidelli}, {Morel},
  {Morgenthaler}, {Morley}, {Morris}, {Mulone}, {Muraveva}, {Musella},
  {Narbonne}, {Nelemans}, {Nicastro}, {Noval}, {Ord{\'e}novic},
  {Ordieres-Mer{\'e}}, {Osborne}, {Pagani}, {Pagano}, {Pailler}, {Palacin},
  {Palaversa}, {Parsons}, {Paulsen}, {Pecoraro}, {Pedrosa}, {Pentik{\"a}inen},
  {Pereira}, {Pichon}, {Piersimoni}, {Pineau}, {Plachy}, {Plum}, {Poujoulet},
  {Pr{\v{s}}a}, {Pulone}, {Ragaini}, {Rago}, {Rambaux}, {Ramos-Lerate},
  {Ranalli}, {Rauw}, {Read}, {Regibo}, {Renk}, {Reyl{\'e}}, {Ribeiro},
  {Rimoldini}, {Ripepi}, {Riva}, {Rixon}, {Roelens}, {Romero-G{\'o}mez},
  {Rowell}, {Royer}, {Rudolph}, {Ruiz-Dern}, {Sadowski}, {Sagrist{\`a}
  Sell{\'e}s}, {Sahlmann}, {Salgado}, {Salguero}, {Sarasso}, {Savietto},
  {Schnorhk}, {Schultheis}, {Sciacca}, {Segol}, {Segovia}, {Segransan},
  {Serpell}, {Shih}, {Smareglia}, {Smart}, {Smith}, {Solano}, {Solitro},
  {Sordo}, {Soria Nieto}, {Souchay}, {Spagna}, {Spoto}, {Stampa}, {Steele},
  {Steidelm{\"u}ller}, {Stephenson}, {Stoev}, {Suess}, {S{\"u}veges}, {Surdej},
  {Szabados}, {Szegedi-Elek}, {Tapiador}, {Taris}, {Tauran}, {Taylor},
  {Teixeira}, {Terrett}, {Tingley}, {Trager}, {Turon}, {Ulla}, {Utrilla},
  {Valentini}, {van Elteren}, {Van Hemelryck}, {van Leeuwen}, {Varadi},
  {Vecchiato}, {Veljanoski}, {Via}, {Vicente}, {Vogt}, {Voss}, {Votruba},
  {Voutsinas}, {Walmsley}, {Weiler}, {Weingrill}, {Werner}, {Wevers},
  {Whitehead}, {Wyrzykowski}, {Yoldas}, {{\v{Z}}erjal}, {Zucker}, {Zurbach},
  {Zwitter}, {Alecu}, {Allen}, {Allende Prieto}, {Amorim},
  {Anglada-Escud{\'e}}, {Arsenijevic}, {Azaz}, {Balm}, {Beck}, {Bernstein},
  {Bigot}, {Bijaoui}, {Blasco}, {Bonfigli}, {Bono}, {Boudreault}, {Bressan},
  {Brown}, {Brunet}, {Bunclark}, {Buonanno}, {Butkevich}, {Carret}, {Carrion},
  {Chemin}, {Ch{\'e}reau}, {Corcione}, {Darmigny}, {de Boer}, {de Teodoro}, {de
  Zeeuw}, {Delle Luche}, {Domingues}, {Dubath}, {Fodor}, {Fr{\'e}zouls},
  {Fries}, {Fustes}, {Fyfe}, {Gallardo}, {Gallegos}, {Gardiol}, {Gebran},
  {Gomboc}, {G{\'o}mez}, {Grux}, {Gueguen}, {Heyrovsky}, {Hoar}, {Iannicola},
  {Isasi Parache}, {Janotto}, {Joliet}, {Jonckheere}, {Keil}, {Kim},
  {Klagyivik}, {Klar}, {Knude}, {Kochukhov}, {Kolka}, {Kos}, {Kutka}, {Lainey},
  {LeBouquin}, {Liu}, {Loreggia}, {Makarov}, {Marseille}, {Martayan},
  {Martinez-Rubi}, {Massart}, {Meynadier}, {Mignot}, {Munari}, {Nguyen},
  {Nordlander}, {Ocvirk}, {O'Flaherty}, {Olias Sanz}, {Ortiz}, {Osorio},
  {Oszkiewicz}, {Ouzounis}, {Palmer}, {Park}, {Pasquato}, {Peltzer}, {Peralta},
  {P{\'e}turaud}, {Pieniluoma}, {Pigozzi}, {Poels}, {Prat}, {Prod'homme},
  {Raison}, {Rebordao}, {Risquez}, {Rocca-Volmerange}, {Rosen}, {Ruiz-Fuertes},
  {Russo}, {Sembay}, {Serraller Vizcaino}, {Short}, {Siebert}, {Silva},
  {Sinachopoulos}, {Slezak}, {Soffel}, {Sosnowska}, {Strai{\v{z}}ys}, {ter
  Linden}, {Terrell}, {Theil}, {Tiede}, {Troisi}, {Tsalmantza}, {Tur},
  {Vaccari}, {Vachier}, {Valles}, {Van Hamme}, {Veltz}, {Virtanen}, {Wallut},
  {Wichmann}, {Wilkinson}, {Ziaeepour}, \& {Zschocke}}]{Gaia+2016}
{Gaia Collaboration}, {Prusti}, T., {de Bruijne}, J.~H.~J., {et~al.} 2016,
  \aap, 595, A1

\bibitem[{{Garrison} \& {Gray}(1994)}]{Garrison+Gray-1994}
{Garrison}, R.~F. \& {Gray}, R.~O. 1994, \aj, 107, 1556

\bibitem[{{Godoy} {et~al.}(2024){Godoy}, {Choquet}, {Serabyn}, {Danielski},
  {Stolker}, {Charnay}, {Hinkley}, {Lagage}, {Ressler}, {Tremblin}, \&
  {Vigan}}]{Godoy+2024}
{Godoy}, N., {Choquet}, E., {Serabyn}, E., {et~al.} 2024, \aap, 689, A185

\bibitem[{{Gratton} {et~al.}(2024){Gratton}, {Bonavita}, {Mesa}, {Zurlo},
  {Marino}, {Desidera}, {D'Orazi}, {Rigliaco}, {Squicciarini}, \&
  {Nogueira}}]{Gratton+2024}
{Gratton}, R., {Bonavita}, M., {Mesa}, D., {et~al.} 2024, \aap, 684, A69

\bibitem[{{Groff} {et~al.}(2014){Groff}, {Kasdin}, {Limbach}, {Galvin}, {Carr},
  {Knapp}, {Brandt}, {Loomis}, {Jarosik}, {Mede}, {McElwain}, {Janson},
  {Guyon}, {Jovanovic}, {Takato}, {Martinache}, \& {Hayashi}}]{Groff+2014}
{Groff}, T.~D., {Kasdin}, N.~J., {Limbach}, M.~A., {et~al.} 2014, in Society of
  Photo-Optical Instrumentation Engineers (SPIE) Conference Series, Vol. 9147,
  Ground-based and Airborne Instrumentation for Astronomy V, ed. S.~K.
  {Ramsay}, I.~S. {McLean}, \& H.~{Takami}, 91471W

\bibitem[{{Groff} {et~al.}(2015){Groff}, {Kasdin}, {Limbach}, {Galvin}, {Carr},
  {Knapp}, {Brandt}, {Loomis}, {Jarosik}, {Mede}, {McElwain}, {Leviton},
  {Miller}, {Quijada}, {Guyon}, {Jovanovic}, {Takato}, \&
  {Hayashi}}]{Groff+2015}
{Groff}, T.~D., {Kasdin}, N.~J., {Limbach}, M.~A., {et~al.} 2015, in Society of
  Photo-Optical Instrumentation Engineers (SPIE) Conference Series, Vol. 9605,
  Techniques and Instrumentation for Detection of Exoplanets VII, ed.
  S.~{Shaklan}, 96051C

\bibitem[{{Hauschildt} {et~al.}(1999){Hauschildt}, {Allard}, {Ferguson},
  {Baron}, \& {Alexander}}]{Hauschildt+1999}
{Hauschildt}, P.~H., {Allard}, F., {Ferguson}, J., {Baron}, E., \& {Alexander},
  D.~R. 1999, \apj, 525, 871

\bibitem[{{Helling} {et~al.}(2008){Helling}, {Dehn}, {Woitke}, \&
  {Hauschildt}}]{Helling+2008}
{Helling}, C., {Dehn}, M., {Woitke}, P., \& {Hauschildt}, P.~H. 2008, \apjl,
  675, L105

\bibitem[{{Helou} \& {Walker}(1988)}]{IRAS+1}
{Helou}, G. \& {Walker}, D.~W., eds. 1988, {Infrared Astronomical Satellite
  (IRAS) Catalogs and Atlases.Volume 7: The Small Scale Structure Catalog.},
  Vol.~7

\bibitem[{{Hinkley} {et~al.}(2008){Hinkley}, {Oppenheimer}, {Brenner}, {Parry},
  {Sivaramakrishnan}, {Soummer}, \& {King}}]{Hinkley+2008}
{Hinkley}, S., {Oppenheimer}, B.~R., {Brenner}, D., {et~al.} 2008, in Society
  of Photo-Optical Instrumentation Engineers (SPIE) Conference Series, Vol.
  7015, Adaptive Optics Systems, ed. N.~{Hubin}, C.~E. {Max}, \& P.~L.
  {Wizinowich}, 701519

\bibitem[{{Hinkley} {et~al.}(2011){Hinkley}, {Oppenheimer}, {Zimmerman},
  {Brenner}, {Parry}, {Crepp}, {Vasisht}, {Ligon}, {King}, {Soummer},
  {Sivaramakrishnan}, {Beichman}, {Shao}, {Roberts}, {Bouchez}, {Dekany},
  {Pueyo}, {Roberts}, {Lockhart}, {Zhai}, {Shelton}, \&
  {Burruss}}]{Hinkley+2011c}
{Hinkley}, S., {Oppenheimer}, B.~R., {Zimmerman}, N., {et~al.} 2011, \pasp,
  123, 74

\bibitem[{{Hinkley} {et~al.}(2013){Hinkley}, {Pueyo}, {Faherty}, {Oppenheimer},
  {Mamajek}, {Kraus}, {Rice}, {Ireland}, {David}, {Hillenbrand}, {Vasisht},
  {Cady}, {Brenner}, {Veicht}, {Nilsson}, {Zimmerman}, {Parry}, {Beichman},
  {Dekany}, {Roberts}, {Roberts}, {Baranec}, {Crepp}, {Burruss}, {Wallace},
  {King}, {Zhai}, {Lockhart}, {Shao}, {Soummer}, {Sivaramakrishnan}, \&
  {Wilson}}]{Hinkley+2013}
{Hinkley}, S., {Pueyo}, L., {Faherty}, J.~K., {et~al.} 2013, \apj, 779, 153

\bibitem[{{Hodapp} {et~al.}(2008){Hodapp}, {Suzuki}, {Tamura}, {Abe}, {Suto},
  {Kandori}, {Morino}, {Nishimura}, {Takami}, {Guyon}, {Jacobson},
  {Stahlberger}, {Yamada}, {Shelton}, {Hashimoto}, {Tavrov}, {Nishikawa},
  {Ukita}, {Izumiura}, {Hayashi}, {Nakajima}, {Yamada}, \&
  {Usuda}}]{Hodapp+2008}
{Hodapp}, K.~W., {Suzuki}, R., {Tamura}, M., {et~al.} 2008, in Society of
  Photo-Optical Instrumentation Engineers (SPIE) Conference Series, Vol. 7014,
  Ground-based and Airborne Instrumentation for Astronomy II, ed. I.~S.
  {McLean} \& M.~M. {Casali}, 701419

\bibitem[{{H{\o}g} {et~al.}(2000){H{\o}g}, {Fabricius}, {Makarov}, {Bastian},
  {Schwekendiek}, {Wicenec}, {Urban}, {Corbin}, \& {Wycoff}}]{TYCHO+0}
{H{\o}g}, E., {Fabricius}, C., {Makarov}, V.~V., {et~al.} 2000, \aap, 357, 367

\bibitem[{{Ishihara} {et~al.}(2010){Ishihara}, {Onaka}, {Kataza}, {Salama},
  {Alfageme}, {Cassatella}, {Cox}, {Garc{\'\i}a-Lario}, {Stephenson}, {Cohen},
  {Fujishiro}, {Fujiwara}, {Hasegawa}, {Ita}, {Kim}, {Matsuhara}, {Murakami},
  {M{\"u}ller}, {Nakagawa}, {Ohyama}, {Oyabu}, {Pyo}, {Sakon}, {Shibai},
  {Takita}, {Tanab{\'e}}, {Uemizu}, {Ueno}, {Usui}, {Wada}, {Watarai},
  {Yamamura}, \& {Yamauchi}}]{AKARI+2}
{Ishihara}, D., {Onaka}, T., {Kataza}, H., {et~al.} 2010, \aap, 514, A1

\bibitem[{{Jones} {et~al.}(2016){Jones}, {White}, {Quinn}, {Ireland},
  {Boyajian}, {Schaefer}, \& {Baines}}]{Jones+2016}
{Jones}, J., {White}, R.~J., {Quinn}, S., {et~al.} 2016, \apjl, 822, L3

\bibitem[{{Kammerer} {et~al.}(2022){Kammerer}, {Girard}, {Carter}, {Perrin},
  {Cooper}, {Thatte}, {Vandal}, {Leisenring}, {Wang}, {Balmer},
  {Sivaramakrishnan}, {Pueyo}, {Ward-Duong}, {Sunnquist}, \& {Adams
  Redai}}]{Kammerer+2022}
{Kammerer}, J., {Girard}, J., {Carter}, A.~L., {et~al.} 2022, in Society of
  Photo-Optical Instrumentation Engineers (SPIE) Conference Series, Vol. 12180,
  Space Telescopes and Instrumentation 2022: Optical, Infrared, and Millimeter
  Wave, ed. L.~E. {Coyle}, S.~{Matsuura}, \& M.~D. {Perrin}, 121803N

\bibitem[{{Karalidi} {et~al.}(2021){Karalidi}, {Marley}, {Fortney}, {Morley},
  {Saumon}, {Lupu}, {Visscher}, \& {Freedman}}]{Karalidi+2021}
{Karalidi}, T., {Marley}, M., {Fortney}, J.~J., {et~al.} 2021, \apj, 923, 269

\bibitem[{{Kirkpatrick}(2005)}]{Kirkpatrick-2005}
{Kirkpatrick}, J.~D. 2005, \araa, 43, 195

\bibitem[{{Kirkpatrick} {et~al.}(1999){Kirkpatrick}, {Reid}, {Liebert},
  {Cutri}, {Nelson}, {Beichman}, {Dahn}, {Monet}, {Gizis}, \&
  {Skrutskie}}]{Kirkpatrick+1999}
{Kirkpatrick}, J.~D., {Reid}, I.~N., {Liebert}, J., {et~al.} 1999, \apj, 519,
  802

\bibitem[{{K{\"u}hn} {et~al.}(2018){K{\"u}hn}, {Serabyn}, {Lozi}, {Jovanovic},
  {Currie}, {Guyon}, {Kudo}, {Martinache}, {Liewer}, {Singh}, {Tamura},
  {Mawet}, {Hagelberg}, \& {Defrere}}]{Kuhn+2018}
{K{\"u}hn}, J., {Serabyn}, E., {Lozi}, J., {et~al.} 2018, \pasp, 130, 035001

\bibitem[{{Kurucz}(2013)}]{ATLAS12}
{Kurucz}, R.~L. 2013, {ATLAS12: Opacity sampling model atmosphere program},
  Astrophysics Source Code Library, record ascl:1303.024

\bibitem[{{Kurucz}(2014)}]{ATLAS12+code}
{Kurucz}, R.~L. 2014, in Determination of Atmospheric Parameters of B, ed.
  E.~{Niemczura}, B.~{Smalley}, \& W.~{Pych}, 39--51

\bibitem[{{Lagrange} {et~al.}(2010){Lagrange}, {Bonnefoy}, {Chauvin}, {Apai},
  {Ehrenreich}, {Boccaletti}, {Gratadour}, {Rouan}, {Mouillet}, {Lacour}, \&
  {Kasper}}]{Lagrange+2010}
{Lagrange}, A.~M., {Bonnefoy}, M., {Chauvin}, G., {et~al.} 2010, Science, 329,
  57

\bibitem[{{Lagrange} {et~al.}(2009){Lagrange}, {Gratadour}, {Chauvin}, {Fusco},
  {Ehrenreich}, {Mouillet}, {Rousset}, {Rouan}, {Allard}, {Gendron}, {Charton},
  {Mugnier}, {Rabou}, {Montri}, \& {Lacombe}}]{Lagrange+2009}
{Lagrange}, A.~M., {Gratadour}, D., {Chauvin}, G., {et~al.} 2009, \aap, 493,
  L21

\bibitem[{{Lajoie} {et~al.}(2016){Lajoie}, {Soummer}, {Pueyo}, {Hines},
  {Nelan}, {Perrin}, {Clampin}, \& {Isaacs}}]{Lajoie+2016}
{Lajoie}, C.-P., {Soummer}, R., {Pueyo}, L., {et~al.} 2016, in Society of
  Photo-Optical Instrumentation Engineers (SPIE) Conference Series, Vol. 9904,
  Space Telescopes and Instrumentation 2016: Optical, Infrared, and Millimeter
  Wave, ed. H.~A. {MacEwen}, G.~G. {Fazio}, M.~{Lystrup}, N.~{Batalha},
  N.~{Siegler}, \& E.~C. {Tong}, 99045K

\bibitem[{{Larkin} {et~al.}(2006){Larkin}, {Barczys}, {Krabbe}, {Adkins},
  {Aliado}, {Amico}, {Brims}, {Campbell}, {Canfield}, {Gasaway}, {Honey},
  {Iserlohe}, {Johnson}, {Kress}, {LaFreniere}, {Magnone}, {Magnone},
  {McElwain}, {Moon}, {Quirrenbach}, {Skulason}, {Song}, {Spencer}, {Weiss}, \&
  {Wright}}]{Larkin+2006}
{Larkin}, J., {Barczys}, M., {Krabbe}, A., {et~al.} 2006, \nar, 50, 362

\bibitem[{{Lebzelter} {et~al.}(2023){Lebzelter}, {Mowlavi}, {Lecoeur-Taibi},
  {Trabucchi}, {Audard}, {Garc{\'\i}a-Lario}, {Gavras}, {Holl}, {Jevardat de
  Fombelle}, {Nienartowicz}, {Rimoldini}, \& {Eyer}}]{Gaia-variables+2023}
{Lebzelter}, T., {Mowlavi}, N., {Lecoeur-Taibi}, I., {et~al.} 2023, \aap, 674,
  A15

\bibitem[{{Leggett} {et~al.}(2010){Leggett}, {Burningham}, {Saumon}, {Marley},
  {Warren}, {Smart}, {Jones}, {Lucas}, {Pinfield}, \& {Tamura}}]{Leggett+2010L}
{Leggett}, S.~K., {Burningham}, B., {Saumon}, D., {et~al.} 2010, \apj, 710,
  1627

\bibitem[{{Leggett} {et~al.}(2002){Leggett}, {Golimowski}, {Fan}, {Geballe},
  {Knapp}, {Brinkmann}, {Csabai}, {Gunn}, {Hawley}, {Henry}, {Hindsley},
  {Ivezi{\'c}}, {Lupton}, {Pier}, {Schneider}, {Smith}, {Strauss}, {Uomoto}, \&
  {York}}]{Leggett+2002}
{Leggett}, S.~K., {Golimowski}, D.~A., {Fan}, X., {et~al.} 2002, \apj, 564, 452

\bibitem[{{Linder} {et~al.}(2019){Linder}, {Mordasini}, {Molli{\`e}re},
  {Marleau}, {Malik}, {Quanz}, \& {Meyer}}]{Linder+2019}
{Linder}, E.~F., {Mordasini}, C., {Molli{\`e}re}, P., {et~al.} 2019, \aap, 623,
  A85

\bibitem[{{Lodders} \& {Fegley}(2002)}]{Lodders+2002}
{Lodders}, K. \& {Fegley}, B. 2002, \icarus, 155, 393

\bibitem[{{Lodieu} {et~al.}(2018){Lodieu}, {Zapatero Osorio}, {B{\'e}jar}, \&
  {Pe{\~n}a Ram{\'\i}rez}}]{Lodieu+2018}
{Lodieu}, N., {Zapatero Osorio}, M.~R., {B{\'e}jar}, V.~J.~S., \& {Pe{\~n}a
  Ram{\'\i}rez}, K. 2018, \mnras, 473, 2020

\bibitem[{{M{\^a}lin} {et~al.}(2025){M{\^a}lin}, {Boccaletti}, {Perrot},
  {Baudoz}, {Rouan}, {Lagage}, {Waters}, {G{\"u}del}, {Henning},
  {Vandenbussche}, {Absil}, {Barrado}, {Charnay}, {Choquet}, {Cossou},
  {Danielski}, {Decin}, {Glauser}, {Pye}, {Olofsson}, {Glasse}, {Patapis},
  {Royer}, {Scheithauer}, {Serabyn}, {Tremblin}, {Whiteford}, {van Dishoeck},
  {Ostlin}, {Ray}, \& {Wright}}]{Malin+2025}
{M{\^a}lin}, M., {Boccaletti}, A., {Perrot}, C., {et~al.} 2025, \aap, 693, A315

\bibitem[{{M{\^a}lin} {et~al.}(2024){M{\^a}lin}, {Boccaletti}, {Perrot},
  {Baudoz}, {Rouan}, {Lagage}, {Waters}, {G{\"u}del}, {Henning},
  {Vandenbussche}, {Absil}, {Barrado}, {Cossou}, {Decin}, {Glauser}, {Pye},
  {Olofsson}, {Glasse}, {Lahuis}, {Patapis}, {Royer}, {Scheithauer},
  {Whiteford}, {Serabyn}, {Choquet}, {Colina}, {Ostlin}, {Ray}, \&
  {Wright}}]{Malin+2024}
{M{\^a}lin}, M., {Boccaletti}, A., {Perrot}, C., {et~al.} 2024, \aap, 690, A316

\bibitem[{{Marleau} \& {Cumming}(2014)}]{Marleau+Cumming-2014}
{Marleau}, G.~D. \& {Cumming}, A. 2014, \mnras, 437, 1378

\bibitem[{{Marley} {et~al.}(2012){Marley}, {Saumon}, {Cushing}, {Ackerman},
  {Fortney}, \& {Freedman}}]{Marley+2012}
{Marley}, M.~S., {Saumon}, D., {Cushing}, M., {et~al.} 2012, \apj, 754, 135

\bibitem[{{Marley} {et~al.}(2021){Marley}, {Saumon}, {Visscher}, {Lupu},
  {Freedman}, {Morley}, {Fortney}, {Seay}, {Smith}, {Teal}, \&
  {Wang}}]{Marley+2021-son}
{Marley}, M.~S., {Saumon}, D., {Visscher}, C., {et~al.} 2021, \apj, 920, 85

\bibitem[{{Marley} {et~al.}(2002){Marley}, {Seager}, {Saumon}, {Lodders},
  {Ackerman}, {Freedman}, \& {Fan}}]{Marley+2002}
{Marley}, M.~S., {Seager}, S., {Saumon}, D., {et~al.} 2002, \apj, 568, 335

\bibitem[{{Marois} {et~al.}(2008){Marois}, {Macintosh}, {Barman}, {Zuckerman},
  {Song}, {Patience}, {Lafreni{\`e}re}, \& {Doyon}}]{Marois+2008}
{Marois}, C., {Macintosh}, B., {Barman}, T., {et~al.} 2008, Science, 322, 1348

\bibitem[{{Matthews} {et~al.}(2024){Matthews}, {Carter}, {Pathak}, {Morley},
  {Phillips}, {P.~M.}, {Feng}, {Bonse}, {Boogaard}, {Burt}, {Crossfield},
  {Douglas}, {Henning}, {Hom}, {Ko}, {Kasper}, {Lagrange}, {Petit dit de la
  Roche}, \& {Philipot}}]{Matthews+2024}
{Matthews}, E.~C., {Carter}, A.~L., {Pathak}, P., {et~al.} 2024, \nat, 633, 789

\bibitem[{{Mawet} {et~al.}(2014){Mawet}, {Milli}, {Wahhaj}, {Pelat}, {Absil},
  {Delacroix}, {Boccaletti}, {Kasper}, {Kenworthy}, {Marois}, {Mennesson}, \&
  {Pueyo}}]{Mawet+2014}
{Mawet}, D., {Milli}, J., {Wahhaj}, Z., {et~al.} 2014, \apj, 792, 97

\bibitem[{{Mermilliod}(1997)}]{Mermilliod-1997}
{Mermilliod}, J.~C. 1997, {VizieR Online Data Catalog: Homogeneous Means in the
  UBV System (Mermilliod 1991)}, VizieR On-line Data Catalog: II/168.
  Originally published in: Institut d'Astronomie, Universite de Lausanne (1991)

\bibitem[{{Mesa} {et~al.}(2018){Mesa}, {Baudino}, {Charnay}, {D'Orazi},
  {Desidera}, {Boccaletti}, {Gratton}, {Bonnefoy}, {Delorme}, {Langlois},
  {Vigan}, {Zurlo}, {Maire}, {Janson}, {Antichi}, {Baruffolo}, {Bruno},
  {Cascone}, {Chauvin}, {Claudi}, {De Caprio}, {Fantinel}, {Farisato}, {Feldt},
  {Giro}, {Hagelberg}, {Incorvaia}, {Lagadec}, {Lagrange}, {Lazzoni}, {Lessio},
  {Salasnich}, {Scuderi}, {Sissa}, \& {Turatto}}]{Mesa+2018}
{Mesa}, D., {Baudino}, J.~L., {Charnay}, B., {et~al.} 2018, \aap, 612, A92

\bibitem[{{Miles} {et~al.}(2023){Miles}, {Biller}, {Patapis}, {Worthen},
  {Rickman}, {Hoch}, {Skemer}, {Perrin}, {Whiteford}, {Chen}, {Sargent},
  {Mukherjee}, {Morley}, {Moran}, {Bonnefoy}, {Petrus}, {Carter}, {Choquet},
  {Hinkley}, {Ward-Duong}, {Leisenring}, {Millar-Blanchaer}, {Pueyo}, {Ray},
  {Sallum}, {Stapelfeldt}, {Stone}, {Wang}, {Absil}, {Balmer}, {Boccaletti},
  {Bonavita}, {Booth}, {Bowler}, {Chauvin}, {Christiaens}, {Currie},
  {Danielski}, {Fortney}, {Girard}, {Grady}, {Greenbaum}, {Henning}, {Hines},
  {Janson}, {Kalas}, {Kammerer}, {Kennedy}, {Kenworthy}, {Kervella}, {Lagage},
  {Lew}, {Liu}, {Macintosh}, {Marino}, {Marley}, {Marois}, {Matthews},
  {Matthews}, {Mawet}, {McElwain}, {Metchev}, {Meyer}, {Molliere}, {Pantin},
  {Quirrenbach}, {Rebollido}, {Ren}, {Schneider}, {Vasist}, {Wyatt}, {Zhou},
  {Briesemeister}, {Bryan}, {Calissendorff}, {Cantalloube}, {Cugno}, {De
  Furio}, {Dupuy}, {Factor}, {Faherty}, {Fitzgerald}, {Franson}, {Gonzales},
  {Hood}, {Howe}, {Kraus}, {Kuzuhara}, {Lagrange}, {Lawson}, {Lazzoni}, {Liu},
  {Llop-Sayson}, {Lloyd}, {Martinez}, {Mazoyer}, {Quanz}, {Redai}, {Samland},
  {Schlieder}, {Tamura}, {Tan}, {Uyama}, {Vigan}, {Vos}, {Wagner}, {Wolff},
  {Ygouf}, {Zhang}, {Zhang}, \& {Zhang}}]{Miles+2023}
{Miles}, B.~E., {Biller}, B.~A., {Patapis}, P., {et~al.} 2023, \apjl, 946, L6

\bibitem[{{Morley} {et~al.}(2012){Morley}, {Fortney}, {Marley}, {Visscher},
  {Saumon}, \& {Leggett}}]{Morley+2012}
{Morley}, C.~V., {Fortney}, J.~J., {Marley}, M.~S., {et~al.} 2012, \apj, 756,
  172

\bibitem[{{Morley} {et~al.}(2014){Morley}, {Marley}, {Fortney}, \&
  {Lupu}}]{Morley+2014}
{Morley}, C.~V., {Marley}, M.~S., {Fortney}, J.~J., \& {Lupu}, R. 2014, \apjl,
  789, L14

\bibitem[{{Morris} {et~al.}(2024){Morris}, {Wang}, {Hsu}, {Ruffio}, {Xuan},
  {Delorme}, {Hood}, {Bryan}, {Martin}, {Pezzato}, {Mawet}, {Skemer}, {Baker},
  {Bartos}, {Calvin}, {Cetre}, {Doppmann}, {Echeverri}, {Finnerty},
  {Fitzgerald}, {Jovanovic}, {Liberman}, {Lopez}, {Sappey}, {Schofield},
  {Wallace}, \& {Wang}}]{Morris+2024}
{Morris}, E.~C., {Wang}, J.~J., {Hsu}, C.-C., {et~al.} 2024, \aj, 168, 144

\bibitem[{{Murakami} {et~al.}(2007){Murakami}, {Baba}, {Barthel}, {Clements},
  {Cohen}, {Doi}, {Enya}, {Figueredo}, {Fujishiro}, {Fujiwara}, {Fujiwara},
  {Garcia-Lario}, {Goto}, {Hasegawa}, {Hibi}, {Hirao}, {Hiromoto}, {Hong},
  {Imai}, {Ishigaki}, {Ishiguro}, {Ishihara}, {Ita}, {Jeong}, {Jeong},
  {Kaneda}, {Kataza}, {Kawada}, {Kawai}, {Kawamura}, {Kessler}, {Kester},
  {Kii}, {Kim}, {Kim}, {Kobayashi}, {Koo}, {Kwon}, {Lee}, {Lorente}, {Makiuti},
  {Matsuhara}, {Matsumoto}, {Matsuo}, {Matsuura}, {M{\"U}ller}, {Murakami},
  {Nagata}, {Nakagawa}, {Naoi}, {Narita}, {Noda}, {Oh}, {Ohnishi}, {Ohyama},
  {Okada}, {Okuda}, {Oliver}, {Onaka}, {Ootsubo}, {Oyabu}, {Pak}, {Park},
  {Pearson}, {Rowan-Robinson}, {Saito}, {Sakon}, {Salama}, {Sato}, {Savage},
  {Serjeant}, {Shibai}, {Shirahata}, {Sohn}, {Suzuki}, {Takagi}, {Takahashi},
  {Tanab{\'E}}, {Takeuchi}, {Takita}, {Thomson}, {Uemizu}, {Ueno}, {Usui},
  {Verdugo}, {Wada}, {Wang}, {Watabe}, {Watarai}, {White}, {Yamamura},
  {Yamauchi}, \& {Yasuda}}]{AKARI+1}
{Murakami}, H., {Baba}, H., {Barthel}, P., {et~al.} 2007, \pasj, 59, S369

\bibitem[{{Nasa / Esa /}(2004)}]{IEU}
{Nasa / Esa /}, S. 2004, {VizieR Online Data Catalog: Final Merged Log of IUE
  Observations (NASA-ESA, 2000)}, VizieR On-line Data Catalog: VI/110.
  Originally published in: IUE Newsletters (1978-2000)

\bibitem[{{Nasedkin} {et~al.}(2023){Nasedkin}, {Molli{\`e}re}, {Wang},
  {Cantalloube}, {Kreidberg}, {Pueyo}, {Stolker}, \& {Vigan}}]{Nasedkin+2023}
{Nasedkin}, E., {Molli{\`e}re}, P., {Wang}, J., {et~al.} 2023, \aap, 678, A41

\bibitem[{{Nielsen} {et~al.}(2019){Nielsen}, {De Rosa}, {Macintosh}, {Wang},
  {Ruffio}, {Chiang}, {Marley}, {Saumon}, {Savransky}, {Ammons}, {Bailey},
  {Barman}, {Blain}, {Bulger}, {Burrows}, {Chilcote}, {Cotten}, {Czekala},
  {Doyon}, {Duch{\^e}ne}, {Esposito}, {Fabrycky}, {Fitzgerald}, {Follette},
  {Fortney}, {Gerard}, {Goodsell}, {Graham}, {Greenbaum}, {Hibon}, {Hinkley},
  {Hirsch}, {Hom}, {Hung}, {Dawson}, {Ingraham}, {Kalas}, {Konopacky},
  {Larkin}, {Lee}, {Lin}, {Maire}, {Marchis}, {Marois}, {Metchev},
  {Millar-Blanchaer}, {Morzinski}, {Oppenheimer}, {Palmer}, {Patience},
  {Perrin}, {Poyneer}, {Pueyo}, {Rafikov}, {Rajan}, {Rameau}, {Rantakyr{\"o}},
  {Ren}, {Schneider}, {Sivaramakrishnan}, {Song}, {Soummer}, {Tallis},
  {Thomas}, {Ward-Duong}, \& {Wolff}}]{Nielsen+2019}
{Nielsen}, E.~L., {De Rosa}, R.~J., {Macintosh}, B., {et~al.} 2019, \aj, 158,
  13

\bibitem[{{{\"O}berg} {et~al.}(2011){{\"O}berg}, {Murray-Clay}, \&
  {Bergin}}]{Oberg+2011}
{{\"O}berg}, K.~I., {Murray-Clay}, R., \& {Bergin}, E.~A. 2011, \apjl, 743, L16

\bibitem[{{Ochsenbein} {et~al.}(2000){Ochsenbein}, {Bauer}, \&
  {Marcout}}]{Vizier-data}
{Ochsenbein}, F., {Bauer}, P., \& {Marcout}, J. 2000, \aaps, 143, 23

\bibitem[{{Onaka} {et~al.}(2007){Onaka}, {Matsuhara}, {Wada}, {Fujishiro},
  {Fujiwara}, {Ishigaki}, {Ishihara}, {Ita}, {Kataza}, {Kim}, {Matsumoto},
  {Murakami}, {Ohyama}, {Oyabu}, {Sakon}, {Tanab{\'e}}, {Takagi}, {Uemizu},
  {Ueno}, {Usui}, {Watarai}, {Cohen}, {Enya}, {Ootsubo}, {Pearson}, {Takeyama},
  {Yamamuro}, \& {Ikeda}}]{AKARI+0}
{Onaka}, T., {Matsuhara}, H., {Wada}, T., {et~al.} 2007, \pasj, 59, S401

\bibitem[{{Oppenheimer} {et~al.}(2012){Oppenheimer}, {Beichman}, {Brenner},
  {Burruss}, {Cady}, {Crepp}, {Hillenbrand}, {Hinkley}, {Ligon}, {Lockhart},
  {Parry}, {Pueyo}, {Rice}, {Roberts}, {Roberts}, {Shao}, {Sivaramakrishnan},
  {Soummer}, {Vasisht}, {Vescelus}, {Wallace}, {Zhai}, \&
  {Zimmerman}}]{Oppenheimer+2012}
{Oppenheimer}, B.~R., {Beichman}, C., {Brenner}, D., {et~al.} 2012, in Society
  of Photo-Optical Instrumentation Engineers (SPIE) Conference Series, Vol.
  8447, Adaptive Optics Systems III, ed. B.~L. {Ellerbroek}, E.~{Marchetti}, \&
  J.-P. {V{\'e}ran}, 844720

\bibitem[{{Paunzen}(2015)}]{Paunzen+2015P}
{Paunzen}, E. 2015, \aap, 580, A23

\bibitem[{{Peters-Limbach} {et~al.}(2012){Peters-Limbach}, {Groff}, {Kasdin},
  {McElwain}, {Galvin}, {Carr}, {Lupton}, {Gunn}, {Knapp}, {Gong}, {Carlotti},
  {Brandt}, {Janson}, {Guyon}, {Martinache}, {Hayashi}, \&
  {Takato}}]{Peters+2012}
{Peters-Limbach}, M.~A., {Groff}, T., {Kasdin}, N.~J., {et~al.} 2012, in
  Society of Photo-Optical Instrumentation Engineers (SPIE) Conference Series,
  Vol. 8446, Ground-based and Airborne Instrumentation for Astronomy IV, ed.
  I.~S. {McLean}, S.~K. {Ramsay}, \& H.~{Takami}, 84467U

\bibitem[{{Petrus} {et~al.}(2023){Petrus}, {Chauvin}, {Bonnefoy}, {Tremblin},
  {Charnay}, {Delorme}, {Marleau}, {Bayo}, {Manjavacas}, {Lagrange},
  {Molli{\`e}re}, {Palma-Bifani}, {Biller}, {Jenkins}, {Goyal}, \&
  {Hoch}}]{Petrus+2023}
{Petrus}, S., {Chauvin}, G., {Bonnefoy}, M., {et~al.} 2023, \aap, 670, L9

\bibitem[{{Phillips} {et~al.}(2020){Phillips}, {Tremblin}, {Baraffe},
  {Chabrier}, {Allard}, {Spiegelman}, {Goyal}, {Drummond}, \&
  {H{\'e}brard}}]{Phillips+2020}
{Phillips}, M.~W., {Tremblin}, P., {Baraffe}, I., {et~al.} 2020, \aap, 637, A38

\bibitem[{{Pickles}(1998)}]{Pickles-1998}
{Pickles}, A.~J. 1998, \pasp, 110, 863

\bibitem[{{Radigan}(2014)}]{Radigan+2014b}
{Radigan}, J. 2014, \apj, 797, 120

\bibitem[{{Rajan} {et~al.}(2017){Rajan}, {Rameau}, {De Rosa}, {Marley},
  {Graham}, {Macintosh}, {Marois}, {Morley}, {Patience}, {Pueyo}, {Saumon},
  {Ward-Duong}, {Ammons}, {Arriaga}, {Bailey}, {Barman}, {Bulger}, {Burrows},
  {Chilcote}, {Cotten}, {Czekala}, {Doyon}, {Duch{\^e}ne}, {Esposito},
  {Fitzgerald}, {Follette}, {Fortney}, {Goodsell}, {Greenbaum}, {Hibon},
  {Hung}, {Ingraham}, {Johnson-Groh}, {Kalas}, {Konopacky}, {Lafreni{\`e}re},
  {Larkin}, {Maire}, {Marchis}, {Metchev}, {Millar-Blanchaer}, {Morzinski},
  {Nielsen}, {Oppenheimer}, {Palmer}, {Patel}, {Perrin}, {Poyneer},
  {Rantakyr{\"o}}, {Ruffio}, {Savransky}, {Schneider}, {Sivaramakrishnan},
  {Song}, {Soummer}, {Thomas}, {Vasisht}, {Wallace}, {Wang}, {Wiktorowicz}, \&
  {Wolff}}]{Rajan+2017}
{Rajan}, A., {Rameau}, J., {De Rosa}, R.~J., {et~al.} 2017, \aj, 154, 10

\bibitem[{{Ressler} {et~al.}(2015){Ressler}, {Sukhatme}, {Franklin}, {Mahoney},
  {Thelen}, {Bouchet}, {Colbert}, {Cracraft}, {Dicken}, {Gastaud}, {Goodson},
  {Eccleston}, {Moreau}, {Rieke}, \& {Schneider}}]{Ressler+2015}
{Ressler}, M.~E., {Sukhatme}, K.~G., {Franklin}, B.~R., {et~al.} 2015, \pasp,
  127, 675

\bibitem[{{Rieke} {et~al.}(2015){Rieke}, {Ressler}, {Morrison}, {Bergeron},
  {Bouchet}, {Garc{\'\i}a-Mar{\'\i}n}, {Greene}, {Regan}, {Sukhatme}, \&
  {Walker}}]{Rieke+2015}
{Rieke}, G.~H., {Ressler}, M.~E., {Morrison}, J.~E., {et~al.} 2015, \pasp, 127,
  665

\bibitem[{{Rigby} {et~al.}(2023){Rigby}, {Perrin}, {McElwain}, {Kimble},
  {Friedman}, {Lallo}, {Doyon}, {Feinberg}, {Ferruit}, {Glasse}, {Rieke},
  {Rieke}, {Wright}, {Willott}, {Colon}, {Milam}, {Neff}, {Stark}, {Valenti},
  {Abell}, {Abney}, {Abul-Huda}, {Acton}, {Adams}, {Adler}, {Aguilar}, {Ahmed},
  {Albert}, {Alberts}, {Aldridge}, {Allen}, {Altenburg},
  {{\'A}lvarez-M{\'a}rquez}, {Alves de Oliveira}, {Andersen}, {Anderson},
  {Anderson}, {Argyriou}, {Armstrong}, {Arribas}, {Artigau}, {Arvai},
  {Atkinson}, {Bacon}, {Bair}, {Banks}, {Barrientes}, {Barringer}, {Bartosik},
  {Bast}, {Baudoz}, {Beatty}, {Bechtold}, {Beck}, {Bergeron}, {Bergkoetter},
  {Bhatawdekar}, {Birkmann}, {Blazek}, {Blome}, {Boccaletti}, {B{\"o}ker},
  {Boia}, {Bonaventura}, {Bond}, {Bosley}, {Boucarut}, {Bourque}, {Bouwman},
  {Bower}, {Bowers}, {Boyer}, {Bradley}, {Brady}, {Braun}, {Breda},
  {Bresnahan}, {Bright}, {Britt}, {Bromenschenkel}, {Brooks}, {Brooks},
  {Brown}, {Brown}, {Brown}, {Bunker}, {Burger}, {Bushouse}, {Cale}, {Cameron},
  {Cameron}, {Canipe}, {Caplinger}, {Caputo}, {Cara}, {Carey}, {Carniani},
  {Carrasquilla}, {Carruthers}, {Case}, {Catherine}, {Chance}, {Chapman},
  {Charlot}, {Charlow}, {Chayer}, {Chen}, {Cherinka}, {Chichester}, {Chilton},
  {Chonis}, {Clampin}, {Clark}, {Clark}, {Coe}, {Coleman}, {Comber}, {Comeau},
  {Connolly}, {Cooper}, {Cooper}, {Coppock}, {Correnti}, {Cossou}, {Coulais},
  {Coyle}, {Cracraft}, {Curti}, {Cuturic}, {Davis}, {Davis}, {Dean}, {DeLisa},
  {deMeester}, {Dencheva}, {Dencheva}, {DePasquale}, {Deschenes}, {Hunor
  Detre}, {Diaz}, {Dicken}, {DiFelice}, {Dillman}, {Dixon}, {Doggett},
  {Donaldson}, {Douglas}, {DuPrie}, {Dupuis}, {Durning}, {Easmin}, {Eck},
  {Edeani}, {Egami}, {Ehrenwinkler}, {Eisenhamer}, {Eisenhower}, {Elie},
  {Elliott}, {Elliott}, {Ellis}, {Engesser}, {Espinoza}, {Etienne}, {Etxaluze},
  {Falini}, {Feeney}, {Ferry}, {Filippazzo}, {Fincham}, {Fix}, {Flagey},
  {Florian}, {Flynn}, {Fontanella}, {Ford}, {Forshay}, {Fox}, {Franz}, {Fu},
  {Fullerton}, {Galkin}, {Galyer}, {Garc{\'\i}a Mar{\'\i}n}, {Gardner},
  {Gardner}, {Garland}, {Garrett}, {Gasman}, {Gaspar}, {Gaudreau}, {Gauthier},
  {Geers}, {Geithner}, {Gennaro}, {Giardino}, {Girard}, {Giuliano},
  {Glassmire}, {Glauser}, {Glazer}, {Godfrey}, {Golimowski}, {Gollnitz},
  {Gong}, {Gonzaga}, {Gordon}, {Gordon}, {Goudfrooij}, {Greene}, {Greenhouse},
  {Grimaldi}, {Groebner}, {Grundy}, {Guillard}, {Gutman}, {Ha}, {Haderlein},
  {Hagedorn}, {Hainline}, {Haley}, {Hami}, {Hamilton}, {Hammel}, {Hansen},
  {Harkins}, {Harr}, {Hart}, {Hart}, {Hartig}, {Hashimoto}, {Haskins},
  {Hathaway}, {Havey}, {Hayden}, {Hecht}, {Heller-Boyer}, {Henriques}, {Henry},
  {Hermann}, {Hernandez}, {Hesman}, {Hicks}, {Hilbert}, {Hines}, {Hoffman},
  {Holfeltz}, {Holler}, {Hoppa}, {Hott}, {Howard}, {Howard}, {Hunter},
  {Hunter}, {Hurst}, {Husemann}, {Hustak}, {Ilinca Ignat}, {Illingworth},
  {Irish}, {Jackson}, {Jahromi}, {Jakobsen}, {James}, {James}, {Januszewski},
  {Jenkins}, {Jirdeh}, {Johnson}, {Johnson}, {Jones}, {Jones}, {Jones},
  {Jones}, {Jordan}, {Jordan}, {Jurczyk}, {Jurling}, {Kaleida}, {Kalmanson},
  {Kammerer}, {Kang}, {Kao}, {Karakla}, {Kavanagh}, {Kelly}, {Kendrew},
  {Kennedy}, {Kenny}, {Keski-kuha}, {Keyes}, {Kidwell}, {Kinzel}, {Kirk},
  {Kirkpatrick}, {Kirshenblat}, {Klaassen}, {Knapp}, {Knight}, {Knollenberg},
  {Koehler}, {Koekemoer}, {Kovacs}, {Kulp}, {Kumari}, {Kyprianou}, {La Massa},
  {Labador}, {Labiano}, {Lagage}, {Lajoie}, {Lallo}, {Lam}, {Lamb}, {Lambros},
  {Lampenfield}, {Langston}, {Larson}, {Law}, {Lawrence}, {Lee}, {Leisenring},
  {Lepo}, {Leveille}, {Levenson}, {Levine}, {Levy}, {Lewis}, {Lewis},
  {Libralato}, {Lightsey}, {Link}, {Liu}, {Lo}, {Lockwood}, {Logue}, {Long},
  {Long}, {Loomis}, {Lopez-Caniego}, {Lorenzo Alvarez}, {Love-Pruitt}, {Lucy},
  {Luetzgendorf}, {Maghami}, {Maiolino}, {Major}, {Malla}, {Malumuth},
  {Manjavacas}, {Mannfolk}, {Marrione}, {Marston}, {Martel}, {Maschmann},
  {Masci}, {Masciarelli}, {Maszkiewicz}, {Mather}, {McKenzie}, {McLean},
  {McMaster}, {Melbourne}, {Mel{\'e}ndez}, {Menzel}, {Merz}, {Meyett}, {Meza},
  {Miskey}, {Misselt}, {Moller}, {Morrison}, {Morse}, {Moseley}, {Mosier},
  {Mountain}, {Mueckay}, {Mueller}, {Mullally}, {Murphy}, {Murray}, {Murray},
  {Mustelier}, {Muzerolle}, {Mycroft}, {Myers}, {Myrick}, {Nanavati}, {Nance},
  {Nayak}, {Naylor}, {Nelan}, {Nickson}, {Nielson}, {Nieto-Santisteban},
  {Nikolov}, {Noriega-Crespo}, {O'Shaughnessy}, {O'Sullivan}, {Ochs}, {Ogle},
  {Oleszczuk}, {Olmsted}, {Osborne}, {Ottens}, {Owens}, {Pacifici}, {Pagan},
  {Page}, {Park}, {Parrish}, {Patapis}, {Paul}, {Pauly}, {Pavlovsky}, {Pedder},
  {Peek}, {Pena-Guerrero}, {Penanen}, {Perez}, {Perna}, {Perriello},
  {Phillips}, {Pietraszkiewicz}, {Pinaud}, {Pirzkal}, {Pitman}, {Piwowar},
  {Platais}, {Player}, {Plesha}, {Pollizi}, {Polster}, {Pontoppidan},
  {Porterfield}, {Proffitt}, {Pueyo}, {Pulliam}, {Quirt}, {Quispe Neira},
  {Ramos Alarcon}, {Ramsay}, {Rapp}, {Rapp}, {Rauscher}, {Ravindranath},
  {Rawle}, {Regan}, {Reichard}, {Reis}, {Ressler}, {Rest}, {Reynolds}, {Rhue},
  {Richon}, {Rickman}, {Ridgaway}, {Ritchie}, {Rix}, {Robberto}, {Robinson},
  {Robinson}, {Robinson}, {Rock}, {Rodriguez}, {Rodriguez Del Pino}, {Roellig},
  {Rohrbach}, {Roman}, {Romelfanger}, {Rose}, {Roteliuk}, {Roth}, {Rothwell},
  {Rowlands}, {Roy}, {Royer}, {Royle}, {Rui}, {Rumler}, {Runnels}, {Russ},
  {Rustamkulov}, {Ryden}, {Ryer}, {Sabata}, {Sabatke}, {Sabbi}, {Samuelson},
  {Sapp}, {Sappington}, {Sargent}, {Sauer}, {Scheithauer}, {Schlawin},
  {Schlitz}, {Schmitz}, {Schneider}, {Schreiber}, {Schulze}, {Schwab}, {Scott},
  {Sembach}, {Shanahan}, {Shaughnessy}, {Shaw}, {Shawger}, {Shay}, {Sheehan},
  {Shen}, {Sherman}, {Shiao}, {Shih}, {Shivaei}, {Sienkiewicz}, {Sing},
  {Sirianni}, {Sivaramakrishnan}, {Skipper}, {Sloan}, {Slocum}, {Slowinski},
  {Smith}, {Smith}, {Smith}, {Smith}, {Snyder}, {Soh}, {Sohn}, {Soto},
  {Spencer}, {Stallcup}, {Stansberry}, {Starr}, {Starr}, {Stewart},
  {Stiavelli}, {Straughn}, {Strickland}, {Stys}, {Summers}, {Sun}, {Sunnquist},
  {Swade}, {Swam}, {Swaters}, {Swoish}, {Taylor}, {Taylor}, {Te Plate}, {Tea},
  {Teague}, {Telfer}, {Temim}, {Thatte}, {Thompson}, {Thompson}, {Thomson},
  {Tikkanen}, {Tippet}, {Todd}, {Toolan}, {Tran}, {Trejo}, {Truong},
  {Tsukamoto}, {Tustain}, {Tyra}, {Ubeda}, {Underwood}, {Uzzo}, {Van Campen},
  {Vandal}, {Vandenbussche}, {Vila}, {Volk}, {Wahlgren}, {Waldman}, {Walker},
  {Wander}, {Warfield}, {Warner}, {Wasiak}, {Watkins}, {Weaver}, {Weilert},
  {Weiser}, {Weiss}, {Weissman}, {Welty}, {West}, {Wheate}, {Wheatley},
  {Wheeler}, {White}, {Whiteaker}, {Whitehouse}, {Whiteleather}, {Whitman},
  {Williams}, {Willmer}, {Willoughby}, {Wilson}, {Wirth}, {Wislowski}, {Wolf},
  {Wolfe}, {Wolff}, {Workman}, {Wright}, {Wu}, {Wu}, {Wymer}, {Yates},
  {Yeager}, {Yeates}, {Yerger}, {Yoon}, {Young}, {Yu}, {Zak}, {Zeidler},
  {Zhou}, {Zielinski}, {Zincke}, \& {Zonak}}]{Rigby+2023}
{Rigby}, J., {Perrin}, M., {McElwain}, M., {et~al.} 2023, \pasp, 135, 048001

\bibitem[{{Rodrigo} \& {Solano}(2020)}]{Rodrigo+2020}
{Rodrigo}, C. \& {Solano}, E. 2020, in XIV.0 Scientific Meeting (virtual) of
  the Spanish Astronomical Society, 182

\bibitem[{{Rodrigo} {et~al.}(2012){Rodrigo}, {Solano}, \&
  {Bayo}}]{Rodrigo+2012}
{Rodrigo}, C., {Solano}, E., \& {Bayo}, A. 2012, {SVO Filter Profile Service
  Version 1.0}, IVOA Working Draft 15 October 2012

\bibitem[{{Rouan} {et~al.}(2000){Rouan}, {Riaud}, {Boccaletti}, {Cl{\'e}net},
  \& {Labeyrie}}]{Rouan+2000}
{Rouan}, D., {Riaud}, P., {Boccaletti}, A., {Cl{\'e}net}, Y., \& {Labeyrie}, A.
  2000, \pasp, 112, 1479

\bibitem[{{Ruffio} {et~al.}(2024){Ruffio}, {Perrin}, {Hoch}, {Kammerer},
  {Konopacky}, {Pueyo}, {Madurowicz}, {Rickman}, {Theissen}, {Agrawal},
  {Greenbaum}, {Miles}, {Barman}, {Balmer}, {Llop-Sayson}, {Girard},
  {Rebollido}, {Soummer}, {Allen}, {Anderson}, {Beichman}, {Bellini}, {Bryden},
  {Espinoza}, {Glidden}, {Huang}, {Lewis}, {Libralato}, {Louie}, {Sohn},
  {Seager}, {van der Marel}, {Wakeford}, {Watkins}, {Ygouf}, \&
  {Mountain}}]{Ruffio+2024}
{Ruffio}, J.-B., {Perrin}, M.~D., {Hoch}, K. K.~W., {et~al.} 2024, \aj, 168, 73

\bibitem[{{Samland} {et~al.}(2017){Samland}, {Molli{\`e}re}, {Bonnefoy},
  {Maire}, {Cantalloube}, {Cheetham}, {Mesa}, {Gratton}, {Biller}, {Wahhaj},
  {Bouwman}, {Brandner}, {Melnick}, {Carson}, {Janson}, {Henning}, {Homeier},
  {Mordasini}, {Langlois}, {Quanz}, {van Boekel}, {Zurlo}, {Schlieder},
  {Avenhaus}, {Beuzit}, {Boccaletti}, {Bonavita}, {Chauvin}, {Claudi}, {Cudel},
  {Desidera}, {Feldt}, {Fusco}, {Galicher}, {Kopytova}, {Lagrange}, {Le
  Coroller}, {Martinez}, {Moeller-Nilsson}, {Mouillet}, {Mugnier}, {Perrot},
  {Sevin}, {Sissa}, {Vigan}, \& {Weber}}]{Samland+2017}
{Samland}, M., {Molli{\`e}re}, P., {Bonnefoy}, M., {et~al.} 2017, \aap, 603,
  A57

\bibitem[{{Saumon} \& {Marley}(2008{\natexlab{a}})}]{Saumon+Marley-2008}
{Saumon}, D. \& {Marley}, M.~S. 2008{\natexlab{a}}, \apj, 689, 1327

\bibitem[{{Saumon} \& {Marley}(2008{\natexlab{b}})}]{Saumon+2008}
{Saumon}, D. \& {Marley}, M.~S. 2008{\natexlab{b}}, \apj, 689, 1327

\bibitem[{{Skemer} {et~al.}(2014){Skemer}, {Marley}, {Hinz}, {Morzinski},
  {Skrutskie}, {Leisenring}, {Close}, {Saumon}, {Bailey}, {Briguglio},
  {Defrere}, {Esposito}, {Follette}, {Hill}, {Males}, {Puglisi}, {Rodigas}, \&
  {Xompero}}]{Skemer+2014}
{Skemer}, A.~J., {Marley}, M.~S., {Hinz}, P.~M., {et~al.} 2014, \apj, 792, 17

\bibitem[{{Smith} \& {Terrile}(1984)}]{Smith+Terrile-1984}
{Smith}, B.~A. \& {Terrile}, R.~J. 1984, Science, 226, 1421

\bibitem[{{Snellen} \& {Brown}(2018)}]{Snellen+Brown-2018}
{Snellen}, I.~A.~G. \& {Brown}, A.~G.~A. 2018, Nature Astronomy, 2, 883

\bibitem[{{Spiegel} \& {Burrows}(2012)}]{Spiegel+Burrows-2012}
{Spiegel}, D.~S. \& {Burrows}, A. 2012, \apj, 745, 174

\bibitem[{{Stone} {et~al.}(2020){Stone}, {Barman}, {Skemer}, {Briesemeister},
  {Brock}, {Hinz}, {Leisenring}, {Woodward}, {Skrutskie}, \&
  {Spalding}}]{Stone+2020}
{Stone}, J.~M., {Barman}, T., {Skemer}, A.~J., {et~al.} 2020, \aj, 160, 262

\bibitem[{{Su{\'a}rez} \& {Metchev}(2022)}]{Suarez+Metchev-2022}
{Su{\'a}rez}, G. \& {Metchev}, S. 2022, \mnras, 513, 5701

\bibitem[{{Sutlieff} {et~al.}(2021){Sutlieff}, {Bohn}, {Birkby}, {Kenworthy},
  {Morzinski}, {Doelman}, {Males}, {Snik}, {Close}, {Hinz}, \&
  {Charbonneau}}]{Sutlieff+2021}
{Sutlieff}, B.~J., {Bohn}, A.~J., {Birkby}, J.~L., {et~al.} 2021, \mnras, 506,
  3224

\bibitem[{{Todorov} {et~al.}(2016){Todorov}, {Line}, {Pineda}, {Meyer},
  {Quanz}, {Hinkley}, \& {Fortney}}]{Todorov+2016}
{Todorov}, K.~O., {Line}, M.~R., {Pineda}, J.~E., {et~al.} 2016, \apj, 823, 14

\bibitem[{{Tokunaga} {et~al.}(1998){Tokunaga}, {Kobayashi}, {Bell}, {Ching},
  {Hodapp}, {Hora}, {Neill}, {Onaka}, {Rayner}, {Robertson}, {Warren}, {Weber},
  \& {Young}}]{Tokunaga+1998}
{Tokunaga}, A.~T., {Kobayashi}, N., {Bell}, J., {et~al.} 1998, in Society of
  Photo-Optical Instrumentation Engineers (SPIE) Conference Series, Vol. 3354,
  Infrared Astronomical Instrumentation, ed. A.~M. {Fowler}, 512--524

\bibitem[{{Tremblin} {et~al.}(2016){Tremblin}, {Amundsen}, {Chabrier},
  {Baraffe}, {Drummond}, {Hinkley}, {Mourier}, \& {Venot}}]{Tremblin+2016}
{Tremblin}, P., {Amundsen}, D.~S., {Chabrier}, G., {et~al.} 2016, \apjl, 817,
  L19

\bibitem[{{Uyama} {et~al.}(2020){Uyama}, {Currie}, {Hori}, {De Rosa}, {Mede},
  {Brandt}, {Kwon}, {Guyon}, {Lozi}, {Jovanovic}, {Martinache}, {Kudo},
  {Tamura}, {Kasdin}, {Groff}, {Chilcote}, {Hayashi}, {McElwain},
  {Asensio-Torres}, {Janson}, {Knapp}, \& {Serabyn}}]{Uyama+2020}
{Uyama}, T., {Currie}, T., {Hori}, Y., {et~al.} 2020, \aj, 159, 40

\bibitem[{{Vigan} {et~al.}(2021){Vigan}, {Fontanive}, {Meyer}, {Biller},
  {Bonavita}, {Feldt}, {Desidera}, {Marleau}, {Emsenhuber}, {Galicher}, {Rice},
  {Forgan}, {Mordasini}, {Gratton}, {Le Coroller}, {Maire}, {Cantalloube},
  {Chauvin}, {Cheetham}, {Hagelberg}, {Lagrange}, {Langlois}, {Bonnefoy},
  {Beuzit}, {Boccaletti}, {D'Orazi}, {Delorme}, {Dominik}, {Henning}, {Janson},
  {Lagadec}, {Lazzoni}, {Ligi}, {Menard}, {Mesa}, {Messina}, {Moutou},
  {M{\"u}ller}, {Perrot}, {Samland}, {Schmid}, {Schmidt}, {Sissa}, {Turatto},
  {Udry}, {Zurlo}, {Abe}, {Antichi}, {Asensio-Torres}, {Baruffolo}, {Baudoz},
  {Baudrand}, {Bazzon}, {Blanchard}, {Bohn}, {Brown Sevilla}, {Carbillet},
  {Carle}, {Cascone}, {Charton}, {Claudi}, {Costille}, {De Caprio},
  {Delboulb{\'e}}, {Dohlen}, {Engler}, {Fantinel}, {Feautrier}, {Fusco},
  {Gigan}, {Girard}, {Giro}, {Gisler}, {Gluck}, {Gry}, {Hubin}, {Hugot},
  {Jaquet}, {Kasper}, {Le Mignant}, {Llored}, {Madec}, {Magnard}, {Martinez},
  {Maurel}, {M{\"o}ller-Nilsson}, {Mouillet}, {Moulin}, {Orign{\'e}}, {Pavlov},
  {Perret}, {Petit}, {Pragt}, {Puget}, {Rabou}, {Ramos}, {Rickman}, {Rigal},
  {Rochat}, {Roelfsema}, {Rousset}, {Roux}, {Salasnich}, {Sauvage}, {Sevin},
  {Soenke}, {Stadler}, {Suarez}, {Wahhaj}, {Weber}, \& {Wildi}}]{Vigan+2021}
{Vigan}, A., {Fontanive}, C., {Meyer}, M., {et~al.} 2021, \aap, 651, A72

\bibitem[{{Visscher} {et~al.}(2010){Visscher}, {Lodders}, \&
  {Fegley}}]{Visscher+2010}
{Visscher}, C., {Lodders}, K., \& {Fegley}, Jr., B. 2010, \apj, 716, 1060

\bibitem[{{Wilcomb} {et~al.}(2020){Wilcomb}, {Konopacky}, {Barman}, {Theissen},
  {Ruffio}, {Brock}, {Macintosh}, \& {Marois}}]{Hoch+2020}
{Wilcomb}, K.~K., {Konopacky}, Q.~M., {Barman}, T.~S., {et~al.} 2020, \aj, 160,
  207

\bibitem[{{Wright} {et~al.}(2010){Wright}, {Eisenhardt}, {Mainzer}, {Ressler},
  {Cutri}, {Jarrett}, {Kirkpatrick}, {Padgett}, {McMillan}, {Skrutskie},
  {Stanford}, {Cohen}, {Walker}, {Mather}, {Leisawitz}, {Gautier}, {McLean},
  {Benford}, {Lonsdale}, {Blain}, {Mendez}, {Irace}, {Duval}, {Liu}, {Royer},
  {Heinrichsen}, {Howard}, {Shannon}, {Kendall}, {Walsh}, {Larsen}, {Cardon},
  {Schick}, {Schwalm}, {Abid}, {Fabinsky}, {Naes}, \& {Tsai}}]{WISE}
{Wright}, E.~L., {Eisenhardt}, P. R.~M., {Mainzer}, A.~K., {et~al.} 2010, \aj,
  140, 1868

\bibitem[{{Wright} {et~al.}(2015){Wright}, {Wright}, {Goodson}, {Rieke},
  {Aitink-Kroes}, {Amiaux}, {Aricha-Yanguas}, {Azzollini}, {Banks},
  {Barrado-Navascues}, {Belenguer-Davila}, {Blommaert}, {Bouchet}, {Brandl},
  {Colina}, {Detre}, {Diaz-Catala}, {Eccleston}, {Friedman},
  {Garc{\'\i}a-Mar{\'\i}n}, {G{\"u}del}, {Glasse}, {Glauser}, {Greene},
  {Groezinger}, {Grundy}, {Hastings}, {Henning}, {Hofferbert}, {Hunter},
  {Jessen}, {Justtanont}, {Karnik}, {Khorrami}, {Krause}, {Labiano}, {Lagage},
  {Langer}, {Lemke}, {Lim}, {Lorenzo-Alvarez}, {Mazy}, {McGowan}, {Meixner},
  {Morris}, {Morrison}, {M{\"u}ller}, {rgaard-Nielson}, {Olofsson},
  {O'Sullivan}, {Pel}, {Penanen}, {Petach}, {Pye}, {Ray}, {Renotte}, {Renouf},
  {Ressler}, {Samara-Ratna}, {Scheithauer}, {Schneider}, {Shaughnessy},
  {Stevenson}, {Sukhatme}, {Swinyard}, {Sykes}, {Thatcher}, {Tikkanen}, {van
  Dishoeck}, {Waelkens}, {Walker}, {Wells}, \& {Zhender}}]{Wright+2015}
{Wright}, G.~S., {Wright}, D., {Goodson}, G.~B., {et~al.} 2015, \pasp, 127, 595

\bibitem[{{Xuan} {et~al.}(2024){Xuan}, {Hsu}, {Finnerty}, {Wang}, {Ruffio},
  {Zhang}, {Knutson}, {Mawet}, {Mamajek}, {Inglis}, {Wallack}, {Bryan},
  {Blake}, {Molli{\`e}re}, {Hejazi}, {Baker}, {Bartos}, {Calvin}, {Cetre},
  {Delorme}, {Doppmann}, {Echeverri}, {Fitzgerald}, {Jovanovic}, {Liberman},
  {L{\'o}pez}, {Morris}, {Pezzato}, {Sappey}, {Schofield}, {Skemer}, {Wallace},
  {Wang}, {Agrawal}, \& {Horstman}}]{Xuan+2024}
{Xuan}, J.~W., {Hsu}, C.-C., {Finnerty}, L., {et~al.} 2024, \apj, 970, 71

\bibitem[{{Xuan} {et~al.}(2022){Xuan}, {Wang}, {Ruffio}, {Knutson}, {Mawet},
  {Molli{\`e}re}, {Kolecki}, {Vigan}, {Mukherjee}, {Wallack}, {Wang}, {Baker},
  {Bartos}, {Blake}, {Bond}, {Bryan}, {Calvin}, {Cetre}, {Chun}, {Delorme},
  {Doppmann}, {Echeverri}, {Finnerty}, {Fitzgerald}, {Horstman}, {Inglis},
  {Jovanovic}, {L{\'o}pez}, {Martin}, {Morris}, {Pezzato}, {Ragland}, {Ren},
  {Ruane}, {Sappey}, {Schofield}, {Skemer}, {Venenciano}, {Wallace}, \&
  {Wizinowich}}]{Xuan+2022}
{Xuan}, J.~W., {Wang}, J., {Ruffio}, J.-B., {et~al.} 2022, \apj, 937, 54

\bibitem[{{Xuan} {et~al.}(2018){Xuan}, {Mawet}, {Ngo}, {Ruane}, {Bailey},
  {Choquet}, {Absil}, {Alvarez}, {Bryan}, {Cook}, {Femen{\'\i}a Castell{\'a}},
  {Gomez Gonzalez}, {Huby}, {Knutson}, {Matthews}, {Ragland}, {Serabyn}, \&
  {Zawol}}]{Xuan+2018}
{Xuan}, W.~J., {Mawet}, D., {Ngo}, H., {et~al.} 2018, \aj, 156, 156

\bibitem[{{Zhang} {et~al.}(2023){Zhang}, {Duch{\^e}ne}, {De Rosa}, {Ansdell},
  {Konopacky}, {Esposito}, {Chiang}, {Rice}, {Matthews}, {Kalas}, {Macintosh},
  {Marchis}, {Metchev}, {Patience}, {Rameau}, {Ward-Duong}, {Wolff},
  {Fitzgerald}, {Bailey}, {Barman}, {Bulger}, {Chen}, {Chilcotte}, {Cotten},
  {Doyon}, {Follette}, {Gerard}, {Goodsell}, {Graham}, {Greenbaum}, {Hibon},
  {Hung}, {Ingraham}, {Maire}, {Marley}, {Marois}, {Millar-Blanchaer},
  {Nielsen}, {Oppenheimer}, {Palmer}, {Perrin}, {Poyneer}, {Pueyo}, {Rajan},
  {Rantakyr{\"o}}, {Ruffio}, {Savransky}, {Schneider}, {Sivaramakrishnan},
  {Song}, {Soummer}, {Thomas}, {Wang}, \& {Wiktorowicz}}]{Zhang+2023}
{Zhang}, S.~Y., {Duch{\^e}ne}, G., {De Rosa}, R.~J., {et~al.} 2023, \aj, 165,
  219

\bibitem[{{Zhang} {et~al.}(2018{\natexlab{a}}){Zhang}, {Galvez-Ortiz},
  {Pinfield}, {Burgasser}, {Lodieu}, {Jones}, {Mart{\'\i}n}, {Burningham},
  {Homeier}, {Allard}, {Zapatero Osorio}, {Smith}, {Smart}, {L{\'o}pez
  Mart{\'\i}}, {Marocco}, \& {Rebolo}}]{Zhang+2018b}
{Zhang}, Z.~H., {Galvez-Ortiz}, M.~C., {Pinfield}, D.~J., {et~al.}
  2018{\natexlab{a}}, \mnras, 480, 5447

\bibitem[{{Zhang} {et~al.}(2018{\natexlab{b}}){Zhang}, {Pinfield},
  {G{\'a}lvez-Ortiz}, {Homeier}, {Burgasser}, {Lodieu}, {Mart{\'\i}n},
  {Zapatero Osorio}, {Allard}, {Jones}, {Smart}, {L{\'o}pez Mart{\'\i}},
  {Burningham}, \& {Rebolo}}]{Zhang+2018a}
{Zhang}, Z.~H., {Pinfield}, D.~J., {G{\'a}lvez-Ortiz}, M.~C., {et~al.}
  2018{\natexlab{b}}, \mnras, 479, 1383

\bibitem[{{Zuckerman} {et~al.}(2011){Zuckerman}, {Rhee}, {Song}, \&
  {Bessell}}]{Zuckerman+2011}
{Zuckerman}, B., {Rhee}, J.~H., {Song}, I., \& {Bessell}, M.~S. 2011, \apj,
  732, 61

\end{thebibliography}

\appendix


\onecolumn 

\section{Removing prominent and energetic cosmic rays}\label{apx:cosmic}

The pipeline \texttt{spaceKLIP} (which uses JWST pipeline routines) cannot remove some cosmic rays efficiently. For example, the most energetic cosmic rays are shown in the reduced frame on the left side in Figure\,\ref{fig:cosmic_example}. These energetic cosmic rays are challenging, and despite the initial parameters for the data reduction, they remain in the final image. Our approach is based on the nature of MIRI observations, which use nondestructive readout techniques (see \citealt{Ressler+2015}), and works as a pre-reduction step at the raw data level. Essentially, the detector continuously collects data without interrupting the observation process or stopping the receipt of photons. As a result, the data consists of a series of images, each with progressively increasing flux measurements. From the perspective of a single pixel, this process resembles a ramp, where the slope of the ramp represents the increase in detected signal over time.

\begin{figure}[htb!]
\centering
\sidecaption
\includegraphics[width=10.2cm]{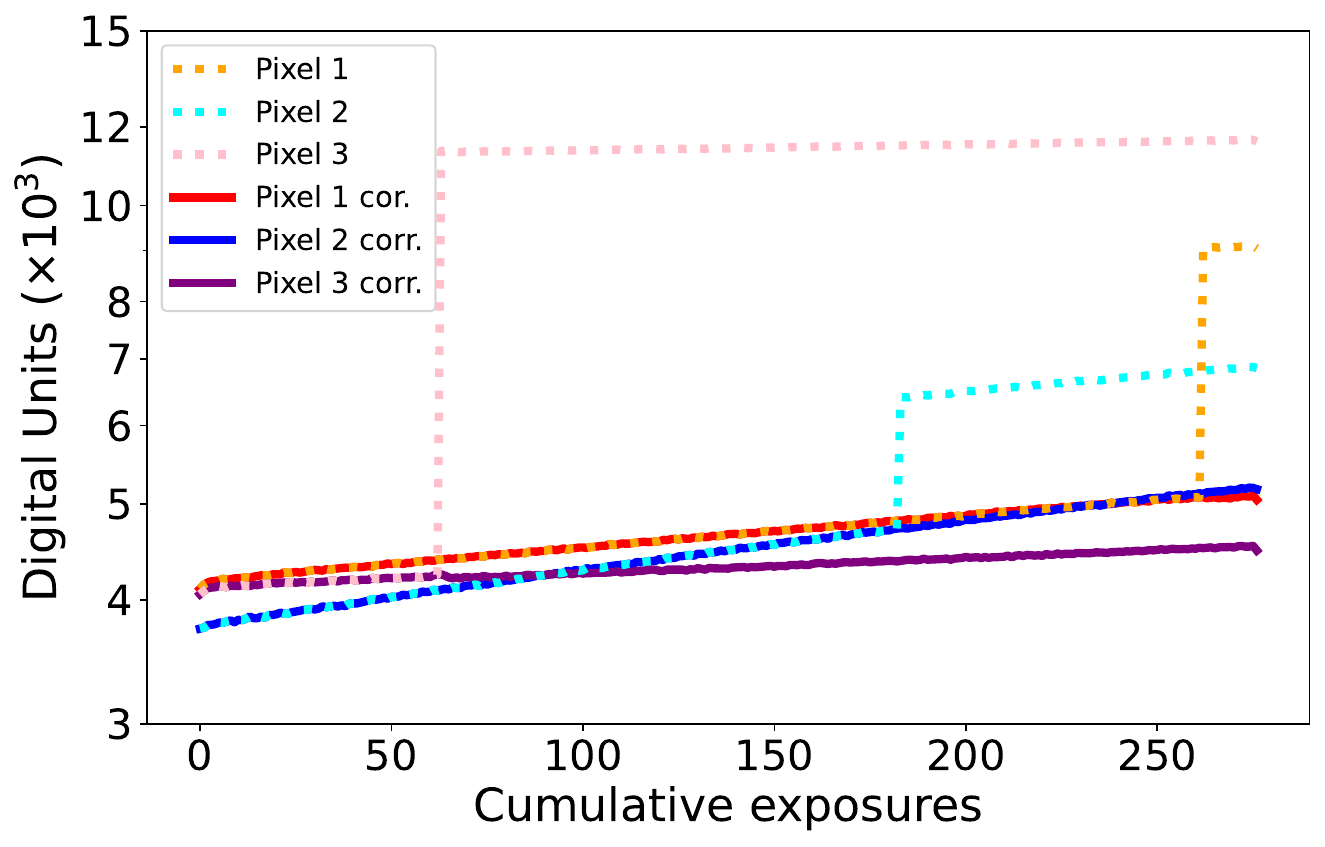}
\caption{Ramp in one integration for MIRI observations at \texttt{F1550C} filter for three specific pixels. The dotted colored lines are the original ramp for three pixels at the location of three different cosmic rays that impacted at different times during the integration. The solid-colored lines are the corrected ramp for these three pixels.  }
\label{fig:pixel_ramp}%
\end{figure}

\begin{figure*}[htb!]
    \centering
    \sidecaption
    \begin{tabular}{ccc}
        \includegraphics[width=2.8cm]{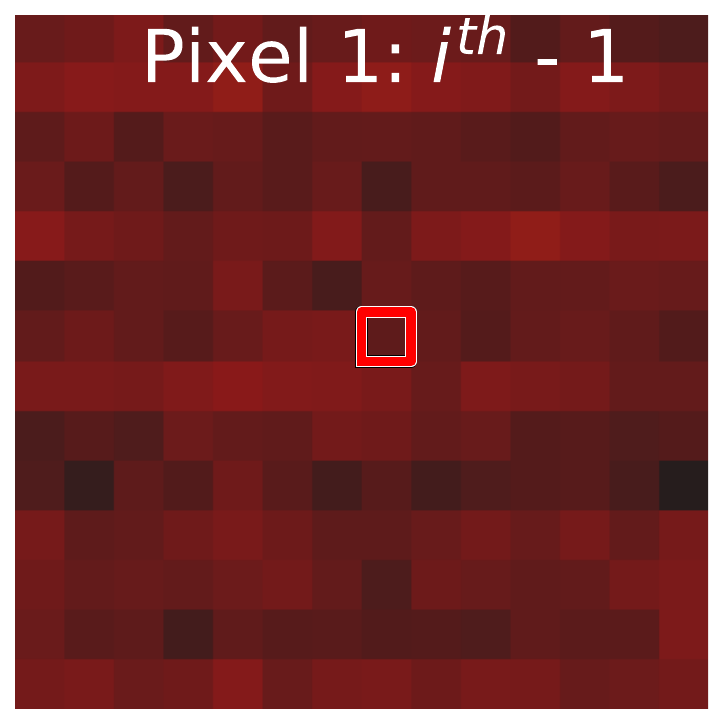} &
        \includegraphics[width=2.8cm]{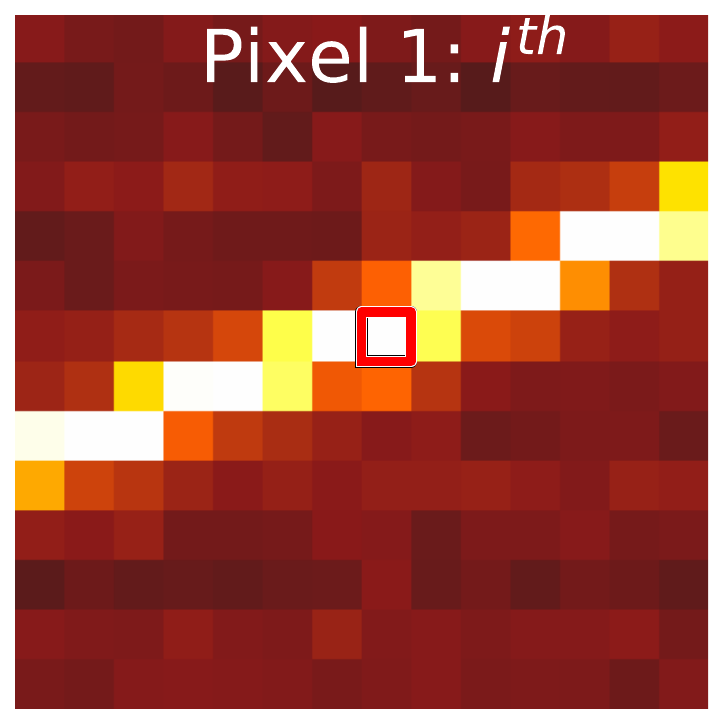} &
        \includegraphics[width=2.8cm]{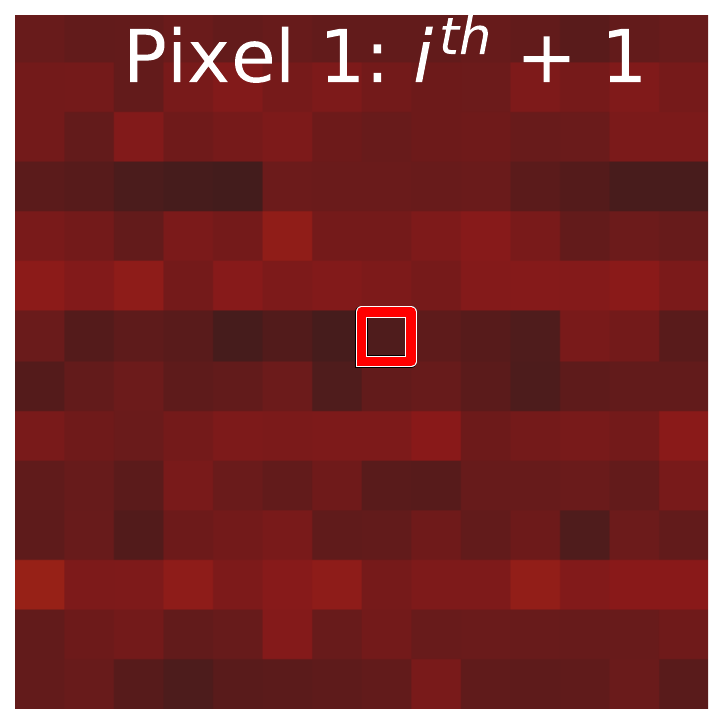} \\
        \includegraphics[width=2.8cm]{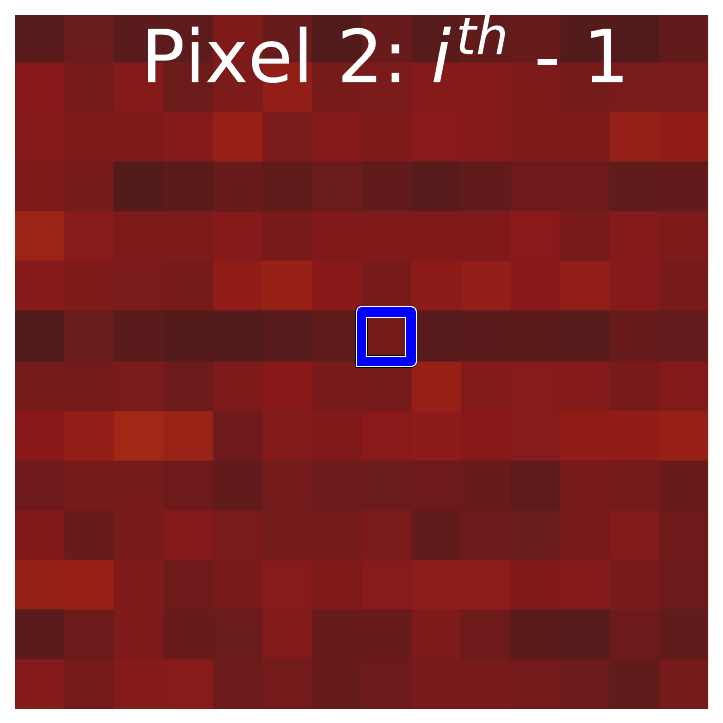} &
        \includegraphics[width=2.8cm]{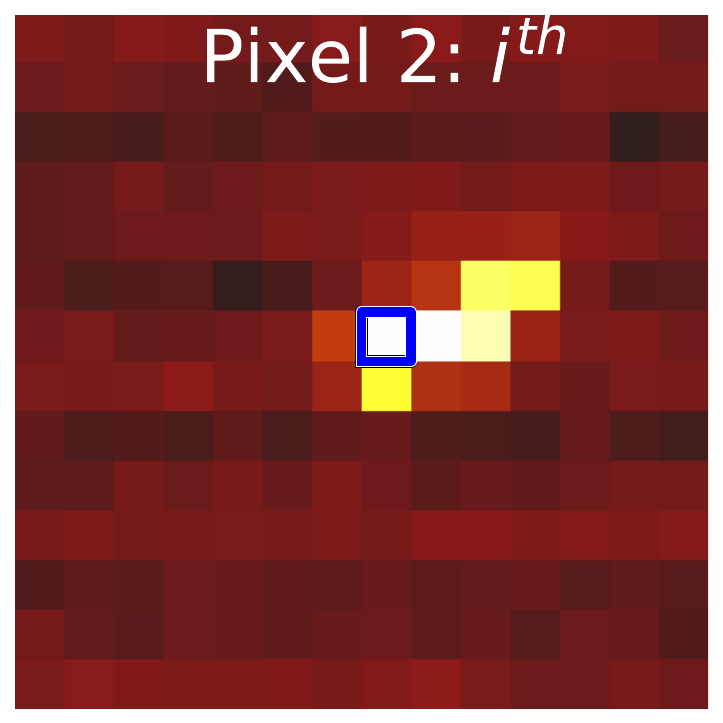} &
        \includegraphics[width=2.8cm]{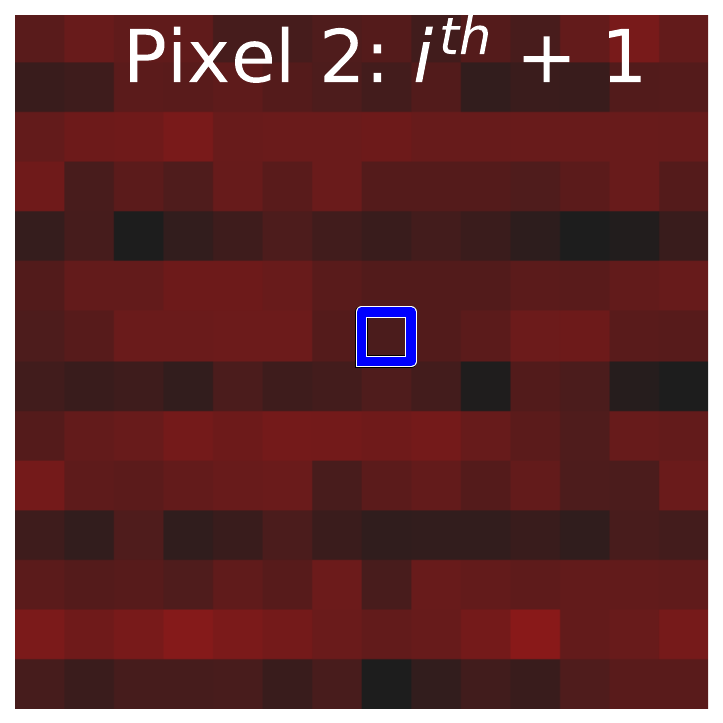} \\
        \includegraphics[width=2.8cm]{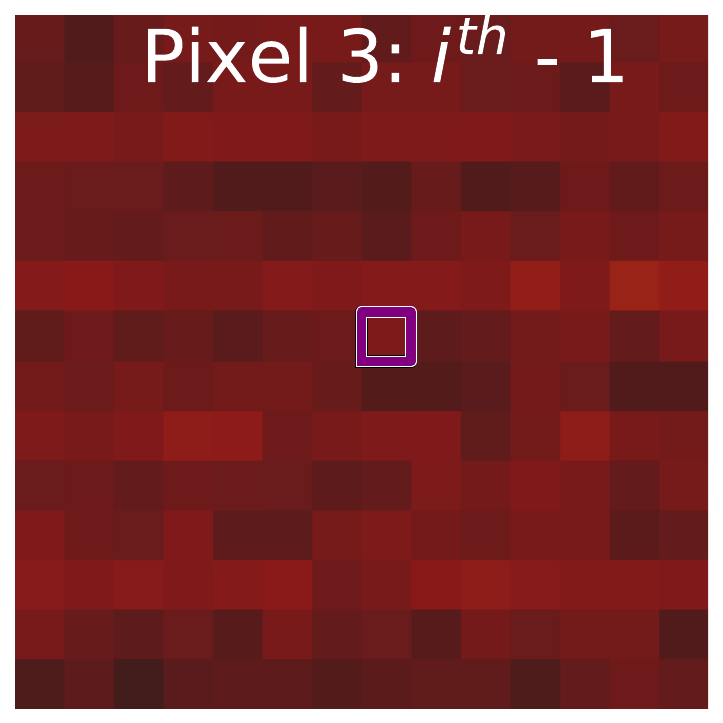} &
        \includegraphics[width=2.8cm]{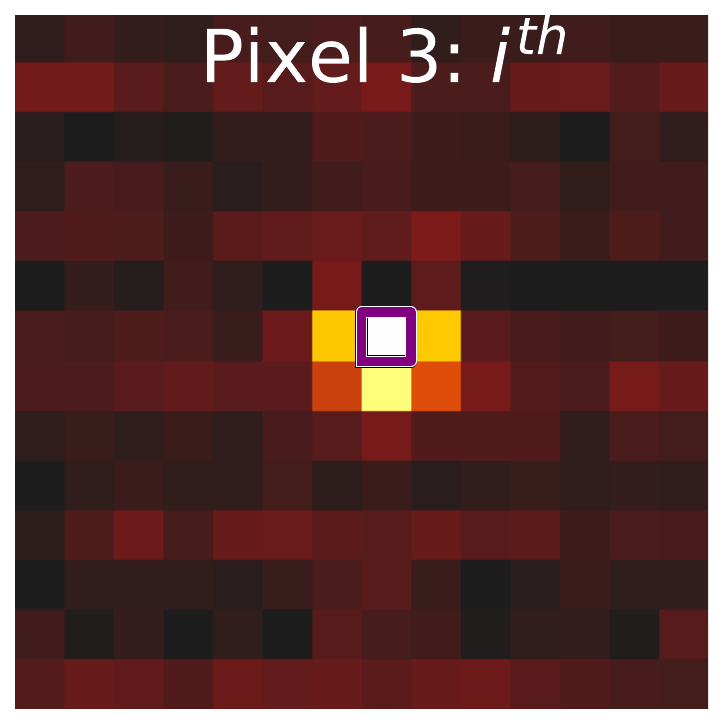} &
        \includegraphics[width=2.8cm]{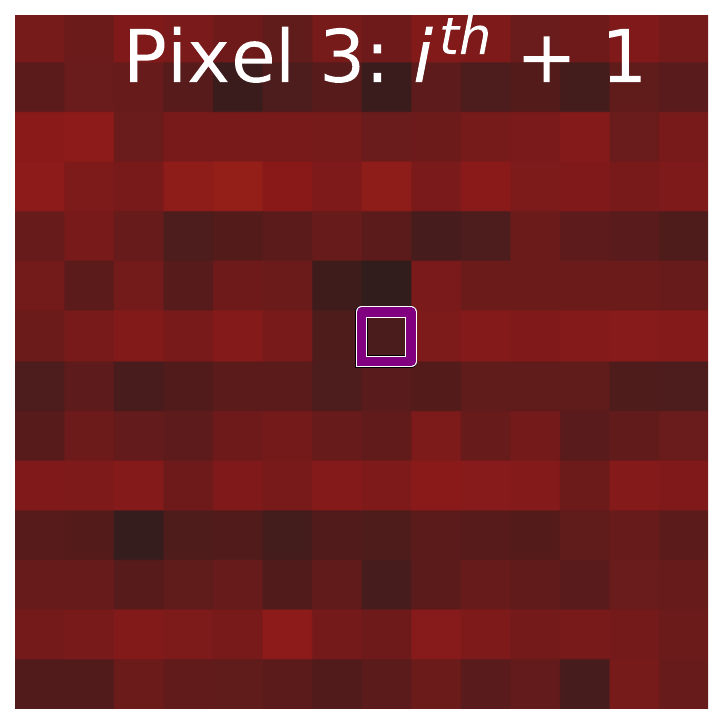} \\
    \end{tabular}
    \caption{
        Images showing the ramp separated into individual integrations and ordered in time, centered at the cosmic ray hit (middle column). The first and third columns are the moments before and after the hit of the cosmic ray. The colored squares in each panel highlight the pixel selected in Figure\,\ref{fig:pixel_ramp} for each cosmic ray (different rows). The images have the same color scale.
    }
    \label{fig:serie_images}
\end{figure*}

\begin{figure}[htb!]
\centering
\sidecaption
\includegraphics[width=8.9cm]{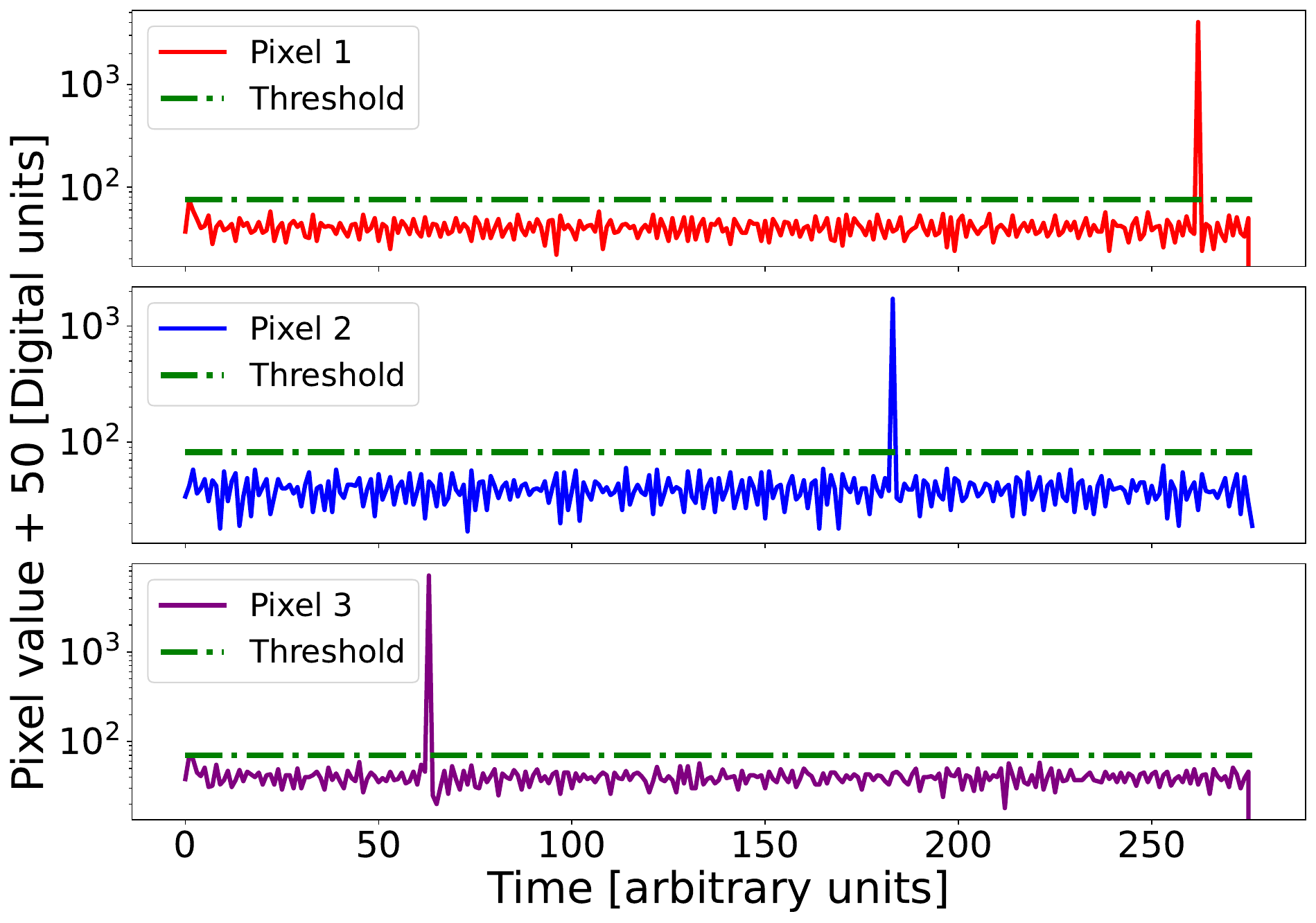}
\caption{Flux difference between consecutive groups for each selected pixel affected by a cosmic ray. The red, blue, and purple curves correspond to each pixel flux as a function of time. The dashed green line corresponds to the threshold used to select outliers. The cosmic ray hit appears as a peak in each flux-differentiation ramp. }
\label{fig:LC_pix}%
\end{figure}

When a cosmic ray strikes a pixel, it causes a sudden increase in the signal, appearing as a sharp jump in the ramp (\citealt{Anderson+Gordon-2011}). This jump is distinct from the steady increase caused by incoming photons and can be detected and corrected during data processing. Figure\,\ref{fig:pixel_ramp} shows one pixel centered at three cosmic rays received at different times in dashed, colored lines. The image shows the ``jump'' in the so-called digital units (flux measurement unit in the detector). The correction used by the pipeline works well for low-energy cosmic rays but results in more masked pixels and residual structures with higher-energy ones. Our procedure breaks the ramp into individual images, generating a series of exposures. As a result, a cosmic ray impact will be captured in one of these images, corresponding to the moment of the cosmic-ray arrival. Figure\,\ref{fig:serie_images} presents three consecutive images of the same three pixels shown in Figure\,\ref{fig:pixel_ramp}. The middle image contains the cosmic ray, which is absent in the subsequent image. Figure\,\ref{fig:LC_pix} shows the signal of these three single-pixels across the image series (flux difference between each consecutive pair of groups within the ramp). The cosmic ray appears as an outlier in the curves.

\FloatBarrier

\FloatBarrier

\FloatBarrier

\begin{figure}[htb!]
\centering
\sidecaption
\includegraphics[width=4.8cm]{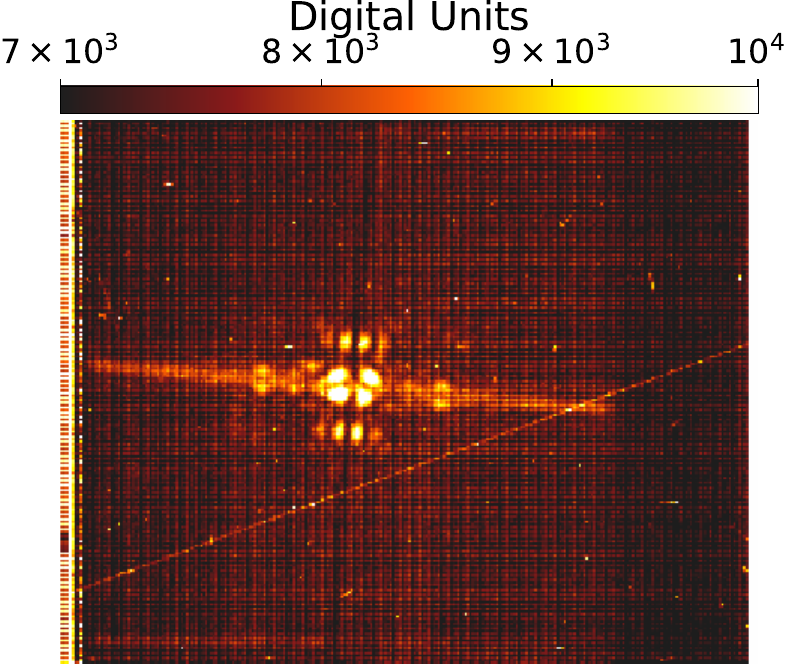}
\includegraphics[width=4.8cm]{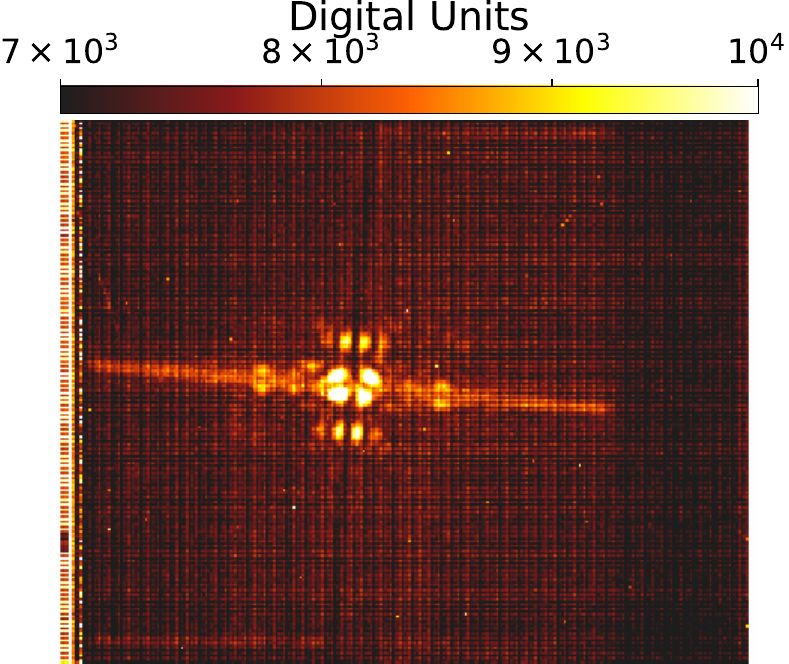}
\caption{Second row ramp integration of the third dither position of the reference star HD\,222389 at \texttt{F1550C} filter. \textit{Left}: Raw, last image in the ramp integration without applying any correction (raw or \texttt{*uncal*} image). \textit{Right}: Same frame but applying the cosmic rays and basic bad pixel corrections directly in the raw image.}
\label{fig:cosmic_raw_example}%
\end{figure}

By analyzing the differentiated-ramp flux across all images, we can create a representative image by computing the median to avoid outlier effects. The outliers can be identified using a threshold of six times the MAD (green dashed line in Figure\,\ref{fig:LC_pix}) and then replaced with the median value. This method effectively identifies high-energy cosmic rays but may miss low-energy ones and bad pixels, as they do not produce significant variations in the differentiated ramp. We then restore the ramp structure using the corrected differentiated ramp flux. Figure\,\ref{fig:pixel_ramp} shows the ramp before (dashed colored lines) and after (solid colored lines) cosmic ray removal, while Figure\,\ref{fig:cosmic_raw_example} shows the last image in the ramp before (left) and after (right) cosmic ray correction.

\FloatBarrier

\renewcommand{\thefootnote}{\fnsymbol{footnote}}
\section{Fringing structure\protect\footnotemark[1]}\label{Apx:fringing}

\footnotetext[1]{Since we processed the data, this ``fringing'' has been associated with the 390\,Hz pattern noise seen in many MIRI subarray observations. The effect is described at \url{https://jwst-docs.stsci.edu/known-issues-with-jwst-data/miri-known-issues\#MIRIKnownIssues-emiElectromagneticinterference(EMI)patternnoise}. A pipeline correction is now available as part of the \texttt{\textit{emicorr}} step of the \texttt{calwebb\_detector1} module of the pipeline. }
\renewcommand{\thefootnote}{\arabic{footnote}}

\begin{figure*}[htb!]
    \centering
    \begin{minipage}[c]{0.48\textwidth}
        \centering
        \includegraphics[width=5.8cm]{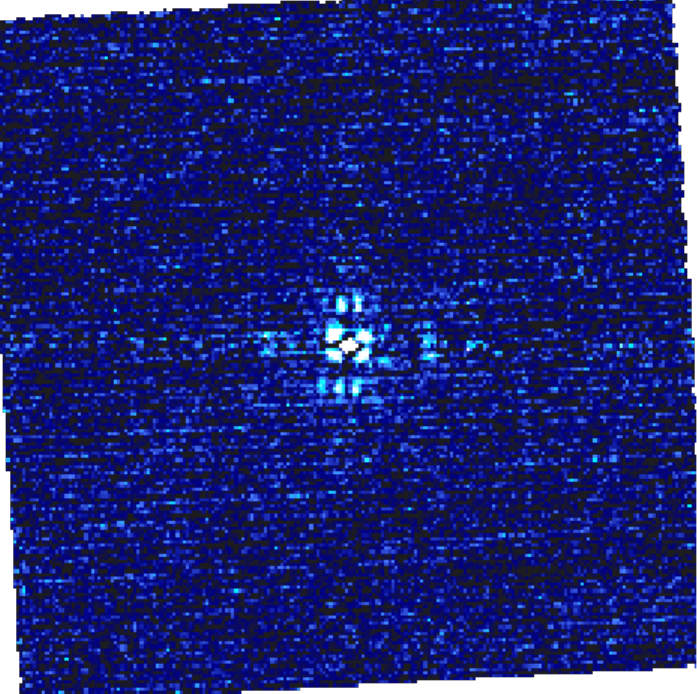}
        \caption{Post-processing, PCA subtraction of $\kappa$\,And using RDI and 15 components. In this case, the reference frames were not corrected for the fringing structure.  }
\label{fig:bad_PCAsub}%
    \end{minipage}
    \hfill
    \begin{minipage}[c]{0.48\textwidth}
        \centering
        \includegraphics[width=8cm]{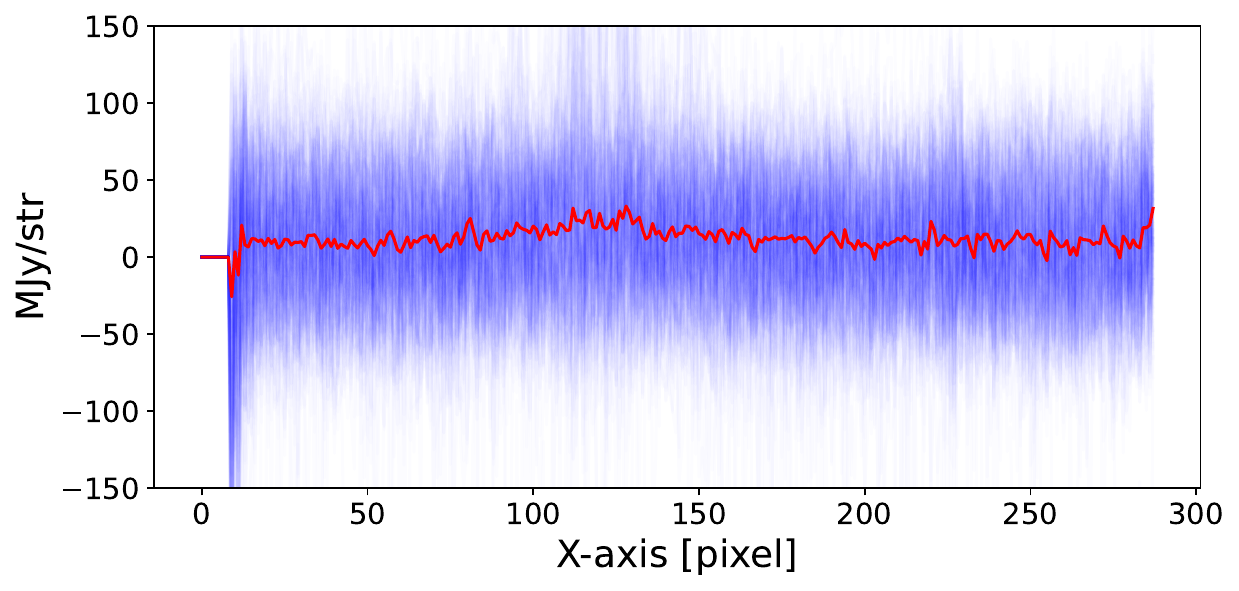}
        \includegraphics[width=8cm]{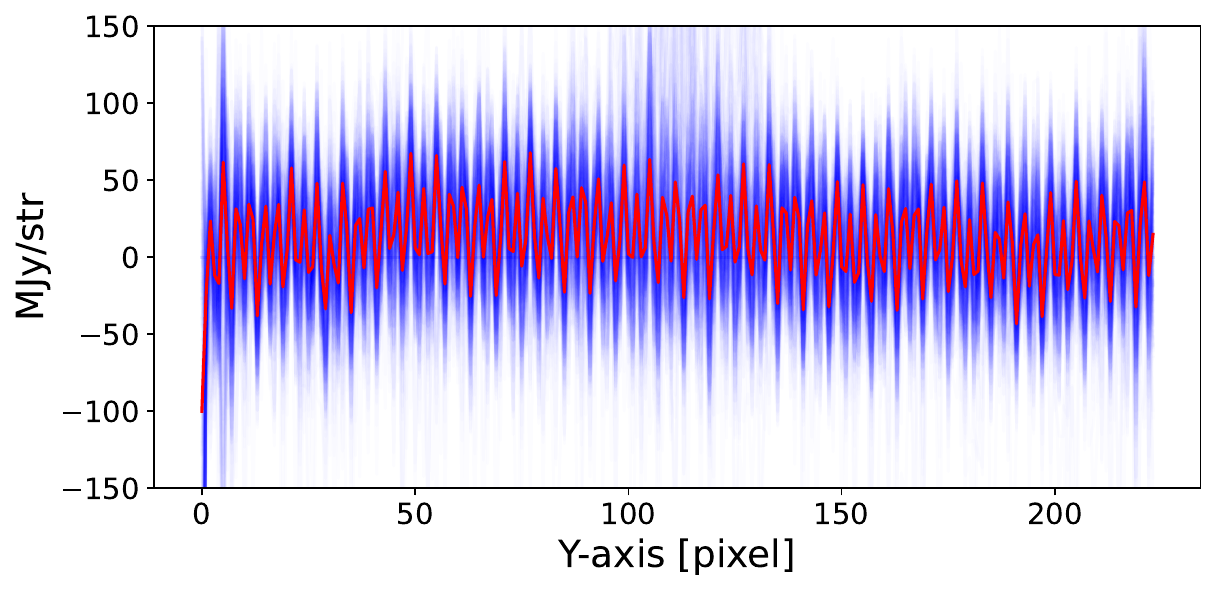}
        \caption{Main structures and flux dispersion in the X and Y axes. \texttt{Top:} Median value along the x-axis corresponds to the median value per column. \texttt{Bottom:} Same but for the y-axis, and along the rows. The blue lines correspond to the values per column (left) and row (right), while the red line corresponds to the respective median. }
\label{fig:raw_structures}%
    \end{minipage}
\end{figure*}

The fringing structure is present in the reference star integrations in the \texttt{F1065C} and \texttt{F1140C} filters. The fringing directly impacts post-processing, resulting in poor (or even no) correction of the coronagraphic stellar PSF in the science target. Figure\,\ref{fig:bad_PCAsub} shows this effect in the PCA subtraction in $\kappa$\,And at \texttt{F1065C} using $15$ components. We proceed as follows to remove and correct the fringing in all the reference integrations. To exemplify the procedure, we show here the first integration of the first dither position of the reference star at \texttt{F1065C}. First, we identify the main structure by computing the median value in the X (along the columns) and Y (along the rows) axes. Figure\,\ref{fig:raw_structures} shows the values per column (top) and rows (bottom) and the respective median (red curve) values. The fringing structure is visible on the bottom side. We also noted the dispersion along both axes, which reached a mean upper value of 100 MJy/str.

We proceed as follows to correct this structure presented in all the reference star integrations in \texttt{F1065C} and \texttt{F1140C} filters:

\textbf{1)} We computed the median $N_{j0}$ along the Y-axis (median of each row) in the image $I_{i,j}$. Then we compute the median level $n_{0}$ of $N_{j}$.

\textbf{2)} We calculated, per column $I_{i}$, the corrected image $I'_{i}$ using the following equation $I'_{i} = I_{i} - N_{j} + n_{0}$.

\begin{figure*}[htb!]
    \centering
    \begin{minipage}[c]{0.48\textwidth}  
        \centering
        \includegraphics[width=6.0cm]{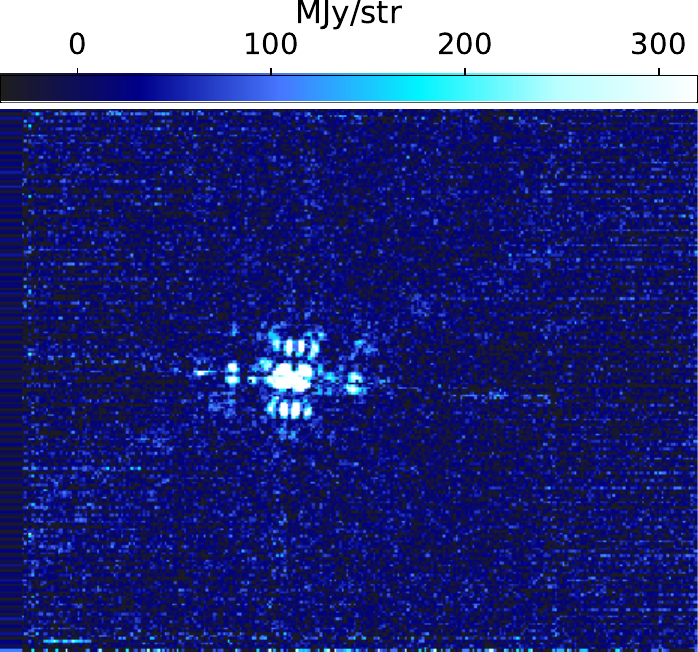}
        \caption{Resulting image after the first iterations to correct the fringing structure. The remnant fringing is still present on the left and right sides of the image.}
        \label{fig:First_step}%
    \end{minipage}
    \hfill
    \begin{minipage}[c]{0.48\textwidth}  
        \centering
        \includegraphics[width=8cm]{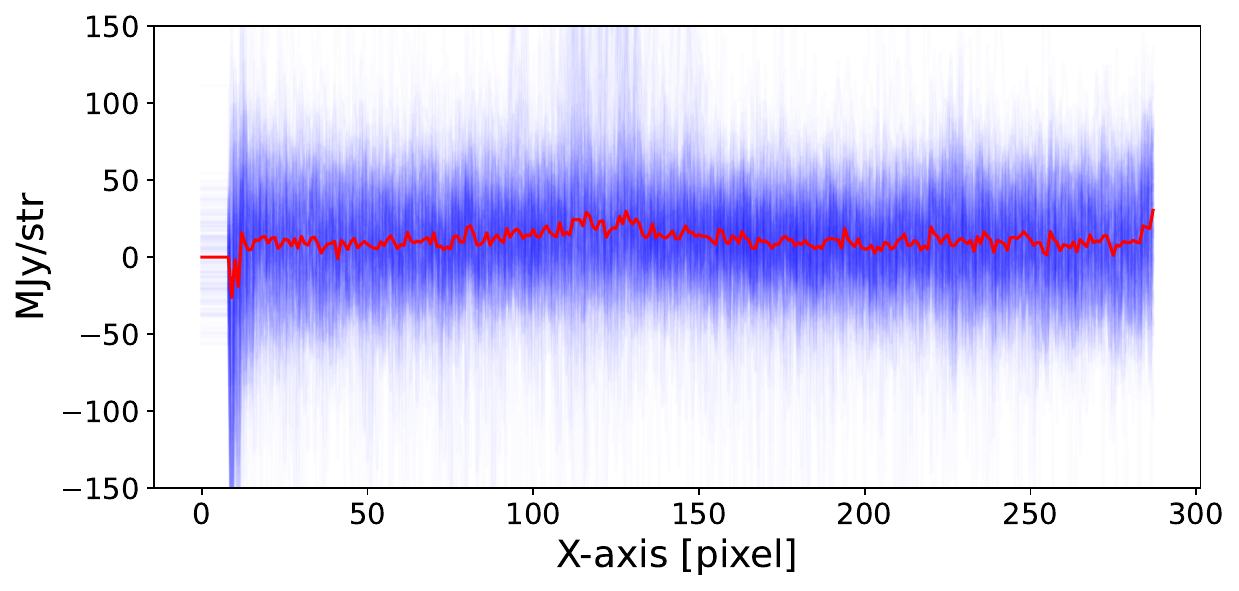}
        \includegraphics[width=8cm]{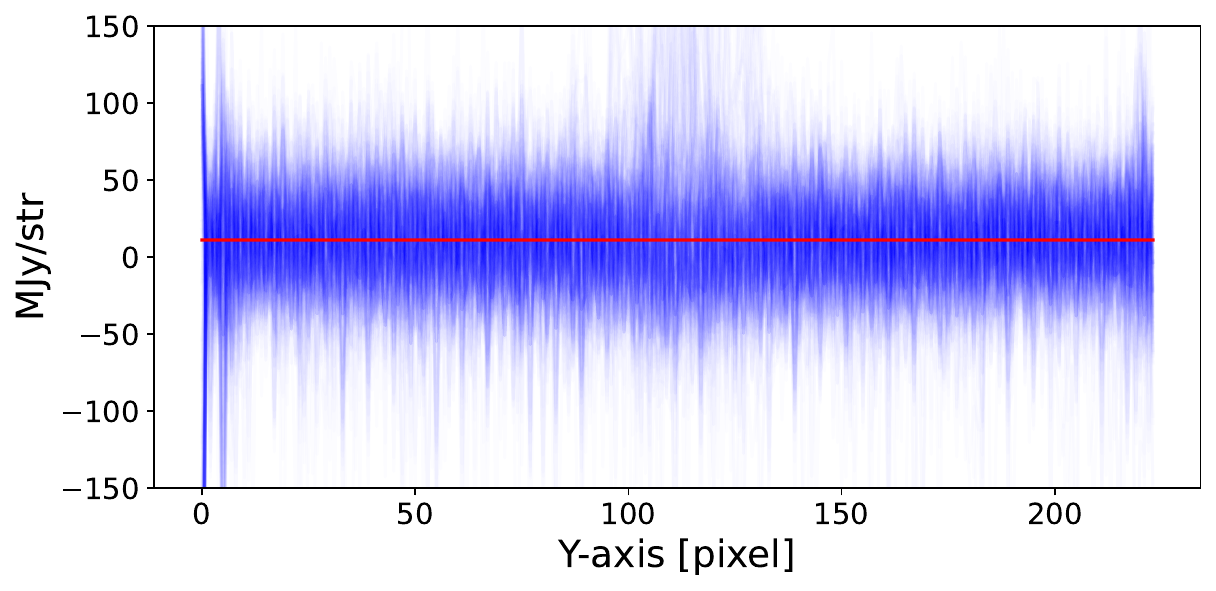}
        \caption{Same as Figure\,\ref{fig:raw_structures} but now using the first fringing corrections. }
        \label{fig:first_structures}%
        \label{fig:A4}
    \end{minipage}
\end{figure*}

\textbf{3)} We repeated steps 1) and 2) two more times using the resulting $I'$ as input each time to obtain a final $I''$. Figure\,\ref{fig:First_step} shows the resulting image at the top and the Y-axis structure at the bottom. Figure\,\ref{fig:first_structures} shows the same as Figure\,\ref{fig:raw_structures} but now using the corrected image $I''$. We note that we still have remnant fringing structures that were not corrected using the previous steps in both figures. Since we still have these remnant fringing structures at the left and right sides of the image, we proceed to fix them individually.

\textbf{4)} We computed the MAD $M_{i}$ along the X-axis. The MAD values will be higher when we still have the presence of the fringing structure and minimum in the regions where we successfully removed the fringing in the previous steps. 

\begin{figure*}[htb!]
    \centering
    \begin{minipage}[c]{0.48\textwidth}  
        \centering
        \includegraphics[width=6.7cm]{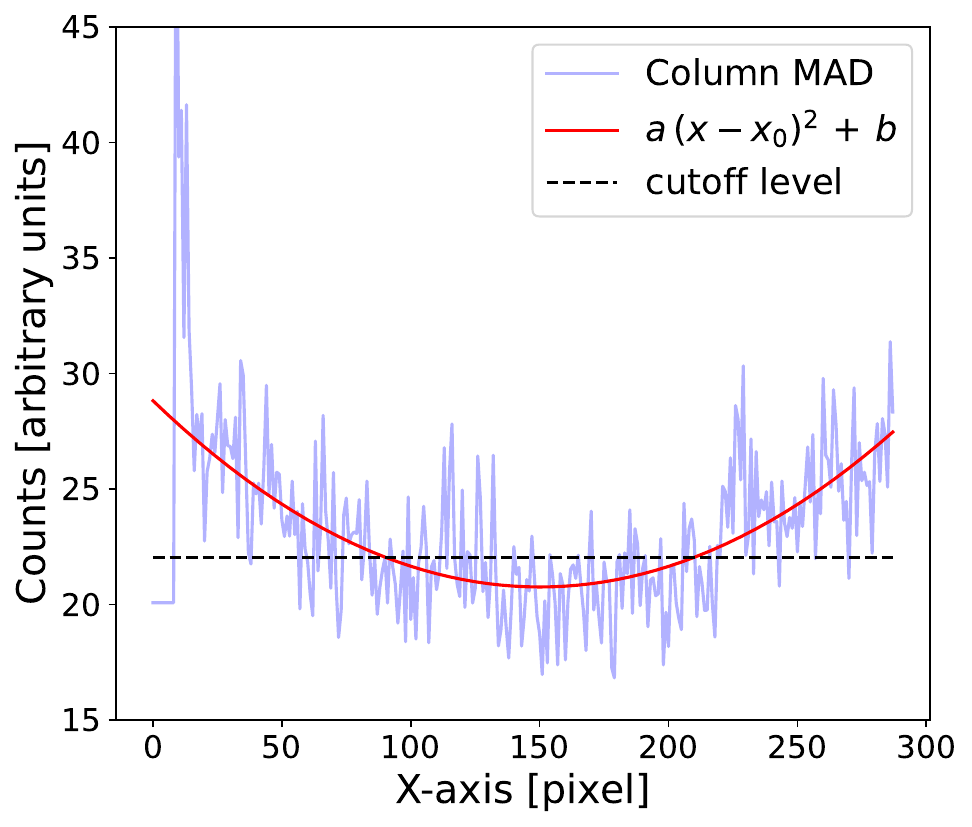}
        \caption{MAD along the columns of the first fringing corrections applied to the image (transparent blue curve). The red curve corresponds to the fitted polynomial $P_{i}$, and the dashed black line to the cutoff level.  }
        \label{fig:cutoff}%
    \end{minipage}
    \hfill
    \begin{minipage}[c]{0.48\textwidth}  
        \centering
        \includegraphics[width=6.5cm]{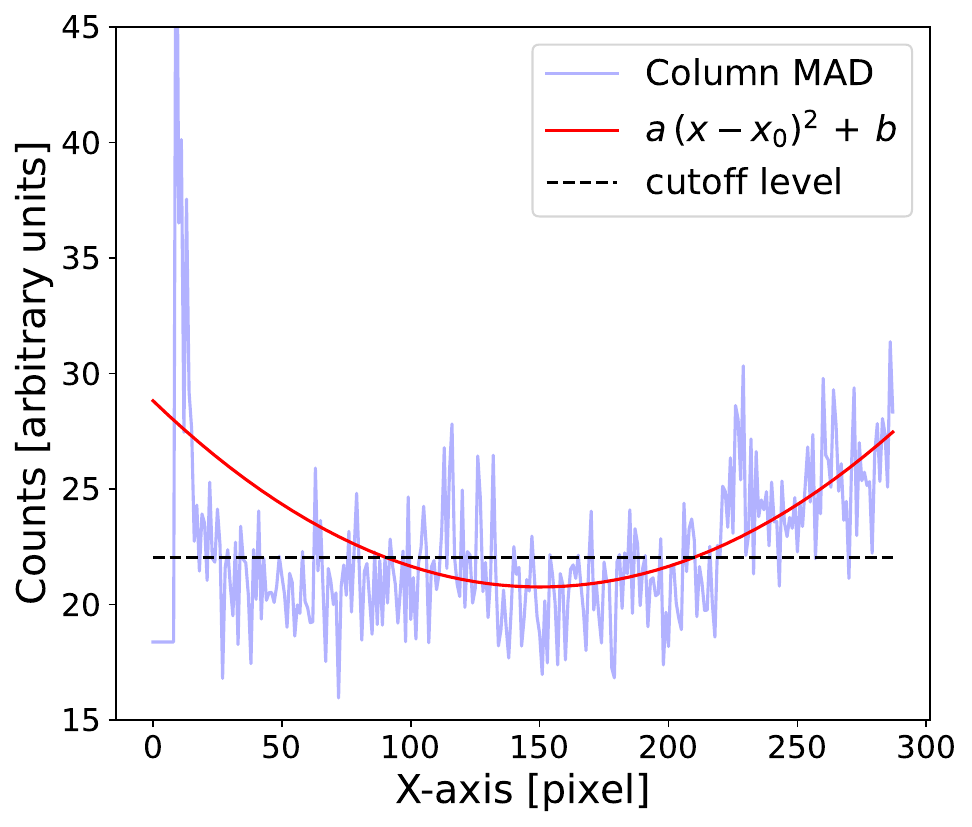}
        \includegraphics[width=6.5cm]{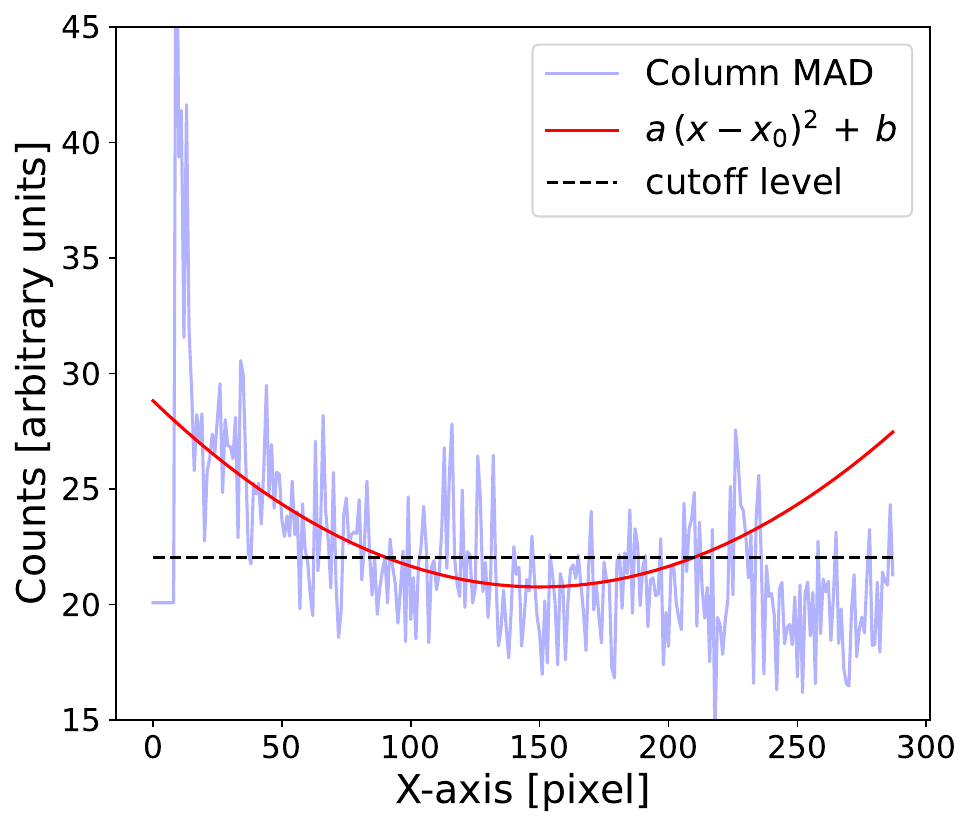}
        \caption{Same as Figure\,\ref{fig:cutoff} but after applying the fringing corrections. \texttt{Top:} Corrections applied only to the left region in step 6). \texttt{Bottom}: Correction applied only to the right region.}
        \label{fig:both_structures}
    \end{minipage}
\end{figure*}

\textbf{5)} To identify the significant remnant fringing structure (i.e., above the noise level), we fit a polynomial of degree 2, $P_{i}$, to get the main structure in $M_{i}$. Then, we normalized $M_{i}$ using $P_{i}$. We used the normalized $M_{i}$ to compute the standard deviation, which represents the main noise level and out cutoff level. Figure\,\ref{fig:cutoff} shows $M_{i}$, $P_{i}$, and the cutoff level.


\textbf{6)} We calculated the intersection between the fit model $P_{i}$ and the cutoff level. This defines the left and right regions still affected by the fringing structure.

\begin{figure*}[htb!]
    \centering
    \begin{minipage}[c]{0.48\textwidth}  
        \centering
        \includegraphics[width=4.1cm]{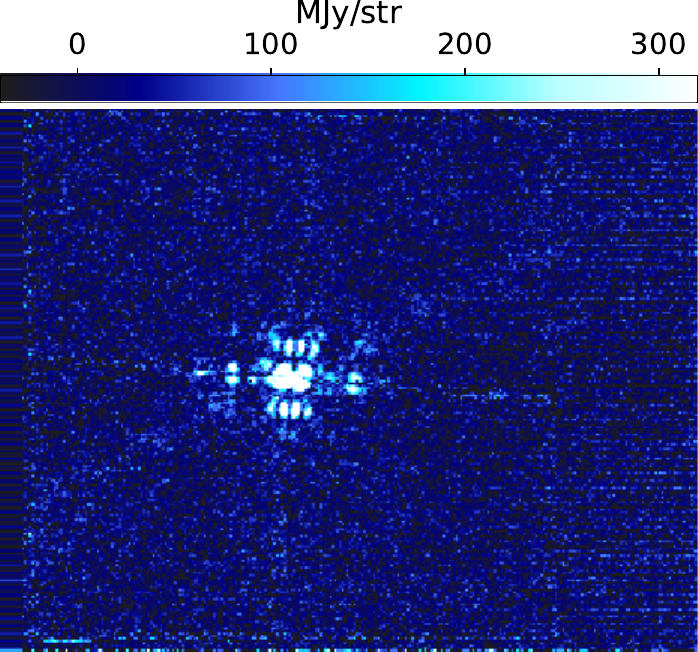}
        \includegraphics[width=4.1cm]{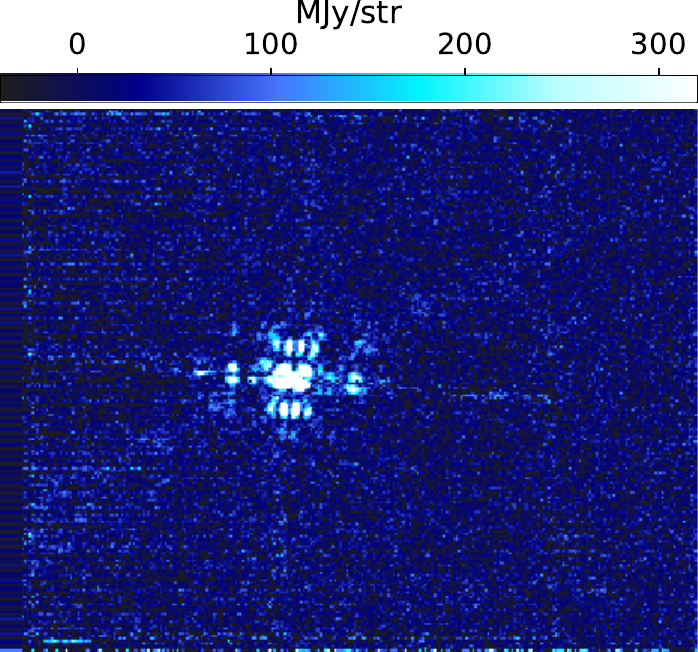}
        \caption{Image after the first and second fringing corrections for the left (\textit{left}) and right (\textit{right}) regions.}
        \label{fig:both_structures_corr}
    \end{minipage}
    \hfill
    \begin{minipage}[c]{0.48\textwidth}  
        \centering
        \includegraphics[width=6.4cm]{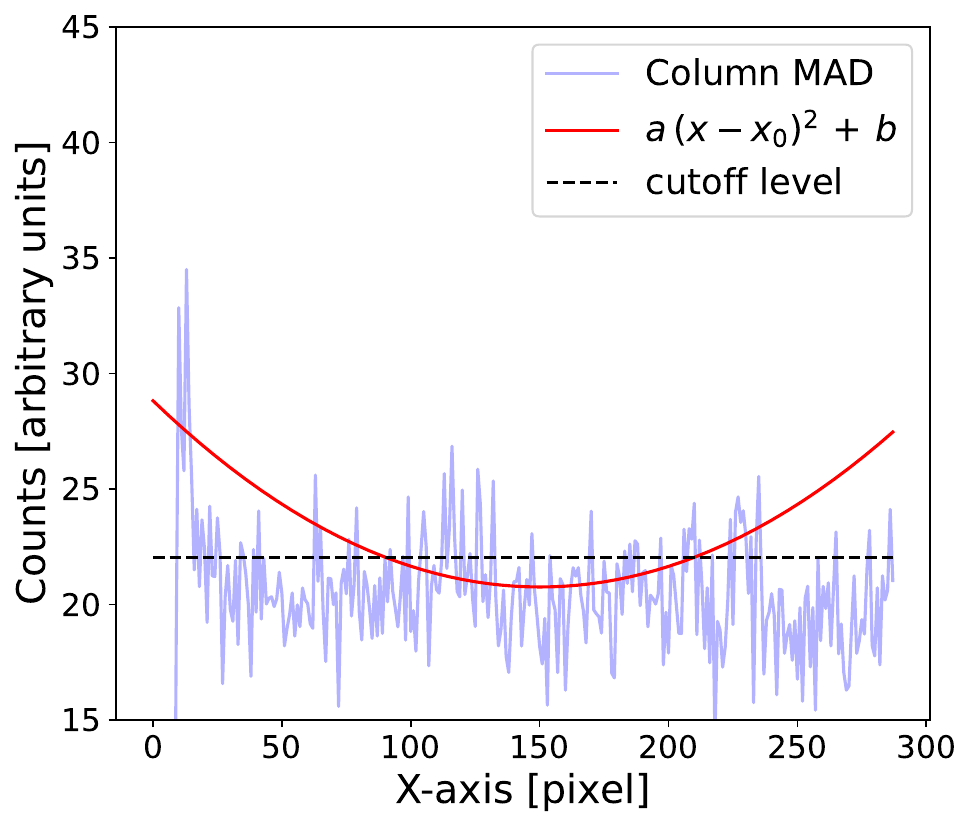}
        \caption{Same as Figure\,\ref{fig:cutoff} but using the image after all the fringing corrections.  }
        \label{fig:cutoff_cor}
    \end{minipage}
\end{figure*}

\textbf{7)} We repeated steps 1) and 2) but now used the left and right areas separately. Figure\,\ref{fig:both_structures} shows the point 4) after applying the correction to each region. Figures\,\ref{fig:both_structures_corr} show the image $I_{i,j}$ after the correction on the left (top) and right (bottom) regions, separately.

\textbf{8)} Finally, we corrected the most prominent negative outliers in the image using the routine \texttt{cube\_correct\_nan} from the \texttt{VIP} python package and $\sigma$-clipping with $\sigma=5$ and five iterations.

\begin{figure}[htb!]
\centering
\includegraphics[width=8.0cm]{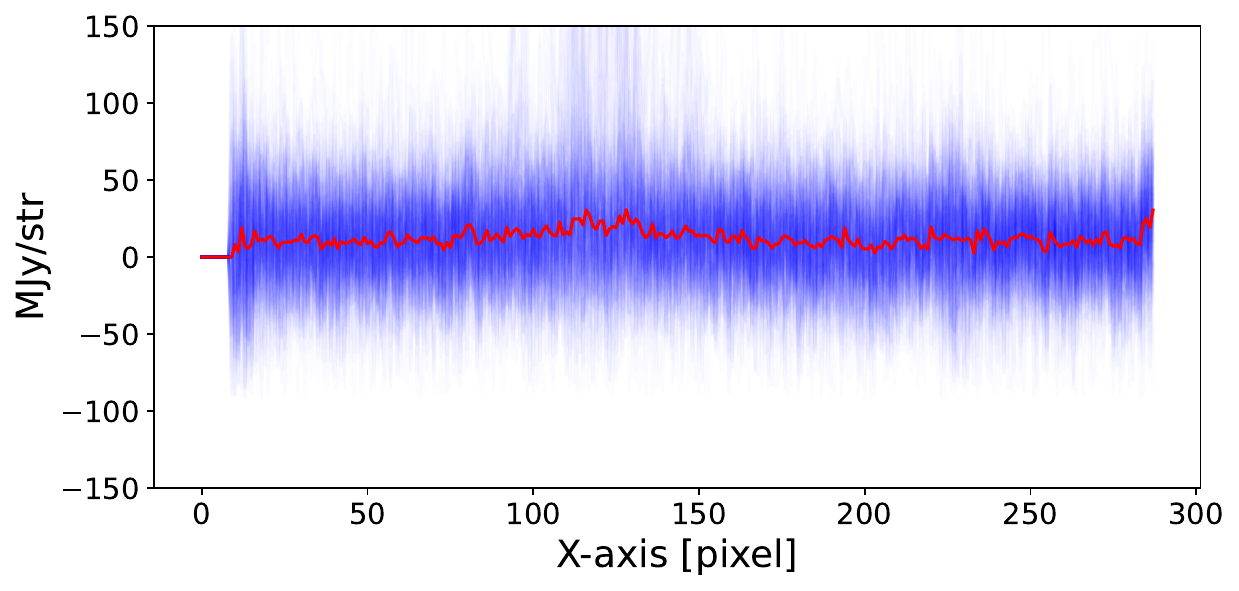}
\includegraphics[width=8.0cm]{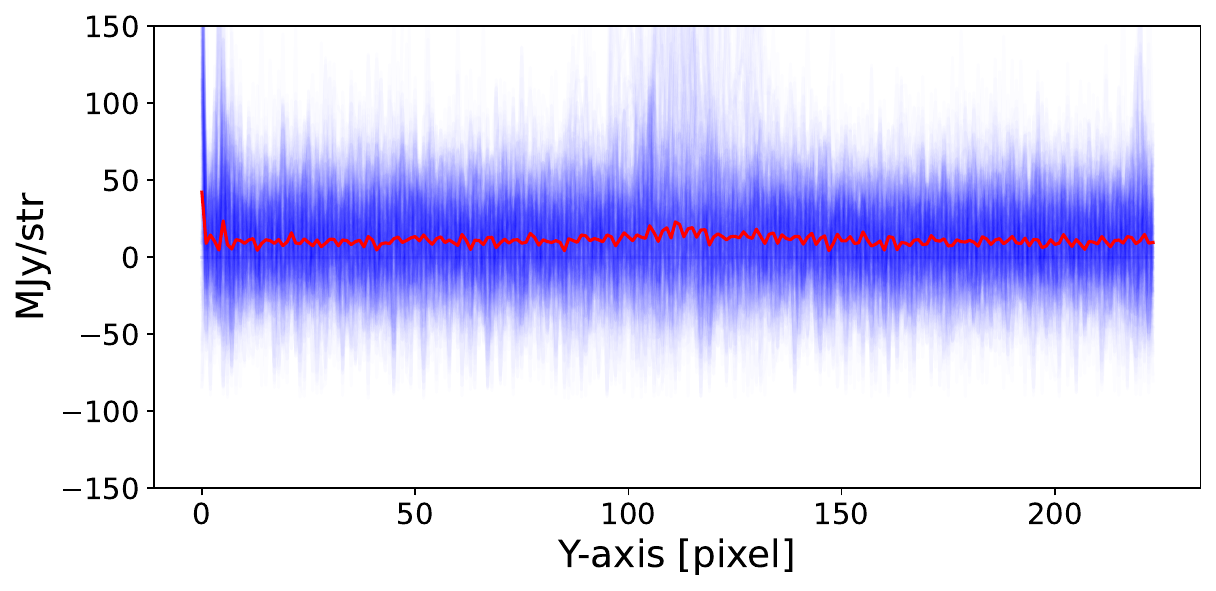}
\caption{Same as Figure\,\ref{fig:raw_structures} but after applying all the fringing corrections. }
\label{fig:final_structures}
\end{figure}

Figure\,\ref{fig:final_structures} shows the main structure in the X and Y axes as in Figure\,\ref{fig:raw_structures}. Figure\,\ref{fig:cutoff_cor} shows the same as Figure\,\ref{fig:cutoff} but after applying the fringing corrections in the left and right regions. We note that the fringing structure disappeared, and the noise level was reduced from $\sim 100$ to $\sim 60$ MJy/sr. Figure\,\ref{fig:fringing_example} shows the image before and after applying these corrections. 

\FloatBarrier

\section{Best extraction parameters}\label{Apx:best_param}

\begin{figure*}[htb!]
\centering
\sidecaption
\includegraphics[width=17.0cm]{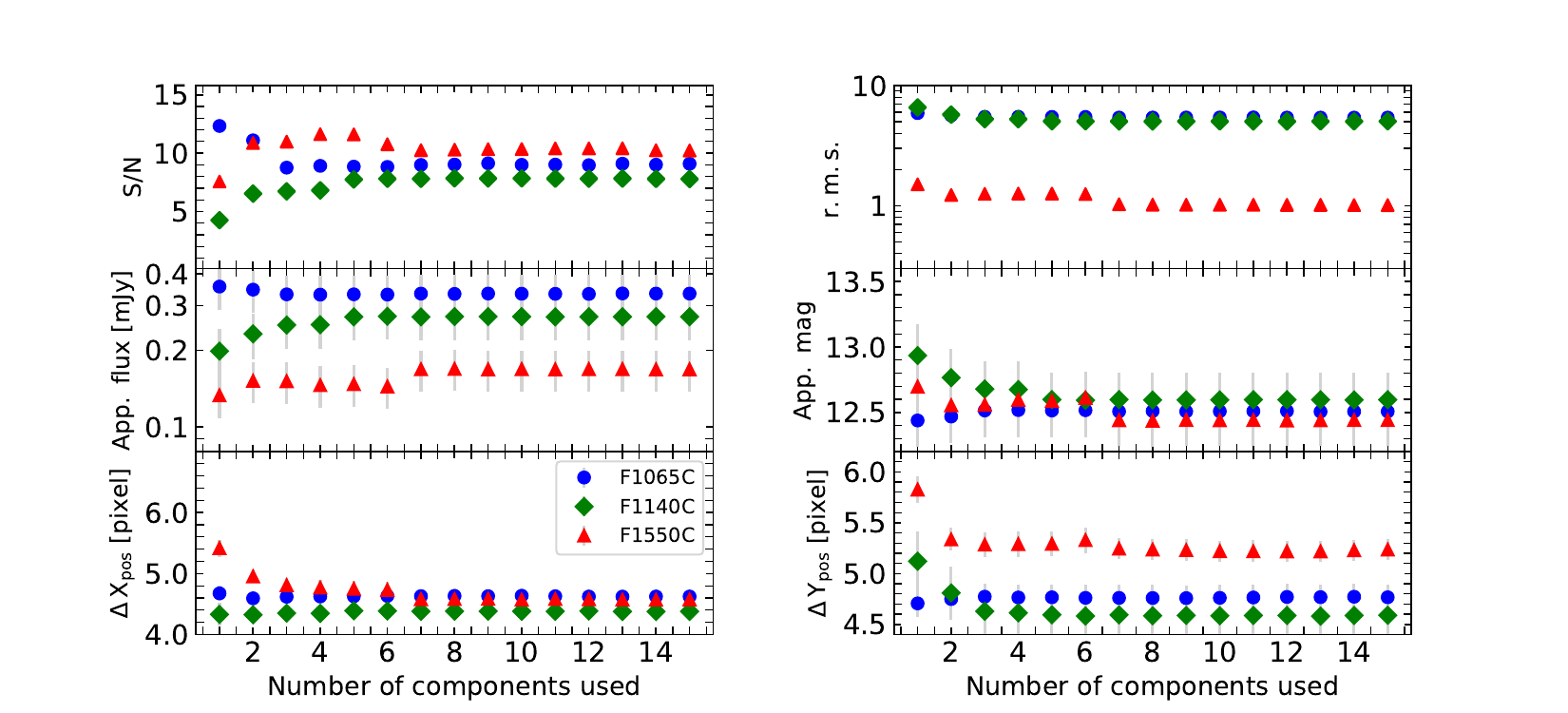}
\caption{General extraction quality for $\kappa$\,And\,b without smoothing the frames. \textit{Left-top}: Signal-to-noise ratio as a function of the number of components used (NPC). \textit{Left-middle}: Apparent flux vs. NPC. \textit{Left-bottom}: Relative position (with respect to the coronagraph) along the X-axis vs. NPC (in pixels). \textit{Right-top}: Root-mean-square in the residual image after subtracting the companion vs. NPC. \textit{Right-middle}: Apparent magnitude in Vega system vs. NPC. \textit{Right-bottom}: Relative position along the Y-axis vs. NPC (in pixels). Blue circles, red triangles, and green diamonds correspond to the bandpasses \texttt{F1065C}, \texttt{F1140C}, and \texttt{F1550C}, respectively. Vertical gray lines are the uncertainties. }
\label{fig:Comp_stat_NS}
\end{figure*}

\begin{figure*}[htb!]
\centering
\sidecaption
\includegraphics[width=17.0cm]{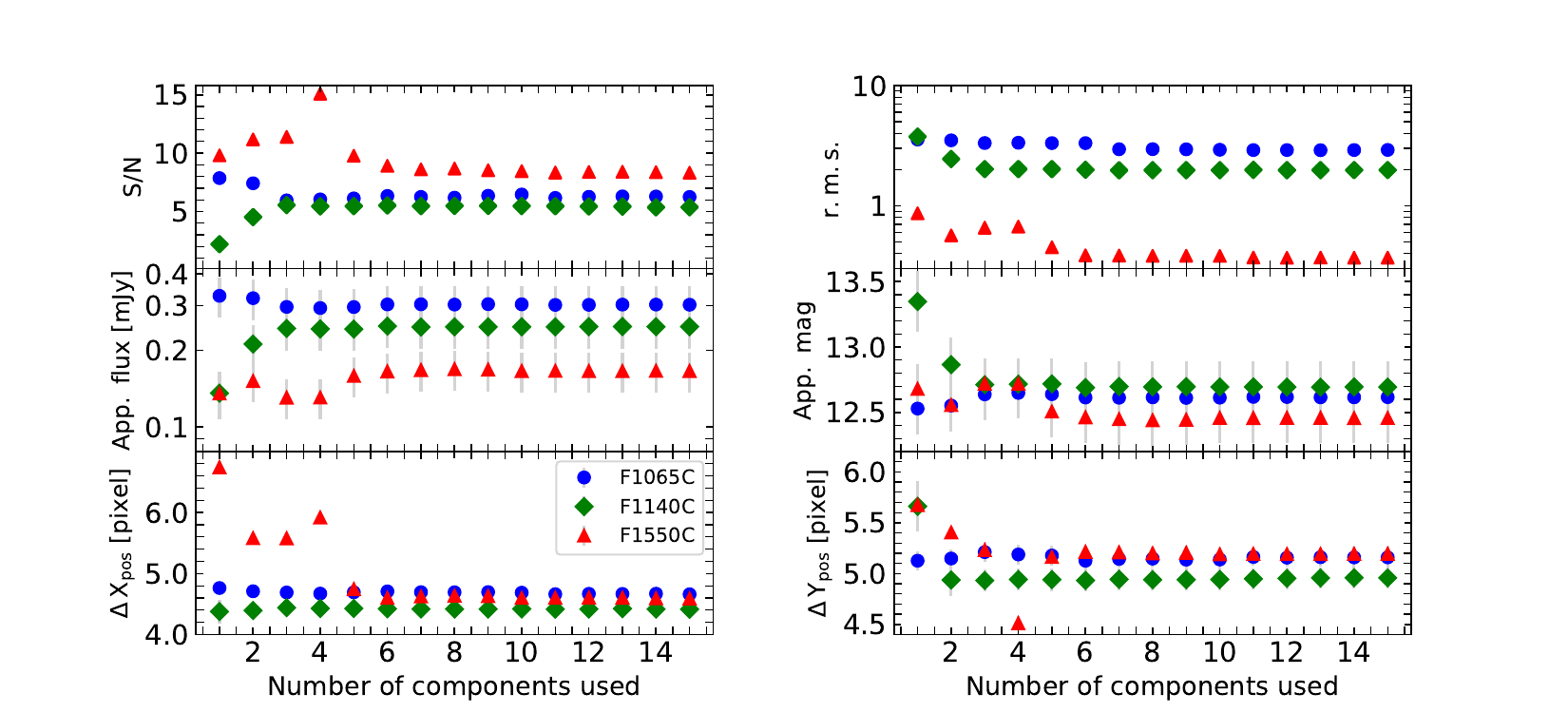}
\caption{Same as Figure\,\ref{fig:Comp_stat_NS} but smoothing the frames. }
\label{fig:Comp_stat_SS}
\end{figure*}

To determine the flux and astrometric position of $\kappa$\,And\,b with minimal uncertainty and bias, we conducted several tests using different numbers of principal components in post-processing. These components were crucial for effectively removing the coronagraphic PSF and speckles, which in turn impacted the signal, flux measurement, and astrometric accuracy of the companion. We utilized the fake planet injection method from \texttt{spaceKLIP} across 1 to 15 components and calculated the S/N with \texttt{VIP}. Additionally, we assessed the post-extraction residuals, flux, magnitude, and relative position in pixels. Since the companion has a low signal, we evaluate the option to blur (to smooth) the integrations. We perform the same analysis with smoothed and unsmoothed frames. Figures\,\ref{fig:Comp_stat_SS} and \ref{fig:Comp_stat_NS} show these values as a function of the number of components subtracted for each of the three filters. 

The optimal results are summarized in Table\,\ref{table:General_ext_param} for unsmoothed and smoothed frames. These selections were based on the highest S/N, stable relative position, consistent flux values with low uncertainty, and minimized r.m.s., indicating efficient starlight removal. The differences between the two approaches are within the uncertainties, but the case of the smoothed frame has better-extracted parameters in terms of r.m.s. and uncertainties. The best number of components are 10, 7, and 8 for the \texttt{F1065C}, \texttt{F1140C}, and \texttt{F1550C} filters, respectively.

\FloatBarrier

\begin{table*}[htb!]
\caption{Best extraction parameters for the unsmoothed and smoothed data for $\kappa$\,And\,b.}
\centering
\begin{tabular}{lcccccccccccc}
    \toprule  \toprule
    & \multicolumn{6}{c}{No smoothing} & \multicolumn{6}{c}{Smoothing} \\
    \cmidrule(lr){2-7} \cmidrule(lr){8-13}
    & \# PC & S/N & Flux & $\Delta$X  & $\Delta$Y  & rms 
    & \# PC & S/N & Flux & $\Delta$X  & $\Delta$Y  & rms \\
    &       &     & [$10^{-4}$\,Jy] & [pixel] & [pixel] & 
    &       &     & [$10^{-4}$\,Jy] & [pixel] & [pixel] &  \\
    \midrule
    \texttt{F1065C} & $7$  & $9$  & $3.35\pm0.63$ & $4.6\pm0.2$ & $4.8\pm0.3$ & $5.4$ 
                    & $10$ & $6$  & $3.04\pm0.55$ & $4.7\pm0.1$ & $5.1\pm0.1$ & $2.9$ \\
    \texttt{F1140C} & $7$  & $8$  & $2.71\pm0.52$ & $4.4\pm0.2$ & $4.6\pm0.2$ & $5.0$ 
                    & $7$  & $6$  & $2.47\pm0.45$ & $4.4\pm0.1$ & $4.9\pm0.1$ & $2.0$ \\
    \texttt{F1550C} & $8$  & $10$ & $1.68\pm0.31$ & $4.6\pm0.1$ & $5.2\pm0.1$ & $1.0$ 
                    & $8$  & $9$  & $1.69\pm0.30$ & $4.6\pm0.1$ & $5.2\pm0.1$ & $0.4$ \\
    \bottomrule
\end{tabular}
\tablefoot{The root-mean-square (rms) is in MJy/str. }
\label{table:General_ext_param}
\end{table*}

\FloatBarrier

\section{Bad-pixel effects in companion extraction}\label{Apx:Bad_Pix}

\begin{figure}[htb!]
\centering
\sidecaption
\includegraphics[width=7.8cm]{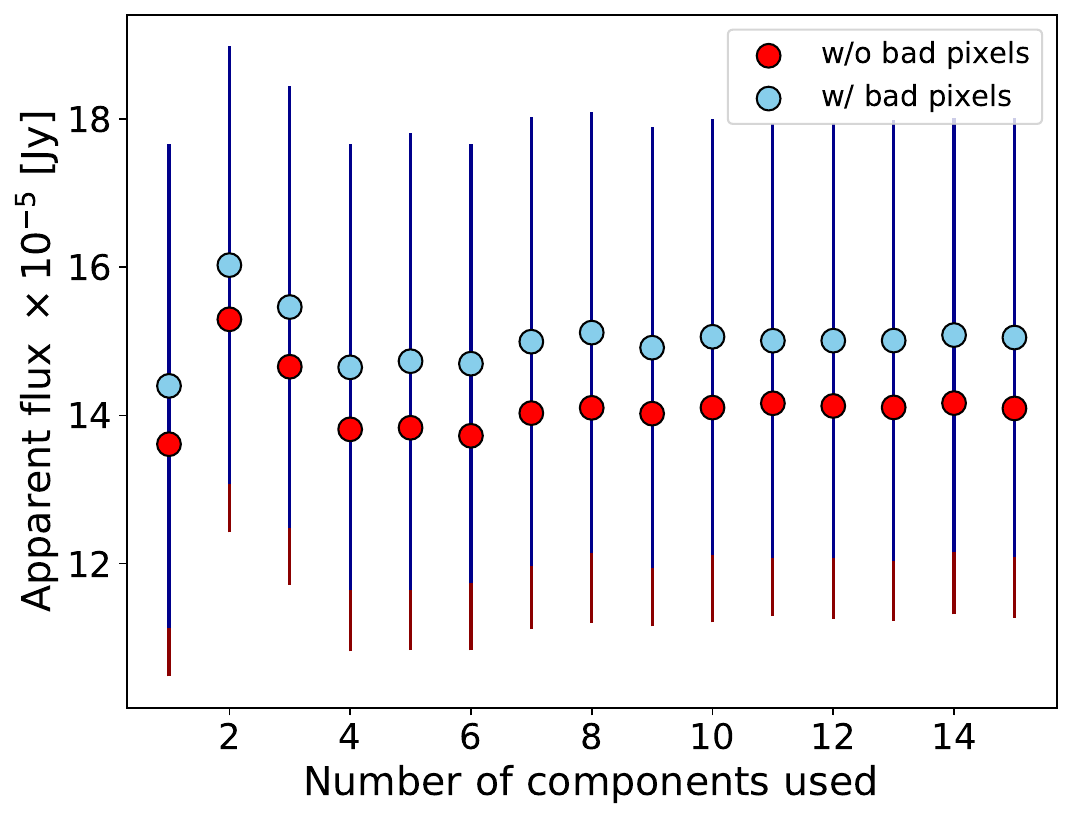}
\caption{Flux measured from the PSF modeling of an injected fake planet at the same angular separation but at a position angle shifted by $\sim180\degree$ relative to $\kappa$\,And\,b. The red dots represent the flux extraction of the injected fake planet without any bad or hot pixels. The light blue dots correspond to the same fake planet but include bad and hot pixels. The vertical line in each colored point corresponds to the uncertainty. }
\label{fig:BP_check}
\end{figure}

We note the presence of persistent hot pixels at the location of $\kappa$\,And\,b in the \texttt{F1550C} filter. These pixels may or may not bias the measurements of the flux and astrometry of $\kappa$\,And\,b. To investigate this possibility, we injected a fake planet at the same separation as $\kappa$\,And\,b on the opposite side of the companion (i.e., at PA+$\sim180$\degree). The fake planet was injected using a simulated PSF and renormalized to approximately match the flux of $\kappa$\,And\,b. We individually injected the same bad and hot pixels in the residuals (see Fig.\,\ref{fig:Extraction_comp}). Then, we proceeded with the post-processing and companion extraction as done for $\kappa$\,And\,b. We repeated the same procedure without injecting the bad and hot pixels to compare the flux extraction results. Figure\,\ref{fig:BP_check} shows the extracted flux for the case without (red) and with (blue) bad pixels. We found that the injected fake planet with bad pixels has $\sim 6\%$ ($\sim 0.09\times 10^{-4}$) more flux than the one without bad pixels. Because the uncertainties are larger ($\sim 0.28\times 10^{-4}$), however, the contribution of this $6\%$ difference can be considered within the actual uncertainties: $0.29\times 10^{-4}$ vs. $0.28\times 10^{-4}$. We determine that the bad and hot pixels within the $\kappa$\,And\,b PSF are not biasing the measured flux of the companion. 

\FloatBarrier

\section{Archival data correction}\label{Apx:data}

We compute our stellar spectrum model using ATLAS/SYNTHE models based on the physical parameters derived by \cite{Jones+2016}. We collected all the archival data and used stellar model. We used the stellar model from the literature and our model to compute a corrector factor in all the datasets. 

\begin{figure}[htb!]
\centering
\includegraphics[width=6.8cm]{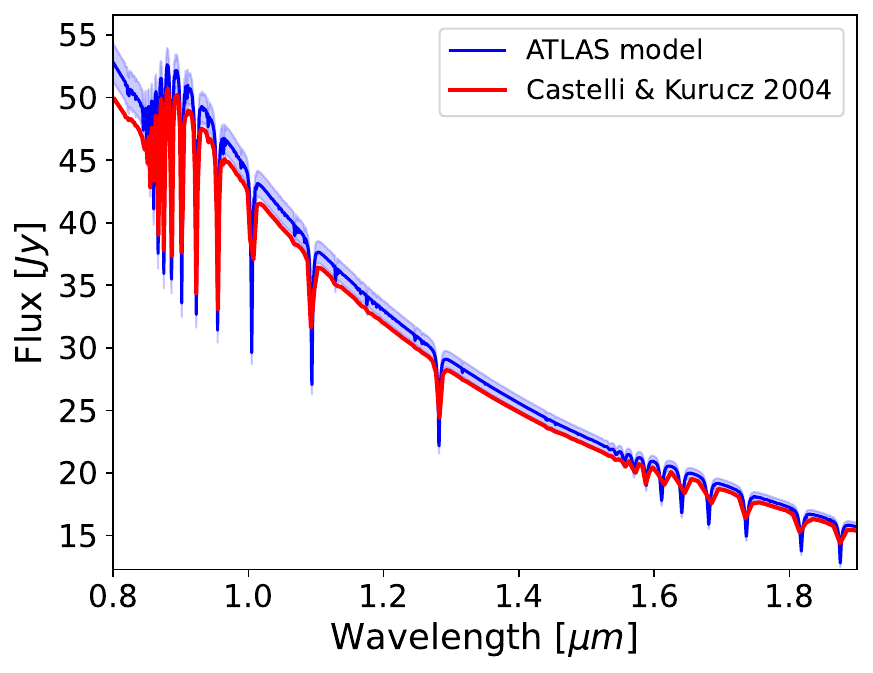}
\includegraphics[width=6.9cm]{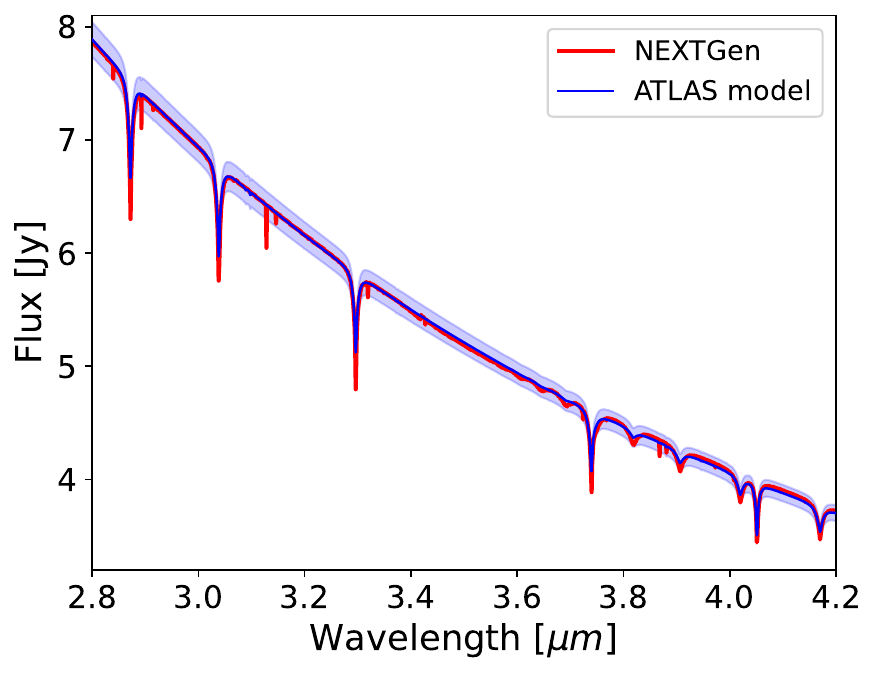}
\caption{Stellar model spectrum used in this study (blue line) and the model used by \cite{Currie+2018} in the left panel, and by \cite{Stone+2020} in the right panel (red line).}
\label{fig:Spec_star_1}
\end{figure}

\begin{figure}[htb!]
\centering
\includegraphics[width=6.8cm]{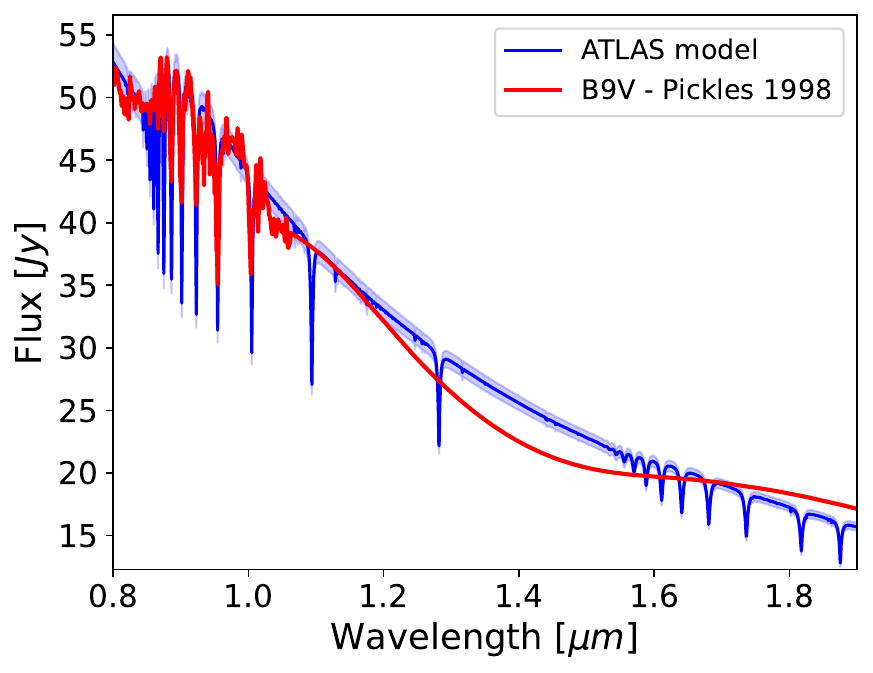}
\includegraphics[width=6.9cm]{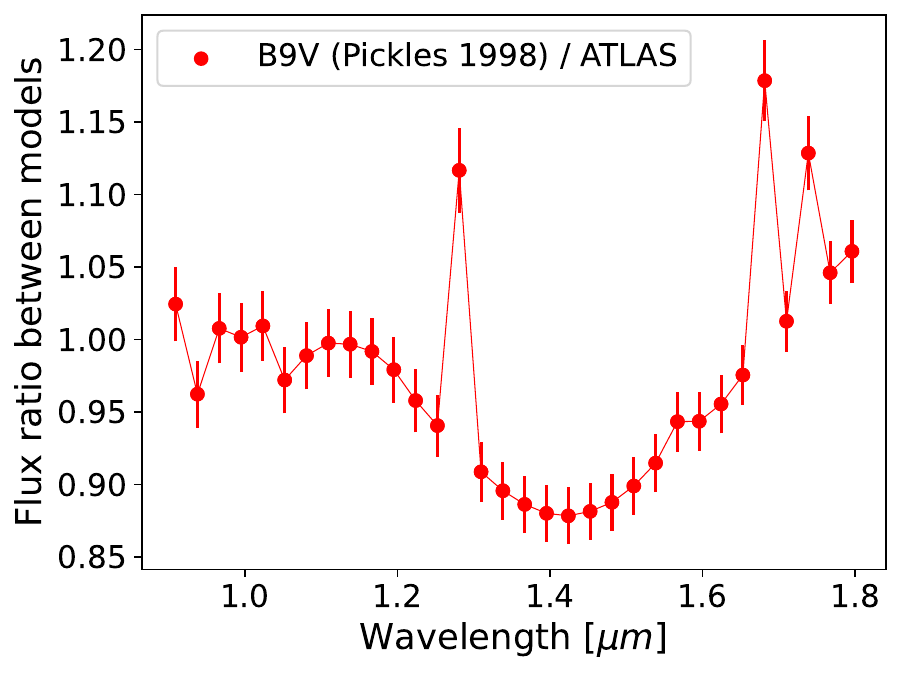}
\caption{Synthetic stellar spectrum (blue line) and the B9V Pickles library spectrum (red line) used to convert the companion contrast into flux (\citealt{Hinkley+2013}). \textit{Left:} Two stellar model spectra in the wavelength range of the IFS observation. \textit{Right:} Ratio of the two model spectra at the resolution of the IFS observation. This ratio corresponds to the corrector factor. }
\label{fig:Spec_star_2}
\end{figure}

The correction is more notorious in spectroscopic observations since we are able to see the changes in the shape of the companion spectrum. We first reduced the resolution of the two stellar model spectra to the resolution of the IFS/medium-resolution observation. Then, we divided them to compute a corrector factor as a function of wavelength. These values are then directly multiplied in the IFS/medium-resolution observations of the companion $\kappa$\,And\,b. Figure\,\ref{fig:Spec_star_1} shows the comparison between our stellar models and those used by \cite{Currie+2018} and \cite{Stone+2020}. Figure\,\ref{fig:Spec_star_2} shows on the left the stellar models of our ATLAS and the B9V Pickles library model used by \cite{Hinkley+2013}, and on the right, the ratio of the two models at the resolution of the IFS observation.

\FloatBarrier

\FloatBarrier

\begin{figure*}[htb!]
\centering
\includegraphics[width=7.5cm]{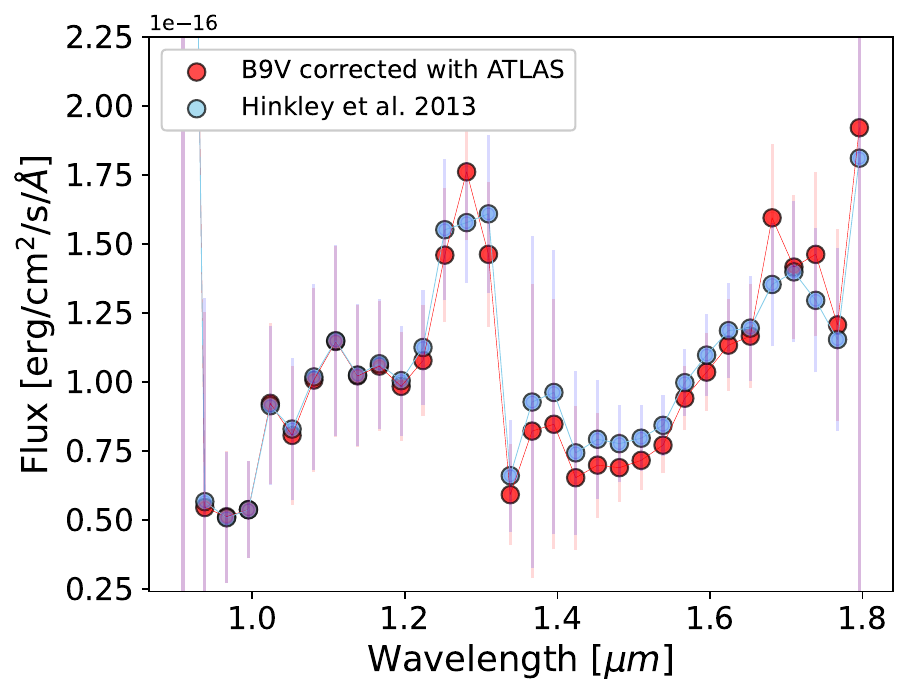}
\includegraphics[width=7.5cm]{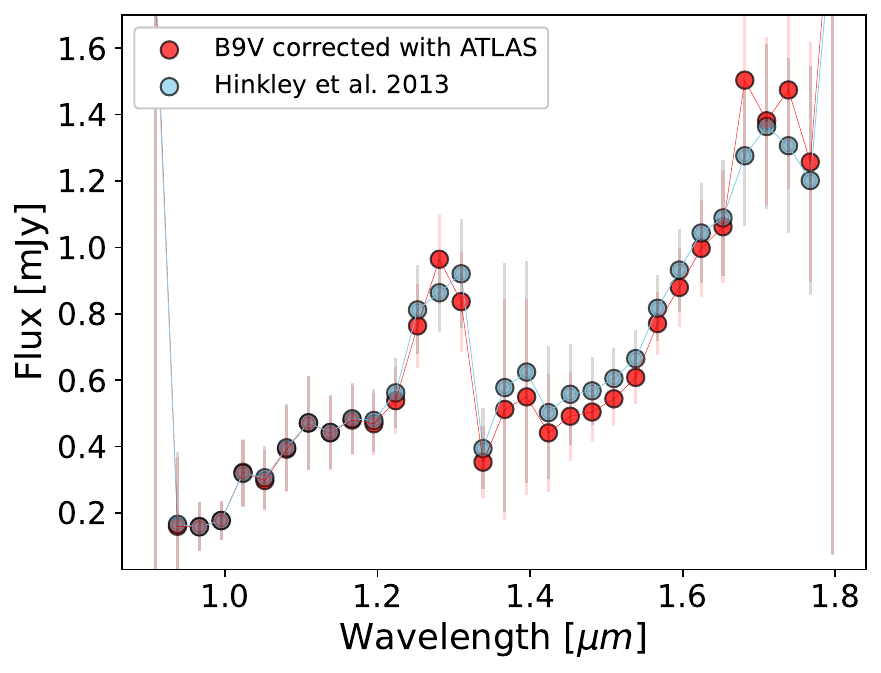}
\includegraphics[width=7.5cm]{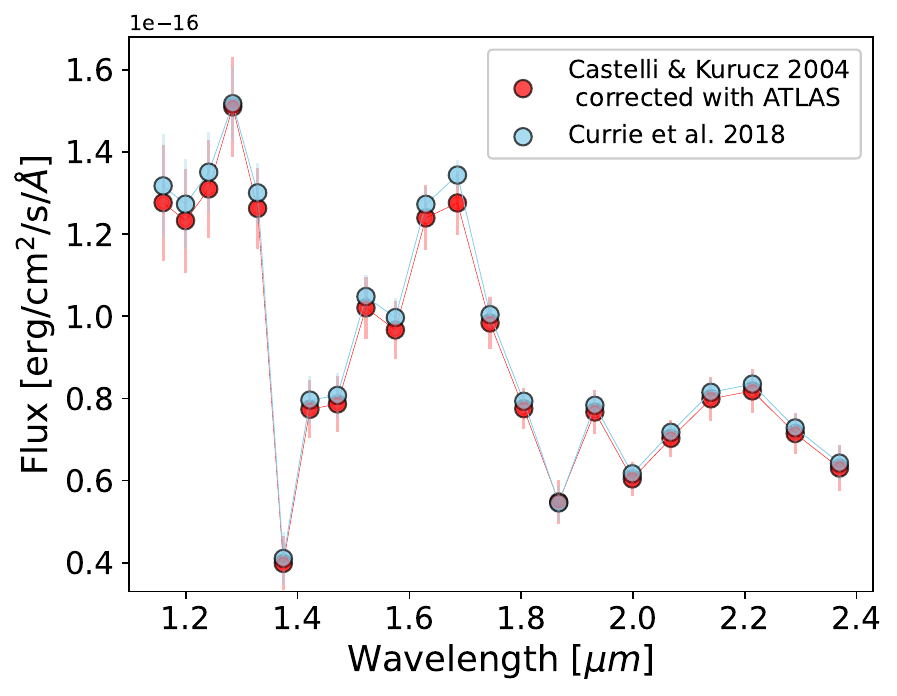}
\includegraphics[width=7.5cm]{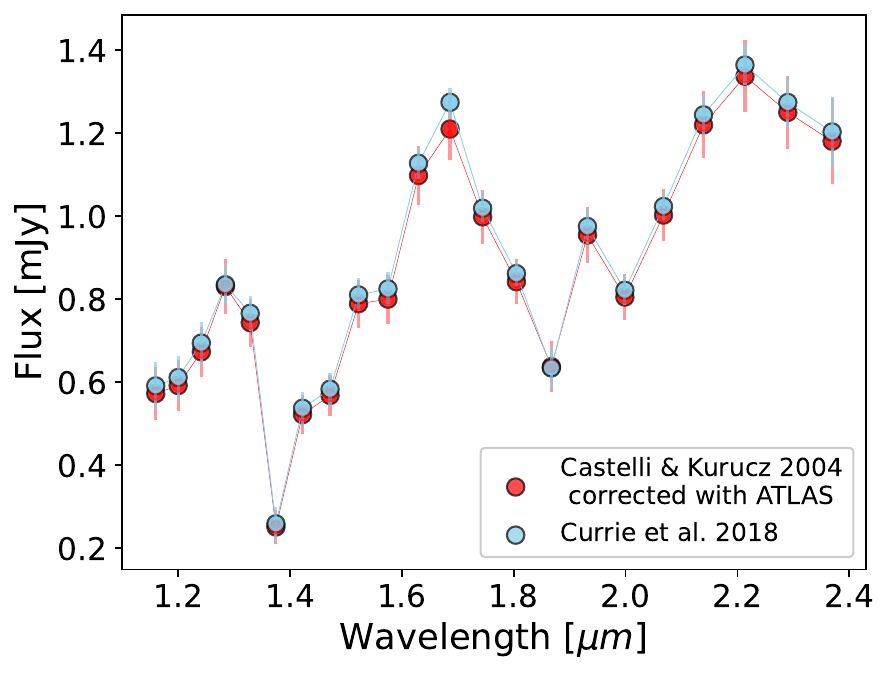}
\includegraphics[width=7.5cm]{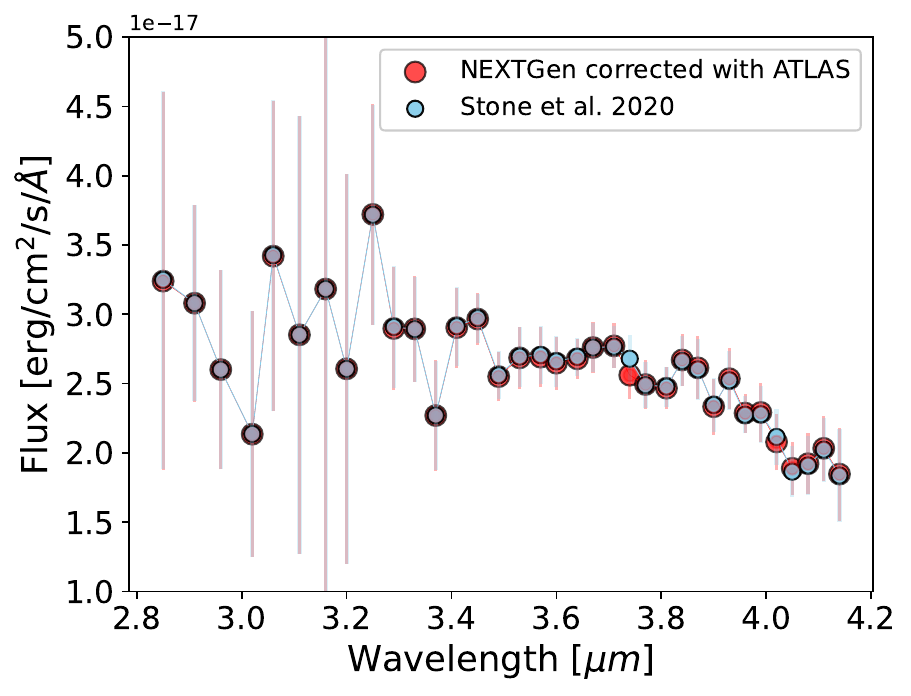}
\includegraphics[width=7.5cm]{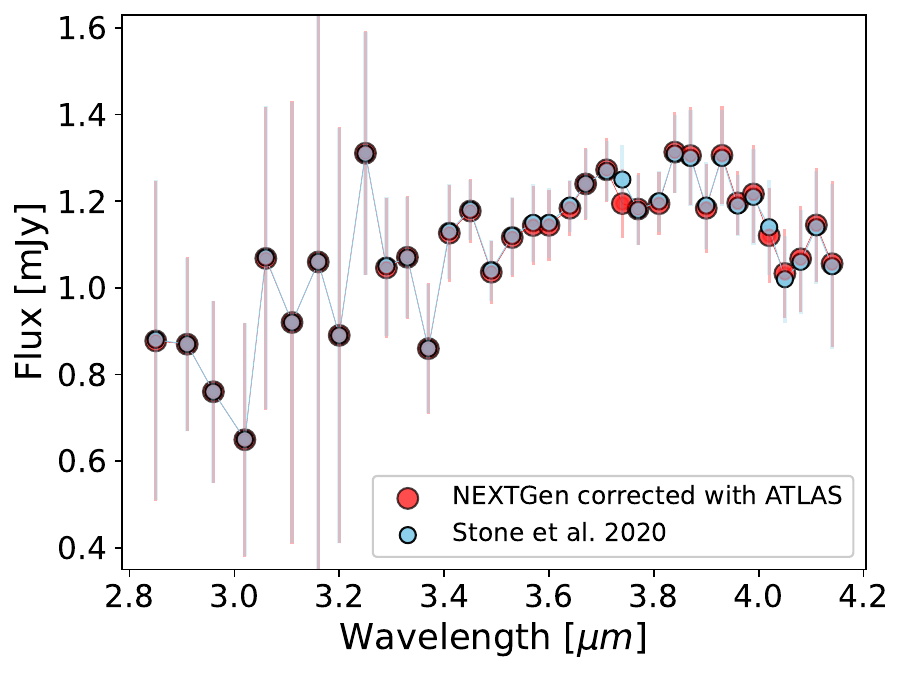}
\includegraphics[width=7.5cm]{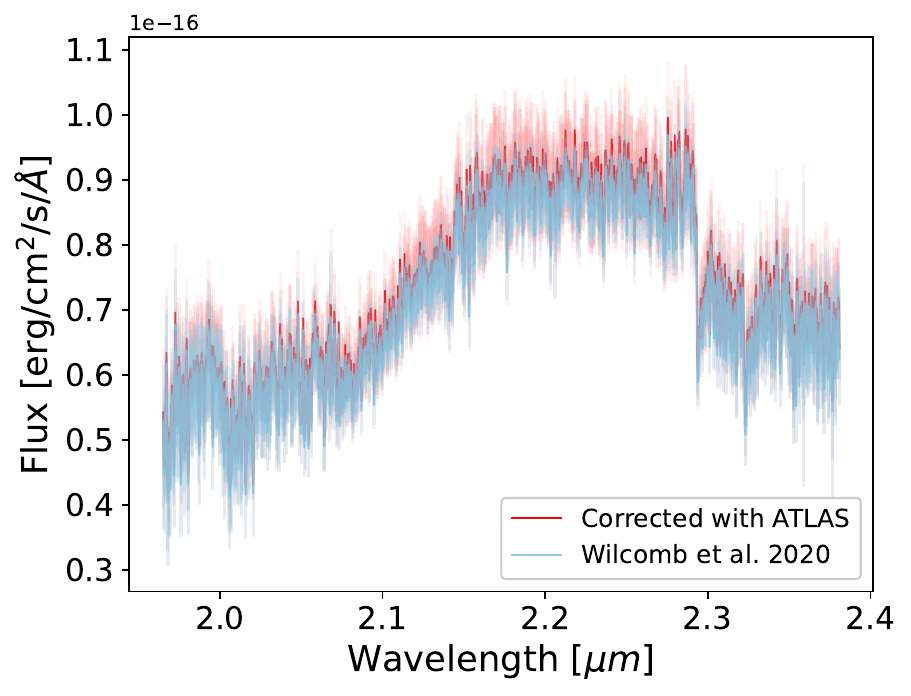}
\includegraphics[width=7.5cm]{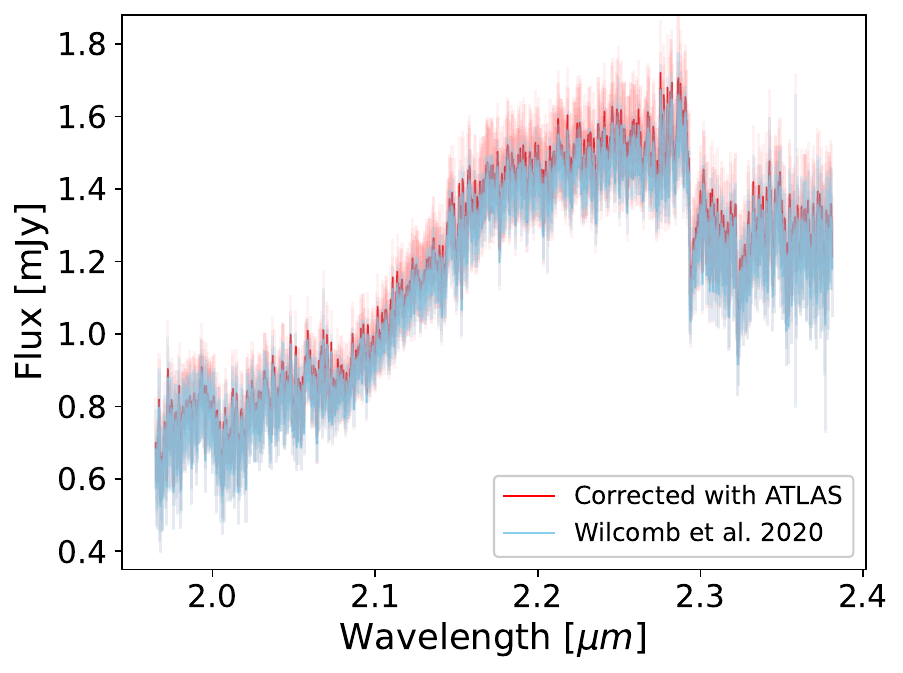}
\caption{$\kappa$\,And\,b archival spectroscopic data (light blue dots) and corrected data (red dots). The left panels show the flux in $\mathrm{erg/cm^2/s/A}$, while the right panels shows the flux in mJy.}
\label{fig:Spec_comp_corr}
\end{figure*}

We corrected all the spectra observations of $\kappa$\,And\,b following the procedure describe above. Figure\,\ref{fig:Spec_comp_corr} shows all the archival spectroscopic data of $\kappa$\,And\,b (black dots and lines), and our correction using our ATLAS stellar model spectrum (red dots and lines) for (from top to bottom):\cite{Hinkley+2013}, \cite{Currie+2018}, \cite{Stone+2020}, and \cite{Hoch+2020}.

\FloatBarrier

\newpage

\section{Age and mass estimates from CMD+isochrones}\label{Apx:cmd-tracks}

\begin{figure*}[htb!]
\centering
\sidecaption
\includegraphics[width=7.8cm]{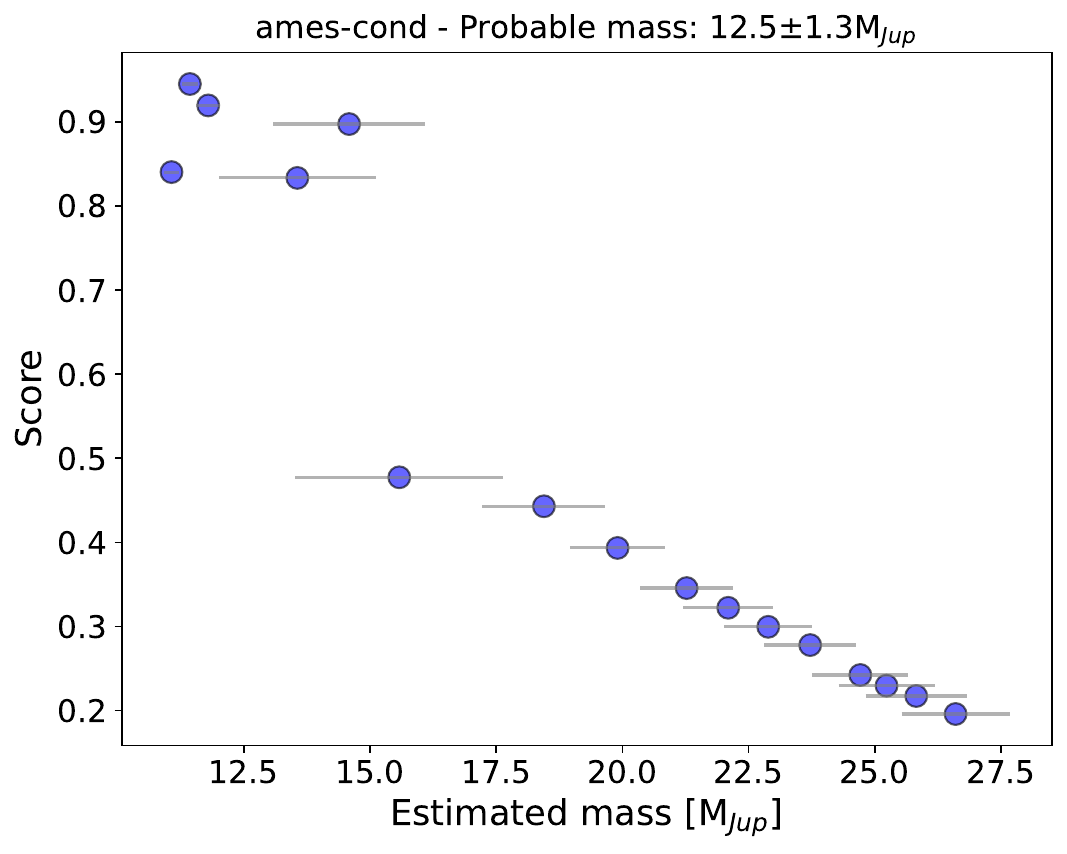} 
\includegraphics[width=7.8cm]{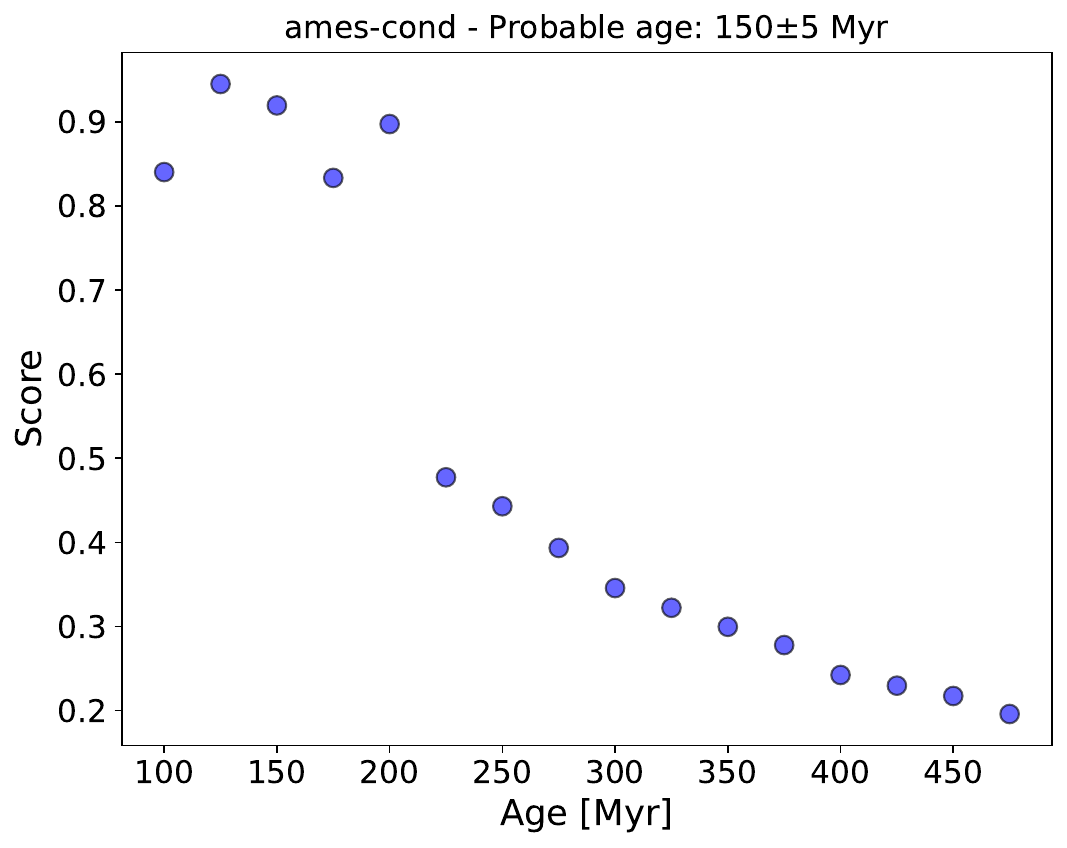}
\caption{Distribution of mass and age for VHS\,1256\,b using AMES-COND evolutionary tracks. \textit{Left:} Best score and probable mass for each isochrone. The masses (x-axis) were sorted from younger to older associated age from the isochrone. The error bar corresponds to the mass-dispersion measurement in each isochrone. \textit{Right:} Isochronal age and the best evaluated score. The final values for each parameter (mass and age) are presented at the top of each figure.}
\label{fig:vhs_mass_age_dist_ames-cond}
\end{figure*}

\begin{figure*}[htb!]
\centering
\sidecaption
\includegraphics[width=7.8cm]{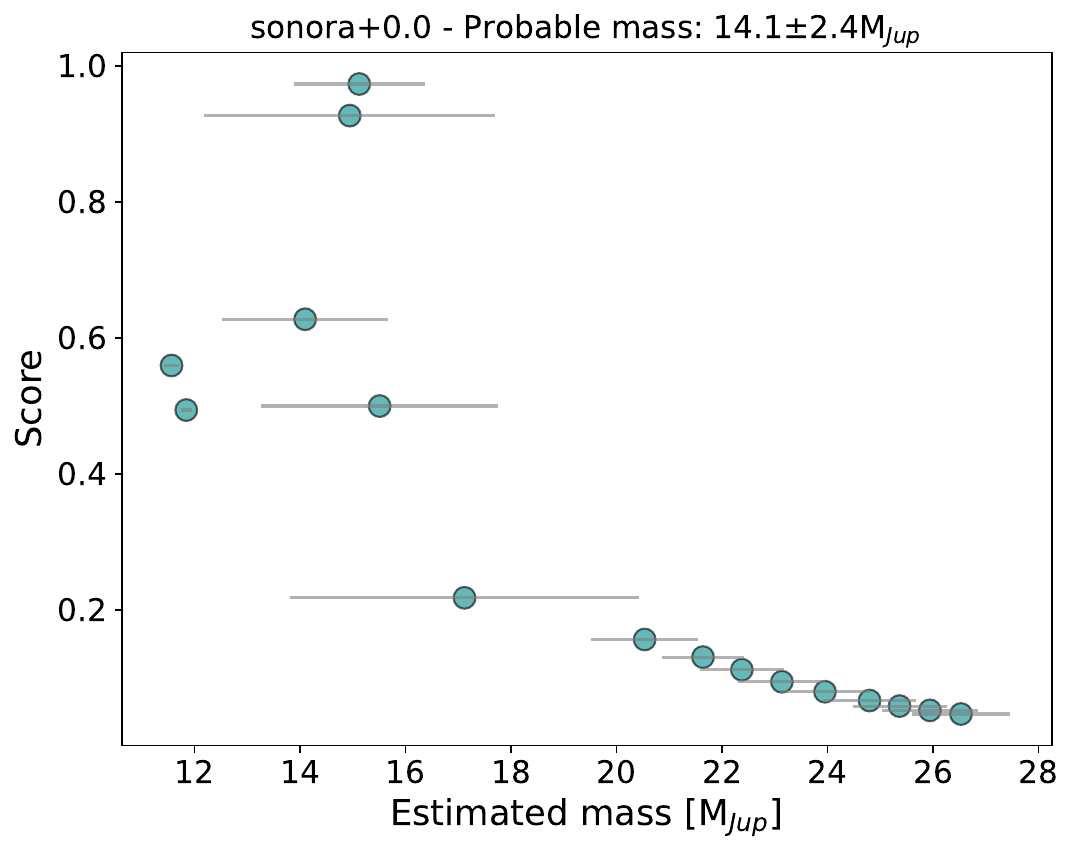}
\includegraphics[width=7.8cm]{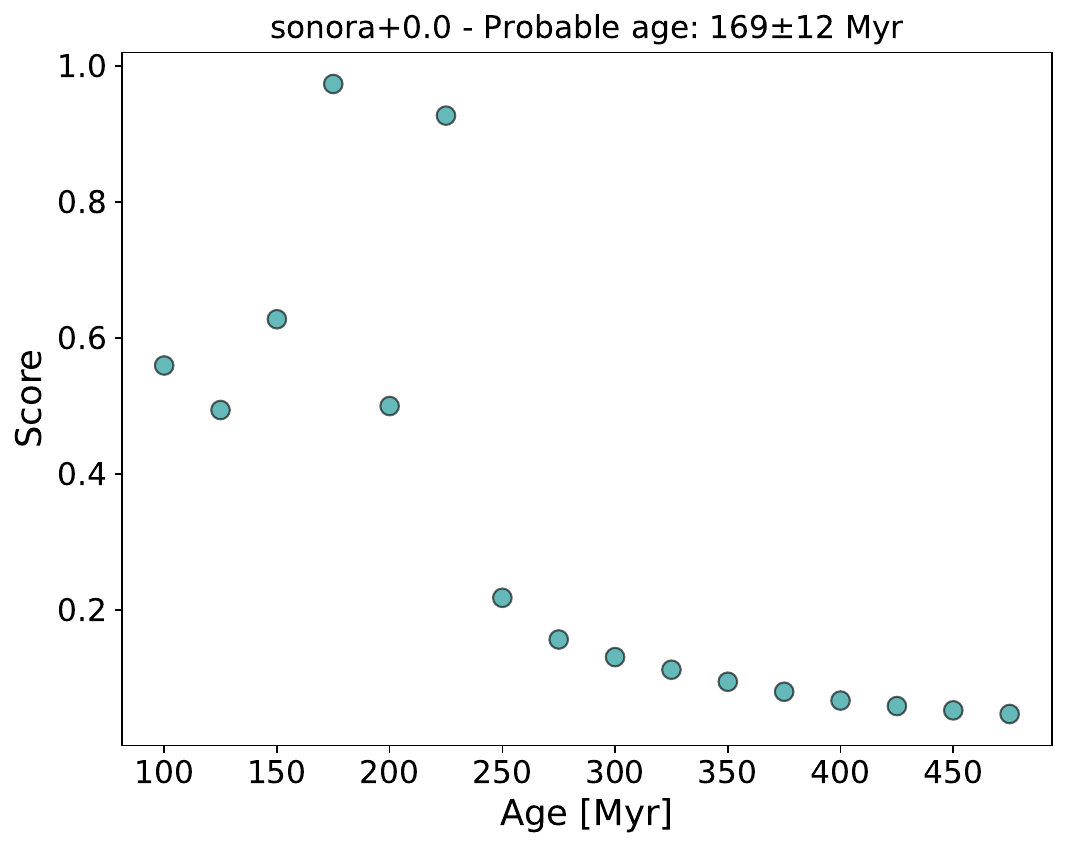}
\caption{Same as Figure\,\ref{fig:vhs_mass_age_dist_ames-cond} but using Sonora solar metallicity evolutionary tracks.}
\label{fig:vhs_mass_age_dist_sonora}
\end{figure*}

We employed our method to estimate the mass and age from the CMD plus isochrones, of the VHS\,1256\,b companion. How was mentioned before, each isochrone has an associated score and a representative mass. This comes from the evaluation of the isochrone in the 2D Gaussian profile, normalized to 1 at the maximum. The 2D Gaussian corresponds to the photometric measurement of the companion, i.e., the color and absolute magnitude, and related uncertainties. For each isochrone, we evaluated the mass track into the 2D Gaussian, whose values can range from 0 to 1. This is equivalent to estimating how good the color-match of a specific mass is in the isochrone with respect to our companion fluxes. When the isochrone matches perfectly the color-magnitude of the companion, the score is 1, while the score will be zero if the isochrone completely mismatches it. Then, for each isochrone, we have a mass distribution (score vs. mass), for which we can estimate the representative mass and the dispersion as uncertainty. This mass is calculated using a weighted mean, with the weight being the associated score for each mass. We kept the maximum score associated with each isochrone to make the age versus score and mass versus score distribution. This distribution in mass corresponds to the weighted mean mass for each isochrone and the best evaluated score. The final age and mass estimates were calculated using the weighted mean and a score threshold of $0.6$. Figures\,\ref{fig:vhs_mass_age_dist_ames-cond} and \ref{fig:vhs_mass_age_dist_sonora} show the mass and age (x-axis) with the respective score (y-axis) for the AMES-COND and Sonora-solar isochrones, respectively. The probable mass and age from the distributions were presented at the top of each figure. 

\begin{figure*}[htbp]
\centering
\sidecaption
\includegraphics[width=8.5cm]{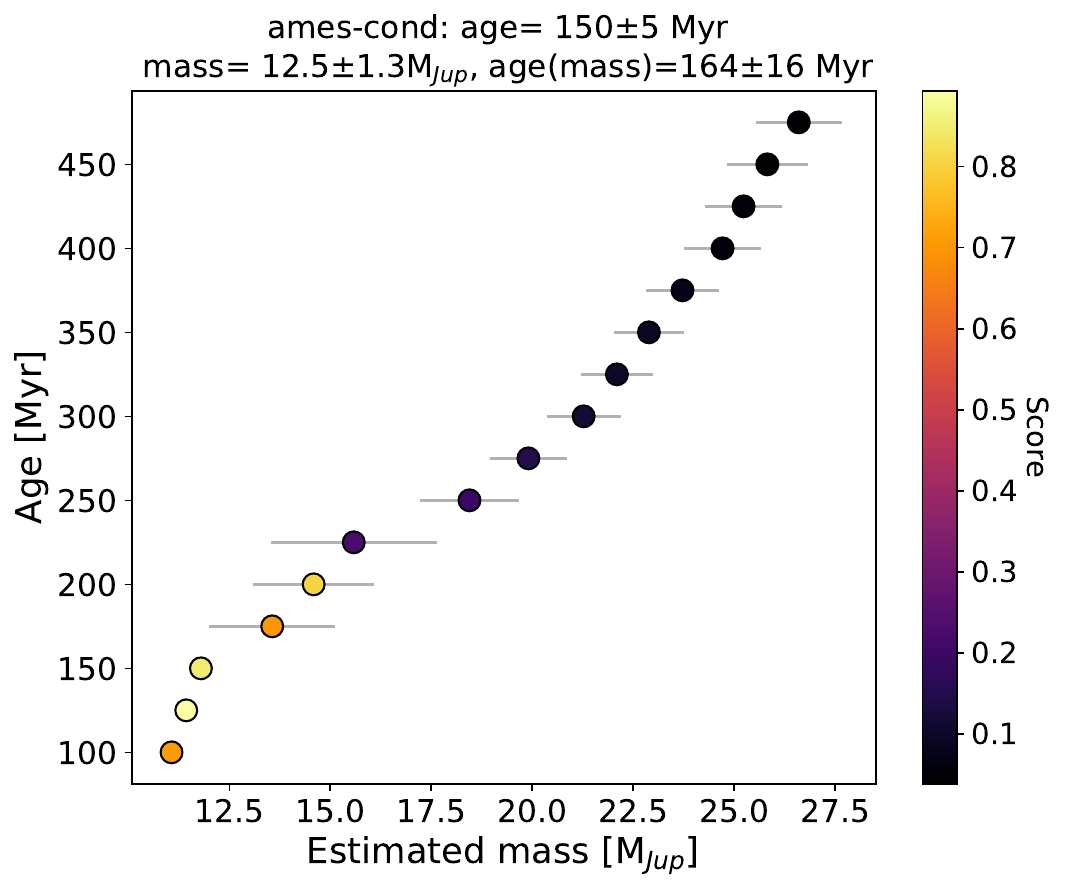}
\includegraphics[width=8.5cm]{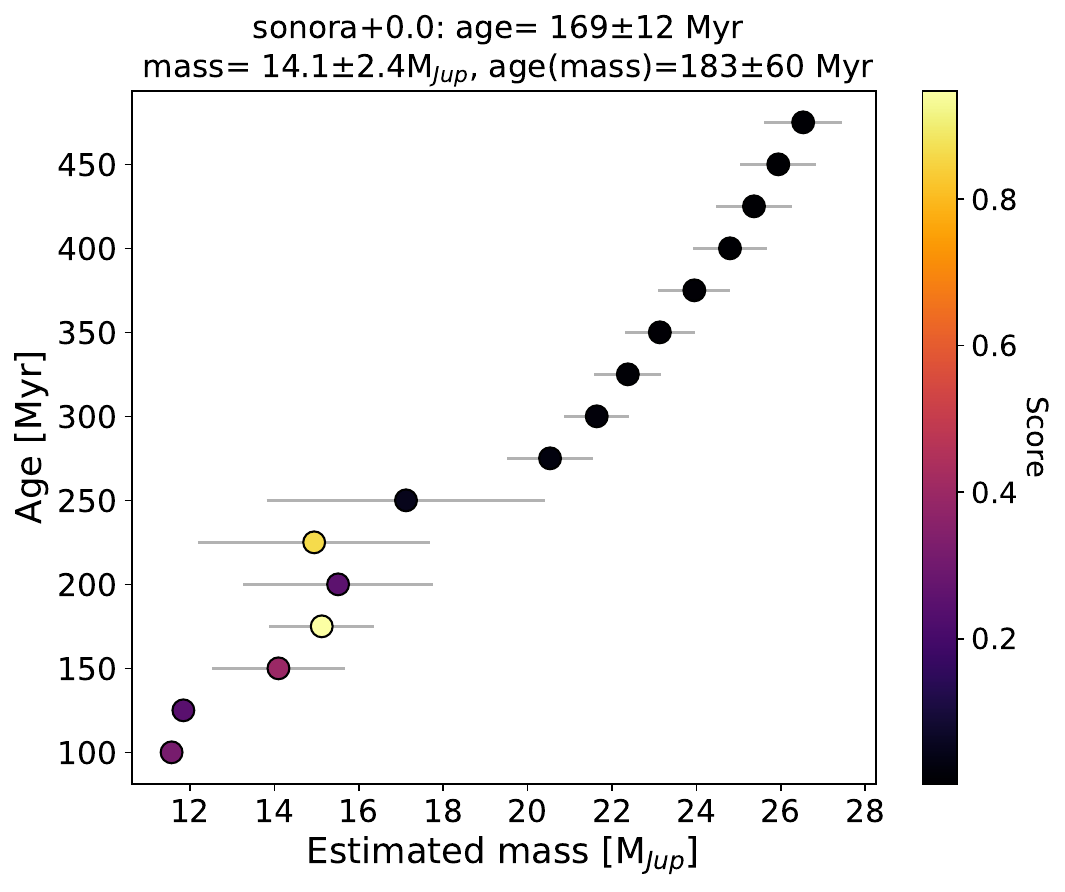}
\caption{Age as a function of the most probable mass using AMES-COND (left) and Sonora-solar (right) evolutionary tracks. The color-scale corresponds to the associated score. The final calculated values for age and mass are presented at the top of each figure.
}
\label{fig:vhs_mass_age_dist}
\end{figure*}

Figure\,\ref{fig:vhs_mass_age_dist} shows the age as a function of mass with the respective associated score in color for the two evolutionary tracks. We calculated the probable age from the most probable mass and uncertainty via Monte Carlo (i.e., age as a function of mass-age(mass)). At the top of each figure, we show the two probable ages and masses. From AMES-COND we estimated an age(mass) of $164\pm16$\,Myr and mass of $12.5\pm1.3\,\mathrm{M_{Jup}}$, while for Sonora-solar we obtained an age(mass) of $176\pm59$\,Myr and $14.1\pm2.3\,\mathrm{M_{Jup}}$, respectively. \cite{Dupuy+2023} reported an age for VHS\,1256 of $140\pm20$\,Myr and a companion mass between $12\,\mathrm{M_{Jup}}$ and $16\,\mathrm{M_{Jup}}$. Our age and mass estimates are in agreement with the one reported by \cite{Dupuy+2023} at $1$$\sigma$ level, further supporting our procedure to derive those parameters for $\kappa$\,And\,b. Notoriously, in Figure\,\ref{fig:vhs_mass_age_dist} (more clearly in the left panel), we observe a double peak in the mass axis, which is in agreement with the bi-modal distribution of mass reported by \citet[Fig. 10]{Miles+2023}, with both peaks at $\sim12\mathrm{M_{Jup}}$ and $\sim16\mathrm{M_{Jup}}$.

\end{document}